\documentclass[a4paper,11pt]{article}
\usepackage{jheppubMod}
\usepackage{amsmath,amssymb}
\usepackage[bottom]{footmisc}
\usepackage{graphicx}
\usepackage[mathlines,running]{lineno}
\usepackage{mathtools,slashed}
\usepackage{dsfont}
\usepackage{physics}
\numberwithin{equation}{section}
\usepackage{pdfpages}
\usepackage[makeroom]{cancel}
\usepackage{slashed}
\usepackage{verbatim}
\usepackage{soul}
\usepackage{subfigure} 
\usepackage[normalem]{ulem}
\usepackage[svgnames,dvipsnames]{xcolor}
\usepackage{hyperref} 
\usepackage{numprint}

\graphicspath{{./Plots/}}

\hypersetup{colorlinks=true,
    urlcolor=blue,
    citecolor=blue,
    linkcolor=blue,
    pdffitwindow=false,
    pdfstartview={FitH},
    pdftitle={Basu \& Ruiz, The Four Polarizations of the W [arXiv:2512.10015]},
    pdfauthor={Names},
    pdfsubject={Subject},
    pdfcreator={Creator},
    pdfproducer={Producer},
    pdfkeywords={Multiboson Scattering}{Helicity Polarization}{Electroweak Interactions}{Large Hadron Collider},
    pdfnewwindow=true
}

\newcommand{\orcid}[1]{\,\href{https://orcid.org/#1}{\includegraphics[width=9pt]{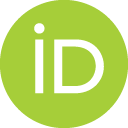}}}
\newcommand{\orcidTB}{0009-0003-1414-0282} 
\newcommand{\orcidRR}{0000-0002-3316-2175} 

\def\GeV{{\rm\ GeV}}
\def\TeV{{\rm\ TeV}}

\definecolor{magenta}{HTML}{FF00FF}
\definecolor{cornflowerblue}{HTML}{6495ED}
\definecolor{turquoise}{HTML}{40E0D0}

\definecolor{darkgreen}{rgb}{0.0, 0.2, 0.13}
\definecolor{darkmagenta}{rgb}{0.55, 0.0, 0.55}
\definecolor{amber}{rgb}{1.0, 0.6, 0.0}

\title{The Four Polarizations of the $W$ at High Energies}

\author{Trina Basu\orcid{\orcidTB} and}
\author{Richard Ruiz\orcid{\orcidRR}}

\affiliation{Institute of Nuclear Physics -- 
Polish Academy of Sciences {\rm (IFJ PAN)},\\ 
ul. Radzikowskiego, 31-342 Krak{\'o}w, Poland}

\emailAdd{trina.basu@ifj.edu.pl}
\emailAdd{rruiz@ifj.edu.pl}

\abstract{We investigate 
polarization-induced interference, off-shell effects, 
and gauge cancellations 
in predictions for high-energy, multi-leg processes
with (near) resonant weak bosons.
Building on the ``polarized propagator'' paradigm,
we carry out our analysis
at the level of helicity amplitudes and squared amplitudes,
computing polarization interference directly.
We introduce analytical decompositions 
of polarized propagators, 
valid for covariant and axial gauges, 
that simplify the organization and evaluation of polarized amplitudes,
and make power counting of mass-over-energy factors manifest.
We show:
(i) For the fully massive case, polarization interference 
can exceed $\mathcal{O}(\Gamma_V^2/M_V^2)$ width-over-mass corrections, 
limiting the applicability of the narrow width and pole approximation at low energies.
(ii) At the fully differential level, 
interference can naturally be larger 
than squared longitudinal amplitudes  
but can also vanish when bosons are emitted by unpolarized sources.
(iii) When weak bosons decay to massless fermions,
the non-interference of polarization 
after angular integration extends  
to the off-shell regime 
but remains approximate due to $V-A$ couplings.
Guided by BRST invariance, we propose a simple scheme 
for grouping together polarizations that reduces gauge 
ambiguities in predictions for polarized scattering rates
and is applicable to the fully massive case.
As case studies, 
we examine polarization interference in
$W$(+jets), top quark decay, 
and neutrino deep-inelastic scattering.
For decays of unpolarized top quarks, 
interference exactly cancels at the 
totally unintegrated level.
}

\preprint{IFJPAN-IV-2025-23, COMETA-2025-45}
\arxivnumber{2512.10015}

\begin{document}
\maketitle
\setcounter{page}{2}

\newpage
\section{Introduction}\label{sec:intro}

In the Standard Model (SM) of particle physics 
the gauge forces are complementary.
In different kinematical regimes 
the different helicity polarizations $(\lambda)$
of the gauge bosons naturally probe different dynamics~\cite{Lam:1978pu,
Hagiwara:1986vm,Mirkes:1994eb,Bern:2011ie,Stirling:2012zt,Azatov:2016sqh,
Panico:2017frx,Covarelli:2021gyz,BuarqueFranzosi:2021wrv,Carrivale:2025mjy}.
Of particular interest are the longitudinal 
polarizations of the $W$ and $Z$,
which can contribute to physical scattering amplitudes 
in on-shell limits due to electroweak (EW) symmetry breaking.
This should be contrasted 
with longitudinally polarized photons and gluons, 
which contribute to physical processes but only when they are 
off-shell~\cite{Budnev:1975poe,Halzen:1984mc,Sterman:1993hfp,
Kroll:1995pv,Ji:1996nm,OPAL:1999rcd,HERMES:2011yno,Coleman:2018mew}.
In this sense, the novelty of studying multiboson processes 
at the Large Hadron Collider (LHC), 
processes like vector boson scattering and triboson production, 
is the ability to directly observe longitudinal $W$ and $Z$ bosons 
with energies above the EW scale.
Because of this novelty, 
there is motivation 
to have a firm theoretical understanding 
of weak boson polarization in the high-energy regime.

Unfortunately, 
helicity polarization is complicated
in gauge quantum field theories.
Since gauge vector bosons belong 
to the 4-vector representation of the 
Lorenz group, 
the states are formally described 
by four polarizations,
i.e., four polarization vectors $\varepsilon^\mu(q,\lambda)$.
These are the wave functions that normalize 
creation and annihilation operators of 
spin-1 quantum fields in Fourier 
space.
And by construction, the $\varepsilon^\mu(q,\lambda)$ 
are related to gauge boson propagators 
though completeness relationships.
That and gauge fixing.

Through gauge fixing, gauge boson propagators,
polarization vectors, and other Feynman rules
are adjusted so that redundant and unphysical (gauge) artifacts
are removed from amplitudes for physical processes.
Depending on the gauge and process,
amplitudes (or graphs or Green's functions)
may feature more or fewer polarized contributions 
in predictions for cross sections and decay 
rates~\cite{Kunszt:1987tk,Bassetto:1991ue,Capper:1981rd,
Leibbrandt:1994wj,Sterman:1993hfp,Ellis:1996mzs,Dams:2004vi,
Dreiner:2008tw,Coleman:2018mew}.
Above all, Refs.~\cite{tHooft:1971qjg,
Becchi:1974md,Becchi:1974xu,Becchi:1975nq,
Tyutin:1975qk,Becchi:2014lsa} 
are clear that 
longitudinally polarized gauge bosons 
should not be considered in isolation 
but in conjunction with scalar polarizations,
Goldstone bosons, and Faddeev-Popov ghosts.

Given the success of the ``polarized propagator''
(or ``spin-truncated propagator'') paradigm~\cite{Ballestrero:2017bxn,
Ballestrero:2019qoy,BuarqueFranzosi:2019boy,Denner:2020bcz,Hoppe:2023uux}
in describing polarized weak-boson data from the 
LHC~\cite{CMS:2011kaj,CMS:2020etf,CMS:2021icx,
ATLAS:2022oge,ATLAS:2023zrv,ATLAS:2024qbd},
in anticipation of the precision-polarization program
at the high-luminosity LHC~\cite{BuarqueFranzosi:2021wrv,Carrivale:2025mjy},
and with the aim of putting the formalism and its practical application
on firmer theoretical footing,
we investigate polarization interference, off-shell effects, 
and gauge cancellations
in processes featuring (near)-resonant, helicity-polarized $W$ bosons.

To carry out this work, we start in Sec.~\ref{sec:polvector} 
by introducing analytical decompositions 
for polarized propagators of EW bosons
in covariant (Sec.~\ref{sec:polvector_covariant}) 
and axial (Sec.~\ref{sec:polvector_axial}) gauges.
These decompositions are exact, covariant,
simplify the organization and evaluation 
of polarized amplitudes,
and make power counting 
of mass-over-energy factors manifest.

In Sec.~\ref{sec:interference}, 
we then build general expressions 
for polarization interference in covariant and axial gauges.
Our ability to compute polarization interference directly
and
in a systematic fashion goes beyond many contemporary 
works on weak boson polarization, 
which typically estimate interference indirectly from closure.
Importantly, our construction relies on the observation 
that helicity polarization can be treated 
diagrammatically~\cite{Javurkova:2024bwa}.
This essentially puts the polarizations of gauge bosons 
on the same footing as Goldstones and ghosts
when computing helicity amplitudes,
in accordance with BRST invariance~\cite{Becchi:1974md,
Becchi:1974xu,Becchi:1975nq,Tyutin:1975qk}.

In Sec.~\ref{sec:interference_independence},
we propose a simple scheme 
for including unphysical scalar degrees of freedom (dof)
in realistic Monte Carlo predictions for polarized processes.
Inspired by BRST invariance, 
the scheme reduces some of the inherent gauge dependence
(ambiguities) in polarized predictions,
is applicable to processes with massive external states,
and helps explain the success of the 
``pole'' and ``narrow width'' approximations 
in polarization studies.

In Sec.~\ref{sec:noninterference}, 
we derive some conditions under which polarization interference vanishes,
effectively extending the findings 
of Refs.~\cite{Azatov:2016sqh,Panico:2017frx,Ballestrero:2017bxn}
for polarization interference in diboson production 
and vector boson scattering to a broader class of processes.
We are able to give new insights into the origin 
of ``accidentally'' small polarization 
interference in $V$+jets~\cite{Belyaev:2013nla},
the interplay between kinematical cuts 
in multiboson production~\cite{Stirling:2012zt,Belyaev:2013nla,
Panico:2017frx,Ballestrero:2019qoy,Maina:2020rgd,Denner:2020eck},
and the observed differences in polarization interference 
between $W$ and $Z$ production~\cite{Ballestrero:2019qoy}.

In Sec.~\ref{sec:case_studies}, 
we further apply our organization and power-counting techniques  
to compute or estimate polarization interference 
in several realistic charged-current processes.

We conclude in Sec.~\ref{sec:conclusion} 
with an outlook for further applications of our work.

A pedagogical construction of polarization vectors
is given in Appendix~\ref{app:polvectors}.
In Appendix~\ref{app:properties} we list additional properties 
 and relationships for polarized propagators
 not covered in Sec.~\ref{sec:polvector}.
 Finally, Appendix~\ref{app:mgpolar} documents extended support 
 for simulated polarized matrix elements 
 with the simulation framework 
\texttt{MadGraph5\_aMC@NLO}~\cite{Stelzer:1994ta,
Alwall:2014hca,BuarqueFranzosi:2019boy}.

\section{Polarized Propagators and Power Counting}
\label{sec:polvector}

In this section we introduce exact, analytical decompositions 
for the outer products of helicity polarization vectors, 
$\varepsilon^\mu(q,\lambda)\varepsilon^\mu(q,\lambda)$,
for EW boson
in both covariant (Sec.~\ref{sec:polvector_covariant})
and axial (Sec.~\ref{sec:polvector_axial}) gauges.
Such outer products appear in the definitions of 
helicity-polarized propagators for gauge bosons,
and hence are needed 
to construct polarized scattering rates.
Our decompositions are inspired 
by power-counting devices used 
in quantum chromodynamics (QCD)~\cite{Sterman:1978bi,
Libby:1978bx,Kunszt:1987tk,Sterman:1995fz}
and accommodates the nonzero masses of the $W$ and $Z$.

In order to map our conventions and notation 
onto popular works on polarization, 
e.g., Refs.~\cite{Ballestrero:2017bxn,
Ballestrero:2019qoy,BuarqueFranzosi:2019boy,Denner:2020bcz,Hoppe:2023uux},
we briefly summarize our organization of 
polarized and unpolarized contributions 
at the diagrammatic $(\mathcal{M})$ 
and squared-diagrammatic $(\vert\mathcal{M}\vert^2)$ 
levels in Sec.~\ref{sec:polvector_organization}.
Our kinematical conventions and notation are defined 
in Sec.~\ref{sec:polvector_kinematics}.

Importantly, predictions for polarized scattering rates 
have inherent gauge ambiguities due 
to the interplay between 
the longitudinal and scalar poalrizations 
with Goldstones and Faddeev-Popov ghosts.
The ambiguity is precisely BRST invariance.
Throughout our work we strive to keep track of this interplay, 
particularly in the (near) on-shell regime 
where finite-width effects become relevant.
As a consequence, our expressions for polarized propagators
in Sec.~\ref{sec:polvector_covariant} and Sec.~\ref{sec:polvector_axial}
are (slight) generalizations of those used 
in Refs.~\cite{Ballestrero:2017bxn,Ballestrero:2019qoy}.

\subsection{Organization of Polarized and Unpolarized Subamplitudes}
\label{sec:polvector_organization}

At the LHC, fiducial cross sections for processes mediated 
by resonant weak bosons with fixed helicities are measured 
using the template method~\cite{Bern:2011ie,
Stirling:2012zt,CMS:2011kaj,CMS:2020etf,CMS:2021icx,
ATLAS:2022oge,ATLAS:2023zrv,ATLAS:2024qbd}.
According to this method, an unpolarized process is 
first measured and a set of templates are then fit to 
its kinematical distributions in the fiducial region 
$(d\sigma_{\rm unpol}^{\rm fid.})$.
Each template corresponds to the original process
but mediated by a helicity-polarized weak boson.
During the fitting procedure, the normalizations 
of helicity-polarized cross sections $(\sigma_\lambda)$ 
are allowed to vary, resulting in a measurement 
of polarization fractions in the fiducial region  
$f_\lambda^{\rm fid.} = 
\sigma_\lambda^{\rm fid.}/\sigma_{\rm unpol}^{\rm fid.}$.

The template method draws on the fact that the distribution 
of unpolarized events $(dN_{\rm unpol})$ are determined 
by the squared matrix elements 
$(\vert \mathcal{M}\vert^2)$
for unpolarized processes,
$dN_{\rm unpol}\ \propto d\sigma_{\rm unpol}\ \propto\
    \vert\mathcal{M}_{\rm unpol}\vert^2$,
with the later related to matrix elements 
for helicity-polarized processes $(\mathcal{M}_\lambda)$
by completeness relationships.
This is expressed by~\cite{Ballestrero:2017bxn,Ballestrero:2019qoy}
\begin{align}
\label{eq:unpol_def}
\vert\mathcal{M}_{\rm unpol}\vert^2\ &=\
\vert\mathcal{M}^{\rm res}_{\rm unpol}\ 
+\ 
\mathcal{M}_{\rm non-res}\vert^2\
\\
\label{eq:nonres_int_def}
&\equiv\
\vert\mathcal{M}^{\rm res}_{\rm unpol}\vert^2\ 
+\ 
\mathcal{I}_{\rm non-res}\
\\
\label{eq:polint_def_prelim}
&=\
\sum_{\lambda\in\{\pm1,0,S\}}\ 
\vert\mathcal{M}_\lambda\vert^2\ 
+\ 
\mathcal{I}_{\rm pol}\  
+\ 
\mathcal{I}_{\rm non-res}\
,\quad\text{where}
\\
\mathcal{I}_{\rm pol}\ &\equiv\ 
\sum_{\lambda\neq\lambda'}
\mathcal{M}^*_{\lambda}\mathcal{M}_{\lambda'}\ .
\label{eq:polint_def}
\end{align}

\begin{figure}[!t]
\includegraphics[width=\textwidth]{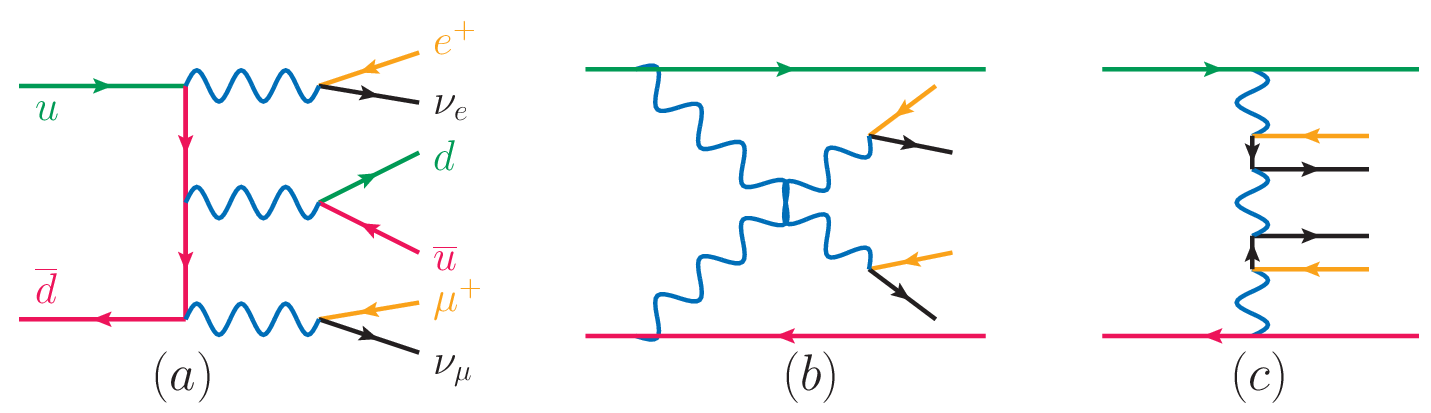}
\caption{(a) Resonant,
(b) mixed resonant-non-resonant,
and 
(c) non-resonant 
contributions to the Born-level,
partonic process 
$u\overline{d}\to e^+ \nu_e \mu^+\nu_\mu d\overline{u}$
at $\mathcal{O}(\alpha^6 \alpha_s^0)$.
Drawn with \texttt{JaxoDraw}~\cite{Binosi:2003yf,Binosi:2008ig}.}
\label{fig:wPolar_res_nores_example}
\end{figure}

In Eq.~\eqref{eq:unpol_def},
the full, unpolarized matrix element $\mathcal{M}_{\rm unpol}$
is assumed to be gauge invariant and consists of both 
resonant, unpolarized subamplitudes
$(\mathcal{M}^{\rm res}_{\rm unpol})$
and 
non-resonant subamplitudes $(\mathcal{M}_{\rm non-res})$.
Resonant, unpolarized contributions  
are understood to contain $s$-channel 
exchanges of weak bosons
that are on shell or nearly on shell.
Non-resonant contributions are understood 
to contain chains of $t$-channel exchanges.
Such subamplitudes are illustrated 
in Fig.~\ref{fig:wPolar_res_nores_example}
for the tree-level process 
$u\overline{d}\to e^+\nu_e \mu^+\nu_\mu d\overline{u}$
at $\mathcal{O}(\alpha^6 \alpha_s^0)$,
which shows 
(a) triply resonant contributions 
(triboson production),
(b) mixed resonant-non-resonant contributions 
(vector bosons scattering),
and (c) non-resonant contributions.

The precise division of $\mathcal{M}_{\rm unpol}$
into two (or more) categories depends on several factors.
These include: process definitions, phase space cuts, 
gauge choice, and kinematical assumptions, 
such as the massless of external fermions,
the narrow width approximation or the pole approximation.
Our work is not at odds with these approximations
since they are usually applied at the level of 
phase space integration while we are working (at least) one step before,
at the level of matrix elements and squared matrix elements.

At the squared level, 
the non-resonant ``interference'' term 
$\mathcal{I}_{\rm non-res}$ 
in Eq.~\eqref{eq:nonres_int_def}
consists of squared, non-resonant contributions, 
$\vert\mathcal{M}_{\rm non-res}\vert^2$,
as well as the interference 
between resonant and non-resonant contributions, 
$\Re[\mathcal{M}_{\rm unpol}^{\rm res} \mathcal{M}_{\rm non-res}^*]$.
Again, non-resonant contributions must be carefully considered 
to avoid spoiling large cancellations~\cite{Ballestrero:2017bxn,
Ballestrero:2019qoy,Denner:2020bcz}.

The ``polarization interference'' $(\mathcal{I}_{\rm pol})$
in Eqs.~\eqref{eq:polint_def_prelim}-\eqref{eq:polint_def}
is the interference among the resonant, helicity-polarized amplitudes 
$\mathcal{M}_\lambda$.
It is the focus of our work.
In experimental measurements~\cite{CMS:2011kaj,
CMS:2020etf,CMS:2021icx,ATLAS:2022oge,ATLAS:2023zrv,ATLAS:2024qbd}, 
$\mathcal{I}_{\rm pol}$ is assumed to be small (or negligible)
due to the smallness of polarization interference 
in inclusive cross sections 
for some multiboson processes~\cite{Mirkes:1994eb,
Bern:2011ie,Stirling:2012zt,Azatov:2016sqh,
Panico:2017frx,Ballestrero:2017bxn}.
However, $\mathcal{I}_{\rm pol}$ is also known 
to reach $\mathcal{I}_{\rm pol}\sim\mathcal{O}(10\%)$ 
in exclusive regions of phase space~\cite{Belyaev:2013nla,
Ballestrero:2019qoy,Maina:2020rgd,
Denner:2020eck,Denner:2021csi,Denner:2023ehn}
and
in low-energy limits~\cite{Panico:2017frx,Accomando:2006mc,
Borel:2012by,Chen:2016wkt,Ruiz:2021tdt,Bigaran:2025rvb}.
However, most of these studies only 
estimated $\mathcal{I}_{\rm pol}$ numerically 
via completeness/closure,
assumed strict on-shell conditions,
and/or worked in the context of high-energy factorization.
As a result, a precise understanding of $\mathcal{I}_{\rm pol}$ 
has remained obfuscated and motivates our investigation.

\subsection{Notation and Kinematics}
\label{sec:polvector_kinematics}

For covariant gauges in Sec.~\ref{sec:polvector_covariant}
and axial gauges in Sec.~\ref{sec:polvector_axial},
we consider a spin-1 state $V_\lambda$ with mass $M_V$, width $\Gamma_V$, 
virtuality $\sqrt{q^2}$, and 4-momentum $q^\mu$ given by
\begin{align}
\label{eq:mom_def}
 q^\mu\ &=\ (E_V, q_x, q_y, q_z) =\ 
 (E_V, 
 \vert\vec{q}\vert \sin\theta_V\cos\phi_V, 
 \vert\vec{q}\vert \sin\theta_V\sin\phi_V, 
 \vert\vec{q}\vert \cos\theta_V)\ .
\end{align}
Here and throughout
$\vec{q}=(q_x,q_y,q_z)$ 
is the 3-momentum of $V(q)$,
$\vert\vec{q}\vert$ is the corresponding magnitude,
and $\hat{q}=\vec{q}/\vert\vec{q}\vert=(\sin\theta_V\cos\phi_V, 
 \sin\theta_V\sin\phi_V, 
 \cos\theta_V)$ is the unit 3-vector 
that points in the direction of $\vec{q}$.
We denote the transverse momentum 2-vector 
by $\vec{q}_T = (q_x,q_y)$,
its magnitude by $q_T=\vert\vec{q}_T\vert$,
and its 4-vector embedment by $q_T^\mu = (0,q_x,q_y,0)$.

Given Eq.~\eqref{eq:mom_def}, 
we are able to build the auxiliary momentum vectors 
\begin{subequations}
\label{eq:mom_def_perp}
\begin{align}
\hat{q}_{\perp}^\mu &= 
\frac{1}{q_T\vert\vec{q}\vert} (0, q_x q_z, q_y q_z, -q_T^2 )\ 
=\
(0,\cos\theta_V\cos\phi_V,\cos\theta_V\sin\phi_V,-\sin\theta_V)\ ,
\\
\hat{q}_{T\perp}^\mu &= 
\frac{1}{q_T} (0, - q_y, q_x, 0 )\ =\
(0,-\sin\phi_V,\cos\phi_V,0)\ . 
\end{align}    
\end{subequations}
The spatial parts of these momenta are the two directions that are 
mutually perpendicular to $\hat{q}$, 
with $\hat{q}_{T\perp}^\mu$ also being orthogonal to $q_T^\mu$.
By construction, they satisfy a number of orthogonality relationships: 
$(\hat{q}_{\perp}\cdot q)$, 
$(\hat{q}_{T\perp}\cdot q)$, 
$(\hat{q}_{\perp}\cdot\hat{q}_{T\perp})$,
and 
$(\hat{q}_{T\perp}\cdot q_T)=0$.

Following the \texttt{HELAS} convention~\cite{Hagiwara:1985yu,Murayama:1992gi},
we quantize the spin axis of $V$ along the $\hat{z}$ direction
and build Lorentz-covariant quantities through boosts and rotations.
As a result, expressions throughout this section 
and Sec.~\ref{sec:interference} 
do not assume a particular reference frame.
We denote the helicity of $V$ by $\lambda\in\{\pm1,0,S\}$, 
where $\lambda=+1\ (-1)$ is the right (left) transverse polarization,
$\lambda=0$ is the longitudinal polarization, and 
$\lambda=S$ is the scalar (or auxiliary) polarization vector. 
The coherent sum over the two transverse polarizations is denoted
by $\lambda=T$.
In expressions for currents and amplitudes,
$P_{L/R} = (1\mp\gamma^5)$ and $\gamma^\mu$ 
are the usual chiral projection operators 
and gamma matrices in the chiral basis.
%

\subsection{Polarized Propagators in the 
\texorpdfstring{$R_\xi$}{Rx} and Unitary Gauges}
\label{sec:polvector_covariant}

In the generalized renormalizable $(R_\xi)$ gauge,
the propagator of an intermediate weak boson
with finite-width corrections
and its decomposition 
into the outer product 
of helicity polarization vectors 
for arbitrary momentum $q$ are given by
\begin{align}
\label{eq:prop_unpol_rx}
    \Pi_{\mu\nu}^V(q)\ &=\
  \cfrac{-i\left[g_{\mu\nu} + (\xi-1)\cfrac{q_\mu q_\nu}{q^2 - \xi M_V^2 +i\xi M_V\Gamma_V}\right]}{q^2-M_V^2 + iM_V\Gamma_V}\ 
  \\
  &=\
    \sum_{\lambda=\pm1,0,S}\
    \cfrac{i \eta_\lambda\ \varepsilon_\mu(q,\lambda)\varepsilon^*_\nu(q,\lambda)
}{q^2 - M_V^2 + iM_V\Gamma_V}\ 
\label{eq:completeness}
\\
&\equiv\ 
\sum_{\lambda=\pm1,0,S}\ 
\Pi_{\mu\nu}^V(q,\lambda)\ , \quad\ \text{where}\
\label{eq:polprop_predef}
\\
-(\eta_{\lambda=S})\
&
=\ \eta_{\lambda=+1}\ 
=\ \eta_{\lambda=-1} 
=\ \eta_{\lambda=0}\ 
=\ +1\ .
\label{eq:polprop_sign_convention}
\end{align}

The $iM_V\Gamma_V$ term in the (lower) denominator of 
Eq.~\eqref{eq:prop_unpol_rx}
is the textbook result of summing over one-particle irreducible diagrams,
generating a Breit-Wigner distribution when $V$ is nearly on shell.
The $i\xi M_V\Gamma_V$ term in the numerator (upper denominator)
of Eq.~\eqref{eq:prop_unpol_rx}
is tied to gauge invariance, 
namely satisfying Ward identities when the $iM_V\Gamma_V$ term 
(lower denominator) is present~\cite{Willenbrock:1991hu,
Stuart:1991xk,LopezCastro:1991nt,LopezCastro:1995kg,
Denner:1999gp,Denner:2005fg}.
In the Complex Mass Scheme, the $iM_V\Gamma_V$ term 
and the $i\xi M_V\Gamma_V$ term are both generated 
when real-valued masses $M_V$ 
are replaced by complex-valued masses 
$\tilde{M}_V=\sqrt{M_V^2-iM_V\Gamma_V}$
at the Lagrangian level~\cite{Denner:1999gp,Denner:2005fg}.

Equation \eqref{eq:completeness} is the completeness relationship 
between the propagator and polarization vectors
for arbitrary momenta (not just on-shell momenta).
The completeness relationship also defines the so-called 
helicity-polarized propagator $\Pi_{\mu\nu}^V(q,\lambda)$ 
in Eq.~\eqref{eq:polprop_predef}, 
\begin{align}
\label{eq:polprop_def}
    \Pi_{\mu\nu}^V(q,\lambda)\ &=\ 
    \cfrac{i\eta_\lambda\ \varepsilon_\mu(q,\lambda)\varepsilon^*_\nu(q,\lambda)
}{q^2 - M_V^2 + iM_V\Gamma_V}\ 
\equiv\ 
\cfrac{i\eta_\lambda}{D_V(q^2)}\ 
\varepsilon_\mu(q,\lambda)\varepsilon^*_\nu(q,\lambda)\ .
\end{align}
Here, we also define our shorthand $D_V(q^2)=q^2 - M_V^2 + iM_V\Gamma_V$ 
for the pole structure of $V$ at $q^2$.
For the values of $\eta_\lambda$, we follow\footnote{\label{foot:haber}See 
also the lecture notes available at 
\href{http://scipp.ucsc.edu/~haber/ph218/polsum.pdf}{http://scipp.ucsc.edu/~haber/ph218/polsum.pdf}.} 
the convention of Ref.~\cite{Dreiner:2008tw} 
and adopt\footnote{The simulation 
framework \texttt{MadGraph5\_aMC@NLO}~\cite{Stelzer:1994ta,Alwall:2014hca} uses $\eta_S=+1$
with $\sqrt{-1}\times\varepsilon(q,\lambda=S)$~\cite{BuarqueFranzosi:2019boy}.} 
a form  that mirrors the 
\textit{negative} signature of the Minkowski metric: 
$-g_{\mu\nu}={\rm diag}(-1,+1,+1,+1)$. 
This convention simplifies the completeness relationship 
for polarization vectors in the Cartesian and helicity bases 
(see App.~\ref{app:polvectors}),
but other conventions can be found in the literature.

The quantity $\xi$ in Eq.~\eqref{eq:prop_unpol_rx}
is the gauge-fixing parameter of the theory 
and is implicit for some helicities in Eq.~\eqref{eq:polprop_def}.
The Unitary gauge is obtained by taking $\xi\to\infty$, 
\begin{align}
\label{eq:prop_unpol_unitary}
    \Pi_{\mu\nu}^V(q)\ {\Big\vert}_{\rm Unitary} &=\
  \cfrac{-i\left[g_{\mu\nu} - \cfrac{q_\mu q_\nu}{M_V^2 -i M_V\Gamma_V}\right]}{q^2-M_V^2 + iM_V\Gamma_V}\ .
\end{align}
Other gauges are obtain by taking the appropriate limits,
e.g., $\xi\to1$ for 'tHooft-Feynman and $\xi\to0$ for the Landau gauge.
For finite $\xi$ and with finite-width corrections, 
the propagators for the EW Goldstone bosons 
$G\in\{G^\pm, G^0\}$ are given by
\begin{align}
\label{eq:prop_goldstone}
\Pi_G(q)\ &=\ 
\frac{i}{q^2 - \xi M_V^2 + i\xi M_V\Gamma_V}\ 
\equiv\ \frac{i}{D_V(q^2,\xi)}\ .
\end{align}
Here, $M_G=\sqrt{\xi} M_V$ and $\Gamma_G=\sqrt{\xi} \Gamma_V$ 
are the $\xi$-dependent mass and width of $G$,
with $M_V$ and $\Gamma_V$ corresponding to those 
of the massive gauge boson $V$.
We also define our shorthand notation 
$D_V(q^2,\xi)=q^2 - \xi M_V^2 + i\xi M_V\Gamma_V$ 
for the pole structure of $G$.

\subsubsection{Transverse Polarized Propagators} 
\label{sec:polvector_covariant_trans}

For transverse helicities $(\lambda=\pm1)$ and 
momentum $q$ as given in Eq.~\eqref{eq:mom_def},
we use the following polarization vectors, valid for all $\xi$ in the $R_\xi$ gauge:
\begin{align}
\label{eq:polvec_transverse_alt}
\varepsilon_\mu (q,&\lambda=\pm1)\ 
=\ -\lambda\ \hat{q}_{\perp\mu}\ -\ i\hat{q}_{T\perp\mu}\ \\ 
&= \frac{1}{\sqrt{2}}\big(0,
\pm\cos\theta_V\cos\phi_V - i\sin\phi_V,
\pm\cos\theta_V\sin\phi_V +  i\cos\phi_V,
\mp\sin\theta_V \big)\ 
\label{eq:polvec_transverse}
\end{align}
The expressions here coincide with those of the \texttt{HELAS} 
convention~\cite{Hagiwara:1985yu,Murayama:1992gi}
as well as Ref.~\cite{Aivazis:1993kh} for $\theta_V\to0$.
They satisfy the helicity-operator relationship  
$h^{\mu\nu}(\hat{q})\varepsilon_\nu(q,\lambda)=
\lambda \varepsilon^\mu(q,\lambda)$ 
[see Eq.~\eqref{eq:polvectors_hel_helicity_op} of App.~\ref{app:polvectors}],
and are valid for massless spin-1 states
in both on-shell and off-shell regimes.
Since the temporal $(\mu=0)$ components are zero, 
$\vec{q}\cdot \vec{\varepsilon}(q,\lambda=\pm1)=0$
in addition to $q\cdot \varepsilon (q,\lambda=\pm1)=0$.
This last point is guaranteed by the gauge-fixing condition
for convariant gauges, $(\partial^\mu A_\mu)=0$,
which stipulates that physical states are orthogonal to $q^\mu$.

The outer product of polarization vectors, 
summed over both transverse helicities, is 
\begin{align}
\label{eq:spinsum_def_trans}
 &   
 \sum_{\lambda=\pm1}\ \varepsilon_\mu(q,\lambda)\varepsilon^*_\nu(q,\lambda)\
 \nonumber\\
 =&
\begin{pmatrix}
0 & 0 & 0 & 0 \\
0 & \cos^2\theta_V\cos^2\phi_V + \sin^2\phi_V & -\cos\phi_V\sin^2\theta_V\sin\phi_V & -\cos\theta_V\sin\theta_V\cos\phi_V \\
0 & -\cos\phi_V\sin^2\theta_V\sin\phi_V & 
\cos^2\phi_V+ \cos^2\theta_V\sin^2\phi_V
& -\cos\theta_V\sin\theta_V\sin\phi_V \\
0 & -\cos\theta_V\sin\theta_V\cos\phi_V & -\cos\theta_V\sin\theta_V\sin\phi_V 
& \sin^2\theta_V \\
\end{pmatrix} \ 
\\
\equiv&\ -g_{\mu\nu} - \Theta_{\mu\nu}(\theta_V,\phi_V) \ .
\label{eq:spinsum_trans}
\end{align}
Here, 
$\Theta_{\mu\nu}$ is our first bookkeeping device;
its negative is defined as the sum of 
the polarization sum and 
the (Minkowski) spacetime metric $g_{\mu\nu}$.
Explicitly, it is given by 
\begin{align}
\label{eq:theta_def_angles}
 \Theta_{\mu\nu}\
=&\
\begin{pmatrix}
-1 & 0 & 0 & 0 \\
0 & \sin^2\theta_V\cos^2\phi_V  & \cos\phi_V\sin^2\theta_V\sin\phi_V & \cos\theta_V\sin\theta_V\cos\phi_V \\
0 & \cos\phi_V\sin^2\theta_V\sin\phi_V & 
\sin^2\theta_V\sin^2\phi_V
& \cos\theta_V\sin\theta_V\sin\phi_V \\
0 & \cos\theta_V\sin\theta_V\cos\phi_V & \cos\theta_V\sin\theta_V\sin\phi_V 
& \cos^2\theta_V \\
\end{pmatrix}  
\\
 =&\ 
\begin{pmatrix}
-1 & 0 & 0 & 0 \\
0 & \hat{q}_x^2  & \hat{q}_x\ \hat{q}_y & \hat{q}_x\ \hat{q}_z \\
0 & \hat{q}_x\ \hat{q}_y & \hat{q}_y^2 & \hat{q}_y\ \hat{q}_z \\
0 & \hat{q}_x\ \hat{q}_z & \hat{q}_y\ \hat{q}_z & \hat{q}_z^2 \\
\end{pmatrix} \ .
\end{align}
In terms of this device, 
the transverse helicity propagator can be written as
\begin{align}
\label{eq:prop_trans}
\boxed{
    \Pi_{\mu\nu}^V(q,\lambda=T)\ =\ 
    \sum_{\lambda=\pm1}
    \frac{i\eta_\lambda\ \varepsilon_\mu(q,\lambda)\varepsilon^*_\nu(q,\lambda)
}{q^2 - M_V^2 + iM_V\Gamma_V}\ 
=\ 
\frac{-i\left(g_{\mu\nu} + \Theta_{\mu\nu}\right)}{q^2 - M_V^2 + iM_V\Gamma_V} 
}\ .
\end{align}

We draw attention to the relative positive sign
preceding $\Theta_{\mu\nu}$.
Since the spatial diagonal elements 
$(\mu=\nu=1,2,3)$
of $\Theta_{\mu\nu}$ are  positive-definite, 
the sign na\"ively 
suggests 
constructive interference with $g_{\mu\nu}$.
However, the spatial elements of $g_{\mu\nu}$ are negative, 
indicating cancellations and destructive interference. 
For off-diagonal spatial components $(\mu\neq\nu=1,2,3)$,
the metric is zero while $\Theta_{\mu\nu}$ 
can take on both positive and negative values 
over the full $4\pi$ domain of $\theta_V$ and $\phi_V$.
Since the transverse polarization vectors 
describe transverse polarization 
relative to $V$'s propagation,
and since $g_{\mu\nu}$ contains temporal and longitudinal 
components, 
then $\Theta_{\mu\nu}$ encodes  
time-like and space-like propagation. 
This is clearer when $q^\mu$ 
is aligned with the $\hat{z}$ axis $(\theta_V\to0,\pi)$:
\begin{align}
\label{eq:theta_limits}
\lim_{\theta_V\to0} 
 \sum_{\lambda=\pm1} 
 \varepsilon_\mu(q,\lambda)\varepsilon^*_\nu(q,\lambda)
    &= 
\begin{pmatrix}
0 & 0 & 0 & 0 \\
0 & 1 & 0 & 0 \\
0 & 0 & 1 & 0 \\
0 & 0 & 0 & 0 \\
\end{pmatrix} \ ,\
\lim_{\theta_V\to0,\pi} 
\Theta_{\mu\nu}(\theta_V,\phi_V) 
=
\begin{pmatrix}
-1 & 0 & 0 & 0 \\
0 & 0  & 0 & 0 \\
0 & 0 & 0 & 0 \\
0 & 0 & 0 & 1 \\
\end{pmatrix}\   .
\end{align}

Now, by using Eq.~\eqref{eq:polvec_transverse_alt},
the polarization sum in Eq.~\eqref{eq:spinsum_trans}
can also be written as  
\begin{align}
    \Phi(\theta_V,\phi_V)\ \equiv\ 
    \sum_{\lambda=\pm1}\ 
    \varepsilon_\mu(q,\lambda)\varepsilon^*_\nu(q,\lambda)\
    =\ 
    \hat{q}_{\perp\mu}\hat{q}_{\perp\nu}\ 
    +\ 
    \hat{q}_{T\perp\mu}\hat{q}_{T\perp\nu}\ .    
\label{Phi_matrix}
\end{align}
In this alternative but equivalent form, the transverse helicity propagator is
\begin{align}
\label{eq:prop_trans_alt}
\boxed{    
\Pi_{\mu\nu}^V(q,\lambda=T)\ =\ 
\frac{i\ \left(\hat{q}_{\perp\mu}\hat{q}_{\perp\nu}\ 
+\ 
\hat{q}_{T\perp\mu}\hat{q}_{T\perp\nu}\right)}{q^2 - M_V^2 + iM_V\Gamma_V}
}\    ,
\end{align}
which makes  directional and orthogonality relationships
of the propagator manifest.

To make these same relationships more explicit in Eq.~\eqref{eq:prop_trans},
we introduce the reference vector $n^\mu$ as a second bookkeeping device.
For the following choices of $n^\mu$
\begin{subequations}
\label{eq:ref_vector_def}
\begin{align}
     \text{light-like\ (LL)}\ :\ &\ n_{\rm LL}^\mu\ 
     =\ (1,-\hat{q}),\quad 
     \text{with}\  n_{\rm LL}^2=\ 0\ ,\\
     \text{time-like\ (TL)}\ :\ &\ n_{\rm TL}^\mu\ 
     =\ (1,\ 0),\quad          
     \text{with}\  n_{\rm TL}^2=+1\  ,\\
     \text{space-like\ (SL)}\ :\ &\ n_{\rm SL}^\mu\ 
     =\ (0,-\hat{q}),\quad   
     \text{with}\  n_{\rm SL}^2=-1\  ,
\end{align}
\end{subequations}
our first bookkeeping device 
$\Theta_{\mu\nu}$ admits the following decomposition
\begin{align}
    \Theta_{\mu\nu}\ &=\ 
    \frac{(n\cdot q)}{(n\cdot q)^2 - q^2 n^2}
    \left[-n_\mu q_\nu - q_\mu n_\nu + \frac{q_\mu q_\nu n^2}{(n\cdot q)}
    + \frac{n_\mu n_\nu q^2}{(n\cdot q)}\right]\ .
    \label{Theta_matrix}
\end{align}

Expressing $\Theta_{\mu\nu}$, 
and hence the sum of transverse polarization vectors, 
in this manner in the $R_\xi$ gauge 
is notable as the Lorentz structure of Eq.~\eqref{Theta_matrix}
is manifest in gauge boson propagators in \textit{axial} gauges. 
 In axial gauges,  unphysical dof are removed from the theory 
 by introducing a reference axis $n^\mu_{\rm axial}$
 and imposing the gauge-fixing condition $(n^\mu A_\mu)=0$ 
 on physical gauge states.
In well-known applications of the axial gauge 
in QCD~\cite{Sterman:1978bi,Libby:1978bx,Kunszt:1987tk,Sterman:1995fz},
 the structure of Eq.~\eqref{Theta_matrix}
 makes mass-over-energy power counting manifest.

 In axial gauges, common choices  for the gauge-fixing \textit{vector} 
 $n^\mu_{\rm axial}$  include those  
 in Eq.~\eqref{eq:ref_vector_def}~\cite{Capper:1981rd,Leibbrandt:1994wj}.
 Taking $n^\mu_{\rm LL}=(1,-\hat{q})$ 
 is sometimes called the ``parton shower gauge,''
and leads to softer gauge cancellations among 
 diagrams~\cite{Nagy:2007ty,Nagy:2014mqa,Hagiwara:2020tbx,Chen:2022gxv}.
In general, the reference vectors in Eq.~\eqref{eq:ref_vector_def} 
do not respect Lorentz transformations, i.e., they are not Lorentz covariant. 
However, when using $n^\mu_{\rm LL}=(1,-\hat{q})$,
rotational invariance is respected when $n^\mu_{\rm LL}$ is contracted with 
Lorentz-covariant vectors, e.g., currents and momenta~\cite{Hagiwara:2020tbx,
Chen:2022gxv}.

Similar decompositions as Eq.~\eqref{Theta_matrix}
have been used in covariant gauges~\cite{Beenakker:1993yr,
Aivazis:1993kh,Bohm:2001yx,Dittmaier:2025htf},
but in the context of gluon~\cite{Beenakker:1993yr,Bohm:2001yx}
and 
photon~\cite{Aivazis:1993kh} propagation.
In Ref.~\cite{Dittmaier:2025htf}, a structure similar to 
Eq.~\eqref{eq:prop_trans} and Eq.~\eqref{Theta_matrix} 
were identified for weak boson propagation 
but assumed on-shell momentum $(q^2=M_V^2)$ and $n^2=0$,
and therefore has somewhat narrower applicability.
 
We stress that $n^\mu$ in Eq.~\eqref{Theta_matrix} 
is an unphysical bookkeeping device.
It is not a gauge-fixing parameter.
The vector $n^\mu$ appears because we insist on writing 
the polarization sum in Eq.~\eqref{eq:spinsum_def_trans}
in terms of the spacetime metric $g_{\mu\nu}$.
However, like gauge-fixing parameters,  
physical matrix elements $\mathcal{M}_{\rm unpol}$ 
must be independent\footnote{In the context of 
factorization~\cite{Contopanagos:1996nh}, this leads 
to the constraint equation $d\mathcal{M}_{\rm unpol}/d n^\mu=0$.}
of $n^\mu$.

Finally, 
for  the choices of $n^\mu$ in Eq.~\eqref{eq:ref_vector_def}
we can also take the difference between 
right-handed and left-handed outer products
to recover the identity~\cite{Aivazis:1993kh}
\begin{align}
\label{eq:poldiff_transverse}
\varepsilon_\mu(q,\lambda=+1)\varepsilon^*_\nu(q,\lambda=+1)\
-\ 
\varepsilon_\mu(q,\lambda=-1)\varepsilon^*_\nu(q,\lambda=-1)\ 
=\ 
\frac{i\epsilon_{\mu\nu\alpha\beta}\ q^\alpha n^\beta}{\sqrt{(n\cdot q)^2 - q^2 n^2}}\ .
\end{align}
Here, the antisymmetric tensor is normalized to 
$\epsilon^{\mu\nu\alpha\beta}=-\epsilon_{\mu\nu\alpha\beta}=+1$.
Its contraction with $q^\alpha$ and $n^\beta$ 
can be evaluated using trace relationships.
The result is 
\begin{align}
\frac{i\epsilon_{\mu\nu\alpha\beta}\ q^\alpha n^\beta}{\sqrt{(n\cdot q)^2 - q^2 n^2}}\ 
&=\ 
\frac{1}{\sqrt{(n\cdot q)^2 - q^2 n^2}}\ 
\left(\frac{-1}{4}\right)\
{\rm Tr}\left[\gamma^5\gamma_\mu \gamma_\nu \not\!q\not\!n\right]
\\
= 
\begin{pmatrix}
0 & 0 & 0 & 0 \\
0 & 0 & i\hat{q}_z & -i\hat{q}_y \\
0 & -i\hat{q}_z & 0 & i\hat{q}_x \\
0 & i\hat{q}_y & -i\hat{q}_x & 0 \\
\end{pmatrix} 
&=
\begin{pmatrix}
0 & 0 & 0 & 0 \\
0 & 0 & i\cos\theta_V & -i \sin\theta_V\sin\phi_V \\
0 & -i\cos\theta_V & 0 & i \sin\theta_V\cos\phi_V \\
0 & i \sin\theta_V\sin\phi_V & -i \sin\theta_V\cos\phi_V & 0 \\
\end{pmatrix}\ .
\end{align}

For $\lambda=\pm1$,
the outer product for 
individual polarization vectors can be written as 
\begin{align}
\varepsilon_\mu(q,\lambda)\varepsilon^*_\nu(q,\lambda)\
&=\ 
\frac{1}{2}\hat{q}_{\perp\mu}\hat{q}_{\perp\nu}\ 
    +\ 
\frac{1}{2}\hat{q}_{T\perp\mu}\hat{q}_{T\perp\nu}\ 
    +\ 
\frac{\lambda}{2}\frac{i \epsilon_{\mu\nu\alpha\beta}\ q^\alpha n^\beta}{\sqrt{(n\cdot q)^2 - q^2 n^2}}\ 
\\
&=\ 
-\frac{1}{2}g_{\mu\nu}\ 
-\frac{1}{2}\Theta_{\mu\nu}\
+\ 
\frac{\lambda}{2}
\frac{i \epsilon_{\mu\nu\alpha\beta}\ q^\alpha n^\beta}{\sqrt{(n\cdot q)^2 - q^2 n^2}}\ ,
\end{align}
which follows from $-g_{\mu\nu}=\Phi_{\mu\nu}+\Theta_{\mu\nu}$,
i.e., the equivalence of Eq.~\eqref{eq:spinsum_trans} and Eq.~\eqref{Phi_matrix}.
These lead to the right-handed $(\lambda=+1)$ and left-handed $(\lambda=-1)$ helicity propagators
\begin{align}
\label{eq:prop_trans_nosum}
\boxed{
    \Pi_{\mu\nu}^V(q,\lambda=\pm1)\ =\  
\cfrac{\cfrac{-i}{2}
\left(g_{\mu\nu} + \Theta_{\mu\nu}
-\lambda\cfrac{\epsilon_{\mu\nu\alpha\beta}\ q^\alpha n^\beta}{\sqrt{(n\cdot q)^2 - q^2 n^2}}
\right)}{q^2 - M_V^2 + iM_V\Gamma_V} 
}\ .
\end{align}
While $g_{\mu\nu}$ and $\Theta_{\mu\nu}$ are symmetric 
in $\mu\leftrightarrow\nu$ exchange,
$\epsilon_{\mu\nu\alpha\beta}$ is antisymmetric.
This means that the propagators 
in Eq.~\eqref{eq:prop_trans_nosum}
are neither symmetric or antisymmetric. 

Additional properties of the transverse propagators
are given in App.~\ref{app:properties}.

\subsubsection{Longitudinal Polarized Propagator} 
\label{sec:polvector_covariant_long}

For the longitudinal helicity $(\lambda=0)$ and 
momentum $q$ as given in Eq.~\eqref{eq:mom_def},
we use the following polarization vector, valid for all $\xi$ in the $R_\xi$ gauge:
\begin{align}
\label{eq:polvec_long}
\varepsilon^\mu (q,\lambda=0)\ &=\ 
\frac{E_V}{\sqrt{q^2}}\ 
\left(\frac{\vert\vec{q}\vert}{E_V},\ 
\sin\theta_V \cos\phi_V,\ 
\sin\theta_V \sin\phi_V,\ 
\cos\theta_V \right)
\\
&=\ 
\frac{1}{\sqrt{(n\cdot q)^2 - q^2 n^2}}\ 
\left[\frac{(n\cdot q)}{\sqrt{q^2}}q^\mu\ -\ n^\mu\sqrt{q^2}\right]\ .
\label{eq:polvec_long_alt}
\end{align}

Here, $n^\mu$ can be any of those listed in Eq.~\eqref{eq:ref_vector_def}.
The decomposition into $q^\mu$ and $n^\mu$ is exact and draws attention 
to the polarization vector having both a forward-like component 
$[\varepsilon^\mu(\lambda=0) \sim q^\mu]$ and 
a backward-like (or stationary for $n_{\rm TL}$) component 
$[\varepsilon^\mu(\lambda=0)\sim n^\mu]$, relative to $V$'s motion.
For the reference vector $n_{\rm LL}$, our decomposition maps to those in 
Refs.~\cite{Dawson:1984gx,Chen:2016wkt,Ruiz:2021tdt,Bigaran:2025rvb}
when $q^2\to M_V^2$.
Similarly, setting $q^2\to -q^2$ recovers the longitudinal polarization vector 
used in Eq.~\cite{Aivazis:1993kh} for an off-shell, $t$-channel photon.

We stress the factor of $1/\sqrt{q^2}$ in Eq.~\eqref{eq:polvec_long}.
A factor of $1/M_V$ is only appropriate for massive spin-1 states with on-shell momenta.
The $1/\sqrt{q^2}$ factor is necessary for consistent application 
of Eq.~\eqref{eq:polvec_long} to massive vector bosons with arbitrary momentum
and
massless, off-shell vector bosons, e.g., longitudinally polarized photons.
The $1/\sqrt{q^2}$ factor is also necessary 
to recover the completeness relationship of Eq.~\eqref{eq:completeness}.
Using $1/M_V$ in Eq.~\eqref{eq:polvec_long} but allowing $q^2\neq M_V^2$ 
can lead to $\mathcal{O}\left((q^2-M_V^2)/M_V^2\right)$ 
miscancellations in Eq.~\eqref{eq:completeness}.

With Eq.~\eqref{eq:polvec_long_alt}, 
the outer product of polarization vectors is easily found to be
\begin{align}
\label{eq:spinsum_def_long}
\varepsilon_\mu (q,\lambda=0)\varepsilon_\nu (q,\lambda=0)
&=\  
\cfrac{q_\mu q_\nu}{q^2} +
\cfrac{(n\cdot q)\
\left[-n_\mu q_\nu - q_\mu n_\nu + \cfrac{q_\mu q_\nu n^2}{(n\cdot q)}
+ \cfrac{n_\nu n_\mu q^2}{(n\cdot q)}\right]}{(n\cdot q)^2 - q^2 n^2}\
\\
&=\ \frac{q_\mu q_\nu}{q^2}\ +\ \Theta_{\mu\nu}\ .
\label{eq:spinsum_long}
\end{align}
This leads to the longitudinal helicity propagator
in terms of our bookkeeping devices:
\begin{align}
\label{eq:prop_long}
\boxed{
\Pi_{\mu\nu}^V(q,\lambda=0)\ =\   
\frac{i\eta_{\lambda=0}\ \varepsilon_\mu(q,\lambda=0)\varepsilon_\nu(q,\lambda=0)
}{q^2 - M_V^2 + iM_V\Gamma_V}\ 
=\ 
\cfrac{i\left(\Theta_{\mu\nu} + \cfrac{q_\mu q_\nu}{q^2}\right)}{q^2 - M_V^2 + iM_V\Gamma_V} 
}\ .
\end{align}

We draw attention to the $1/q^2$ pole in Eq.~\eqref{eq:prop_long}.
The divergence is spurious and an artifact of working 
in covariant gauges~\cite{Capper:1981rd,Bassetto:1991ue,Leibbrandt:1994wj}.
In covariant constructions of the EW theory
and depending on the precise choice of $\xi$, 
it is canceled by the scalar polarization and/or Goldstone bosons 
in gauge-invariant processes~\cite{Becchi:1974xu,Becchi:1975nq,Tyutin:1975qk}.
In quantum electrodynamics, the term decouples from amplitudes due 
to current conservation $q\cdot J_{\rm QED}=0$~\cite{Bohm:2001yx,Coleman:2018mew}.

In App.~\ref{app:properties} we list some properties 
of Eq.~\eqref{eq:prop_long} and the longitudinal polarization vector.

\subsubsection{Scalar Polarized Propagator for Weak Bosons and Photons} 
\label{sec:polvector_covariant_scalar}

Polarization vectors for the ``scalar'' helicity carry the remnants of gauge fixing
and spontaneous symmetry breaking.
Since gauge fixing fixes the form of the unpolarized propagator, 
the completeness relationship of Eq.~\eqref{eq:completeness}
fixes the form for scalar polarization vector.

For weak bosons in the $R_\xi$ and Unitary gauges, the scalar polarization vectors are
\begin{subequations}
\label{eq:polvec_scalar}
\begin{align}
\label{eq:polvec_scalar_rxi}
\varepsilon^\mu (q,\lambda=S)\ 
=&\ \sqrt{\frac{1}{q^2} + \frac{(\xi-1)}{q^2-\xi M_V^2 + i\xi M_V\Gamma_V}}\ 
q^\mu 
=
\sqrt{\frac{1}{q^2} + \frac{(\xi-1)}{D_V(q^2,\xi)}} 
q^\mu\ ,
\\
\label{eq:polvec_scalar_uni}
\varepsilon^\mu (q,\lambda=S)\ {\Big\vert}_{\rm Unitary}
=&\ \sqrt{\frac{1}{q^2} - \frac{1}{M_V^2 - i M_V\Gamma_V}}\ 
q^\mu\ 
= 
\sqrt{\frac{-D_V(q^2)}{(q^2)\ (M_V^2 - i M_V\Gamma_V)}}\ 
q^\mu\ .
\end{align}
\end{subequations}
The rightmost equalities follow from the definitions 
for $D_V(q^2)$ in Eq.~\eqref{eq:polprop_def}
and $D_V(q^2,\xi)$ in Eq.~\eqref{eq:prop_goldstone}.
For all choices of $\xi$, the scalar polarization vector carries 
zero helicity, as shown in Eq.~\eqref{eq:helicityOp_scalar_pol}, 
and behaves as a scalar, as discussed in
Ref.~\cite{Sterman:1993hfp,Weinberg:1995mt,Coleman:2018mew}.

In the $R_\xi$ and Unitary gauges, the outer product of scalar polarization vectors are 
\begin{align}
\label{eq:spinsum_def_scalar}
 \varepsilon_\mu (q,\lambda=S)\varepsilon_\nu (q,\lambda=S)
 =&\ 
 \left(\frac{1}{q^2} + 
 \frac{(\xi-1)}{q^2-\xi M_V^2 + i\xi M_V\Gamma_V}\right)\
q_\mu q_\nu \ 
 \\
 =&\
 \frac{\xi\ (q^2 - M_V^2 + iM_V\Gamma_V)}
 {(q^2)\ (q^2-\xi M_V^2 + i\xi M_V\Gamma_V)}\
q_\mu q_\nu \ ,
\\
\label{eq:spinsum_def_scalar_uni}
\varepsilon_\mu (q,\lambda=S)\varepsilon_\nu (q,\lambda=S)\ {\Big\vert}_{\rm Unitary}\
 =&\
 \left(\frac{1}{q^2} -
 \frac{1}{M_V^2 - i M_V\Gamma_V}\right)\
q_\mu q_\nu \
 \\
 =&\
 \frac{-(q^2 - M_V^2 + iM_V\Gamma_V)}{(q^2)\ (M_V^2 - i M_V\Gamma_V)}\
q_\mu q_\nu \ .
\end{align}
It follows that the scalar helicity propagator 
in the $R_\xi$ gauge is given by
\begin{align}
\label{eq:prop_scalar}  
\boxed{
    \Pi_{\mu\nu}^V(q,\lambda=S) =
\cfrac{-i\left(\cfrac{q_\mu q_\nu}{q^2} + \cfrac{(\xi-1)\ q_\mu q_\nu}{q^2-\xi M_V^2 + i\xi M_V\Gamma_V}\right)}{q^2 - M_V^2 + iM_V\Gamma_V} 
},
\end{align}
with the minus sign originating from $\eta_{\lambda=S}=-1$,
and in the Unitary gauge by
\begin{align}
\Pi_{\mu\nu}^V(q,\lambda=S)
{\Big\vert}_{\rm Unitary}
=
\cfrac{-i\
\left(\cfrac{q_\mu q_\nu}{q^2} -
 \cfrac{q_\mu q_\nu}{M_V^2 - i M_V\Gamma_V}\right)
}{q^2 - M_V^2 + iM_V\Gamma_V}
\ .
\label{eq:prop_scalar_uni}
\end{align}
Explicitly summing the polarized propagators 
of Eqs.~\eqref{eq:prop_trans},
\eqref{eq:prop_long},
and 
\eqref{eq:prop_scalar}
recovers the unpolarized propagators
in accordance with the completeness relationship 
of Eq.~\eqref{eq:completeness}.

For photons, 
the scalar polarization vector and propagator 
in the $R_\xi$ gauge are
\begin{align}
\label{eq:polvec_scalar_photon}
\varepsilon_\mu^\gamma (q,\lambda=S)\ 
&=\ \sqrt{\frac{\xi}{q^2}}\ q_\mu 
\\
\Pi_{\mu\nu}^\gamma(q,\lambda=S)\ &=\   
\cfrac{i\ \eta_{\lambda=S}}{q^2} 
\varepsilon_\mu^\gamma(q,\lambda=S)
\varepsilon_\nu^\gamma(q,\lambda=S)\
=\ 
-i \xi\ \frac{q_\mu q_\nu}{(q^2)^2} \ .
\end{align}
Combining this with the transverse and longitudinal propagators in 
Eqs.~\eqref{eq:prop_trans} and \eqref{eq:prop_long},
one recovers the usual unpolarized propagator for the photon
in the $R_\xi$ gauge:
\begin{align}
\label{eq:prop_unpol_rx_photon}
    \Pi_{\mu\nu}^\gamma(q)\ &=\ 
    \sum_{\lambda=T,0,S}\ \Pi_{\mu\nu}^\gamma(q,\lambda)\ 
=
  \frac{-i}{q^2}\ \left[g_{\mu\nu} + (\xi-1)\frac{q_\mu q_\nu}{q^2}\right]\ . 
\end{align}

The expressions in Eq.~\eqref{eq:prop_scalar} 
and Eq.~\eqref{eq:prop_scalar_uni}
are notable as they simplify to
\begin{align}
\label{eq:prop_scalar_reduced}
\Pi_{\mu\nu}^V(q,\lambda=S) &=
\cfrac{-i\ \xi\ 
\cfrac{q_\mu q_\nu}{q^2}}{D_V(q^2,\xi)}
\quad\text{and}\quad
\Pi_{\mu\nu}^V(q,\lambda=S)
{\Big\vert}_{\rm Unitary}\
=\
\frac{+i\ \cfrac{q_\mu q_\nu}{q^2}}{M_V^2 - i M_V\Gamma_V}\ ,
\end{align}
where poles at $q^2=M_V^2-iM_V\Gamma_V$ have been canceled,
leaving only poles at $q^2=0$ and $q^2=\xi M_V^2-i\xi M_V\Gamma_V$.
This is a manifestation of BRST invariance 
and indicative that we are propagating gauge artifacts correctly
into polarized propagators.
For example:
In the Unitary gauge there are no Goldstone bosons
and the scalar polarization of $V$
propagates like a massless particle
in order to cancel the spurious $1/q^2$ pole 
in the longitudinal propagator of Eq.~\eqref{eq:prop_long}.
In the Landau gauge $(\xi\to0)$ the situation is reversed:
there are no scalar polarizations of $V$
and the $1/q^2$ pole in Eq.~\eqref{eq:prop_long} 
is canceled by a Goldstone boson propagating 
like a massless particle $(M_G^2 = \xi M_V^2=0)$.
Maintaining the correspondence among masses of gauge bosons, scalar polarizations, 
Goldstones, and Faddeev-Poppov ghosts for arbitrary choices of $\xi$
is subtle and demonstrates the importance 
of consistently maintaining $\mathcal{O}(M_V\Gamma_V)$ terms 
in predictions for processes with polarized weak bosons.

We stress that Eq.~\eqref{eq:prop_scalar} and Eq.~\eqref{eq:prop_scalar_uni}
differ from Refs.~\cite{Ballestrero:2017bxn,Ballestrero:2019qoy,
BuarqueFranzosi:2019boy,Denner:2020bcz,Hoppe:2023uux}.
In those works scalar propagators are obtained by taking
$\Gamma_V\to0$ in Eq.~\eqref{eq:polvec_scalar} 
but keeping the Breit-Wigner propagator
as in Eq.~\eqref{eq:prop_scalar}.
This configuration does not respect BRST invariance,
and breaks electromagnetic gauge invariance at tree-level~\cite{Willenbrock:1991hu,
Stuart:1991xk,LopezCastro:1991nt,LopezCastro:1995kg}.
The numerical impact of retaining all $\mathcal{O}(M_V\Gamma_V)$ 
is explored in Sec.~\eqref{sec:top}.

We note that in our convention 
the dependence of polarization vectors and polarized propagators 
on the gauge-fixing parameter $\xi$ is carried entirely 
by the scalar contribution, i.e., Eqs.~\eqref{eq:polvec_scalar} and \eqref{eq:prop_scalar}.
In other conventions~\cite{Hagiwara:2020tbx,Chen:2022gxv},
the $\xi$ dependence is absorbed into the definition of $\eta_{\lambda=S}$.
In both conventions the transverse [Eq.~\eqref{eq:prop_trans}]
or longitudinal [Eq.~\eqref{eq:prop_long}] propagators are independent of $\xi$.

\subsection{Polarized Propagators in Axial Gauges}
\label{sec:polvector_axial}

In the 4-dimensional EW axial gauge, the unpolarized
propagator of a weak boson is~\cite{Dams:2004vi}
\begin{align}
\label{eq:prop_unpol_axial}
\Pi_{\mu\nu}^V(q)\ 
    {\Big\vert}_{\rm axial}\ &=\
  \cfrac{-i\left[g_{\mu\nu} - 
\cfrac{(n_{\rm axial})_\mu q_\nu + (n_{\rm axial})_\nu q_\mu}
{(n_{\rm axial}\cdot q)} 
+ \cfrac{n^2_{\rm axial}}{(q\cdot n_{\rm axial})^2}\
q_\mu q_\nu\right]}{q^2-M_V^2 + iM_V\Gamma_V}\ .
\end{align}
Here, $n^\mu_{\rm axial}$ is 
a reference vector that sets the gauge condition
on physical states $(n^\mu A_\mu)=0$.
The condition stipulates 
that the physical components of $V$
are orthogonal to the preferred direction, 
or \textit{axis}, $n^\mu_{\rm axial}$.
In axial gauges, $n^\mu_{\rm axial}$ is not a bookkeeping device in the sense of 
Eq.~\eqref{Theta_matrix} but a gauge-fixing 4-vector.
The allowed values for $n^\mu_{\rm axial}$ are more restricted,
and in some sense are defined by the  orthogonality and identities one wants 
in practical calculations~\cite{Capper:1981rd,Leibbrandt:1994wj}.
Common choices for $n^\mu_{\rm axial}$ include those 
given in Eq.~\eqref{eq:ref_vector_def}.

A feature of 4-dimensional axial gauges is that the propagators 
for the photon and gluon can be obtained from Eq.~\eqref{eq:prop_unpol_axial} 
by taking $M_V,\Gamma_V\to0$~\cite{Ellis:1996mzs,Dams:2004vi,Coleman:2018mew}.
The correspondence makes comparing our work on polarized weak bosons 
to works on polarized photons in deeply virtual Compton 
scattering~\cite{Kroll:1995pv,Ji:1996nm}
easier, but is left to future investigations.

As the existence of completeness relationships 
among polarization vectors is independent of gauge fixing,
the propagator in Eq.~\eqref{eq:prop_unpol_axial} 
obeys a similar completeness relationship to Eq.~\eqref{eq:completeness}.
We are therefore able to build helicity polarized propagators 
in this gauge according to Eq.~\eqref{eq:polprop_def}.
For concreteness, we choose the convention 
for $\eta_\lambda$ as in Eq.~\eqref{eq:polprop_sign_convention}.

To build helicity-polarized propagators 
in terms of our power-counting devices in the axial gauge,
we first note that the numerator of 
Eq.~\eqref{eq:prop_unpol_axial} can be written as:
\begin{align}
-g_{\mu\nu} 
    +\frac{(n_\mu q_\nu + q_\mu n_\nu)}{(q\cdot n)}
    -\frac{n^2 q_\mu q_\nu}{(q\cdot n)^2}
    &=\ 
-g_{\mu\nu} 
    - 
    \left[\frac{(q\cdot n)^2 - q^2 n^2}{(q\cdot n)^2}\right]
    \Theta_{\mu\nu}
    +
    \frac{q^2 n_\mu n_\nu}{(q\cdot n)^2}
\\
&=\ \left[-g_{\mu\nu} - \Theta_{\mu\nu}\right]
    +
    \left[
    \frac{q^2 n^2}{(q\cdot n)^2}
    \Theta_{\mu\nu}
    +
    \frac{q^2 n_\mu n_\nu}{(q\cdot n)^2}
    \right]\ .
\label{eq:polprop_axial_decomp}
\end{align}
This rewrite allows one to read-off the helicity-polarized propagators.
Additional properties of polarization vectors and 
polarized propagators in this gauge are given in App.~\ref{app:properties}.

\paragraph{Transverse polarization} 
In axial gauges, the polarization vectors for transverse 
helicities are the same as those given in 
Sec.~\ref{sec:polvector_covariant_trans},
both for massive and massless $V(q)$.
This means that the 
transverse $(\lambda=T)$, 
right-handed $(\lambda=+1)$, and
left-handed $(\lambda=-1)$
helicity-polarized propagators 
are the same as those given in 
Eqs.~\eqref{eq:prop_trans}, 
\eqref{Phi_matrix}, and 
\eqref{eq:prop_trans_nosum}.

\paragraph{Longitudinal polarization} 
For the longitudinal helicity and 
momentum $q$ as given in Eq.~\eqref{eq:mom_def},
we use the following polarization vector~\cite{Kunszt:1987tk},
which is valid for $q^2\neq M_V^2$
\begin{align}
\varepsilon^\mu (q,\lambda=0)\ 
    {\Big\vert}_{\rm axial}\ &=\ 
\frac{\sqrt{q^2}}{\sqrt{(q\cdot n)^2-q^2 n^2 }}\ 
\left[\frac{ n^2}{(q\cdot n)}q^\mu\ -\ n^\mu\right] \ .
\end{align}
As in the $R_\xi$ gauge [see Eq.~\eqref{eq:polvec_long_alt}],
the longitudinal polarization vector 
in axial gauges carries a forward component 
$[\varepsilon^\mu(\lambda=0) \sim q^\mu]$ 
and a backward (or stationary for $n_{\rm TL}$) 
component $[\varepsilon^\mu(\lambda=0)\sim n^\mu]$, 
relative to $V$'s motion.
The expression here is valid for off-shell photons and gluons,
and consistently vanishes when $q^2\to0$.

The outer product of polarization vectors 
for $\lambda=0$ in this class of gauges is then 
\begin{align}
    \varepsilon_\mu (q,\lambda=0)\varepsilon_\nu (q,\lambda=0) 
    =&
    \frac{q^2 n^2}{(q\cdot n)^2-q^2 n^2 } 
    \left[
    \frac{n^2}{(q\cdot n)^2} q_\mu q_\nu 
    +  
    \frac{n_\mu n_\nu}{n^2}
    - 
    \frac{(q_\mu n_\nu +  q_\nu n_\mu)}{(q\cdot n)}
    \right] 
    \\
    =&
\cfrac{q^2 n^2}{(q\cdot n)^2}
    \Theta_{\mu\nu}\ 
    +
    \cfrac{q^2}{(q\cdot n)^2}\
    n_\mu n_\nu\ .
\end{align}
In terms of our bookkeeping devices,
the longitudinal helicity propagator in axial gauges is 
\begin{align}
\label{eq:prop_long_axial}
\boxed{
    \Pi_{\mu\nu}^V(q,\lambda=0)\ 
    {\Big\vert}_{\rm axial}\ =\      
\cfrac{i\left(
\cfrac{q^2 n^2}{(q\cdot n)^2}
    \Theta_{\mu\nu}\ 
    +
    \cfrac{q^2}{(q\cdot n)^2}\
    n_\mu n_\nu
\right)}{q^2 - M_V^2 + iM_V\Gamma_V}\ 
}\ .
\end{align}

\paragraph{Scalar polarization} 
Given the decomposition of Eq.~\eqref{eq:polprop_axial_decomp}
and longitudinal propagator in Eq.~\eqref{eq:prop_long_axial},
the scalar polarization vector in this gauge is simply the null vector,
\begin{align}
\label{eq:polvec_scalar_axial}
    \varepsilon^\mu (q,\lambda=S) &= 0^\mu\ .
\end{align}
Similarly, the 
scalar helicity propagator in the axial gauge is the null tensor
\begin{align}
\label{eq:prop_scalar_axial}  
\boxed{
    \Pi_{\mu\nu}^V(q,\lambda=S)\ 
    {\Big\vert}_{\rm axial}\
    =\   0_{\mu\nu}\ 
}\ .
\end{align}

We note that in 5-dimensional constructions 
of the EW axial gauge~\cite{Kunszt:1987tk,
Accomando:2006mc,Borel:2012by,Chen:2016wkt}, 
Goldstone bosons and gauge bosons are embedded 
into 5-dimensional multiplets.
In those frameworks, the ``scalar component''
of the 5-vector gauge field is the Goldstone boson.
The $n\cdot A=0$ gauge conditions 
forbids polarization vectors of the form
$\varepsilon^\mu (q,\lambda=S)\propto q^\mu$.

\section{Power Counting Polarization Interference}
\label{sec:interference}

\begin{figure}[!t]
    \centering
    \includegraphics[width=\textwidth]{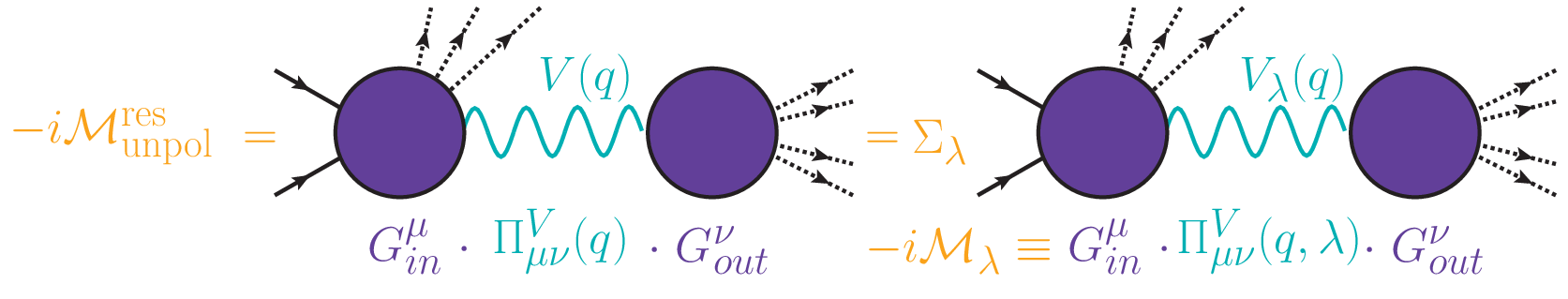}
    \caption{Graphical depiction of
    the matrix element
    for a resonant, unpolarized process
    $\mathcal{M}_{\rm unpol}^{\rm res}$,
    in terms of incoming/outgoing graphs
    $G_{in}^\mu$/$G_{out}^\nu$
    and unpolarized propagator $\Pi^V_{\mu\nu}$,
    and its expansion (right-most expression) in terms of
    polarized matrix elements
    and propagators
    $\mathcal{M}_{\lambda}$ and
    $\Pi^V_{\mu\nu}(\lambda)$.
    }
    \label{fig:wPolar_MatrixElementSum}
\end{figure}

With the expressions for polarized propagators in Sec.~\ref{sec:polvector},
we are in position to organize and estimate, in a generic way, 
the polarization interference $\mathcal{I}_{\rm pol}$, 
as defined in Eq.~\eqref{eq:polint_def}.
The starting point is the observation~\cite{Javurkova:2024bwa}
that helicity polarizations at the level of helicity amplitudes 
can be treated diagrammatically. 
In other words, 
interpret the completeness relationship
of Eq.~\eqref{eq:completeness}
as a sum over interfering diagrams,
where each subamplitude is mediated 
by a weak boson in a fixed helicity.
Doing so puts the polarizations of gauge bosons 
on the same footing as Goldstones and ghosts, in accordance with BRST invariance.

\subsection{Strategy for Power Counting}
\label{sec:interference_strategy}

The first step of our analysis strategy is illustrated 
 in Fig.~\ref{fig:wPolar_MatrixElementSum}.
We start from a collection of subamplitudes 
$\mathcal{M}_{\rm unpol}^{\rm res}$,
as defined in Eq.~\eqref{eq:unpol_def},
that constitute the resonant part 
of a full, gauge-invariant amplitude $\mathcal{M}_{\rm unpol}$.
The unpolarized propagator $\Pi^V_{\mu\nu}(q)$ 
of the intermediate gauge boson $V(q)$
is sandwiched between a collection of 
incoming and outgoing graphs\footnote{For example: 
for $u\overline{d}\to W^+\to \tau^+\nu_\tau$,
as shown in Fig.~\ref{fig:polexpand_dy} of Sec.~\ref{sec:dy},
$G_{in}^\mu$ and $G_{out}^\nu$ each contain one graph,
while in $u\overline{d}\to W^+g\to \tau^+\nu_\tau g$,
as shown in Fig.~\ref{Wjet} in Sec.~\ref{sec:dy},
$G_{in}^\mu$ contains two graphs
and $G_{out}^\nu$ one.} 
(or subamplitudes or Green's functions)
that we collectively 
label as $G_{in}^\mu$ and $G_{out}^\nu$.

From the completeness relationship 
of Eq.~\eqref{eq:completeness}
we generate a collection 
of helicity-polarized amplitudes $\mathcal{M}_\lambda$
in terms of graphs $G_{in}^\mu$ and $G_{out}^\nu$
and the helicity-polarized propagator $\Pi^{V}_{\mu\nu}(q,\lambda)$.
Using the expressions for $\Pi^{V}_{\mu\nu}(q,\lambda)$
given in Sec.~\ref{sec:polvector},
we then build squared polarized amplitudes 
$\vert\mathcal{M}_\lambda\vert^2$
and the 
polarization interference $\mathcal{I}_{\rm pol}$
in terms of our power-counting devices 
($\Theta_{\mu\nu}$ and $\Phi_{\mu\nu}$)
and incoming/outgoing graphs.
Finally, we identify the leading contributions 
to $\mathcal{I}_{\rm pol}$ in generic kinematical limits.

We carry out this analysis first 
in the Unitary gauge in Sec.~\ref{sec:interference_uni},
which is free of Goldstone bosons and Faddeev-Popov ghosts.
We then generalize the analysis 
to the $R_\xi$ gauge in Sec.~\ref{sec:interference_rx}.
We move onto axial gauges in Sec.~\ref{sec:interference_axial}.
To make some cancellations more explicit,
throughout this section we use 
the transverse propagator given in Eq.~\eqref{eq:prop_trans},
with $\Pi^{V}_{\mu\nu}(q,\lambda=T)\propto 
-g_{\mu\nu}-\Theta_{\mu\nu}$, 
instead of the equal but alternative 
expression in Eq.~\eqref{eq:prop_trans_alt}.

\subsection{Unitary Gauge}    
\label{sec:interference_uni}

In terms of incoming/outgoing graphs $G_{in}^\nu$ and $G_{out}^\mu$
the resonant, unpolarized amplitude in the Unitary gauge is given by
\begin{align}
\label{eq:inter_uni_me_unpol}
    -i\mathcal{M}_{\rm unpol}^{\rm res}\ &=\  
    G_{out}^\mu 
    i\left[-g_{\mu\nu} + \frac{q_{\mu}q_{\nu}}{M_V^2-i M_V\Gamma_V}\right]
    D_V^{-1}(q^2)
    G_{in}^\nu\
    \equiv\
    -\mathcal{G} + \frac{\mathcal{Q}}{\tilde{M}_V^2}\ ,
    \\
    \tilde{M}_V\ &=\ \sqrt{M_V^2-i M_V\Gamma_V}\ .
\label{eq:def_mvtilde}
\end{align}
$\mathcal{G}$ and $\mathcal{Q}$ are 
defined as the contractions between external graphs 
with the metric $g_{\mu\nu}$ and tensor $q_\mu q_\nu$,
respectively, multiplied by $i/D_V(q^2)$.
To simplify expressions, 
we adopt the notation of the Complex Mass Scheme $\tilde{M}_V$,
noting that $\tilde{M}_V^2 + (\tilde{M}_V^2)^*= 
2 \Re [\tilde{M}_V^2]= 2M_V^2$.

Similar to the unpolarized amplitude, 
the helicity-polarized amplitudes 
in terms of external graphs and 
our bookkeeping devices are 
\begin{subequations}
\label{eq:inter_uni_me_pol}
\begin{align}
\label{eq:inter_uni_me_pol_trans}
    -i\mathcal{M}_{\lambda=T}\ &=\
    G_{out}^\mu 
    i\left[-g_{\mu\nu} - \Theta_{\mu\nu}\right]
    D_V^{-1}(q^2)
    G_{in}^\nu\
    \equiv\ 
    -\mathcal{G}\ - \vartheta\ ,    
    \\
    -i\mathcal{M}_{\lambda=0}\ &=\ 
    G_{out}^\mu 
    i\left[\Theta_{\mu\nu} + \frac{q_{\mu}q_{\nu}}{q^2}\right]
    D_V^{-1}(q^2)
    G_{in}^\nu\
    \equiv\ +\vartheta\ + \frac{\mathcal{Q}}{q^2}\ ,
    \\
    -i\mathcal{M}_{\lambda=S}\ &=\ 
    G_{out}^\mu  
    i\left[\left(\frac{q_{\mu}q_{\nu}}{M_V^2-i M_V\Gamma_V}- 
    \frac{q_{\mu}q_{\nu}}{q^2}\right)
    \right]
    D_V^{-1}(q^2)
    G_{in}^\nu\
    \equiv\ 
    \frac{\mathcal{Q}}{\tilde{M}_V^2}
    - \frac{\mathcal{Q}}{q^2} \ .
\end{align}
\end{subequations}
Here, $\vartheta$ is the contraction of 
$\Theta_{\mu\nu}$ with $G_{in}^\nu$ and $G_{out}^\mu$,
scaled by the pole $i/D_V(q^2)$.
The sign factors $\eta_\lambda$ 
are included via the definitions 
of the polarized propagators.
It is easy to check that the sum of polarized 
amplitudes recovers the unpolarized case.
We focus first on the $\lambda=T$ polarization
and treat individual $\lambda=\pm1$ transverse helicities
in Eq.~\eqref{eq:inter_uni_me_pol_trans_by_helicity}.

At the squared level, 
the unpolarized and polarized contributions
are
\begin{subequations}
\label{eq:inter_uni_me2}
\begin{align}
\label{eq:inter_uni_me2_unpol}
    \vert\mathcal{M}_{\rm unpol}^{\rm res}\vert^2 &=
    \vert\mathcal{G}\vert^2 + 
    \frac{1}{\vert\tilde{M}_V^2\vert^2}
    \vert\mathcal{Q}\vert^2 
    - 2\Re\left[\frac{\mathcal{G}^*\mathcal{Q}}{\tilde{M}_V^2}\right]\ 
    ,\
    \frac{1}{\vert \tilde{M}_V^2\vert^2} = 
    \frac{1}{M_V^4 + (M_V\Gamma_V)^2}\ ,
    \\
\label{eq:inter_uni_me2_trans}     
    \vert\mathcal{M}_{\lambda=T}\vert^2 &= 
    {\color{violet}\vert\mathcal{G}\vert^2} + 
    \vert\vartheta\vert^2 + 
    2\Re[\mathcal{G}^*\vartheta]\ ,
    \\
\label{eq:inter_uni_me2_long}
    \vert\mathcal{M}_{\lambda=0}\vert^2 &= \vert\vartheta\vert^2 + 
    \frac{1}{(q^2)^2}\vert\mathcal{Q}\vert^2 + 
    \frac{2}{q^2}\Re[\vartheta^*\mathcal{Q}]\ ,
    \\
\label{eq:inter_uni_me2_scalar}
    \vert\mathcal{M}_{\lambda=S}\vert^2 &= 
    \left[\frac{1}{(q^2)^2}+
    {\color{violet}\frac{1}{\vert\tilde{M}_V^2\vert^2}}
    -\frac{2 M_V^2}{q^2 \vert \tilde{M}_V^2\vert^2}\right]
    \vert\mathcal{Q}\vert^2 \ 
    =\ 
    \frac{\vert D_V(q)\vert^2}
    {(q^2)^2 \vert \tilde{M}_V^2\vert ^2}
    \vert \mathcal{Q}\vert^2\ .
\end{align}
\end{subequations}
We draw attention ({\color{violet}dark highlight}) 
to the $\mathcal{O}(\mathcal{G}^2)$ term in Eq.~\eqref{eq:inter_uni_me2_trans}
 and the $\mathcal{O}(1/\tilde{M}_V^4)$ term in Eq.~\eqref{eq:inter_uni_me2_scalar}.
In the Unitary gauge, these contribute to the \textit{unpolarized}
squared matrix element in Eq.~\eqref{eq:inter_uni_me2_unpol}
 and survive cancellation against other contributions at this level.
None of the terms in Eq.~\eqref{eq:inter_uni_me2_long}
appear in the unpolarized squared matrix element,
which highlights the argument of Refs.~\cite{tHooft:1971qjg,
Becchi:1974md,Becchi:1974xu,Becchi:1975nq,Tyutin:1975qk,Becchi:2014lsa}
that longitudinal polarizations should not be treated in isolation 
but in conjunction but with gauge-fixing scalars.

The difference between the squared unpolarized amplitude 
and the squared polarized amplitudes gives the net polarization interference.
In the Unitary gauge, this is 
\begin{align}
&\mathcal{I}_{\rm pol}\ =\ 
\vert\mathcal{M}_{\rm unpol}^{\rm res}\vert^2\ 
-\
\sum_{\lambda\in\{T,0,S\}} \vert \mathcal{M}_\lambda\vert^2
\\ 
\label{int_unitary}
& = 
-2\Re\left[\frac{\mathcal{G}^*\mathcal{Q}}{\tilde{M}_V^2}\right]
-2\vert\vartheta\vert^2  
-2\Re[\mathcal{G}^*\vartheta]
-\frac{2}{q^2}\Re[\vartheta^*\mathcal{Q}]
+\frac{2M_V^2 \left( q^2- M_V^2 - \Gamma_V^2\right)}{(q^2)^2\vert\tilde{M}_V^2\vert^2}\vert\mathcal{Q}\vert^2 \ .
\end{align}
Direct computation shows that 
the net interference has multiple sources,
\begin{subequations}
\label{eq:inter_uni_me2_interference}
\begin{align}
&\mathcal{I}_{\rm pol}\ =\ 
\sum_{\lambda\neq\lambda'\in\{T,0,S\}}
\mathcal{M}^*_{\lambda}\mathcal{M}_{\lambda'}\
\nonumber\\
&=\ 
2\Re\left[\mathcal{M}^*_{\lambda=T}\mathcal{M}_{\lambda=0}\right]
+\ 
2\Re\left[\mathcal{M}^*_{\lambda=T}\mathcal{M}_{\lambda=S}\right]\
+\
2\Re\left[\mathcal{M}^*_{\lambda=0}\mathcal{M}_{\lambda=S}\right]\
,\ \text{where}
\end{align}
\begin{align}
\label{eq:inter_uni_me2_interference_tlong}
2\Re\left[\mathcal{M}^*_{\lambda=T}\mathcal{M}_{\lambda=0}\right]\
&=
- 2\vert\vartheta\vert^2
-2\Re[\mathcal{G}^*\vartheta]
{\color{teal}-\frac{2}{q^2}\Re[\vartheta^*\mathcal{Q}]
-\frac{2}{q^2}\Re[\mathcal{G}^*\mathcal{Q}]}\ ,
\\
\label{eq:inter_uni_me2_interference_tscalar}
2\Re\left[\mathcal{M}^*_{\lambda=T}\mathcal{M}_{\lambda=S}\right]\
&= 
{\color{violet}-2\Re\left[\frac{\mathcal{G}^*\mathcal{Q}}{\tilde{M}_V^2}\right]}
{\color{teal}+\frac{2}{q^2}\Re[\vartheta^*\mathcal{Q}]
+\frac{2}{q^2}\Re[\mathcal{G}^*\mathcal{Q}]
-2\Re\left[\frac{\vartheta^*\mathcal{Q}}{\tilde{M}_V^2}\right]}\ ,
\\
\label{eq:inter_uni_me2_interference_longscalar}
2\Re\left[\mathcal{M}^*_{\lambda=0}\mathcal{M}_{\lambda=S}\right]\
&=
-\frac{2}{q^2}\Re[\vartheta^*\mathcal{Q}] 
+\frac{2 M_V^2}{q^2 \vert \tilde{M}_V^2\vert^2}\vert\mathcal{Q}\vert^2
- \frac{2}{(q^2)^2} \vert\mathcal{Q}\vert^2
{\color{teal} +2\Re\left[\frac{\vartheta^*\mathcal{Q}}{\tilde{M}_V^2}\right]}\ .
\end{align}
\end{subequations}

Several features in the polarization interference 
of Eq.~\eqref{int_unitary} are worth noting:

(i) The $\mathcal{O}(\mathcal{G}^*\mathcal{Q}/\tilde{M}_V^2)$ term 
in the net  interference $\mathcal{I}_{\rm pol}$
also appears in the \textit{unpolarized}
squared matrix element in Eq.~\eqref{eq:inter_uni_me2_unpol}.
Here, it originates from the interference between scalar and transverse
polarizations ({\color{violet}dark highlight}) in 
Eq.~\eqref{eq:inter_uni_me2_interference_tscalar}.
This is easy to see in our organization since $\mathcal{G}$ terms appear only
in $\mathcal{M}_{\lambda=T}$ and $1/\tilde{M}_V^2$ factors  appear
only in $\mathcal{M}_{\lambda=S}$.
In this sense, scalar-induced interference
contributes to physical cross sections in the Unitary gauge.
However, we caution that interpreting individual interference terms 
is not a well-defined practice.
Regardless of origin, the $\mathcal{O}(\mathcal{G}^*\mathcal{Q}/\tilde{M}_V^2)$ 
term in the net interference prevents 
the sum of measured polarization fractions 
$f_\lambda = \sigma_\lambda/\sigma_{\rm unpol}$
from ever adding to unity. 
In practice, it is possible to suppress
such interference by suppressing $\mathcal{Q}$.

(ii) There are many exact cancellations among the different
polarization combinations ({\color{teal}light highlight}).
This follows from exact cancellations at the matrix-element level.

(iii) Due to $\vartheta$ and $\mathcal{Q}$ terms, 
the net interference does not vanish in the (near) on-shell limit.
In fact, the last term in Eq.~\eqref{int_unitary} 
only vanishes at $q^2=M_V^2+\Gamma_V^2$, 
and generates constructive (destructive)
interference when $q^2$ is larger (smaller) than $M_V^2+\Gamma_V^2$.
As $\mathcal{Q}$ terms scale quadratically with the momentum of $V$,
justifying the narrow width and pole approximations 
in polarized predictions is difficult 
in the absence of additional assumptions.

(iv) Importantly, all terms appearing in Eq.~\eqref{int_unitary}
are either proportional to $\mathcal{Q}$, 
which are generated by $q_{\mu}q_{\nu}$ terms 
in longitudinal and scalar propagators,
or proportional to $\vartheta$.
For real-life processes at the LHC, 
$\mathcal{Q}$ is naturally suppressed when $V(q)$ couples 
to massless fermions and other conserved currents.
($q\cdot G_{in/out}=0$ is essentially the definition of a conserved current.)
$\vartheta$ can also be suppressed in certain kinematical limits.
However, this will also impact pure longitudinal contributions, 
which scale as $\mathcal{O}(\vert\vartheta\vert^2)$.

For completeness, we note that the net interference 
for $t$-channel exchanges is 
\begin{align}
\label{eq:int_unitary_tchannel}
\mathcal{I}_{\rm pol}^{t-ch.} \overset{\Gamma_V\to0}{=} 
-\frac{2}{M_V^2}\Re\left[\mathcal{G}^*\mathcal{Q}\right]
-2\vert\vartheta\vert^2  
-2\Re[\mathcal{G}^*\vartheta]
-\frac{2}{q^2}\Re[\vartheta^*\mathcal{Q}]
+\frac{2\left(q^2- M_V^2\right)}{(q^2)^2 M_V^2}\vert\mathcal{Q}\vert^2  .
\end{align}
And in the absence of $\mathcal{Q}$ contributions, 
the net polarization interference collapses to
\begin{align}
\label{eq:int_unitary_noq}
\mathcal{I}_{\rm pol}^{\rm no-\mathcal{Q}}\ \overset{\mathcal{Q}\to0}{=}&\ 
-2\vert\vartheta\vert^2  
-2\Re[\mathcal{G}^*\vartheta]\ 
=\ 
-2\Re[(\mathcal{G}+\vartheta)^*\vartheta] \ 
=\ 
2\Re[\varphi^*\vartheta] \ ,
\\
\label{eq:inter_uni_me_pol_trans_alt}
\varphi\ \equiv\ & 
    G_{out}^\mu\  
    \Phi_{\mu\nu}\
    G_{in}^\nu\
    iD_V^{-1}(q^2) 
    =\ 
-i\mathcal{M}_{\lambda=T}\ ,
\end{align}
where $\varphi$ is the transverse polarization 
amplitude in Eq.~\eqref{eq:inter_uni_me_pol_trans}
in terms of $\Phi_{\mu\nu}(\theta_V,\phi_V)$
in Eq.~\eqref{Phi_matrix}.
To obtain the rightmost equality 
in Eq.~\eqref{eq:int_unitary_noq}
we rewrote $\mathcal{G}$ 
as $\mathcal{G}=(\mathcal{G}+\vartheta)-\vartheta=-\varphi-\vartheta$,
which makes transparent that $\mathcal{I}_{\rm pol}$
consists entirely of transverse-longitudinal 
interference in the $\mathcal{Q}\to0$ limit.
We return to this points in Sec.~\ref{sec:noninterference}.

Extending polarization interference 
to the RH $(\lambda=+1)$ and LH $(\lambda=-1)$
helicity polarizations
is a minor complication.
In terms of external graphs,
the amplitudes are 
\begin{subequations}
\label{eq:inter_uni_me_pol_trans_by_helicity}
\begin{align}
    -i\mathcal{M}_{\lambda=+1}\ &=\
    G_{out}^\mu 
    \frac{i}{2}\left[-g_{\mu\nu} - \Theta_{\mu\nu}
    +\cfrac{\epsilon_{\mu\nu\alpha\beta}\ q^\alpha n^\beta}{\sqrt{(n\cdot q)^2 - q^2 n^2}}
    \right]
    D_V^{-1}(q^2)
    G_{in}^\nu\
    \nonumber\\
    &\equiv\ 
    -\frac{\mathcal{G}}{2}\ -\ \frac{\vartheta}{2}\ +\ \frac{\mathcal{E}}{2}\  
    =\ \frac{\varphi}{2} +\frac{\mathcal{E}}{2},
    \\
    -i\mathcal{M}_{\lambda=-1}\ &=\
    G_{out}^\mu 
    \frac{i}{2}\left[-g_{\mu\nu} - \Theta_{\mu\nu}
    -\cfrac{\epsilon_{\mu\nu\alpha\beta}\ q^\alpha n^\beta}{\sqrt{(n\cdot q)^2 - q^2 n^2}}
    \right]
    D_V^{-1}(q^2)
    G_{in}^\nu\
    \nonumber\\
    &\equiv\ 
    -\frac{\mathcal{G}}{2}\ -\ \frac{\vartheta}{2}\ -\ \frac{\mathcal{E}}{2}\ 
     =\ \frac{\varphi}{2} -\frac{\mathcal{E}}{2}.
\end{align}
\end{subequations}
$\mathcal{E}$ encapsulates the antisymmetric tensor, 
sandwiched by the incoming and outgoing graphs,
scaled by $i/D_V(q)^2$. 
At the squared level, 
one generates the polarized contributions
\begin{subequations}
\begin{align}
\label{eq:inter_uni_me2_trans_rh}     
    \vert\mathcal{M}_{\lambda=+1}\vert^2 &= 
    \frac{\vert\mathcal{G}\vert^2}{4} + 
    \frac{\vert\vartheta\vert^2}{4} + 
    \frac{\vert\mathcal{E}\vert^2}{4} + 
    \frac{1}{2}\Re[\mathcal{G}^*\vartheta]\ -
    \frac{1}{2}\Re[\mathcal{E}^*\vartheta]\ -
    \frac{1}{2}\Re[\mathcal{G}^*\mathcal{E}]\ 
    \nonumber\\
    &=\ 
    \frac{\vert\varphi\vert^2}{4} + 
    \frac{\vert\mathcal{E}\vert^2}{4} +
    \frac{1}{2}\Re[\varphi^*\mathcal{E}]\ ,
\\
\label{eq:inter_uni_me2_trans_lh}     
    \vert\mathcal{M}_{\lambda=-1}\vert^2 &= 
    \frac{\vert\mathcal{G}\vert^2}{4} + 
    \frac{\vert\vartheta\vert^2}{4} + 
    \frac{\vert\mathcal{E}\vert^2}{4} + 
    \frac{1}{2}\Re[\mathcal{G}^*\vartheta]\ +
    \frac{1}{2}\Re[\mathcal{E}^*\vartheta]\ +
    \frac{1}{2}\Re[\mathcal{G}^*\mathcal{E}]\ 
    \nonumber\\
    &=\ 
    \frac{\vert\varphi\vert^2}{4} + 
    \frac{\vert\mathcal{E}\vert^2}{4} - 
    \frac{1}{2}\Re[\varphi^*\mathcal{E}]\ ,
\\
2\Re\left[\mathcal{M}^*_{\lambda=+1}\mathcal{M}_{\lambda=-1}\right]\    
&=\
\frac{\vert\mathcal{G}\vert^2}{2} + 
    \frac{\vert\vartheta\vert^2}{2} - 
    \frac{\vert\mathcal{E}\vert^2}{2} + 
    \Re[\mathcal{G}^*\vartheta]\ 
    =\ \frac{|\varphi|^2}{2}-\frac{|\mathcal{E}|^2}{2}.
\label{eq:inter_uni_me2_trans_trans}
\end{align}
\end{subequations}

For the interference generated between 
$\lambda=\pm1$ 
and a different helicity $\lambda'=0,S$,
each contribution
$\mathcal{M}(\lambda=\pm1)\mathcal{M}^*(\lambda')$  
and its conjugate
will generate terms that scale linearly with $\pm\mathcal{E}$, 
and therefore cancel in the net polarization interference.
Explicit computation of the net interference 
when RH and LH helicities 
are treated separately gives
\begin{align}
\mathcal{I}_{\rm pol}\ &=\ 
\vert\mathcal{M}_{\rm unpol}^{\rm res}\vert^2\ 
-
\sum_{\lambda\in\{\pm1,0,S\}} \vert \mathcal{M}_\lambda\vert^2
    \\
& =\ \frac{\vert\mathcal{G}\vert^2}{2}\ 
-\frac{3\vert\vartheta\vert^2}{2}\  
-\frac{\vert\mathcal{E}\vert^2}{2}\  
-\ \Re[\mathcal{G}^*\vartheta]\  
-\ 2\Re\left[\frac{\mathcal{G}^*\mathcal{Q}}{\tilde{M}_V^2}\right]\ 
-\frac{2}{q^2}\Re[\vartheta^*\mathcal{Q}]\     
\nonumber\\
&+\ 
\frac{2 M_V^2 (q^2 - M_V^2 - \Gamma_V^2)}{(q^2)^2 \vert \tilde{M}_V^2\vert^2}
 \vert\mathcal{Q}\vert^2 \  \\
 & =\ \frac{|\varphi|^2}{2} -\frac{|\mathcal{E}|^2}{2} + 2\ {\rm Re} [\varphi^* \vartheta] + 2\ {\rm Re}\left[\frac{\vartheta^* 
 \mathcal{Q}}{\tilde{M}_V^2}\right] + 2\ {\rm Re}\left[\frac{\varphi^*\mathcal{Q}}{\tilde{M}_V^2}\right] -\frac{2}{q^2}\ {\rm Re}[\vartheta^*\mathcal{Q}] \nonumber\\
 &+\ 
\frac{2 M_V^2 (q^2 - M_V^2 - \Gamma_V^2)}{(q^2)^2 \vert \tilde{M}_V^2\vert^2}
 \vert\mathcal{Q}\vert^2 \ .
\label{eq:int_unitary_rh_lh}
\end{align}
The difference between this expression and 
Eq.~\eqref{int_unitary} is that the transverse-transverse 
interference in Eq.~\eqref{eq:inter_uni_me2_trans_trans}
has been moved from the squared transverse contribution 
$\vert\mathcal{M}_{\lambda=T}\vert^2$ to the net interference.
Adding Eq.~\eqref{eq:inter_uni_me2_trans_trans} to 
Eq.~\eqref{int_unitary} gives 
Eq.~\eqref{eq:int_unitary_rh_lh} above.

In the absence of $\mathcal{Q}$ terms, 
one still has a simpler expression
for polarization interference
\begin{align}
\mathcal{I}_{\rm pol}^{\rm no-\mathcal{Q}}\ \overset{\mathcal{Q}\to0}{=}\ & 
\frac{\vert\mathcal{G}\vert^2}{2}\ 
-\frac{3\vert\vartheta\vert^2}{2}\  
-\frac{\vert\mathcal{E}\vert^2}{2}\  
-\ \Re[\mathcal{G}^*\vartheta]\  
=\ \ \frac{|\varphi|^2}{2} -\frac{|\mathcal{E}|^2}{2} + 2\ {\rm Re} [\varphi^* \vartheta] .
\end{align}
The expression suggests a higher likelihood 
of interference vanishing through 
accidental kinematical configurations 
than it vanishing structurally.
Such investigations are outside our present scope
and individual transverse 
polarizations will not be considered further.

\subsection{The \texorpdfstring{$R_\xi$}{Rx} Gauge}    
\label{sec:interference_rx}

Building the polarization interference $\mathcal{I}_{\rm pol}$
in the $R_\xi$ gauge follows the same procedure as 
in the Unitary gauge in Sec.~\ref{sec:interference_uni},
but with the added complication of Goldstone amplitudes 
and $\xi$ dependency.
In the $R_\xi$ gauge, the spurious pole 
in the longitudinal propagator now cancels against the combination 
of the scalar polarization and Goldstone boson,
while $\xi$-factors in the scalar propagator 
cancel against the Goldstone boson.
These cancellations are structural\footnote{See also 
Refs.~\cite{Bohm:2001yx,Dittmaier:2025htf} 
for contemporary constructions of these relationships.} 
and follow from EW Ward/Slavnov-Taylor identities~\cite{tHooft:1971qjg,
Fujikawa:1972fe,Bohm:2001yx}.

In light of these relationships and in accordance with BRST invariance,
we argue Goldstones bosons should be treated effectively 
as ``extra'' polarizations\footnote{In 
5-dimensional constructions of the EW axial gauge~\cite{Kunszt:1987tk,
Accomando:2006mc,Borel:2012by,Chen:2016wkt}, 
cancellations are realized naturally by embedding the 
Goldstone and gauge field into the same multiplet.}.
Guided by this organization,
we write the unpolarized, Goldstone, and polarized 
amplitudes in the $R_\xi$ gauge
in terms of incoming/outgoing graphs $G_{in}^\nu$/$G_{out}^\mu$
and our bookkeeping devices as 
\begin{subequations}
\label{eq:inter_rx_matrix}
\begin{align}
    -i\mathcal{M}_{\rm unpol}^{\rm res} =&\ G_{out}^\mu\ i 
    \left[-g_{\mu\nu}
    -\frac{(\xi-1)q_\mu q_\nu}{D_V(q^2,\xi)}
    \right]
    D_V^{-1}(q^2)
    G_{in}^\nu\ 
    \equiv\ -\mathcal{G} - \mathcal{Q}_\xi
    \\ 
   -i\mathcal{M}_{\rm Gold} =&\ G_{out}\ i 
   \left[D_V^{-1}(q^2,\xi)\right]\ 
   G_{in}\
   \equiv\ 
   \Xi\ ,
    \\
    \label{eq:inter_rx_matrix_total}
    -i\mathcal{M}_{\rm total}^{\rm res} =&\  
    -i\mathcal{M}_{\rm unpol}^{\rm res}-i\mathcal{M}_{\rm Gold}\
    =\ -\mathcal{G}-\mathcal{Q_\xi}+\Xi \ ,
\end{align}
\begin{align}
    -i\mathcal{M}_{\lambda=T} =&\   G_{out}^\mu\ i 
    \left[-g_{\mu\nu}-\Theta_{\mu\nu}\right]
    D_V^{-1}(q^2)
    G_{in}^\nu\
    \equiv\ 
    -\mathcal{G} - \vartheta\ ,
    \\
    -i\mathcal{M}_{\lambda=0}=&\ 
    G_{out}^\mu\ i
    \left[\Theta_{\mu\nu}+\frac{q_\mu q_\nu}{q^2}\right]
    D_V^{-1}(q^2)
    G_{in}^\nu\
    \equiv \vartheta+\frac{\mathcal{Q}}{q^2}\ ,
    \\
    -i\mathcal{M}_{\lambda=S}=&\ G_{out}^\mu\ i 
    \left[
    -\frac{q_\mu q_\nu}{q^2}
    -\frac{(\xi-1)q_\mu q_\nu}{D_V(q^2,\xi)}
    \right]
    D_V^{-1}(q^2)
    G_{in}^\nu\ 
    \equiv\ -\frac{\mathcal{Q}}{q^2}-{\mathcal{Q}_\xi}\ .
\end{align}
\end{subequations}

As in Eq.~\eqref{eq:inter_uni_me_pol} for the Unitary gauge,
the right-most expressions define the various  
contractions between incoming/outgoing graphs,
pole factors $i/D_V(q^2)$ and $i/D_V(q^2,\xi)$, 
and (un)polarized propagators.
For example: $\Xi$ is the product of the 
incoming/outgoing scalar-valued currents $G_{in/out}$
along with the Goldstone propagator $i/D_V(q^2,\xi)$.
$Q_\xi$ is related to $\mathcal{Q}$ by 
$Q_\xi = (\xi-1)\mathcal{Q} D_V^{-1}(q^2,\xi)$.
In the limit $\xi\to\infty$, one has $\Xi\to0$
with $\mathcal{Q}_\xi\to -\mathcal{Q}/\tilde{M}_V^2$,
and subsequently recovers the expressions 
for the Unitary gauge.

In Eq.~\eqref{eq:inter_rx_matrix_total}, 
we define the total resonant matrix element 
$\mathcal{M}_{\rm total}^{\rm res}$
as the sum of the unpolarized resonant amplitude 
$\mathcal{M}_{\rm unpol}^{\rm res}$
and 
the associated Goldstone amplitude
$\mathcal{M}_{\rm Gold}$.
In the absence of non-resonant diagrams
$(\mathcal{M}_{\rm non-res})$, 
$\mathcal{M}_{\rm total}^{\rm res}$ is gauge invariant.
It is easy to check that the sum of polarized
amplitudes recovers the unpolarized case.

As Faddeev-Popov ghosts do not contribute at tree level in the $R_\xi$ gauge,
there is no separate amplitude for ghosts in Eq.~\eqref{eq:inter_rx_matrix}.
Instead, each polarized and unpolarized contribution undergo the same 
radiative corrections, e.g., ghost and Goldstone loops, beyond tree level.
In Sec.~\ref{sec:interference_independence} 
we discuss how ghosts can be incorporated in non-standard gauges.

For simpler processes, 
e.g., exchanges of a single boson,
our bookkeeping allows us identify a correspondences 
between the scalar and Goldstone contributions
across different gauges.
This is because external currents $G_{in/out}^\mu$
are related to the scalar-valued currents $G_{in/out}$
by EW Ward/Slavnov-Taylor identities~\cite{Bohm:2001yx,Dittmaier:2025htf}.
Assuming the following (Ward/Slavnov-Taylor) relationship 
between external scalar- and vector-valued currents
\begin{align}
\label{eq:ward}
G_{in/out} = \eta_{in/out}\ \frac{q_\mu}{\tilde{M}_V} G_{in/out}^\mu\ ,  
\quad\text{with}\quad \eta_{in}\eta_{out}=1,
\end{align}
where $\eta_{in/out}$ are phase factors~\cite{Dittmaier:2025htf},
then we can write the Goldstone amplitude as 
\begin{align}
\label{eq:inter_rx_matrix_scalar_alt}
\Xi\ =&\ G_{out}^\mu\ i
\left[\cfrac{q_\mu q_\nu}{\tilde{M}_V^2\ D_V(q^2,\xi)}\right]\ 
G_{in}^\nu\
\\
=&\ 
G_{out}^\mu\ i
\left[\cfrac{q_\mu q_\nu}{\tilde{M}_V^2} + 
\cfrac{(\xi-1)q_\mu q_\nu}{D_V(q^2,\xi)}\right]\ 
D_V^{-1}(q^2)
G_{in}^\nu\
=\ \frac{\mathcal{Q}}{\tilde{M}_V^2}\ +\ \mathcal{Q}_\xi\ .
\end{align}
Combining this with Eq.~\eqref{eq:inter_rx_matrix}
and comparing with Eq.~\eqref{eq:inter_uni_me_pol}, 
we obtain the following correspondences for 
the polarized amplitudes in the $R_\xi$ and Unitary gauges:
\begin{subequations}
\label{eq:ward_gauge_inv}
\begin{align}
\label{eq:ward_gauge_inv_unpol}
\mathcal{M}_{\rm unpol}^{\rm res}\big\vert_{R_\xi}
+\mathcal{M}_{\rm Gold}\big\vert_{R_\xi} =&\ 
i\left(-\mathcal{G} - \mathcal{Q}_\xi\right) 
+ 
i\left(\frac{\mathcal{Q}}{\tilde{M}_V^2}\ +\ \mathcal{Q}_\xi\ \right)
= 
\mathcal{M}_{\rm unpol}^{\rm res}\big\vert_{\rm Unitary}\ ,
\\
\label{eq:ward_gauge_inv_scalar}
\mathcal{M}_{\lambda=S}\big\vert_{R_\xi}
+\mathcal{M}_{\rm Gold}\big\vert_{R_\xi} =&\ 
i\left(-\frac{\mathcal{Q}}{q^2}-{\mathcal{Q}_\xi}\right)\
+
i\left(\frac{\mathcal{Q}}{\tilde{M}_V^2}\ +\ \mathcal{Q}_\xi\ \right)
= 
\mathcal{M}_{\lambda=S}\big\vert_{\rm Unitary}\ .
\end{align}    
\end{subequations}

A proliferation of non-resonant contributions 
and $\xi$ factors complicate similar identifications 
for multiboson processes.
However, a rigorous, all-orders proof
is not needed to see 
that the $\xi$ -independence of $\mathcal{M}^{\rm res}_{\rm total}$
is tied to the cancellation between scalar and Goldstone contributions.
In the $R_\xi$ gauge, $\xi$ factors only appears in the 
propagators of scalar polarizations, Goldstone bosons, 
and Faddeev-Popov ghosts, 
as well as Goldstone-ghost couplings,
implying relationships among these contributions.
For example: Eq.~\eqref{eq:ward} implies that the 
scalar-valued graphs $G_{in/out}$ are zero (or absent)
when the corresponding current is conserved, 
i.e., $G_{in/out}^\mu\cdot q_\mu =0$.
Conversely, the absence of scalar polarization and Goldstones diagrams
indicates an absence of spurious poles in longitudinal amplitudes.

At the squared level, the unpolarized, Goldstone, 
and polarized contributions are
\begin{subequations}
\begin{align}
\label{eq:inter_rxi_me2_unpol}
|\mathcal{M}_{\rm total}^{\rm res}|^2 =&\ 
|\mathcal{G}|^2+|\mathcal{Q}_\xi|^2+|\Xi|^2+ 
2\ \Re[\mathcal{G}^* \mathcal{Q}_\xi]-  
2\ \Re[\mathcal{Q}^*_\xi \Xi]- 2\ \Re[\mathcal{G}^* \Xi]\ ,
\\
|\mathcal{M}_{\rm Gold}|^2 =&\ |\Xi|^2\ , 
\\
|\mathcal{M}_{\lambda=T}|^2=&\ |\mathcal{G}|^2+|\vartheta|^2 + 2 \ 
\Re [\mathcal{G}^* \vartheta]\ , 
\\
|\mathcal{M}_{\lambda=0}|^2=&\ |\vartheta|^2+\frac{1}{(q^2)^2}|\mathcal{Q}|^2+ \frac{2}{q^2}\ \Re[\vartheta^* \mathcal{Q}]\ ,
\\
|\mathcal{M}_{\lambda=S}|^2 =& \ \frac{1}{(q^2)^2} |\mathcal{Q}|^2 +|\mathcal{Q}_\xi|^2+ \frac{2}{q^2} \Re[\mathcal{Q}^* \mathcal{Q}_\xi]\ .
\end{align}
\end{subequations}
In comparison to the Unitary gauge, 
two differences appear:  
(i) the presence of the Goldstone amplitude
and (ii) the presence of $\mathcal{Q}_\xi$.
The latter term prevents some simplification
but makes clearer the scalar polarization's
contribution to the total, unpolarized process.

The net polarization interference,
including Goldstone contributions, 
is the difference between 
the squared total amplitude 
and the squared polarized amplitudes
\begin{align}
\mathcal{I}_{\rm pol}^{R_\xi}\ =&\ 
\vert\mathcal{M}_{\rm total}^{\rm res}\vert^2\ 
-\
\sum_{\lambda\in\{T,0,S,\phi\}} \vert \mathcal{M}_\lambda\vert^2\\
    =&\ 
    +2\Re[\mathcal{G}^* \mathcal{Q}_\xi] 
    -2|\vartheta|^2 
    -2\Re[\mathcal{G}^*\vartheta]
    -\frac{2}{q^2}\Re[\vartheta^* \mathcal{Q}]
    -\frac{2}{q^4}  |\mathcal{Q}|^2
    -\frac{2}{q^2}\Re[\mathcal{Q^*\mathcal{Q}_\xi}]
    \nonumber \\
    &\ 
    - 2\ \Re[\mathcal{Q}_\xi^* \Xi] 
    - 2\ \Re[\mathcal{G}^* \Xi] \ .
\label{eq:int_rx}
\end{align}
In the $\xi\to\infty$ limit, 
the $\mathcal{O}(\Xi)$ terms vanish  
and the first line maps 
onto the expression in Eq.~\eqref{int_unitary}
for the Unitary gauge.
Again, this equality is clearer with the identification 
$\mathcal{Q}_\xi \to\mathcal{Q}/\tilde{M}_V^2$ when $\xi\to\infty$.
Term-by-term, the contributions to the net interference are
\begin{subequations}
\label{eq:polint_term_by_term_rxi}
\begin{align}
2 \Re[\mathcal{M}_{\lambda=T}^* \mathcal{M}_{\lambda=0}] =&\ 
- 2 |\vartheta|^2 
- 2\Re[\mathcal{G}^*\vartheta]
- \frac{2}{q^2}\Re[\vartheta^* \mathcal{Q}]
{\color{teal}- \frac{2}{q^2}\Re[\mathcal{G}^* \mathcal{Q}]} \ ,
    \\
2 \Re[\mathcal{M}_{\lambda=T}^* \mathcal{M}_{\lambda=S}] =&\ 
{\color{violet}2 \Re[\mathcal{G}^*\mathcal{Q}_\xi]}
{\color{teal}+2 \Re[\vartheta^* \mathcal{Q}_\xi] 
+\frac{2}{q^2}\Re[\vartheta^* \mathcal{Q}]
+\frac{2}{q^2} \Re[\mathcal{G}^*\mathcal{Q}]}\ ,
\\
2 \Re[\mathcal{M}_{\lambda=0}^* \mathcal{M}_{\lambda=S}] =&\ 
- \frac{2}{(q^2)^2}|\mathcal{Q}|^2
- \frac{2}{q^2}\Re[\mathcal{Q}^*\mathcal{Q}_\xi]
{\color{teal}-2 \Re[\vartheta^*\mathcal{Q}_\xi]
- \frac{2}{q^2}\Re[\vartheta^*\mathcal{Q}]} \ ,
\\
2 \Re[\mathcal{M}_{\lambda=T}^* \mathcal{M}_{\rm Gold}] =&\ 
{\color{violet}- 2 \Re[\mathcal{G}^* \Xi]}
{\color{teal}- 2 \Re[\vartheta^* \Xi]}\ ,
\\
2 \Re[\mathcal{M}_{\lambda=0}^* \mathcal{M}_{\rm Gold}] =&\ 
{\color{teal}2 \Re[\vartheta^* \Xi]
+\frac{2}{q^2} \Re[\mathcal{Q}^*\Xi]} \ ,
\\
2 \Re[\mathcal{M}_{\lambda=S}^* \mathcal{M}_{\rm Gold}] =&\ 
{\color{violet}-2 \Re[\mathcal{Q}_\xi^*\Xi]}
{\color{teal}- \frac{2}{q^2}\ \Re[\mathcal{Q}^*\Xi]}\ .
\end{align} 
\end{subequations}

Among all the terms in Eq.~\eqref{eq:polint_term_by_term_rxi}, 
we draw attention to each first term in the
transverse-scalar, 
transverse-Goldstone,
and scalar-Goldstone interference  
({\color{violet}dark highlight}).
These appear in both the net polarization interference 
and the total, \textit{unpolarized} matrix element 
at the squared level,
given in Eq.~\eqref{eq:inter_rxi_me2_unpol}.
As in the Unitary gauge, the presence of these terms 
prevents the sum of measured polarization fractions 
$f_\lambda = \sigma_\lambda/\sigma_{\rm unpol}$
from ever adding to unity.
In practice, $\mathcal{Q}$ and $\Xi$
are suppressed if massless external states are involved.
When summing over Eq.~\eqref{eq:polint_term_by_term_rxi},
about 10 of the individual terms 
({\color{teal}light highlight}) cancel, 
including the entire interference 
between longitudinal and Goldstone amplitudes.

\subsection{Axial Gauges}    
\label{sec:interference_axial}

Constructing polarization interference 
in the axial gauge follows a similar path as above.
For simplicity, we neglect contributions from Goldstone bosons.
In this case, the resonant, unpolarized matrix elements
and the squared polarized matrix elements are 
\begin{subequations}
\begin{align}
 -i\mathcal{M}_{\rm unpol}^{\rm res}\ &=\  
    G_{out}^\mu 
    i\left[ -g_{\mu\nu} 
    - 
    \left[
    \frac{(q\cdot n)^2 - q^2\ n^2}{(q\cdot n)^2}\right]
    \Theta_{\mu\nu}
    +\frac{q^2}{(q\cdot n)^2}
    n_\mu n_\nu
    \right]
    D_V^{-1}(q^2)
    G_{in}^\nu
    \nonumber\\
    &
    \equiv\
    -\mathcal{G}\ 
    -\ 
    \left[
    \frac{(q\cdot n)^2 - q^2\ n^2}{(q\cdot n)^2}\right]
    \vartheta\
    +\
    \frac{q^2}{(q\cdot n)^2}\mathcal{N}\ ,    
    \\ 
    -i\mathcal{M}_{\lambda=T}\ &=\
    G_{out}^\mu\ 
    i\left[-g_{\mu\nu} - \Theta_{\mu\nu}\right]
    D_V^{-1}(q^2)\
    G_{in}^\nu\
    \nonumber\\
    &\equiv\ -\mathcal{G}\ -\ \vartheta\
    \\
    -i\mathcal{M}_{\lambda=0}\ &=\ 
    G_{out}^\mu\ 
    i\left[
    \frac{q^2\ n^2}{(q\cdot n)^2}
    \Theta_{\mu\nu}\ 
    +
    \frac{q^2}{(q\cdot n)^2}\
    n_\mu n_\nu
    \right]
    D_V^{-1}(q^2)\
    G_{in}^\nu\
    \nonumber\\
    &\equiv\ 
    \frac{q^2\ n^2}{(q\cdot n)^2}\vartheta\ 
    +\ 
    \frac{q^2}{(q\cdot n)^2}\ \mathcal{N}
    \\
    -i\mathcal{M}_{\lambda=S}\ &=\ 0\ .
\end{align}
\end{subequations}

Here, we introduce $\mathcal{N}$, 
which encapsulates the $\mathcal{O}(n_\mu n_\nu)$ 
reference-vector tensor.
There is no amplitude for the scalar polarization 
as the polarization vector vanishes in this gauge.
Since the transverse helicity propagator $\Pi^V_{\mu\nu}(q,\lambda=T)$ 
in this gauge is the same as in Unitary and $R_\xi$ gauges, 
the transverse matrix element $\mathcal{M}_{\lambda=T}$ 
is the same, up to possible differences 
in the incoming/outgoing graphs 
due to differences in Feynman rules.

The squared unpolarized and helicity-polarized matrix elements are given by
\begin{subequations}
\begin{align}
    \vert\mathcal{M}_{\rm unpol}\vert^2 &=
    \vert\mathcal{G}\vert^2\
    +\ 
    \left[
    \frac{(q\cdot n)^2 - q^2\ n^2}{(q\cdot n)^2}\right]^2
    \vert\vartheta\vert^2\
    +\
    \left[
    \frac{2[(q\cdot n)^2 - q^2\ n^2]}{(q\cdot n)^2}\right]
    \Re[\mathcal{G}^*\vartheta]\
    \nonumber\\
    &
    +
    \frac{(q^2)^2}{(q\cdot n)^4}\vert\mathcal{N}\vert^2 
    - 
    \frac{2q^2}{(q\cdot n)^2}\Re[\mathcal{G}^*\mathcal{N}]
    -
    \left[
    \frac{2q^2[(q\cdot n)^2 - q^2\ n^2]}{(q\cdot n)^4}\right]
    \Re[\vartheta^*\mathcal{N}]\ ,
    \\
    \vert\mathcal{M}_{\lambda=T}\vert^2 &= 
    \vert\mathcal{G}\vert^2\ 
    +\ 
    \vert\vartheta\vert^2\ 
    +\ 
    2\Re[\mathcal{G}^*\vartheta]\ ,
    \\
    \vert\mathcal{M}_{\lambda=0}\vert^2 &= 
    \frac{(q^2)^2}{(q\cdot n)^4}\vert\mathcal{N}\vert^2\
    +\ 
    \frac{(q^2)^2\ (n^2)^2}{(q\cdot n)^4}
    \vert\vartheta\vert^2
    +\
    2\frac{(q^2)^2\ n^2}{(q\cdot n)^4}
    \Re[\vartheta^*\mathcal{N}]\ ,
    \\
    \vert\mathcal{M}_{\lambda=S}\vert^2 &= 0 \ .
\end{align}
\end{subequations}
The more complicated squared matrix elements reflect 
the more complicated structure of propagators in this gauge.
However, judicious choices of $n^\mu$ can simplify expressions.
For example: with the light-light reference vector $n_{\rm LL}^2=0$ 
many prefactors above reduce to unity or vanish altogether.
We draw particular attention to $\mathcal{O}[q^4/(q\cdot n)^4]\sim \mathcal{O}(q^4/E_V^4)$ 
terms, which become highly suppressed in high-energy limits.

With the absence of a scalar amplitude, 
the net polarization interference reduces to a single source:
transverse-longitudinal interference.
Direct computation shows 
\begin{align}
\mathcal{I}_{\rm pol}^{\rm axial}\ &=\ 
\vert\mathcal{M}_{\rm unpol}^{\rm res}\vert^2\ 
-\
\sum_{\lambda\in\{T,0,S\}}   \vert \mathcal{M}_\lambda\vert^2\ 
=\ 
\sum_{\lambda\neq\lambda'}
\mathcal{M}^*_{\lambda}\mathcal{M}_{\lambda'}\
\\
&=\ 
2\Re\left[\mathcal{M}^*_{\lambda=T}\mathcal{M}_{\lambda'=0}\right]\
\\
&=\ 
    -\frac{2q^2\ n^2}{(q\cdot n)^2}\vert\vartheta\vert^2
    -\frac{2q^2\ n^2}{(q\cdot n)^2}
    \Re[\mathcal{G}^*\vartheta]
    -\frac{2q^2}{(q\cdot n)^2}
    \Re[\mathcal{G}^*\mathcal{N}]
    -\frac{2q^2}{(q\cdot n)^2} 
    \Re[\vartheta^*\mathcal{N}] 
\label{eq:int_def_axial_expression}
\\
&=\ 
    \frac{2q^2\ n^2}{(q\cdot n)^2}
    \Re[\varphi^*\vartheta]
    +\frac{2q^2}{(q\cdot n)^2}
    \Re[\varphi^*\mathcal{N}] \ .
\label{eq:int_def_axial_expression_shift}
\end{align}
In the last line we rewrote $\mathcal{G}$ as 
$\mathcal{G}= -\varphi-\vartheta$,
using the definition for $\varphi$ in Eq.~\eqref{eq:inter_uni_me_pol_trans_alt}.

There are two notable features in this expression.
First is that all terms of $\mathcal{I}_{\rm pol}^{\rm axial}$
scale as $\mathcal{O}[q^2/(q\cdot n)^2]$.
Na\"ively, this suggests a milder high-energy behavior 
than the squared longitudinal amplitude.
However, $\vartheta$ and $\mathcal{N}$ both contain 
$\mathcal{O}[q^2/(q\cdot n)^2]$ terms,
putting some of the squared longitudinal amplitude 
and some of the polarization interference 
 on equal footing at high energies.
The second observation is the dependence 
of $\mathcal{I}_{\rm pol}^{\rm axial}$ 
on $n^\mu$, and hence gauge dependence.
While the sensitivity of polarization interference to choices of $n^\mu$
has been previously reported~\cite{Bigaran:2025rvb},
the analytical structure of Eq.~\eqref{eq:int_def_axial_expression_shift} 
is novel.

\section{Gauge Invariance and Gauge Independence: 
The ``2P'' Scheme}
\label{sec:interference_independence}

Throughout this work we try to give special attention 
to gauge cancellations and gauge invariance 
in predictions for polarized cross sections and interference.
In realistic Monte Carlo calculations, 
gauge invariance is often checked by varying $\xi$,
resulting (hopefully) in a stable answer.
Weak boson polarization introduces  a complication to this practice.

As discussed in Sec.~\ref{sec:polvector_covariant_scalar}
and Sec.~\ref{sec:interference_rx}, in the $R_\xi$ gauge
the $\xi$ dependency in a weak boson's propagator 
is restricted to the scalar polarization,
specifically $\mathcal{O}(q_\mu q_\nu)$ terms.
However, such terms vanish
when the weak boson couples 
to currents $G^{\mu}_{in/out}$ that are conserved,
i.e., when $q\cdot G_{in/out} =0$.
This happens, for example, in decays of 
weak bosons to a pair of massless fermions.
Technically, having no dependence on the gauge-fixing parameter $\xi$ 
in any part of an amplitude still gives a gauge-invariant result, 
but in a shallow sense.

What is desirable in polarization studies is 
to have predictions that are \textit{independent} 
of gauge-fixing altogether.
The fact that axial gauges effectively have 
two helicity polarizations $(\lambda=T,0)$
while covariant gauges generally have three $(\lambda=T,0,S)$,
even in on-shell limits,
makes predictions for helicity-polarized processes 
inherently dependent on gauge choice.
Refs.~\cite{Ballestrero:2017bxn,Ballestrero:2019qoy,
BuarqueFranzosi:2019boy,Denner:2020bcz,Hoppe:2023uux}
and related works avoid the direct treatment of scalar polarizations 
by making additional approximations, 
such as the pole approximation 
or coupling polarized weak bosons to massless fermions.
The treatment of scalar polarizations becomes relevant, for example,
in top quark decays to massive leptons,
$t\to W^{(*)}_\lambda\to b\tau\nu_\tau$ (see Sec.~\ref{sec:top}).

To help ameliorate the gauge ambiguity in predictions 
for polarized processes, we propose a simple modification to 
helicity polarized propagators when working in covariant gauges.
In these gauges, we propose combining 
the longitudinal $(\lambda=0)$, 
scalar $(\lambda=S)$
and Goldstone $(\lambda=G)$ 
contributions into a single contribution $(\lambda=0')$
at the matrix-element level.
This is similar to how the RH $(\lambda=+1)$ and LH $(\lambda=-1)$ 
helicity contributions are summed together 
in a single ``transverse'' polarization $(\lambda=T)$.
By summing over $\lambda=0,S,G$,
the effective number of polarizations
in the $R_\xi$ gauge reduces to two $(\lambda=T,0')$, 
like in the axial gauge,
and hence is dubbed the ``two polarization'' (2P) scheme.

Taking the Goldstone propagator in Eq.~\eqref{eq:prop_goldstone},
the longitudinal and scalar propagators 
in Eq.~\eqref{eq:prop_long} and Eq.~\eqref{eq:prop_scalar}, 
and assuming the (EW Ward/Slavnov-Taylor) identity in Eq.~\eqref{eq:ward},
then the 2P propagators in the $R_\xi$ gauge
for finite $\xi$ is
\begin{align}
\Pi_{\mu\nu}^V(q,\lambda=0')\Big\vert_{R_\xi}\ &=\ 
\sum_{\lambda=0,S}
\frac{i\eta_{\lambda}\ 
\varepsilon_\mu(q,\lambda)\varepsilon_\nu(q,\lambda)
}{D_V(q^2)}\    
+\
\frac{i\ \frac{q_\mu q_\nu}{\tilde{M}_V^2}}{D_V(q^2,\xi)}
\\
\label{eq:2pscheme_rxi_long_scalar_gold}
&=\ 
\cfrac{i\left(\Theta_{\mu\nu}\ 
-\ \cfrac{(\xi-1)\ q_\mu q_\nu}{D_V(q^2,\xi)}\right)
}{D_V(q^2)}\
+\
\frac{i\left(\cfrac{q_\mu q_\nu}{\tilde{M}_V^2} + 
\cfrac{(\xi-1)q_\mu q_\nu}{D_V(q^2,\xi)}\right)}
{D_V(q^2)}\ 
\\
&=\ 
\cfrac{i\left(
\Theta_{\mu\nu}\ +\ \cfrac{q_\mu q_\nu}{M_V^2-iM_V\Gamma_V}
\right)}{q^2 - M_V^2 + iM_V\Gamma_V}\ ,
\label{eq:2pscheme_rxi}
\end{align}
where $\tilde{M}_V$ is defined in Eq.~\eqref{eq:def_mvtilde}.
Similarly, the 2P propagator in the Unitary gauge is
\begin{align}
\Pi_{\mu\nu}^V(q,\lambda=0')\Big\vert_{\rm Unitary} &=\ 
\sum_{\lambda=0,S}
\Pi_{\mu\nu}^V(q,\lambda)\Big\vert_{\rm Unitary} 
=\ 
\cfrac{i\left(\Theta_{\mu\nu}\ +\ \cfrac{q_\mu q_\nu}{M_V^2 - i M_V\Gamma_V}\right)
}{q^2 - M_V^2 + iM_V\Gamma_V}\ .
\label{eq:2pscheme_unitary}
\end{align}

The equivalence of these two propagators 
demonstrates a desirable robustness across different gauge choices.
In some sense, this robustness follows from BRST invariance,
which stipulates that gauge-invariant predictions require 
summing over longitudinal polarizations, scalar polarizations,
Goldstone bosons, and Faddeev-Popov ghosts.
It is straightforward to extend theis scheme 
for non-traditional gauges wherein ghosts contribute at tree level.

Comparing Eq.~\eqref{eq:2pscheme_rxi} above 
to the longitudinal propagator in Eq.~\eqref{eq:prop_long},
one sees that the ``2P propagator'' can be obtained from Eq.~\eqref{eq:prop_long}
by making the ad hoc replacement, $(q_\mu q_\nu / q^2) \to (q_\mu q_\nu / \tilde{M}_V^2)$.
Moreover, comparing the 2P propagator to the ``on-shell'' longitudinal propagator 
in the pole approximation~\cite{Ballestrero:2017bxn,
Ballestrero:2019qoy,Denner:2020bcz} (and adapted to our notation),
\begin{align}
\Pi_{\mu\nu}^V(q,\lambda&=0)\big\vert_{\rm Pole\ Approx.} =\
\nonumber\\
&\cfrac{\cfrac{i(n\cdot q)}{(n\cdot q)^2 - M_V^2 n^2}
\left[-n_\mu q_\nu - q_\mu n_\nu + \cfrac{q_\mu q_\nu n^2}{(n\cdot q)}
+ \cfrac{n_\nu n_\mu M_V^2}{(n\cdot q)}\right]+ i\cfrac{q_\mu q_\nu}{\tilde{M}_V^2}}{q^2 - M_V^2 + iM_V\Gamma_V}\ ,
\end{align}
one sees that differences with the 2P scheme (taking $n^2=0$) scale as 
\begin{align}
\delta\Pi^V_{\mu\nu}\ \equiv\ \Pi_{\mu\nu}^V(q,\lambda=0')\
-\ 
\Pi_{\mu\nu}^V(q,\lambda=0)\big\vert_{\rm Pole\ Approx.}\
\sim \mathcal{O}
\left[\frac{(q^2-M_V^2)}{(n\cdot q)^2}\right]\ .
\end{align}
In the resonance region, this translates to 
$\delta\Pi^V_{\mu\nu} \sim \mathcal{O}(\Gamma_V^2/E_V^2)$, 
which is better than the intrinsic $\mathcal{O}(\Gamma_V^2/M_V^2)$ 
uncertainty of the narrow width approximation.
In other words, the procedure of Ref.~\cite{Ballestrero:2017bxn,
Ballestrero:2019qoy,Denner:2020bcz} is so successful in describing 
LHC data because gauge miscancellations are only 
$\mathcal{O}[(q^2-M_V^2)/E_V^2]$.
The same miscancellations 
are also likely responsible for the small differences between 
the pole approximation and 
Refs.~\cite{BuarqueFranzosi:2019boy,Javurkova:2024bwa,Hoppe:2023uux},
which use Eq.~\eqref{eq:prop_long},
in predictions for polarized $ZZ$ pairs decaying 
to massless leptons~\cite{Carrivale:2025mjy}.
However, we advocate caution when applying Ref.~\cite{Ballestrero:2017bxn,
Ballestrero:2019qoy,Denner:2020bcz} 
to the case of massive external states,
i.e., when Goldstone boson diagrams are present,
so as to avoid double counting
$\mathcal{O}(q_\mu q_\nu/\tilde{M}_V^2)$ contributions,
such as the second term in Eq.~\eqref{eq:2pscheme_rxi_long_scalar_gold}.

Using Eq.~\eqref{eq:2pscheme_rxi},
the unpolarized and polarized amplitudes  
in the 2P scheme are 
\begin{subequations}
\begin{align}
-i\mathcal{M}_{\rm unpol}^{\rm res}\ &=\  
    G_{out}^\mu 
    i\left[-g_{\mu\nu} + \frac{q_{\mu}q_{\nu}}{M_V^2-i M_V\Gamma_V}\right]
    D_V^{-1}(q^2)
    G_{in}^\nu\
    \equiv\
    -\mathcal{G} + \frac{\mathcal{Q}}{\tilde{M}_V^2}\ ,
\\
    -i\mathcal{M}_{\lambda=T}\ &=\
    G_{out}^\mu 
    i\left[-g_{\mu\nu} - \Theta_{\mu\nu}\right]
    D_V^{-1}(q^2)
    G_{in}^\nu\
    \equiv\ -\mathcal{G}\ - \vartheta\ 
    =\ \varphi\ ,    
    \\
    -i\mathcal{M}_{\lambda=0'}\ &=\ 
    G_{out}^\mu 
    i\left[\Theta_{\mu\nu} + \frac{q_{\mu}q_{\nu}}{M_V^2-i M_V\Gamma_V}\right]
    D_V^{-1}(q^2)
    G_{in}^\nu\
    \equiv\ +\vartheta\ + \frac{\mathcal{Q}}{\tilde{M}_V^2}\ .
\end{align}
\end{subequations}
At the squared level, the unpolarized and polarized 
contributions in the 2P scheme are 
\begin{subequations}
\label{eq:inter_uni_me2_2p}
\begin{align}
\label{eq:inter_uni_me2_unpol_2p}
    \vert\mathcal{M}_{\rm unpol}^{\rm res}\vert^2 &=
    \vert\mathcal{G}\vert^2 + 
    \frac{1}{\vert\tilde{M}_V^2\vert^2}
    \vert\mathcal{Q}\vert^2 
    - 2\Re\left[\frac{\mathcal{G}^*\mathcal{Q}}{\tilde{M}_V^2}\right]\ 
    ,
    \\
\label{eq:inter_uni_me2_trans_2p}     
    \vert\mathcal{M}_{\lambda=T}\vert^2 &= 
    \vert\mathcal{G}\vert^2 + 
    \vert\vartheta\vert^2 + 
    2\Re[\mathcal{G}^*\vartheta]\ ,
    \\
\label{eq:inter_uni_me2_long_2p}
    \vert\mathcal{M}_{\lambda=0'}\vert^2 &= \vert\vartheta\vert^2 
    +\frac{1}{\vert\tilde{M}_V^2\vert^2}\vert\mathcal{Q}\vert^2
    +2\Re\left[\frac{\vartheta^*\mathcal{Q}}{\tilde{M}_V^2}\right]\ .
\end{align}
\end{subequations}
And by construction, the net polarization interference 
has only one source: the interference 
between the transverse amplitude and the 2P amplitude.
Direct computation shows 
\begin{align}
\mathcal{I}_{\rm pol}^{\rm 2P} &=\ 
\vert\mathcal{M}_{\rm unpol}^{\rm res}\vert^2\ 
-\
\sum_{\lambda\in\{T,0'\}}   \vert \mathcal{M}_\lambda\vert^2\ 
=\ 
\sum_{\lambda\neq\lambda'}
\mathcal{M}^*_{\lambda}\mathcal{M}_{\lambda'}\
\\
&=\ 
2\Re\left[\mathcal{M}^*_{\lambda=T}\mathcal{M}_{\lambda'=0}\right]\
=\ 
2\Re[\varphi^*\vartheta]
+2\Re\left[\frac{\varphi^*\mathcal{Q}}{\tilde{M}_V^2}\right]\ .
\end{align}

In comparison to Eq.~\eqref{int_unitary}, 
polarization interference in the 2P scheme is 
simpler because the longitudinal-scalar interference 
is contained in the squared 
2P matrix element $\vert\mathcal{M}(\lambda=0')\vert^2$.
In comparison to the axial gauge,
the 2P scheme features similar contributions
as Eq.~\eqref{eq:int_def_axial_expression_shift},
though with less clear high-energy behavior.
Importantly, the 2P scheme puts predictions 
for polarized amplitudes in covariant gauges and axial gauges 
on closer footing as there is now a correspondence 
among polarized amplitudes,
$\mathcal{M}_{\lambda=T,0'}\vert_{\rm Unitary}\leftrightarrow
\mathcal{M}_{\lambda=T,0}\vert_{\rm axial}$, 
and interference terms in the two gauge classes.

\section{Polarization Interference in Inclusive Weak Boson Production}
\label{sec:noninterference}

The organization of polarized amplitudes introduced 
in Sec.~\ref{sec:interference},
and hence our ability to directly compute polarization interference $\mathcal{I}_{\rm pol}$, 
gives new insight at high energies.
The behavior of $\mathcal{I}_{\rm pol}$ is particularly elucidating  
when the outgoing graph $G_{out}^\mu$ in Fig.~\ref{fig:wPolar_MatrixElementSum}
describes the splitting of $V$ into a pair of massless fermions $f$ and $\overline{f'}$.
Small or vanishing polarization interference has long-been reported 
in inclusive cross sections~\cite{Belyaev:2013nla,
Azatov:2016sqh,Ballestrero:2017bxn,Panico:2017frx,
BuarqueFranzosi:2019boy,Ballestrero:2019qoy,Denner:2020bcz,
Ballestrero:2020qgv,Javurkova:2024bwa}, 
with some understanding that helicity selection rules (spin correlation) 
play a role~\cite{Azatov:2016sqh,Panico:2017frx}.
While spin correlation plays an integral role,
we find that kinematics and $V-A$ couplings
of EW bosons drive the suppression (or non-suppression) 
of $\mathcal{I}_{\rm pol}$ at high energies.

To show this, we consider the (near) resonant production\footnote{This 
assumption allows us to neglect 
complications from non-resonant contributions.} 
of an EW boson $V(q)$ with virtuality $\sqrt{q^2} \sim \mathcal{O}(M_V)$
that decays into a pair of massless fermions $f$ and $\overline{f'}$.
For generic $V-A$ couplings $g_L^f$ and $g_R^f$, 
the outgoing current $G_{out}^\mu$ is 
\begin{align}
\label{eq:noninterference_out_current_def}
    G_{out}^\mu\ &=\ J^\mu_{\lambda,\lambda'}(p,p')\ 
    =\ \overline{u}(p,\lambda)
    \gamma^\mu
    \left(g_L^f P_L + g_R^f P_R\right) v(p',\lambda')\ .
\end{align}
Here, $p$ and $p'=q-p$ with $\lambda$ and $\lambda'$ are 
the momenta and helicities of $f$ and $\overline{f'}$.
As $f$ and $\overline{f'}$ are massless, only the 
$(\lambda,\lambda')=(-\frac{1}{2},+\frac{1}{2})$
and
$(\lambda,\lambda')=(+\frac{1}{2},-\frac{1}{2})$
helicity configurations contribute,
and by the Dirac equation, e.g., $\not\!p' v(p')=0$,
current conservation is satisfied 
\begin{align}
\label{eq:current_conservation}
q_\mu \cdot J^\mu_{\lambda,\lambda'}\ 
=\ E_V J^0_{\lambda,\lambda'} - q^i J^i_{\lambda,\lambda'}\ 
=\ E_V J^0_{\lambda,\lambda'} - \vert\vec{q}\vert\hat{q}^i J^i_{\lambda,\lambda'}\ =\ 
0\ .
\end{align}

Current conservation is an essential ingredient to our argument
and follows from working in the high-energy limit.
Assuming Eq.~\eqref{eq:current_conservation} holds
and choosing the reference vector 
$n_{\rm LL}^\mu = (1,-\hat{q})$ with $n_{\rm LL}^2=0$,
then according to Sec.~\ref{sec:interference_rx}
the unpolarized and polarized matrix elements  
for (near) resonant production of $V(q)$ 
in the $R_\xi$ gauge  are 
\begin{subequations}
\label{eq:pol_me_rxi_current_expand}
\begin{align}
    -i\mathcal{M}_{\rm unpol}^{\rm res}\ =& \
     -\mathcal{G}\ 
    =\
    J^\mu_{\lambda,\lambda'}\ 
    g_{\mu\nu}\
    G_{in}^\nu\
    i D_V^{-1}(q^2)\ ,\
    \\
    -i\mathcal{M}_{\lambda=T} \ =&\  
    \varphi\ 
    =\
    J^\mu_{\lambda,\lambda'} 
    \left[
    \hat{q}_{\perp\mu}\hat{q}_{\perp\nu}\ 
    +\ 
    \hat{q}_{T\perp\mu}\hat{q}_{T\perp\nu}\
    \right]
    G_{in}^\nu\
    i D_V^{-1}(q^2)\ ,\
    \\
    -i\mathcal{M}_{\lambda=0}=&\  \vartheta\  
    =\
    \frac{(J_{\lambda,\lambda'}\cdot n_{\rm LL})}{(n_{\rm LL}\cdot q)}
    \left[
    -(q\cdot G_{in})
    + \frac{(n_{\rm LL}\cdot G_{in})q^2}{(n_{\rm LL}\cdot q)}
    \right] 
    i D_V^{-1}(q^2)\ , 
    \\ 
   -i\mathcal{M}_{\rm Gold},\  
   -i\mathcal{M}_{\lambda=S}
   =&\ 0\ .    
\end{align}
\end{subequations}
The scalar polarization and Goldstone boson for $V$ do not contribute 
due to Eq.~\eqref{eq:current_conservation},
so by Eq.~\eqref{eq:int_rx} the net polarization interference 
for a fixed $f\overline{f'}$ helicities is
\begin{align}
\label{sec:non_interference_polint_def}
\mathcal{I}_{\rm pol}^{\lambda,\lambda'}\     
=\ 2\Re\left[\vartheta^*\varphi\right]\ 
=\ 2\Re\left[\mathcal{M}_{\lambda=0}^*\mathcal{M}_{\lambda=T}\right]\ .
\end{align}

The matrix elements above, and hence the helicity of $V$, 
can be evaluated in any reference frame.
Once fixed, 
the momentum $q$ and current $J^\mu_{\lambda,\lambda'}$
in this ``primary'' frame are  related\footnote{In 
event generators, this relationship is used 
when populating the phase space for $2\to n$ processes. 
Starting from some parent state $(ff')$ that splits into $f$ and $f'$, 
the momenta of $f$ and $f'$ in the rest frame of $(ff')$ 
are generated (pseudo)randomly and then are boosted and rotated 
to the primary frame~\cite{Barger:1987nn,Ellis:1996mzs}.}
to those in the $V(q)$'s rest frame by a $z$-boost $\Lambda_\nu^\mu(\gamma=E_V^2/q^2)$ 
and a pair of rotations $R_\nu^\mu(i,\theta)$ [given in Eq.~\eqref{Boost_rotation_matrices}].
While $n^\mu$ is not Lorentz covariant,
its contractions with Lorentz-covariant objects, e.g., $J^\mu_{\lambda,\lambda'}$, 
are invariant under rotations.

Now, $n^\mu$, $q_\perp^\mu$, and $q_{T\perp}^\mu$ are 
all related to $q$ by construction.
Therefore, rotations of these momenta are also related. 
Denoting rotated momentum with $\sim$,
such that $\tilde{a}^\mu \equiv [R^{-1}(y,\theta_V)\cdot R^{-1}(z,\phi_V)\cdot a]^\mu$,
and applying the same rotation to each quantity, one has
\begin{subequations}
\begin{align}
\tilde{q}^\mu\ &=\ 
[R^{-1}(y,\theta_V)\cdot R^{-1}(z,\phi_V)\cdot q]^\mu\ =\ (E_V,0,0,+\vert\vec{q}\vert)\ ,
\\
\tilde{n}_{\rm LL}^\mu\ &=\ (1,0,0,-1)\ ,\quad
\tilde{q}_\perp^\mu\ =\ (0,1,0,0)\ ,\quad 
\tilde{q}_{T\perp}^\mu\ =\ (0,0,1,0)\ .
\end{align}
\end{subequations}
Multiplication of the rotation matrices from the left 
can be moved to acting on the right via a transpose, 
noting also that $R^T=R^{-1}$.
After rotating,  the polarized propagators
in Eq.~\eqref{eq:pol_me_rxi_current_expand}
act as collections 
of orthogonal projection operators\footnote{In this frame,
the lightcone coordinates 
$\tilde{q}^\mu_\pm = \frac{1}{2}(\tilde{q}^\mu\pm \tilde{n}^\mu)$,
$\tilde{q}_{\perp}^\mu$, and $\tilde{q}_{T\perp}^\mu$ 
constitute an orthonomal basis.}.
For example: 
defining 
$\tilde{J}^{\mu}_{\lambda,\lambda'} = 
[R^{-1}(y,\theta_V)\cdot R^{-1}(z,\phi_V)
\cdot {J}_{\lambda,\lambda'}]^\mu$ as the rotated current, 
Eq.~\eqref{eq:current_conservation} becomes
\begin{align}
q_\mu \cdot J^\mu_{\lambda,\lambda'}\ 
&=\ E_V \tilde{J}^0_{\lambda,\lambda'} - \vert\vec{q}\vert \tilde{J}^3_{\lambda,\lambda'}\ 
=\ 0\ .
\label{eq:current_conservation_rotated}
\end{align}
Similarly, $\tilde{n}^\mu$, $\tilde{q}_\perp^\mu$, and $\tilde{q}_{T\perp}^\mu$
project out particular directions of $\tilde{J}^{\mu}_{\lambda,\lambda'}$.

After rotating, the transverse and longitudinal amplitudes in the ``primary frame'' are
\begin{subequations}
\label{eq:pol_me_rxi_current_expand_rotated}
\begin{align}
-i\mathcal{M}_{\lambda=T}\ =&\
\left[
\tilde{J}^{\mu=1}_{\lambda,\lambda'}
\tilde{G}_{in}^{\nu=1}\
+\
\tilde{J}^{\mu=2}_{\lambda,\lambda'} 
\tilde{G}_{in}^{\nu=2}
\right]
\
i D_V^{-1}(q^2)\ ,\
\\
-i\mathcal{M}_{\lambda=0}\ =&\
\tilde{J}_{\lambda,\lambda'}^{\mu=3}
\left[-\frac{\vert\vec{q}\vert}{E_V}\tilde{G}_{in}^{\nu=0}
+\tilde{G}_{in}^{\nu=3}\right]\ 
i D_V^{-1}(q^2)\ ,\
\end{align}
\end{subequations}
where $\tilde{G}^{\nu}_{in}= [R^{-1}(y,\theta_V)\cdot R^{-1}(z,\phi_V)\cdot G_{in}]^\nu$.
Similarly, the net polarization interference is
\begin{align}
\label{sec:non_interference_polint_long_form}
\mathcal{I}_{\rm pol}^{\lambda,\lambda'}\
&=\ 2\Re\left[I^{k=1}_{\lambda,\lambda'}\ +\ I^{k=2}_{\lambda,\lambda'}\right]\ ,
\quad\text{where}\quad 
I^k_{\lambda,\lambda'}=\ 
(\tilde{J}_{\lambda,\lambda'}^{\mu=3})^*\ \tilde{J}^{\mu=k}_{\lambda,\lambda'}\
\tilde{r}_{in}^k\ ,\
\\
\text{with}\quad 
\tilde{r}_{in}^k\ &=\
\tilde{G}_{in}^{\nu=k}
\left[
-\frac{\vert\vec{q}\vert}{E_V}(\tilde{G}_{in}^{\nu=0})^*
+(\tilde{G}_{in}^{\nu=3})^*
\right]
\vert D_V(q^2)\vert^2\ .
\label{eq:def_polint_rk}
\end{align}
The sub-interference $I^k_{\lambda,\lambda'}$ runs over the $k=1,2$ directions 
and does not mix transverse components.
$I^k_{\lambda,\lambda'}$ also sequesters the outgoing current 
$\tilde{J}_{\lambda,\lambda'}^{\mu}$, 
which describes $V(q)\to f\overline{f'}$ splitting, 
from the remainder of the incoming process, 
which is encoded in $\tilde{r}_{in}^k$.

Assuming the full process in Fig.~\ref{fig:wPolar_MatrixElementSum}
is an $n$-body final state with $n>2$, then the phase-space volume element $dPS_n$
can be split into incoming and outgoing components:
\begin{align}
\label{eq:noninterference_dpsn}
    dPS_n(P_{tot.}; k_1,\dots,k_{n-2},p,p')\ =&\ 
    dPS_{n-2}(P_{tot.}; k_1,\dots,k_{n-2})\ 
    \nonumber\\
    &\ \times\ 
    dPS_2(q; p,p')\
    \times\ 
    \frac{dq^2}{2\pi}\ , 
    \\
    dPS_2(q; p,p')
    =& \frac{d\Omega_f^{(f\overline{f'})}}{2(4\pi)^2}\ =\ 
    \frac{d\phi_f\ d\cos\theta_f}{2(4\pi)^2}
    \ .
\label{eq:noninterference_dps2}
\end{align}
Individual phase-space volume elements are Lorentz invariant.
The means that the angular integrals for $f$ 
in Eq.~\eqref{eq:noninterference_dps2}
can be evaluated 
in the rest frame of the $f\overline{f'}$ system, 
independent of the incoming $(n-2)$-body system
and 
independent of the virtuality of $V(q)$.

The outgoing currents in the rest frame of the $f\overline{f'}$ system 
$\bar{J}_{\lambda\lambda'}^\mu$, which we denote with a bar $-$,
are obtained by evaluating the current 
in Eq.~\eqref{eq:noninterference_out_current_def} 
with the momenta
\begin{align}
    \bar{q}^\mu = (\sqrt{q^2},0,0,0)\ ,\  
    \bar{p}^\mu = E_f (1,\sin\theta_f\cos\phi_f,\sin\theta_f\sin\phi_f,\cos\theta_f)\ ,\
    \bar{E}_f\ =\ \frac{\sqrt{q^2}}{2}\ ,
\end{align}
for all allowed helicity configurations of $f\overline{f'}$.
From $\bar{J}_{\lambda\lambda'}^\mu$, the outgoing currents 
after rotation $\tilde{J}_{\lambda\lambda'}^\mu$
are obtained by a Lorentz boost,
$\tilde{J}_{\lambda\lambda'}^\mu 
= [\Lambda(\gamma) \cdot \bar{J}_{\lambda\lambda'}]^\mu$,
along the $\hat{z}$ direction with $\gamma = E_V/\sqrt{q^2}$.
This is the same boost needed to take $\bar{q}^\mu \to \tilde{q}^\mu$.
The result is
\begin{subequations}
\begin{align}
\tilde{J}_{LR}^\mu\ =\ g_L^f\ 
\times\ 
\big[-\vert\vec{q}\vert\sin\theta_f,\ 
&\sqrt{q^2}(\cos\theta_f\cos\phi_f -i\sin\phi_f),\ 
\nonumber\\
&\sqrt{q^2}(\cos\theta_f\sin\phi_f +i\cos\phi_f),\
-E_V\sin\theta_f
\big]\ ,
\\
\tilde{J}_{RL}^\mu\ =\ 
\left(\frac{g_R^f}{g_L^f}\right)\ 
\times\ 
\left(\tilde{J}_{LR}^\mu\right)^*\ .  &
\end{align}
\end{subequations}

Finally, for each helicity combination of $f$ and $\overline{f'}$,
the sub-interference $I^k_{\lambda,\lambda'}$ is
\begin{subequations}
\label{eq:polar_interference_subint}
\begin{align}
I^{1}_{LR}\ &=\ 
-\vert g_L^f\vert^2\ \sqrt{q^2} E_V \ 
\tilde{r}_{in}^{k=1}\
\sin\theta_f\left(\cos\theta_f\cos\phi_f-i\sin\phi_f\right)\ ,
\\
I^{2}_{LR}\ &=\ 
-\vert g_L^f\vert^2\ \sqrt{q^2} E_V\ 
\tilde{r}_{in}^{k=2}\
\sin\theta_f\left(\cos\theta_f\sin\phi_f+i\cos\phi_f\right)\ ,
\\
I^{1}_{RL}\ &=\ 
-\vert g_R^f\vert^2\ \sqrt{q^2} E_V\ 
\tilde{r}_{in}^{k=1}\ 
\sin\theta_f\left(\cos\theta_f\cos\phi_f+i\sin\phi_f\right)\ ,
\\
I^{2}_{RL}\ &=\ 
-\vert g_R^f\vert^2\ \sqrt{q^2} E_V\ 
\tilde{r}_{in}^{k=2}\ 
\sin\theta_f\left(\cos\theta_f\sin\phi_f-i\cos\phi_f\right)\ ,
\end{align}
\end{subequations}
where $\tilde{r}_{in}^{k}$, defined in Eq.~\eqref{eq:def_polint_rk}, 
is some arbitrary but fixed 
configuration for the incoming portion 
of Fig.~\ref{fig:wPolar_MatrixElementSum},
and hence remains the same for both $LR$ and $RL$ of $f$ and $\overline{f'}$.
The structure of Eq.~\eqref{eq:polar_interference_subint} 
is remarkably simple despite the information it contains.

The net polarization interference $\mathcal{I}_{\rm pol}$,
defined in Sec.~\ref{sec:interference} and 
computed in Eq.~\eqref{sec:non_interference_polint_long_form},
is local. It is defined at the totally differential level,
that is, before any integration over phase space or 
summation over discrete multiplicities.
To obtain the net polarization interference 
at the level of cross sections $\mathbb{I}_{\rm pol}$, 
$\mathcal{I}_{\rm pol}$ needs to be integrated 
over the phase-space measure of Eq.~\eqref{eq:noninterference_dpsn}.
However, since the real-part operator is a linear operator,
it commutes with integration, 
we can integrate the two-body phase-space measure for
$f$ and $\overline{f'}$ in Eq.~\eqref{eq:noninterference_dps2} 
directly over the polarization sub-interference $I^k_{\lambda,\lambda'}$
in Eq.~\eqref{eq:polar_interference_subint}.
Symbolically, we mean 
\begin{align}
\label{eq:non_interference_integral}
    \frac{d\mathbb{I}_{\rm pol}}{dq^2 dPS_{n-2}}\ =\ \frac{1}{2\pi}\sum_{\lambda,\lambda'}\int dPS_2\ 
    \mathcal{I}_{\rm pol}^{\lambda,\lambda'}\
    =\ 
    \frac{2}{(4\pi)^3}\Re\left[\sum_{\lambda,\lambda',k}\int
    d\phi_f d\cos\theta_f\ I^{k}_{\lambda,\lambda'} \right]\ .
\end{align}

Integrating over the azimuthal angle $\phi_f$, 
we find that each term in Eq.~\eqref{eq:polar_interference_subint} vanishes:
\begin{align}
\label{eq:non_interference_integral_phi}
    \int_0^{2\pi}d\phi_f\ I^{k}_{\lambda,\lambda'}\ =\ 0\ .
\end{align}
This follows from each $I^{k}_{\lambda,\lambda'}$ being linear 
in $\cos\phi_f$ and $\sin\phi_f$, 
and arguably a manifestation of rotational symmetry 
about the decay axis in $1\to2$-body splittings.
For the polar angle $\theta_f$, after integrating we instead obtain 
the following non-vanishing combinations:
\begin{subequations}
\label{eq:non_interference_integral_theta}
\begin{align}
\int_{-1}^{+1} d\cos\theta_f\ \left(I^{1}_{LR}\ +\ I^{1}_{RL}\right)\ =\ 
+\frac{i\pi}{2}\ \left(\vert g_L^f\vert^2-\vert g_R^f\vert^2\right)\ 
\sqrt{q^2} E_V\ 
\tilde{r}_{in}^{k=1}\ \sin\phi_f\ ,
\\
\int_{-1}^{+1} d\cos\theta_f\ \left(I^{2}_{LR}\ +\ I^{2}_{RL}\right)\ =\ 
-\frac{i\pi}{2}\ \left(\vert g_L^f\vert^2-\vert g_R^f\vert^2\right)\ 
\sqrt{q^2} E_V\ 
\tilde{r}_{in}^{k=2}\ \cos\phi_f\ .
\end{align}    
\end{subequations}

These last expressions indicate that 
the net polarization interference 
for the $V(q)\to f\overline{f'}$ process 
at the level of inclusive cross sections 
is tied to the underlying $V-A$ coupling structure,
with $\mathbb{I}_{\rm pol} \propto (\vert g_L^f\vert^2-\vert g_R^f\vert^2)$.
For parity-conserving theories like QED, 
the left- and right-handed couplings are equal 
and the integrated interference 
of Eq.~\eqref{eq:non_interference_integral_phi} is zero.
For the $W$ boson, $g_R=0$ and the integrated interference is maximal.
For the $Z$ boson, $g_L$ and $g_R$ are asymmetric, 
and the integrated interference is milder than the $W$ case.

This $V-A$ dependence explains the differences in
polarization interference for $W^+W^-$ and $W^\pm Z$ production 
reported in Refs.~\cite{Denner:2020bcz,Denner:2020eck}.
The dependence also explains the observed correlations 
between polarization interference and phase space cuts
reported in Refs.~\cite{Denner:2020bcz,
Denner:2020eck,Denner:2021csi,Denner:2024tlu,Denner:2025xdz}.
Moreover, our general treatment 
of the incoming graph $G_{in}^\nu$ and 
relaxing the assumption that $V(q)$ is on-shell 
shows that the findings of Ref.~\cite{Panico:2017frx} 
for on-shell diboson production hold more broadly, 
e.g., for $V$+jets, resonant triboson production, 
and the off-shell regime.
The implications of Eqs.~\eqref{eq:non_interference_integral_phi} 
and \eqref{eq:non_interference_integral_theta} 
on polarization interference 
at higher orders of perturbation theory 
are reported in a forthcoming work~\cite{Basu:2026toappear}.

\section{Case Studies in Polarization Interference}
\label{sec:case_studies}

Given the power-counting devices introduced in Sec.~\ref{sec:polvector}
a prescription for their application in Sec.~\ref{sec:interference},
and the knowledge about non-interference at the inclusive level
from Sec.~\ref{sec:noninterference},
we finish our work with an estimation
of the polarization interference 
at the full differential level
for several representative processes.
As case studies, we consider at lowest order: 
inclusive Drell-Yan 
$q\overline{q'}\to W^*\to \tau\nu_\tau$ in Sec.~\ref{sec:dy};
its real radiative correction 
$q\overline{q'}\to W^* g\to \tau\nu_\tau g $ 
in Sec.~\ref{sec:w1g};
the top quark decay process 
$t\to bW^*\to b\tau\nu_\tau$ in Sec.~\ref{sec:top};
and 
inclusive neutrino deep-inelastic scattering 
$\nu q \to \ell q'$ in Sec.~\ref{sec:dis}.
These charged-current processes show the 
increasing levels of complication (or lack thereof) when 
longitudinal and scalar polarizations facilitate a process.
In Sec.~\ref{sec:setup} we summarize our computational setup.

\subsection{Computational Setup} 
\label{sec:setup}
For numerical computations, 
unpolarized and polarized helicity amplitudes 
are computed in the \texttt{HELAS} 
basis~\cite{Hagiwara:1985yu,Murayama:1992gi}
and checked against the 
simulation framework 
\texttt{MadGraph5\_aMC@NLO}~\cite{Stelzer:1994ta,
Alwall:2014hca,BuarqueFranzosi:2019boy}.
For numerical integration we use 
the \texttt{Vegas} algorithm~\cite{Lepage:2020tgj}
as implemented in the \texttt{Cuba} libraries~\cite{Hahn:2004fe}.
In Sec.~\ref{sec:top}, 
we use a prerelease of 
\texttt{MadGraph5\_aMC@NLO} \texttt{v3.7.0},
which features extended support
for computing polarized amplitudes;
see App.~\ref{app:mgpolar} for details.

We assume no quark-flavor mixing 
and use the following SM inputs:
\begin{align}
\label{eq:sminputs}
    M_W &= 80.419 \ {\rm GeV}\ ,\ M_Z = 91.188 \ {\rm GeV}\ ,\ \Gamma_W = 2.0476\ ,\
    \alpha_{\rm EM}^{-1}(\mu_f=M_Z) = 132.507
    \nonumber\\
    m_\tau &= 1.777\ {\rm GeV}\ ,\ m_t = 173 {\rm \GeV}\ ,\ m_b = 4.7 {\rm \ GeV}\ ,\
    \alpha_s(\mu_f=M_Z) = 0.118\ .
\end{align}
Throughout this section 
we adopt the notation $\tilde{M}_W=\sqrt{M_W^2 - iM_W\Gamma_W}$.
$g\approx0.64$ is the weak coupling constant 
extracted from the EW inputs above.
For hadron-level computations, we use the 
NNPDF3.1+luxQED NLO parton distribution function (PDF) 
set~\cite{Bertone:2017bme}
(\texttt{lhaid=324900})
with scale evolution handled 
using \texttt{LHAPDF}~\cite{Buckley:2014ana}.

\subsection{\texorpdfstring{$W$}{W}
Polarization in Inclusive Drell-Yan}
\label{sec:dy}

\begin{figure}[!t]
\begin{center}
    \includegraphics[width=\textwidth]{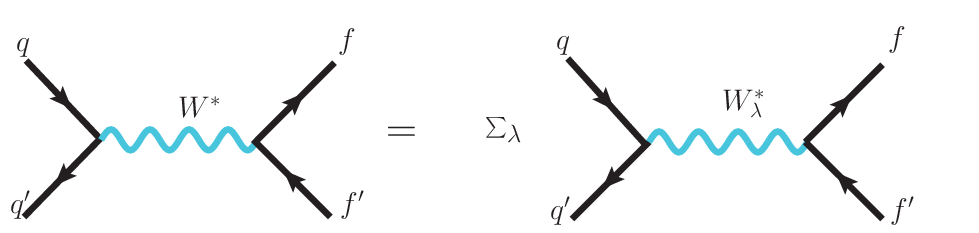}
\end{center}
    \caption{(L) Born-level diagram for the unpolarized, partonic process $q\bar{q}\rightarrow W^{(*)} \rightarrow f \bar{f}$
    and its relationship to (R) the sum of helicity-polarized processes
    $q\bar{q}\rightarrow W^{(*)}_\lambda \rightarrow f \bar{f}$.}
    \label{fig:polexpand_dy}
\end{figure}

As the first case study of our power-counting tools,
we explore $W$ polarization 
in inclusive charged-current Drell-Yan at lowest order.
 Concretely, we consider the partonic process
 $u\bar{d}\rightarrow W^{+(*)}_{(\lambda)}\rightarrow\tau^+\nu_\tau$,
 where external particles are massless except the $\tau$.
 With our bookkeeping devices, we show 
 (i) the scalar polarization  decouples from the process 
 and
 (ii) the longitudinal polarization decouples 
 in the partonic center-of-mass and lab frames.
 
Following the strategy in Sec.~\ref{sec:interference_strategy},
we first make the identification that the unpolarized amplitude for the 
$u\bar{d}\rightarrow W^{+*}\rightarrow\tau^+\nu_\tau$ process is the 
sum of helicity-polarized amplitudes for the processes 
$u\bar{d}\rightarrow W^{+*}_\lambda\rightarrow\tau^+\nu_\tau$.
This is illustrated in Fig.~\ref{fig:polexpand_dy}.
Technically, we work in the $R_\xi$ gauge.
However, due to the masslessness of the incoming quarks
there is no Goldstone contribution,
and by Eq.~\eqref{eq:ward_gauge_inv}
we automatically recover predictions for the Unitary.
At lowest order, there is only one class of resonant diagrams
[Fig.~\ref{fig:polexpand_dy}(L)], no non-resonant diagrams, 
and no non-resonant interference $(\mathcal{I}_{\rm non-res})$ to consider.

Next, we identify the incoming and outgoing graphs 
of Fig.~\ref{fig:wPolar_MatrixElementSum} 
as the incoming $(u\overline{d})$ and outgoing $(\nu_\tau\tau^+)$ 
fermion currents in Fig.~\ref{fig:polexpand_dy}.
We denote these as $J_{in}^\alpha$ and $J_{out}^\beta$, respectively.
In this language, the matrix elements 
for unpolarized and polarized $W^{(*)}$ are 
\begin{subequations}
\label{eq:dy_matrix_elements}
\begin{align}
    -i\mathcal{M}_{\rm unpol}^{\rm res}\ &=\ \frac{-g^2}{2} \frac{-i}{D_W(q^2)} 
    J^\alpha_{\rm in} \ \left(g_{\alpha\beta}+
    \frac{(\xi-1)}{D_W(q^2,\xi)}q_\alpha q_\beta\right) 
    J^\beta_{\rm out}\ 
    \equiv\ -\mathcal{G}\ -\ \mathcal{Q}_\xi\ ,
    \\
    -i\mathcal{M}_{G}\ &=\ 0\ ,
    \\
    -i\mathcal{M}_{\lambda=T}\ &=\ \frac{-g^2}{2} \frac{-i}{D_W(q^2)}
    J^\alpha_{\rm in}\ (g_{\alpha\beta}+\Theta_{\alpha\beta}) 
    J^\beta_{\rm out}\ \equiv\ -\mathcal{G}\ -\ \vartheta\ ,
    \\
    -i\mathcal{M}_{\lambda=0}\ &=\ \frac{-g^2}{2} \frac{+i}{D_W(q^2)} 
    J^\alpha_{\rm in} \ \left(\Theta_{\alpha\beta}+\frac{q_\alpha q_\beta}{q^2}\right) 
    J^\beta_{\rm out}\ \equiv\ 
    \vartheta\ +\ \frac{\mathcal{Q}}{q^2}\ ,
    \\
    \label{MScalar}
    -i\mathcal{M}_{\lambda=S}\ &=\ \frac{-g^2}{2} 
    \frac{-i}{D_W(q^2)}
    J^\alpha_{\rm in} \left(\frac{q_\alpha q_\beta}{q^2}
    +\frac{(\xi-1)}{D_W(q^2,\xi)}q_\alpha q_\beta\right) 
    J^\beta_{\rm out}\ 
    \equiv\ 
    -\frac{\mathcal{Q}}{q^2}\ -\ \mathcal{Q}_\xi\ .
\end{align}
\end{subequations}
We use Eq.~\eqref{eq:prop_trans} for the propagator 
$\Pi_{\mu\nu}^V(q,\lambda=T)$, not Eq.~\eqref{eq:prop_trans_alt}, 
to make cancellations more explicit.
For the following momenta in the partonic center-of-mass frame, 
\begin{subequations}
\begin{align}
p^\mu_u\ &=\ 
\frac{Q}{2}\left( 1 , 0,  0, +1 \right),\  
\quad p^\mu_d = \frac{Q}{2}\left(1, 0,  0, -1 \right)\ , 
\\ 
q^\mu\ &= p^\mu_u + p^\mu_d = p^\mu_\nu + p^\mu_\tau = 
\left(Q,0,  0,  0\right)\ ,
\label{Wmomenta}
\\
p^\mu_\tau\ &=\ 
\left(E_\tau,  \vert p_\tau\vert \sin\theta \cos\phi, 
\vert p_\tau\vert \sin\theta \sin\phi, \vert p_\tau\vert \cos\theta \right),\quad
p^\mu_\nu\ =\ 
\left(\vert p_\tau\vert,  -\vec{p}_\tau \right),
\quad\text{where}
\\
q^2\ &=\ Q^2\ ,\ 
\vert\vec{p}_\tau\vert\ =\ \frac{Q}{2}\left(1 - \frac{m_\tau^2}{Q^2}\right)\ ,
\quad\text{and}\quad 
E_\tau\ =\ \frac{Q}{2}\left(1 + \frac{m_\tau^2}{Q^2}\right)\ ,
\end{align}
\end{subequations}
the incoming and outgoing fermion currents are 
\begin{align}
    J^\alpha_{\rm in} =&\ \bar{v}_{Rj}(p_d) \
    \gamma^\alpha P_L\  \delta_{jk} \ 
    u_{Lk}(p_u)\  =\ [0, Q,-i Q, 0]\ \delta_{jk}
    \label{eq:dy_current_incoming}
    \\
     (J_{\rm out}^{\rm LR})^\beta\ 
     =&\ \bar{u}_{L}(p_\nu) \ \gamma^\beta P_L \ v_{R}(p_\tau) 
     \nonumber\\
     =& \sqrt{2\vert\vec{p}_\tau\vert(E_\tau + \vert\vec{p}_\tau\vert)}\ 
     \left[0, \cos\theta \cos\phi + i \sin\phi, -i\cos\phi+\cos\theta\sin\phi,-\sin\theta \right],
     \label{eq:dy_current_outgoing_LR}
    \\
    (J_{\rm out}^{\rm LL})^\beta =&\ 
    \bar{u}_{L}(p_\nu) \ \gamma^\beta P_L \ v_{L}(p_\tau)  
    \nonumber\\
    =& \sqrt{2 \vert\vec{p}_\tau\vert(E_\tau - \vert\vec{p}_\tau\vert)} 
    \left[ e^{i\phi},-\frac{1}{2}(1+e^{2i\phi})\sin\theta, \frac{i}{2}(-1+e^{2i\phi})\sin\theta, - e^{i\phi} \cos\theta\right]  .
    \label{eq:dy_current_outgoing_LL}
\end{align}

The indices $j,k$ in the incoming 
$(u\overline{d})$ current $J^\alpha_{\rm in}$
are color indices that trivially contract 
in the color-neutral process.
$ (J_{\rm out}^{\rm LR})^\beta$ and $(J_{\rm out}^{\rm LL})^\beta$ 
are the outgoing $(\nu\tau^+)$ current for RH and LH $\tau^+$, respectively.
The LH $\tau^+$ only contributes to the LH chiral current 
through helicity inversion of the $\tau^+$, 
with $(J_{\rm out}^{\rm LL})^\beta$ vanishing when $(m_\tau^2/Q^2) \to 0$.

We note that the temporal and longitudinal components 
of the quark current are both zero in this frame, 
i.e.,  $J^{\alpha=0}_{\rm in}$, $J^{\alpha=3}_{\rm in}=0$. 
 In fact, comparing $J^{\alpha}_{\rm in}$ and $(J_{\rm out}^{\rm LR})^\beta$
 to the definition of transverse polarization vectors 
 in Eq.~\eqref{eq:polvec_transverse},
 one sees that both currents are proportional to the $\lambda=-1$
 polarization vector for the three-momentum directions  
 $\hat{q}=(0,0,1)$ and $\hat{p}_\tau$, respectively.
 The significance of this will be made clear shortly.

Continuing with the strategy,
we evaluate each term in 
$\mathcal{M}_{\rm unpol}^{\rm res}$ and
$\mathcal{M}_\lambda$.
The momentum-tensor terms $\mathcal{Q}$ and $\mathcal{Q}_\xi$
are proportional to  $q_\mu q_\nu$.
By the Dirac equation,  we have
\begin{subequations}
\begin{align}
    J^\alpha_{\rm in}\ q_\alpha\ &\propto\  
    \bar{v}_{R}(p_d) (\not\! p_u + \not\! p_d)P_L u_{L}(p_u)\ 
    =\ \bar{v}_{R}(p_d) (m_u P_R-m_d P_L)  u_{L}(p_u)\ =\ 0\ ,
      \label{JdotQ}
    \\
    J^\beta_{\rm out}\ q_\beta &\propto\   
    \bar{u}_{L}(p_\nu)(\not\! p_\nu + \not\! p_\tau) P_L v_{\lambda}(p_\tau)\   
    =\
    \bar{u}_{L}(p_\nu)(m_\nu P_L - m_\tau P_R) v_{\lambda}(p_\tau)   \  .
    \label{JLdotQ}
\end{align}
\end{subequations}
This means that  $\mathcal{Q}$ and $\mathcal{Q}_\xi$
as well as the scalar polarization matrix element are all zero:
\begin{align}
\label{eq:dy_varq_zero}
\mathcal{Q}\ =&\ \frac{-g^2}{2}
\frac{i}{D_W(q^2)}\ 
J^\alpha_{\rm in}\ \frac{q_\alpha  q_\beta}{q^2}\ J^\beta_{\rm out} = \ 0\ ,
\\
\mathcal{Q}_\xi\ =&\ \frac{-g^2}{2}
\frac{i(\xi-1)D_V(q^2,\xi)}{D_W(q^2)}\ 
J^\alpha_{\rm in}\ q_\alpha  q_\beta\ J^\beta_{\rm out} = \ 0\ ,
\\
\mathcal{M}_{\lambda=S}\ =&\
    -\frac{\mathcal{Q}}{q^2}\ -\ \mathcal{Q}_\xi\ =\ 0\ .
\end{align}

The longitudinal tensor $\vartheta$ 
is proportional to $\Theta_{\alpha\beta}$.
Evaluating each term we have 
\begin{align}
     \vartheta\ =&\ \frac{-g^2}{2}
     \frac{i}{D_W(q^2)}\ 
     J^\alpha_{\rm in}\ \Theta_{\alpha\beta}\ 
     J^\beta_{\rm out}\
     =\ \frac{-g^2}{2}
    \frac{i}{D_W(q^2)}\ 
    \frac{(n\cdot q)}{(n\cdot q)^2 - q^2 n^2}
    \nonumber\\
    &\times
    \left[
    -\underbrace{J^\alpha_{\rm in} n_\alpha  q_\beta J^\beta_{\rm out}}_{\rm term\ 1} 
    - \underbrace{J^\alpha_{\rm in} q_\alpha    n_\beta J^\beta_{\rm out}}_{\rm term\ 2} 
    + \frac{n^2}{(n\cdot q)}
    \underbrace{J^\alpha_{\rm in}q_\alpha   q_\beta J^\beta_{\rm out}}_{\rm term\ 3}
    + \frac{q^2}{(n\cdot q)}
    \underbrace{J^\alpha_{\rm in} n_\alpha    n_\beta J^\beta_{\rm out}}_{\rm term\ 4} 
    \right]\ .
\label{eq:JJdotTheta}
\end{align}
From Eq.~\eqref{JdotQ} above, terms 2 and 3 are zero 
due to current conservation.
In principle, 
using the identities in Eq.~\eqref{eq:refvector_ident},
 $n^\mu$ can be expressed in terms of 
 the time-like reference vector $n^\mu_{\rm TL}$
 and the boson momentum $q^\mu$,
 which will generate more zeros via Eq.~\eqref{JdotQ}.
However, the drawback of working in the rest frame of the $W^{(*)}$
is that the magnitude of its three-momentum is zero.
The makes the $(n_{\rm SL}\cdot q)^{-1} = \vert\vec{q}\vert^{-1}$
factors in Eq.~\eqref{eq:refvector_ident} singular.
Moreover, when $n^\mu=n^\mu_{\rm TL}=(1,\vec{0})$ is chosen at the outset,
the absence of three-momentum introduces spurious (soft) singularities 
in the prefactor
$[(n_{\rm TL}\cdot q)^2 - q^2 n^2_{\rm TL}]^{-1} = [E_V^2 - q^2]^{-1}$.
We stress that the singularities are artifacts, 
i.e., a limitation on choices of $n^\alpha$ 
in particular reference frames and configurations.
$\Theta_{\mu\nu}$ does not contain singular entries [see Eq.~\eqref{eq:theta_def_angles}].

For both light-like and space-like reference vectors 
the same $\Theta_{\alpha\beta}$, and hence $\vartheta$, 
is obtained.
In the frame of $V(q)$, the momentum direction three-vector reduces 
to $\hat{q}\vert_{\rm rest\ frame}=(0,0,\pm1)$, with a twofold ambiguity.
For the light-like case, $\Theta_{\alpha\beta}$ is 
\begin{align}
    \Theta_{\alpha\beta} =&\ 
    \frac{(n_{\rm LL}\cdot q) }{(n_{\rm LL}\cdot q)^2 - q^2n_{\rm LL}^2} 
    \left[
    -(n_{\rm LL})_\alpha q_\beta\ 
    -\ q_\alpha(n_{\rm LL})_\beta\ 
    +\ 0_{\alpha\beta}\
    +\ \frac{q^2}{(n_{\rm LL}\cdot q)} (n_{\rm LL})_\alpha (n_{\rm LL})_\beta 
     \right]
    \nonumber\\
    =&\ -\frac{1}{Q} 
    \begin{pmatrix}
Q & 0 & 0 & 0 \\
0 & 0 & 0 & 0 \\
0 & 0 & 0 & 0 \\
\pm Q & 0 & 0 & 0
\end{pmatrix} -\frac{1}{Q}
   \begin{pmatrix}
Q & 0 & 0 & \pm Q \\
0 & 0 & 0 & 0 \\
0 & 0 & 0 & 0 \\
0 & 0 & 0 & 0
\end{pmatrix}+ 
\begin{pmatrix}
1 & 0 & 0 & \pm 1 \\
0 & 0 & 0 & 0 \\
0 & 0 & 0 & 0 \\
\pm 1 & 0 & 0 & 1
\end{pmatrix}
= \begin{pmatrix}
-1 & 0 & 0 & 0 \\
0 & 0 & 0 & 0 \\
0 & 0 & 0 & 0 \\
0 & 0 & 0 & 1
\end{pmatrix} \ . 
\end{align}
In the first line, term 3 is zero due to the light-like condition $n_{\rm LL}^2=0$.
For the space-like case, 
$(n_{\rm SL}\cdot q)\vert_{\rm rest\ frame}=0$, 
leading to terms 1 and 2 to vanish.
$\Theta_{\alpha\beta}$ is then similarly 
\begin{align}
    \Theta_{\alpha\beta} =&\ 
    \frac{1}{(n_{\rm SL}\cdot q)^2 - q^2n_{\rm SL}^2} \left[0_{\alpha\beta}\ 
    +\ 0_{\alpha\beta}\ +\ 
    q^2 (n_{\rm SL})_\alpha (n_{\rm SL})_\beta\ +\ n_{\rm SL}^2 q_\alpha q_\beta\right]
    \nonumber\\ 
    =&\
\frac{Q^2}{(-1)^2Q^2}
\begin{pmatrix}
0 & 0 & 0 & 0 \\
0 & 0 & 0 & 0 \\
0 & 0 & 0 & 0 \\
0 & 0 & 0 & 1
\end{pmatrix} + \frac{(-1)}{(-1)^2Q^2} 
\begin{pmatrix}
Q^{2} & 0 & 0 & 0 \\
0 & 0 & 0 & 0 \\
0 & 0 & 0 & 0 \\
0 & 0 & 0 & 0
\end{pmatrix}
= \begin{pmatrix}
-1 & 0 & 0 & 0 \\
0 & 0 & 0 & 0 \\
0 & 0 & 0 & 0 \\
0 & 0 & 0 & 1
\end{pmatrix}\ .
\label{thetaSL}
\end{align}

Importantly, regardless of the representation for $n^\mu$,
the longitudinal tensor $\Theta_{\alpha\beta}$
reduces to a diagonal temporal $(\mu=\nu=0)$ component 
and a diagonal longitudinal $(\mu=\nu=3)$ component.
All transverse and off-diagonal components 
of $\Theta_{\alpha\beta}$ vanish in this frame.

Comparing $\Theta_{\alpha\beta}$ to the incoming 
$(u\overline{d})$ current $J^{\alpha}_{in}$
in Eq.~\eqref{eq:dy_current_incoming},
one sees that the two are orthogonal,
$J^{\alpha}_{in}\ \Theta_{\alpha\beta} = 0_\beta$.
This follows from the orthogonality 
of $J^{\alpha}_{in}$ and $n^\alpha$.
$J^{\alpha}_{in}$ contains only transverse components
while $n^\alpha$ contains no transverse components.
Consequentially, $\vartheta$ itself is zero 
and with Eq.~\eqref{eq:dy_varq_zero}
as is  the matrix element for the transverse polarization:
\begin{align}
\label{eq:dy_vartheta_zero}
    \vartheta\ &=\ \frac{-g^2}{2}
     \frac{i}{D_W(q^2)}\ 
     J^\alpha_{\rm in}\ \Theta_{\alpha\beta}\ 
     J^\beta_{\rm out}\ =\ 0\ ,
     \\
     \mathcal{M}_{\lambda=0}\ &=\ 
    \vartheta\ +\ \frac{\mathcal{Q}}{q^2}\ =\ 0\ .
\end{align}

What remains are the $\mathcal{G}$ terms
in the unpolarized matrix element $\mathcal{M}_{\rm unpol}^{\rm res}$
and the matrix element 
for the transverse polarization $\mathcal{M}_{\lambda=T}$.
As no other terms in Eq.~\eqref{eq:dy_matrix_elements} survives
[see Eqs.~\eqref{eq:dy_varq_zero} and \eqref{eq:dy_vartheta_zero}],
the two amplitudes are equal and are given by
\begin{align}
    -i\mathcal{M}_{\rm unpol}^{\rm res}\ &=\ 
    -\mathcal{G}\ -\ \mathcal{Q}_\xi\ =\ -\mathcal{G},
    \\
    -i\mathcal{M}_{\lambda=T}\ &=\ 
    -\mathcal{G}\ -\ \vartheta\ =\ -\mathcal{G}\ .
\end{align}    
For completeness, the $\mathcal{G}$ terms 
for the two $(\nu_\tau\tau^+)$ helicity configurations are
\begin{align}
    \mathcal{G}^{\rm LR}\ =\ & \frac{+g^2}{2} 
    \frac{i}{D_W(q^2)} J^\alpha_{\rm in}\ g_{\alpha\beta}\ (J_{\rm out}^{\rm LR})^\beta 
    = 
    \frac{-g^2}{2} \frac{ie^{-i\phi} Q}{D_W(q^2)} 
    \sqrt{2 \ E_\nu(E_\tau+E_\nu)}\ (1-\cos\theta)\ ,
\\
    \mathcal{G}^{\rm LL}\ =\ & \frac{+g^2}{2} 
    \frac{i}{D_W(q^2)} J^\alpha_{\rm in}\ g_{\alpha\beta}\ (J_{\rm out}^{\rm LL})^\beta 
    = 
    \frac{+g^2}{2} \frac{i Q}{D_W(q^2)} 
    \sqrt{2 \ E_\nu(E_\tau-E_\nu)}\ \sin\theta\ .
\end{align}

Turning to interference, 
due to the absence of $\mathcal{Q}$ terms 
[see Eq.~\eqref{eq:dy_varq_zero}], 
we can use the expression for $\mathcal{I}_{\rm pol}$
given in Eq.~\eqref{eq:int_unitary_noq}.
However, due to the additional absence of $\vartheta$ terms
[see Eq.~\eqref{eq:dy_vartheta_zero}], 
the total matrix element is  purely
the transverse contribution.
Therefore, the polarization interference 
for the charged-current Drell-Yan process 
vanishes 
\begin{align}
\label{eq:int_unitary_noq_dy}
\mathcal{I}_{\rm pol}^{\rm no-\mathcal{Q}}\ \overset{\mathcal{Q}\to0}{=}\ 
-2\Re[(\mathcal{G}+\vartheta)^*\vartheta]\ 
=\ 
-2\Re[(\mathcal{G}+0)^*0]\ 
=\ 0\ .
\end{align}

At the hadronic level, 
$\mathcal{Q}$ terms remain absent due 
to the massless of the incoming quarks.
And in the absence real radiative corrections 
[see Sec.~\ref{sec:w1g}], 
the incoming momenta $p_u$ and $p_d$ are  
only boosted along the $\hat{z}$ direction.
While none of the $n^\mu$  
in Eq.~\eqref{eq:ref_vector_def} is Lorentz covariant, 
$\Theta_{\mu\nu}$ is Lorentz covariant and well-defined 
in 
the lab frame (lab) and 
the rest frame (rest) of $W^{(*)}$.
Subsequently, by boost invariance, one has
\begin{align}
  (J_{in}^{\rm lab})^\alpha \cdot 
  \Theta_{\alpha\beta}\vert^{\rm lab}\
  =\ 
  (J_{in}^{\rm rest})^\alpha\cdot 
  \Theta_{\alpha\beta}\vert^{\rm rest}\
  = 0\ .
\end{align}
The intuition for this follows from longitudinal boosts 
only shuffling $\alpha=0,3$ elements
of $J_{\in}^{\alpha}$, $q^\alpha$, and $n^\alpha$
while the $\alpha=1,2$ elements (transverse) unaltered.

This means that since the transverse components of $q$ remain zero,
the incoming quark current $J_{\rm in}^\alpha$ remains a transverse current,
and the projection of its temporal and longitudinal components 
also remains zero, $J_{\rm in}^\alpha n_{\alpha}=0$.
It then follows that Drell-Yan currents 
at this order are driven entirely  
by the transverse polarization of the intermediate boson,
in accordance with Refs.~\cite{Collins:1977iv,Lam:1978pu}.
The longitudinal polarization ($\vartheta$ terms)
and polarization interference are absent.
At $\mathcal{O}(\alpha_s)$, 
virtual QCD corrections 
to the $Wqq'$ vertex factorize for massless quarks~\cite{Altarelli:1979ub},
and do not alter the outcome.
We now turn to $W+1g$ production.

\subsection{\texorpdfstring{$W$}{W} 
Polarization in \texorpdfstring{$W$}{W}+jets}
\label{sec:w1g}

We now discuss helicity polarization 
and polarization interference in the $W+$jets process.
This process was previously studied at tree-level in Refs.~\cite{Bern:2011ie,
Stirling:2012zt,Belyaev:2013nla,Pellen:2021vpi},
but using the single-particle-inclusive angular decomposition 
of Ref.~\cite{Ellis:1996mzs}, 
which obfuscates polarization interference.
The treatment here employs the polarized propagator methods
of Ref.~\cite{Ballestrero:2017bxn,Ballestrero:2019qoy};
polarization interference 
in $W$+jets at higher orders in QCD is discussed in Ref.~\cite{Basu:2026toappear}.

\begin{figure}[t!]
\begin{center}
\includegraphics[width=\textwidth]{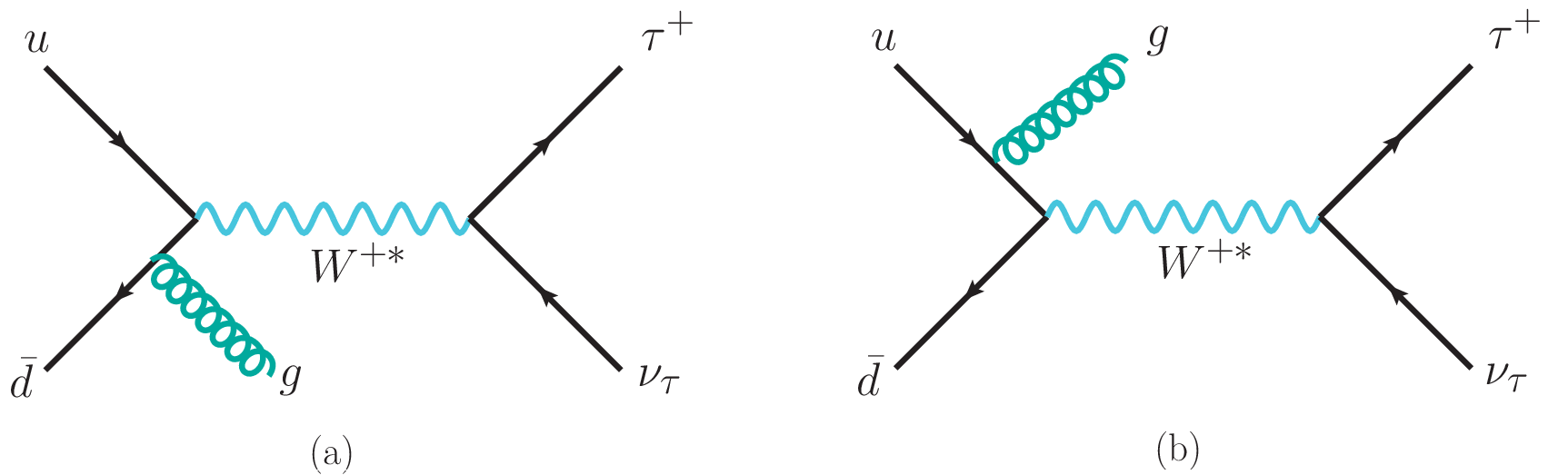}
\end{center}
\caption{Lowest order diagrams for the unpolarized partonic process 
$u\bar{d}\rightarrow W^{+(*)} g \rightarrow \nu_\tau \tau^+  g$,
featuring a $(u\overline{d}g)$ current with
(a) $d\to d^*g$ emission and a $t$-channel $d^*$ 
($\mathcal{D}_{in}^\alpha$ in the main text),
and 
(b) $u\to u^*g$ emission and a $t$-channel $u^*$ 
($\mathcal{U}_{in}^\alpha$  in the main text).}
\label{Wjet}
\end{figure}

For conciseness, we focus on the channel
$u(p_u)\bar{d}(p_d)
\rightarrow W^{(*)}(q) g(k)
\rightarrow \tau^+(p_\tau) \nu_\tau(p_\nu) g(k)$
at lowest order as illustrated in Fig.~\ref{Wjet}.
We again take external particles massless except for the $\tau^+$.
Like the Drell-Yan case in Sec.~\ref{sec:dy}, 
the scalar polarization does not contribute in the $R_\xi$ gauge.
However, unlike the previous case the longitudinal polarization 
is present in the $W+g$ process.
There are no non-resonant diagrams 
at this order $(\mathcal{M}_{\rm non-res}=0)$.

\begin{figure}[!t]
\begin{center}
    \includegraphics[width=\textwidth]{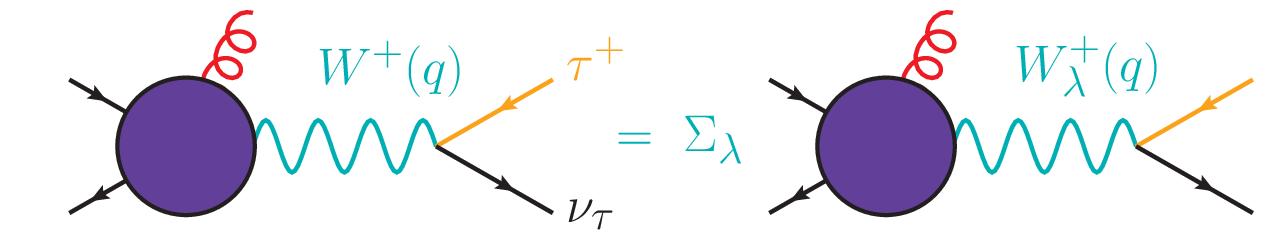}
\end{center}    
    \caption{(L) Born-level diagram for the unpolarized, partonic process $q\bar{q}\rightarrow W^{(*)} g \rightarrow \tau^+ \nu_\tau g$ 
    and its relationship to (R) the sum of helicity-polarized processes
    $q\bar{q}\rightarrow W^{(*)}_\lambda g \rightarrow \tau^+ \nu_\tau g$.}
    \label{fig:wPolar_MEsum_w1Jet}
\end{figure}

Following the strategy for computing polarization interference
in Sec.~\ref{sec:interference_strategy},
we first make the identification that 
the full amplitude 
for an unpolarized intermediate $W^{(*)}$
is the sum of amplitudes 
for helicity-polarized intermediate $W^{(*)}_\lambda$.
This identification is illustrated in Fig.~\ref{fig:wPolar_MEsum_w1Jet}.
Next, we make the identification 
that the outgoing graph $G_{out}^\beta$
is just the $(\nu_\tau\tau^+)$ current $J_{out}^\beta$.
For $W_\lambda$ momentum $q = p_u+p_d - k= p_\tau+p_\nu$, this is given by 
\begin{align}
    J^\beta_{\rm out} =&\ \bar{u}_{L}(p_\nu,\lambda_\nu)
    \left(-\frac{ig}{\sqrt{2}}\gamma^\beta P_L\right)
    v_{R}(p_\tau,\lambda_\tau) 
    =\ \frac{-ig}{\sqrt{2}} 
    \left[\bar{u}_{L}(p_\nu,\lambda_\nu) \ \gamma^\beta P_L\ 
    v_{R}(p_\tau,\lambda_\tau)\right].
\end{align}

The incoming graph 
$G_{in}^\alpha=\mathcal{D}^\alpha_{in}+\mathcal{U}^\alpha_{in}$ 
is composed of two $(u\overline{d}g)$ currents.
The first, labeled $\mathcal{D}^\alpha_{in}$ 
and shown in Fig.~\ref{Wjet}(a), features  
$d\to d^*g$ emission and a $t$-channel $d^*$
with momentum $p_b=p_d -k$.
The second, labeled $\mathcal{U}^\alpha_{in}$ 
and shown in Fig.~\ref{Wjet}(b), features a
$u\to u^*g$ emission and a $t$-channel $u^*$
with momentum $p_a=p_u -k$.
Explicitly, these are given by
\begin{subequations}
\begin{align}
    D^\alpha_{in}  =&\ \frac{ig\ g_s}{\sqrt{2}}\ \delta_{jk} T^A_{lk}\ 
    \left(\frac{+1}{p_b^2}\right) \left[\bar{v}_{lR} (p_d,\lambda_d)\ \gamma^\rho \ \epsilon_\rho^*(k,\lambda_g)\ \slashed{p}_b \gamma^\alpha  P_L \ u_{jL}(p_u,\lambda_u)  \right],\ 
    \\
    U^\alpha_{in}  =&\ \frac{ig\ g_s}{\sqrt{2}}\ \delta_{kl} T^A_{jk}\ 
    \left(\frac{-1}{p_a^2}\right) \left[\bar{v}_{lR} (p_d,\lambda_d)\ \gamma^\alpha P_L\ \slashed{p}_a \gamma^\rho \epsilon_\rho^*(k,\lambda_g)\ u_{jL}(p_u,\lambda_u)  \right]\ .    
\end{align}
\end{subequations}
Here, $g_s= \sqrt{4\pi \alpha_s}$ is the strong coupling constant, 
$T^A_{jk}$ is the color matrix for quark-gluon vertex, 
$\delta_{jk}$ is the (trivial) color matrix for the quark-$W$ vertex,
$\epsilon_\rho^*(k,\lambda_g)$ is the polarization vector 
for the outgoing gluon 
[same expression as given in Eq.~\eqref{eq:polvec_transverse}].

With our bookkeeping, 
the unpolarized and polarized amplitudes  
in the $R_\xi$ gauge are
\begin{subequations}
\label{eq:w1g_matrix_elements}
\begin{align}
\label{eq:w1g_matrix_elements_unpol}
    -i\mathcal{M}_{\rm unpol}\ &=\ 
    -\mathcal{G}\ -\ \mathcal{Q}_\xi\ 
    =\ 
    -\mathcal{G}_U\  
    -\mathcal{G}_D\ 
    -\mathcal{Q}_{\xi U}\ 
    -\mathcal{Q}_{\xi D}\,
\\
\label{eq:w1g_matrix_elements_trans}
-i\mathcal{M}_{\lambda=T}\ &=\  -\mathcal{G}\ -\ \vartheta\
=\ -\mathcal{G}_U - \mathcal{G}_D - \vartheta_U -\vartheta_D\ ,
\\
\label{eq:w1g_matrix_elements_long}
-i\mathcal{M}_{\lambda=0}\ &=\ \vartheta\ +\ \frac{\mathcal{Q}}{q^2}\ 
=\ \vartheta_U + \vartheta_D 
+ \frac{\mathcal{Q}_U}{q^2} + \frac{\mathcal{Q}_D}{q^2}\ ,
\\
-i\mathcal{M}_{\lambda=S}\ &=\  
-\frac{\mathcal{Q}}{q^2}\ -\ \mathcal{Q}_\xi\
=\ 
-\frac{\mathcal{Q}_U}{q^2} -\frac{\mathcal{Q}_D}{q^2} 
- \mathcal{Q}_{\xi U} - \mathcal{Q}_{\xi D}\ ,
    \\
    -i\mathcal{M}_{G}\ &=\ 0\ ,
 \end{align}
\end{subequations}
where in terms of external currents 
$G_{in}^\alpha$ and $J_{out}^\beta$ we have 
\begin{subequations}
\begin{align}
\mathcal{G}_{U} &= \frac{i}{D_W(q^2)}
\left(U^\alpha_{in}\ g_{\alpha\beta} \ J^\beta_{\rm out}\right)
\quad,\quad 
\mathcal{Q}_{\xi U} =\ \frac{i}{D_W(q^2)}
\left(U^\alpha_{in}\ 
\frac{(\xi-1)q_\alpha q_\beta}{D_V(q^2,\xi)}\
J^\beta_{\rm out}\right)\ ,\ 
\\
\vartheta_U &=\ \frac{i}{D_W(q^2)}
\left(U^\alpha_{in}\ \Theta_{\alpha\beta} \ J^\beta_{\rm out}\right)
\quad,\quad 
\mathcal{Q}_U =\ \frac{i}{D_W(q^2)}
\left(U^\alpha_{in}\ q_\alpha q_\beta \ J^\beta_{\rm out}\right)\ .\
\end{align}
\end{subequations}
``$D$''-current terms obtained 
by making the replacement $U_{in}^\alpha\to D_{in}^\alpha$.

We now focus on the $\mathcal{Q}$ and $\mathcal{Q}_\xi$ terms.
Since both incoming quarks are massless,
both sets of quark spinors 
obey an equation of motion of the form
$\!\not p_u u(p_u)=0$ and $\!\not p_d v(p_d)=0$.
After (anti)commuting, 
the contractions of the incoming quark currents 
with $q_\alpha$ are
\begin{subequations}
\begin{align}
    D^\alpha_{in}q_{\alpha} =&\ 
    \frac{ig\ g_s}{\sqrt{2}}\ \delta_{jk} T^A_{lk}\ \frac{(+1)}{p_b^2}\left[\bar{v}_{lR}(p_d)\ \gamma^\rho \epsilon_\rho^*(k,\lambda_g) \  \slashed{p}_b\ \gamma^\alpha\  P_L\ u_{jL}(p_u) \right] (p_{b\alpha}+p_{u\alpha})
    \nonumber\\
    =&\ \frac{ig\ g_s}{\sqrt{2}}\ \delta_{jk} T^A_{lk}\ 
    (+1)
    \left[\bar{v}_{lR}(p_d)\ \gamma^\rho \epsilon_\rho^*(k,\lambda_g)\ P_L\ u_{jL}(p_u)\ \right]
    \label{DinUin}
\\
U^\alpha_{in} q_\alpha = &\ 
    \frac{ig\ g_s}{\sqrt{2}}\ \delta_{jk} T^A_{lk}\ 
    \frac{(-1)}{p_a^2}
    \left[\bar{v}_{lR} (p_d)\ \gamma^\alpha P_L\ \slashed{p}_a \gamma^\rho \epsilon_\rho^*(k,\lambda_g)\ u_{jL}(p_u)  \right] (p_{a\alpha}+p_{d\alpha})
    \nonumber\\
     =&\ \frac{ig\ g_s}{\sqrt{2}}\ \delta_{jk} T^A_{lk}\ 
     (-1)
     \left[\bar{v}_{lR} (p_d)\ \gamma^\rho \epsilon_\rho^*(k,\lambda_g)\ P_L\ u_{jL}(p_u)  \right]\ =\ -D^\alpha_{in}q_{\alpha}.
\\
G_{in}^\alpha q_\alpha = &\ (D^\alpha_{in} + U^\alpha_{in})q_{\alpha}\ =\ 0\ .
\end{align}
\end{subequations}

Since $G_{in}^\alpha$ is a conserved current, it follows  
that $\mathcal{Q}\propto G_{in}^\alpha q_\alpha =0$ 
and 
$\mathcal{Q}_\xi \propto G_{in}^\alpha q_\alpha = 0$,
and subsequently that the scalar-helicity matrix element is zero,
$\mathcal{M}_{\lambda=S}=0$.
Heuristically, this could be anticipated because 
a nonzero scalar amplitude would imply a dependence 
on the gauge-fixing parameter $\xi$ 
in the unpolarized amplitude, 
after summing over contributions. 
However, as there is no Goldstone amplitude
and no $\xi$ dependence 
in either the transverse or longitudinal amplitudes,
then gauge invariance 
requires that $\mathcal{M}_{\lambda=S}=0$.
In other words, when Goldstone bosons are absent 
in a process (but $\xi$ is finite),
then both $\mathcal{Q}_\xi$ and $\mathcal{Q}$ are zero.
This is a manifestation of the Ward identity in Eq.~\eqref{eq:ward}.

More rigorously, the full incoming quark graph 
$G_{in}^\alpha=\mathcal{D}^\alpha_{in}+\mathcal{U}^\alpha_{in}$ 
is a conserved current because 
(a) we are summing over all possible QCD contributions 
to the incoming graph at this order,
(b) the external quarks are both massless,
(c) the quark-gluon vertex is a vector current,
and hence is helicity conserving
Since the operator $\not\!q$ is 
a helicity-inverting operator,
$G_{in}^\alpha q_\alpha$ is only nonzero 
when $u$ or $\overline{d}$ is massive
or when color indices remain uncontracted.
The same argument holds 
for the $(ug)$ and $(\overline{d}g)$ 
scattering configurations.
Moreover, 
attaching additional gluons (or photons) 
to the incoming quark lines in Fig.~\ref{Wjet}
does not alter this  property as each extra 
real emission either (i) leaves the Dirac algebra in 
$\mathcal{D}^\alpha_{in}$ and $\mathcal{U}^\alpha_{in}$
unchanged 
($\varepsilon_\rho(k)$ is replaced 
by a more complicated object 
but remains a scalar in spinor space),
or (ii) leaves the helicity unchanged since 
for each additional $\gamma^m\gamma^n$ pair 
(one for the vertex and one for the propagator)
one has $\gamma^m\gamma^n P_L = P_L \gamma^m\gamma^n$.
Consequentially, 
$G_{in}^\alpha$ remains a conserved current,
i.e., $G_{in}^\alpha q_\alpha=0$.

Moving briefly to the $\vartheta$ term,
$G_{in}^\alpha q_\alpha= 0$
allows us to write $\vartheta$ as
\begin{align}
    \vartheta  =\ \frac{i}{D_W(q^2)}\  \frac{(n\cdot q)}{(n\cdot q)^2 - q^2 n^2}  \left[(U^\alpha_{in}+D^\alpha_{in}) 
    \left(n_\alpha q_\beta + \frac{q^2}{(n\cdot q)}\ n_\alpha n_\beta  \right)
    J^\beta_{\rm out} \right]\ .
    \label{thetaWJ}
\end{align}

Now, with the absence of $\mathcal{Q}$ and $\mathcal{Q}_\xi$ terms,
the polarization interference reduces to
\begin{align}
\label{eq:int_unitary_noq_w1g}
\mathcal{I}_{\rm pol}^{W+1g}\ =\
-2|\vartheta|^2- 2 {\rm \ Re}(\mathcal{G}^* \vartheta)\ =\
2\Re[\varphi^*\vartheta]\ .
\end{align}
Unlike the Drell-Yan process, 
the interference in the $W+1g$ process is non-zero. 
To estimate its magnitude and dependence on scattering energy,
we use na\"ive power counting.

For this analysis, we make some simplifying assumptions 
as we are only interested in the na\"ive scaling 
with hard scattering energy. 
Working in the partonic center-of-mass frame,
we first assume that the $W^{(*)}g$ pair in the 
$u\overline{d}\to W^{(*)}g$ sub-process are produced 
at wide angles and at high $p_T$ such that 
$E_g, E_{W} \sim E_u, E_d = \sqrt{\hat{s}}/2$,
where $\sqrt{\hat{s}}$ is the partonic center-of-mass energy,
and $E_g \lesssim E_W$ due 
to the virtuality of $W^{(*)}$  $(\sqrt{q^2}>0)$.

For $t$-channel $u^*(p_a)$ and $d^*(p_b)$, this implies the scaling
\begin{subequations}
\begin{align}
p_a^2 =&\ p_u^2+k^2 -2(p_u\cdot k) =  -2E_u E_g (1-\cos\theta_{ug}) \sim -\hat{s}\ ,
\\
p_b^2 =&\ p_d^2+k^2 -2(p_d\cdot k) =  -2E_d E_g (1-\cos\theta_{dg}) \sim -\hat{s}\ .
\end{align}
\end{subequations}
Spinors have an energy dependence 
of $u,v\sim\sqrt{E}$, 
but other objects, such as $\gamma$-matrices and 
the gluon's polarization vector $\epsilon$, 
do not carry any explicit energy dependence. 

We use the timelike reference vector $n_{\rm TL}^\alpha$ 
as we can always express $n_{\rm SL}^\alpha$ and $n_{\rm LL}^\alpha$ 
in terms of $n_{\rm TL}^\alpha$ and momentum $q^\alpha$ in this frame.
Now, when incoming currents $D_{in}^\alpha$ and $U_{in}^\alpha$ 
contract with $n^\alpha_{\rm TL}$, 
it returns a quantity that does not na\"ively scale: 
\begin{subequations}
\label{eq:w1g_current_projection_quarks}
\begin{align}
D^\alpha_{in}\cdot (n_{\rm TL})_{\alpha}\sim&\ 
\frac{1}{p_b^2} 
\bar{v}_{lR} (p_d) \gamma^\rho \epsilon_\rho^*(k,\lambda_g) 
\slashed{p}_b \gamma^0  P_L  u_{jL}(p_u) 
\sim\ \frac{\sqrt{E_d} E_d  \sqrt{E_u}}{E_d E_g}\sim (E_u)^0\ ,
\\
U^\alpha_{in}\cdot (n_{\rm TL})_{\alpha} \sim&\ 
\frac{1}{p_a^2} 
\bar{v}_{lR} (p_d) \gamma^0 \slashed{p}_a \gamma^\rho 
\epsilon_\rho^*(k,\lambda_g) P_L u_{jL}(p_u)
\sim\ \frac{\sqrt{E_d} E_u  \sqrt{E_u}}{E_u E_g}\sim (E_d)^0\ .
\end{align}
\end{subequations}
For the outgoing $(\nu_\tau \tau^+)$ current $J_{out}^\beta$ 
we have for different helicity configurations 
\begin{subequations}
\label{eq:w1g_current_projection_leptons}
\begin{align}
q_\beta \cdot J^\beta_{\rm out}(\nu_{\tau L} \tau^+_R)\sim\ 
m_\tau\ \bar{u}_L(p_\nu,\lambda_\nu=-\frac{1}{2}) 
P_R v_R(p_\tau,\lambda_\tau=+\frac{1}{2}) 
&\sim \frac{m_\tau^2 \sqrt{E_\nu}}{\sqrt{E_\tau}}\ ,
\\
q_\beta \cdot J^\beta_{\rm out}(\nu_{\tau L} \tau^+_L)\sim\ 
m_\tau\ \bar{u}_L(p_\nu,\lambda_\nu=-\frac{1}{2}) 
P_R v_R(p_\tau,\lambda_\tau=-\frac{1}{2}) 
&\sim m_\tau \sqrt{E_\nu}\sqrt{E_\tau}\ ,
\\
(n_{\rm TL})_{\beta} \cdot J^\beta_{\rm out}(\nu_{\tau L} \tau^+_R) \sim\ 
\bar{u}_{L}(p_\nu,\lambda_\nu=-\frac{1}{2}) \gamma^0 
P_L v_{R}(p_\tau,\lambda_\tau=+\frac{1}{2})
&\sim\ \sqrt{E_\nu} \sqrt{E_\tau}\ .
\\
(n_{\rm TL})_{\beta} \cdot J^\beta_{\rm out}(\nu_{\tau L} \tau^+_L) \sim\ 
\bar{u}_{L}(p_\nu,\lambda_\nu=-\frac{1}{2}) \gamma^0  
P_L v_{R}(p_\tau,\lambda_\tau=-\frac{1}{2})
&\sim\  \frac{m_\tau \sqrt{E_\nu}}{\sqrt{E_\tau}}\ .
\end{align}
\end{subequations}
In the first lines we used the Dirac equation 
as done in Eq.~\eqref{JdotQ} for the Drell-Yan case.

We draw attention to the different degrees 
to which helicity inversion is present.
For the $q_\beta \cdot J^\beta_{\rm out}$ cases,
the difference is whether one is
(a) inverting the helicity 
of a helicty-preserving vector current $(\nu_L\tau^+_R)$,
which is suppressed by $\mathcal{O}(m_\tau^2/\sqrt{E_\tau})$, or 
(b) inverting the helicity 
of a helicity-flipped vector current $(\nu_L\tau^+_L)$,
which is mildly enhanced by $\mathcal{O}(m_\tau \sqrt{E_\tau})$.
For the $(n_{\rm TL})_{\beta} \cdot J^\beta_{\rm out}$ cases,
we see helicity preservation in the $(\nu_L\tau^+_R)$ current
and $\mathcal{O}(m_\tau/\sqrt{E_\tau})$ helicity inversion 
in the $(\nu_L\tau^+_L)$ current.
In the massless $\tau$ limit, 
only $(n_{\rm TL})_{\beta} \cdot J^\beta_{\rm out}(\nu_L \tau^+_R)$
survives because it is the only helicity-conserving contribution.

Putting these scalings into Eqs.~\eqref{thetaWJ}
and \eqref{eq:w1g_matrix_elements_long}, 
we get for the $\lambda=0$ amplitude
\begin{subequations}
\begin{align}
-i\mathcal{M}_{\lambda=0}(\nu_{\tau L}\tau^+_R)\ =\ 
\vartheta(\nu_{\tau L}\tau^+_R)\ &\sim\ 
\frac{\sqrt{E_\nu} \sqrt{E_\tau}}{D_W(q^2)}
\frac{E_W^2}{E_W^2 - q^2}
\left(
\frac{m_\tau^2}{E_W E_\tau} + 
\frac{q^2}{E_W^2} 
\right)\ ,
\\
-i\mathcal{M}_{\lambda=0}(\nu_{\tau L}\tau^+_L)\ =\
\vartheta(\nu_{\tau L}\tau^+_L)\ &\sim\ 
\frac{\sqrt{E_\nu}\sqrt{E_\tau}}{D_W(q^2)}
\frac{E_W^2}{E_W^2 - q^2}
\left(
\frac{m_\tau}{E_W}  + \frac{m_\tau}{E_\tau}\frac{q^2}{E_W^2}
\right)\ .
\end{align}
\end{subequations}
The longitudinal polarization amplitudes 
are nonzero for the $(\nu_{\tau L}\tau^+_R)$ helicity configuration, 
even for massless $\tau$ leptons, 
while the $(\nu_L\tau^+_L)$ helicity configuration
is zero for massless $\tau$ leptons.
In ultra-low-energy scattering 
where $\mathcal{O}(q^2/E_W^2)$ terms can be neglected
and $\tau$s are replaced by electrons,
then the longitudinal amplitude remains nonzero
due to lepton masses (and likely quark masses).
At ultra-high-energy scattering 
where $\mathcal{O}(q^2/E_W^2)$ 
and $\mathcal{O}(m_\tau/E_W)$ terms can be neglected,
the longitudinal matrix element vanishes.

For the $\mathcal{G}$ term 
in the unpolarized and transverse polarization
amplitudes, we note that the scaling of $g_{\alpha\beta}$ is the same
as $(n_{\rm TL})_\alpha (n_{\rm TL})_\beta$ since 
the $(\alpha,\beta)=(0,0)$ components in the two are the same.
(This is feature built into the definition $\Theta_{\alpha\beta}$ 
and its decomposition, as discussed 
in Sec.~\ref{sec:polvector_covariant_trans}).
Using the scalings 
in Eq.~\eqref{eq:w1g_current_projection_quarks} 
and 
Eq.~\eqref{eq:w1g_current_projection_leptons}
for the quark and lepton currents,
the unpolarized matrix elements 
for the different lepton helicities scale as
\begin{subequations}
\begin{align}
-i\mathcal{M}_{\rm unpol}(\nu_L\tau^+_R) = 
\mathcal{G}(\nu_{\tau L}\tau^+_R)\ 
&\sim\
G_{in}^\alpha
\frac{g_{\alpha\beta}}{D_W(q^2)}
 J^\beta_{\rm out}(\nu_{\tau L} \tau^+_R)
\nonumber\\
&\sim\ 
G_{in}^\alpha
\frac{(n_{\rm TL})_{\alpha}(n_{\rm TL})_{\beta}}{D_W(q^2)}
 J^\beta_{\rm out}(\nu_{\tau L} \tau^+_R) 
\sim \frac{\sqrt{E_\nu} \sqrt{E_\tau}}{D_W(q^2)}\  ,
\\
-i\mathcal{M}_{\rm unpol}(\nu_{\tau L}\tau^+_L) = 
\mathcal{G}(\nu_{\tau L}\tau^+_L) 
&\sim\ 
G_{in}^\alpha
\frac{g_{\alpha\beta}}{D_W(q^2)}
 J^\beta_{\rm out}(\nu_{\tau L} \tau^+_L)
\nonumber\\
\sim&\ 
G_{in}^\alpha  
\frac{(n_{\rm TL})_{\alpha}(n_{\rm TL})_{\beta}}{D_W(q^2)}
 J^\beta_{\rm out}(\nu_{\tau L} \tau^+_L)
\sim
\frac{\sqrt{E_\nu}\sqrt{E_\tau}}{D_W(q^2)}
\frac{m_\tau}{ E_\tau}\ .
\end{align}
\end{subequations}
Using these, the transverse matrix elements 
for the different lepton helicities are given by 
\begin{subequations}
\begin{align}
-i\mathcal{M}_{\lambda=T}(\nu_{\tau L}\tau^+_R) &=
-\mathcal{G}(\nu_{\tau L}\tau^+_R)
-\vartheta(\nu_{\tau L}\tau^+_R) 
\sim \frac{\sqrt{E_\nu} \sqrt{E_\tau}}{D_W(q^2)}  
\frac{E_W^2}{E_W^2 - q^2}
\left(1+
\frac{m_\tau^2}{E_W E_\tau}
\right)\ ,
\\
-i\mathcal{M}_{\lambda=T}(\nu_{\tau L}\tau^+_L) &= 
-\mathcal{G}(\nu_{\tau L}\tau^+_L)
-\vartheta(\nu_{\tau L}\tau^+_L) 
\sim 
\frac{\sqrt{E_\nu}\sqrt{E_\tau}}{D_W(q^2)}
\frac{m_\tau}{E_\tau}
\frac{E_W^2}{E_W^2 - q^2}
\left(1+\frac{E_\tau}{E_W}\right) .
\end{align}
\end{subequations}
In the absence of $\tau$ masses, we see strong resemblance to
the unpolarized matrix element,
with the difference being a factor of 
$\mathcal{M}_{\lambda=T}/\mathcal{M}_{\rm unpol}\sim E_W^2/(E_W^2-q^2)$\ .

Given the expressions for the scaling 
of $\vartheta$ and $\mathcal{G}$ above 
and Eq.~\eqref{eq:int_unitary_noq_w1g},
then the scaling of polarization interference  
for different lepton helicities scale as
\begin{subequations}
\label{eq:int_unitary_noq_w1g_full}
\begin{align}
\mathcal{I}_{\rm pol}^{W+1g}(\nu_{\tau L}\tau^+_R) \sim\
&\frac{E_\tau E_\nu}{\vert D_W(q^2)\vert^2}
\frac{E_W^4}{(E_W^2 - q^2)^2}
\left(1+\frac{m_\tau^2}{E_W E_\tau}\right)
\left(
\frac{m_\tau^2}{E_W E_\tau} + 
\frac{q^2}{E_W^2} 
\right)
\nonumber\\
\overset{m_\tau\to0}{\sim}&
\frac{E_\tau E_\nu}{\vert D_W(q^2)\vert^2}
\frac{q^2\ E_W^2}{(E_W^2 - q^2)^2}
\\
\mathcal{I}_{\rm pol}^{W+1g}(\nu_{\tau L}\tau^+_L)
\sim\
&\frac{E_\tau E_\nu}{\vert D_W(q^2)\vert^2}
\frac{m_\tau^2}{E_\tau^2}
\frac{E_W^4}{(E_W^2 - q^2)^2}
\left(1+\frac{E_\tau}{E_W}\right)
\left(\frac{E_\tau}{E_W}  + \frac{q^2}{E_W^2}\right)\ ,
\nonumber\\
\overset{m_\tau\to0}{\sim}& 0\ .
\end{align}
\end{subequations}
In the second line of both expressions we took the $m_\tau\to0$ 
limit since $m_\tau\approx1.78\GeV$ is small compared to typical 
high-$p_T$ scales at the LHC.
The interference for both helicity permutations 
should be compared to the scaling 
of the squared unpolarized matrix element:
\begin{align}
\label{eq:w1g_sqme}
\vert\mathcal{M}_{\rm unpol}(\nu_{\tau L}\tau^+_R)\vert^2\ +\
\vert\mathcal{M}_{\rm unpol}(\nu_{\tau L}\tau^+_L)\vert^2\
\sim 
\frac{E_\nu E_\tau}{\vert D_W(q^2)\vert^2}
\left(1+\frac{m_\tau^2}{E_\tau^2}\right)\ .
\end{align}
This is essentially the leading factor in 
the  polarization interference.

Taking the ratio of interference and unpolarized
result (and neglecting $m_\tau$),  we get
\begin{align}
\mathcal{R}^{W+1g}_{\rm pol\ int}\ \equiv&\ 
\cfrac{
\mathcal{I}_{\rm pol}^{W+1g}(\nu_{\tau L}\tau^+_R) + 
\mathcal{I}_{\rm pol}^{W+1g}(\nu_{\tau L}\tau^+_L)}{
\vert\mathcal{M}_{\rm unpol}(\nu_{\tau L}\tau^+_R)\vert^2 +
\vert\mathcal{M}_{\rm unpol}(\nu_{\tau L}\tau^+_L)\vert^2}
\\
\sim&\
\frac{E_W^4}{(E_W^2 - q^2)^2}
\left(\frac{E_\tau}{E_\tau + m_\tau}\right)
\nonumber\\
\times& 
\left[
\left(1+\frac{m_\tau^2}{E_W E_\tau}\right)
\left(
\frac{m_\tau^2}{E_W E_\tau} + 
\frac{q^2}{E_W^2} 
\right)
+
\frac{m_\tau^2}{E_\tau^2}
\left(1+\frac{E_\tau}{E_W}\right)
\left(\frac{E_\tau}{E_W}  + \frac{q^2}{E_W^2}\right)
\right]
\\
\overset{m_\tau\to0}\sim&\
\frac{E_W^4}{(E_W^2 - q^2)^2}
\left(\frac{q^2}{E_W^2}\right)\
\sim\
\frac{q^2}{E_W^2}
\left[1+\mathcal{O}\left(\frac{q^2}{E_W^2}\right)\right]^2
\ .
\end{align}
The scaling of the relative size of the polarization 
interference shows that the polarization interference 
$\mathcal{R}^{W+1g}_{\rm pol\ int}$
for the
$u\bar{d}\rightarrow W^{+(*)} g \rightarrow \nu_\tau \tau^+  g$
process is quickly suppressed $(\sim1/\gamma^2)$ for increasing 
$W$ energy in the partonic center-of-mass frame. 
For instance, assuming the $W$ is on-shell and at rest,
then in the absence of $\tau$ lepton masses the polarization interference 
is $\mathcal{R}^{W+1g}_{\rm pol\ int}\sim\mathcal{O}(100\%)$.
Instead, when $p_T^W > 100\GeV\ (250\GeV)$,
which is well within the reach of LHC analyses,
one has $E_W\gtrsim \sqrt{p_T^2 + M_W^2}\sim 125\GeV\ (260\GeV)$,
and the ratio drops to $\mathcal{R}^{W+1g}_{\rm pol\ int}\lesssim 
\mathcal{O}(40\%)\ [\mathcal{O}(10\%)]$ since $1/\gamma_W^2 = q^2/E_W^2\sim0.4\ (0.1)$.

At the same time, the relative size of the
squared longitudinal polarization  
 matrix element carries a stronger mass-over-energy scaling:
\begin{align}
\mathcal{R}^{W+1g}_{\lambda=0}\ \equiv&\ 
\cfrac{
\vert\mathcal{M}_{\lambda=0}(\nu_L\tau^+_R)\vert^2
+ 
\vert\mathcal{M}_{\lambda=0}(\nu_L\tau^+_L)\vert^2
}{
\vert\mathcal{M}_{\rm unpol}(\nu_L\tau^+_R)\vert^2 +
\vert\mathcal{M}_{\rm unpol}(\nu_L\tau^+_L)\vert^2}
\\
\label{eq:ratio_w1g_long_etau}
\sim&\
\frac{E_W^4}{(E_W^2 - q^2)^2}
\left(\frac{E_\tau}{E_\tau + m_\tau}\right)
\left[
\left(
\frac{m_\tau^2}{E_W E_\tau} + 
\frac{q^2}{E_W^2} 
\right)^2
+
\left(
\frac{m_\tau}{E_W}  + \frac{m_\tau}{E_\tau}\frac{q^2}{E_W^2}
\right)^2
\right]
\\
\overset{m_\tau\to0}\sim&\
\frac{E_W^4}{(E_W^2 - q^2)^2}
\left(\frac{q^2}{E_W^2}\right)^2
\ \sim\ 
\left(\frac{q^2}{E_W^2}\right)^2
\left[1+\mathcal{O}\left(\frac{q^2}{E_W^2}\right)\right]^2
\ .
\end{align}
While $\mathcal{R}^{W+1g}_{\lambda=0}\sim\mathcal{O}(100\%)$
when $W$ is at rest,
$\mathcal{R}^{W+1g}_{\lambda=0}\lesssim 
\mathcal{O}(15\%)\ [\mathcal{O}(1\%)]$
for the $p_T^W$ thresholds above. 
In other words, in the high-energy limit,
the longitudinal polarization contribution 
will decouple from the $W+1g$ process 
before the interference. 

\begin{figure}[!t]
\subfigure[]{\includegraphics[width=0.48\textwidth]{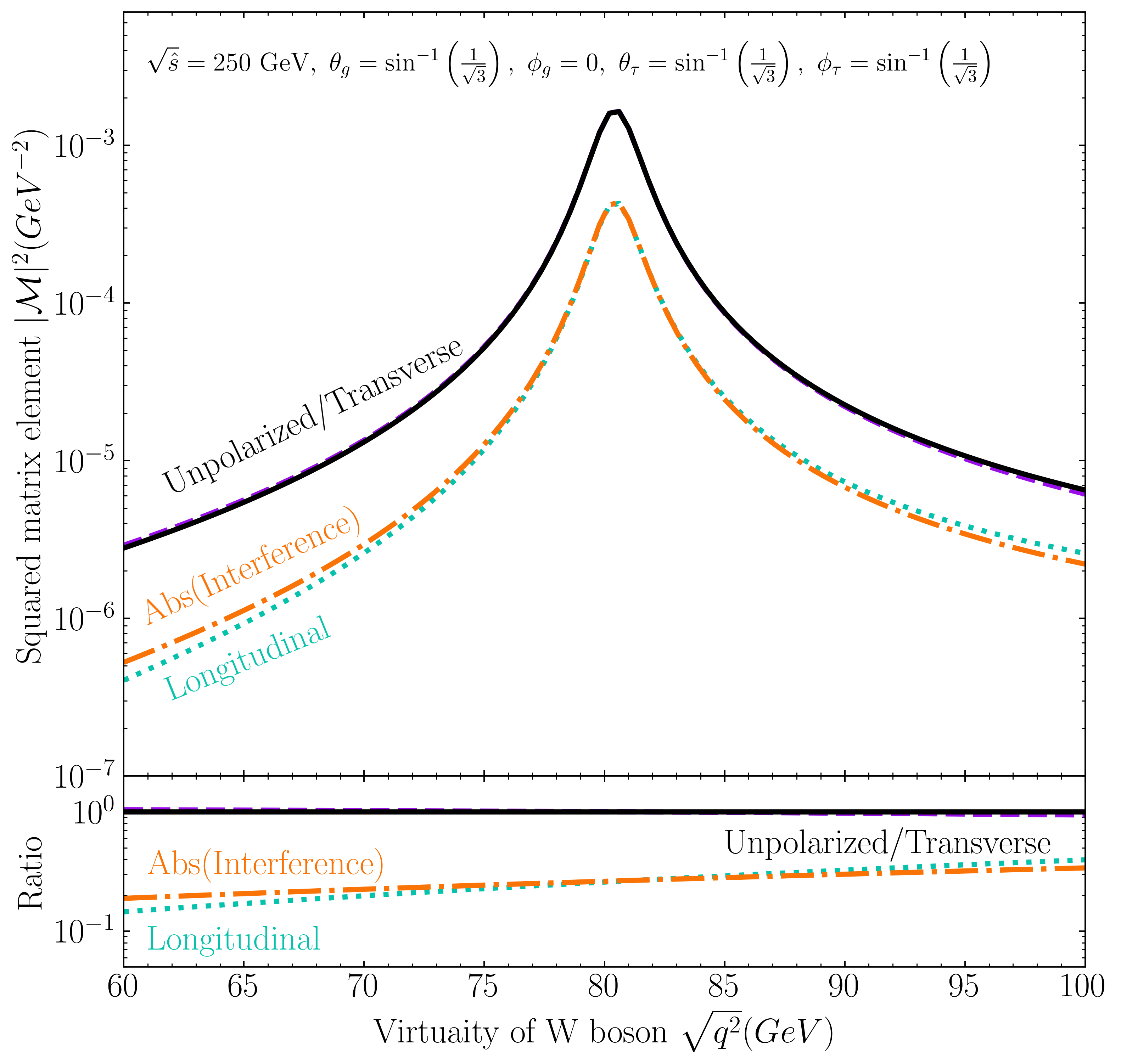}
\label{ME1}}\hfill
\subfigure[]{\includegraphics[width=0.48\textwidth]{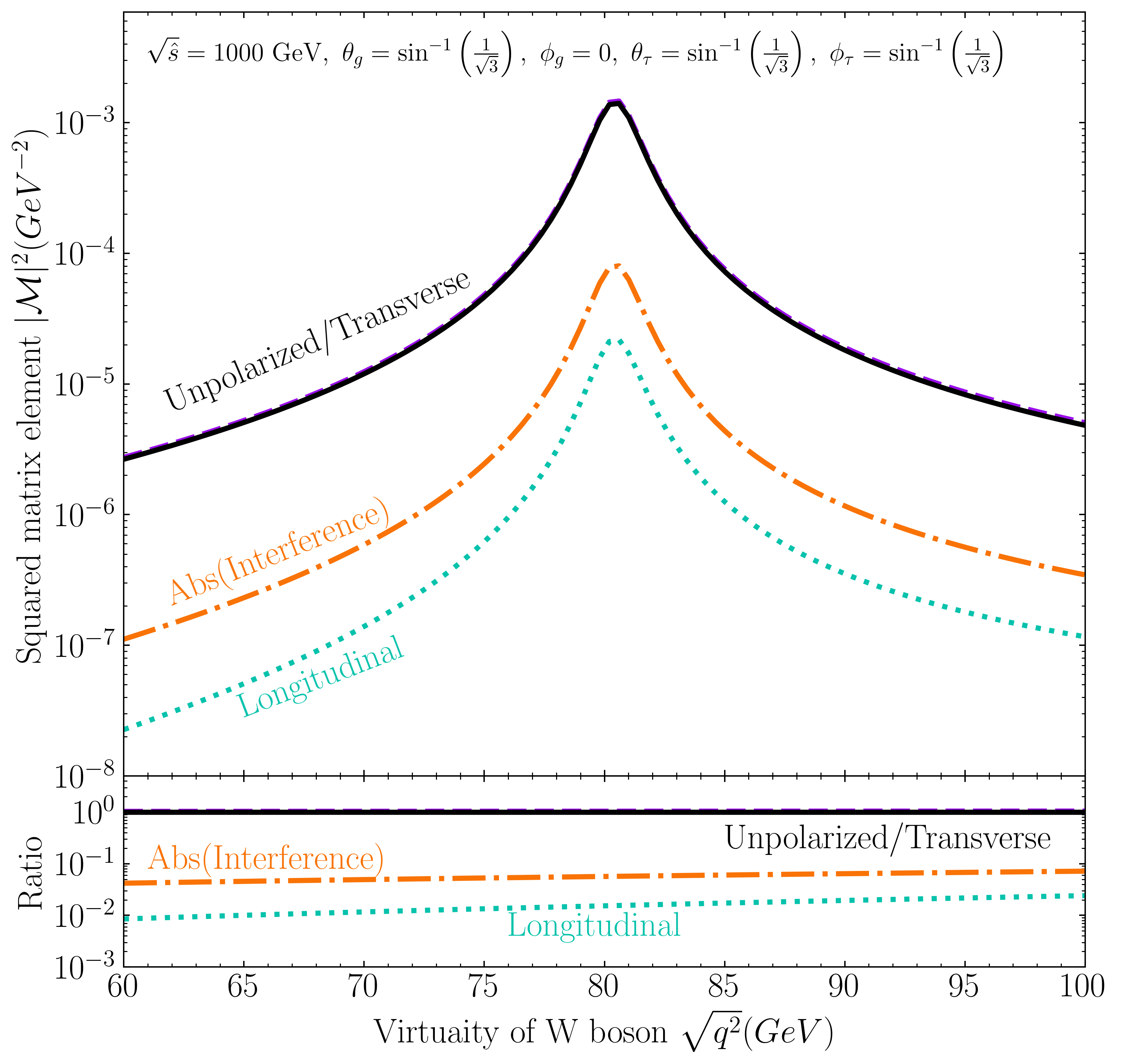}
\label{ME2}}
\subfigure[]{\includegraphics[width=0.48\textwidth]{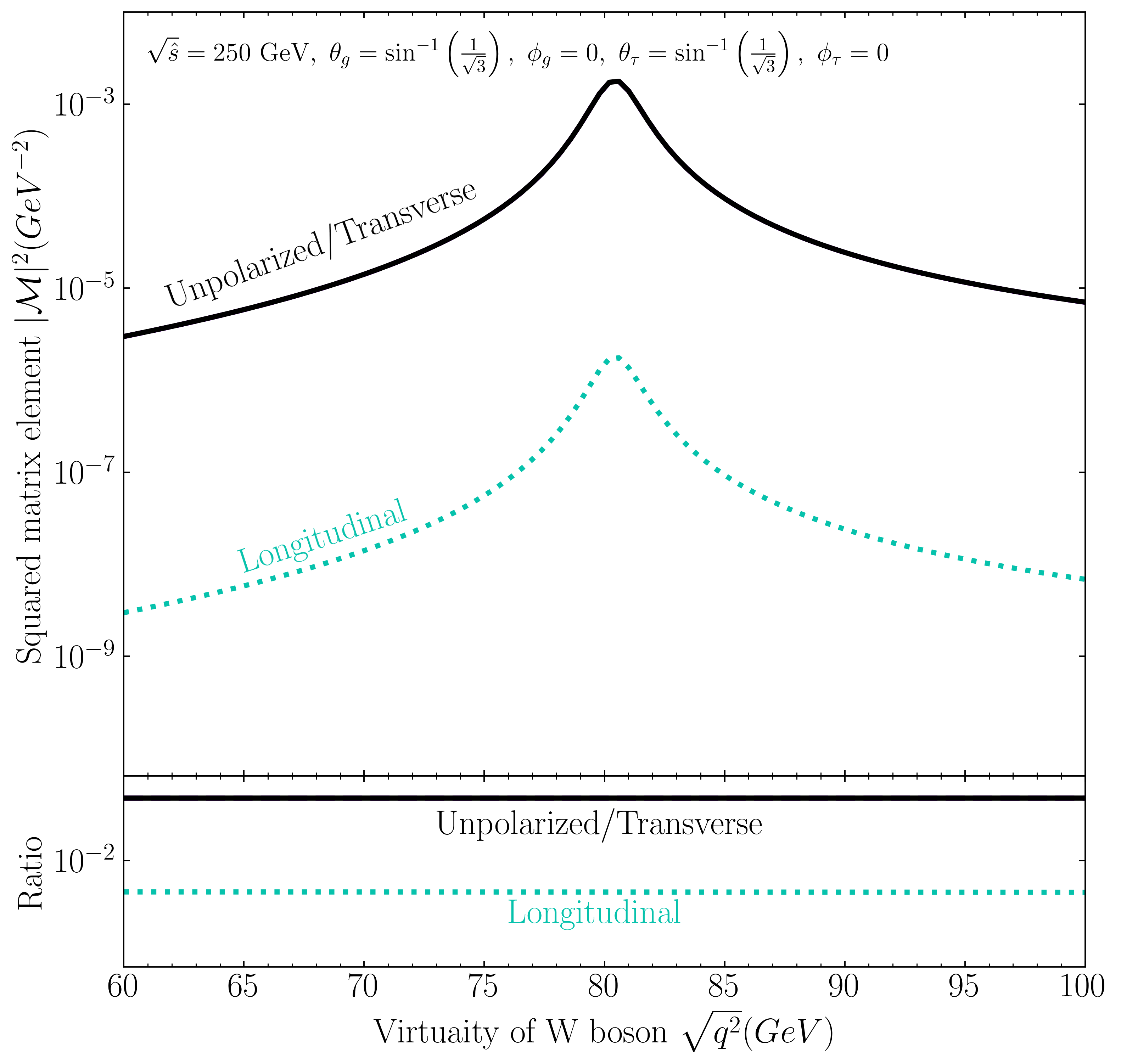}
\label{ME3}}\hfill
\subfigure[]{\includegraphics[width=0.48\textwidth]{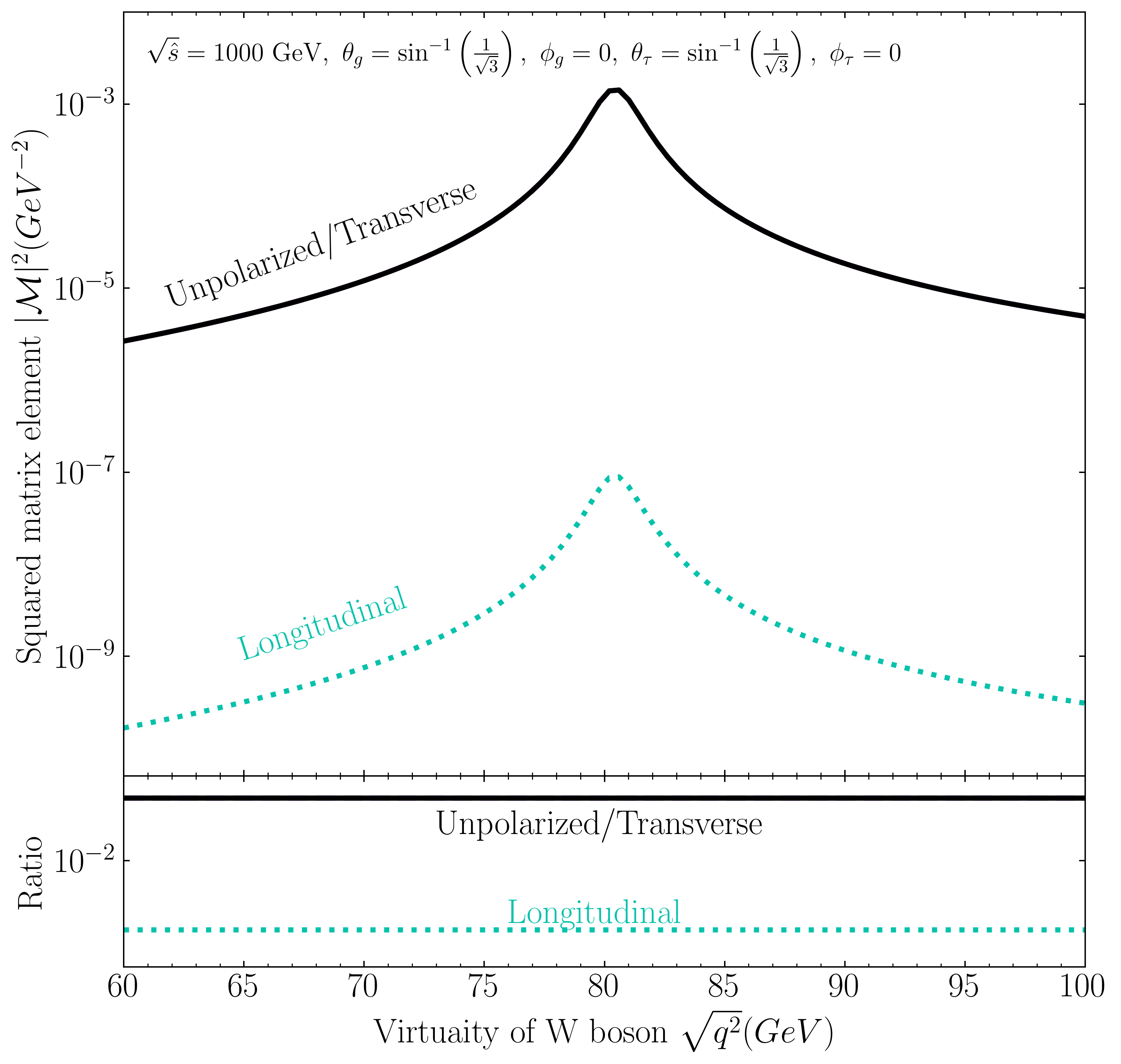}
\label{ME4}}
\caption{As a function of $W^{(*)}$ virtuality $\sqrt{q^2}$
and in the partonic center-of-mass frame at different phase space points,
the squared matrix element 
for the unpolarized process
$u\bar{d}\rightarrow W^{+(*)} g \rightarrow \nu_\tau \tau^+  g$
(solid),
the squared amplitudes 
for transversely polarized $W_{\lambda=T}^{+(*)}$ (dash)
and 
longitudinally polarized $W_{\lambda=T}^{+(*)}$ (dot),
and the absolute value 
of the polarization interference (dash-dot).}
\label{fig:w1g_me2}
\end{figure}

To demonstrate the behavior 
of the polarized matrix elements
and polarization interference,
we plot in Fig.~\ref{fig:w1g_me2}
as a function of $W^{(*)}$ virtuality $\sqrt{q^2}$
and in the partonic center-of-mass frame 
the full squared matrix element for the 
unpolarized $W+1g$ process (solid)
as well as the squared amplitudes 
for the transverse polarization (dash),
longitudinal polarization (dot),
and the absolute value 
of the net polarization interference (dash-dot).
In the first panel are the ratios of the curves
with respect to the unpolarized case.

In Fig.~\ref{ME1} and \ref{ME2}, 
we take a partonic center-of-mass energy 
of $\sqrt{\hat{s}} = 250$ GeV  and $1000$ GeV, respectively, 
for the phase space point corresponding to
\begin{align}
    \theta_g,\  
    \theta_\tau^{W\ \rm frame},\ 
    \phi_\tau^{W\ \rm frame}\ 
    = \sin^{-1}\left(\frac{1}{\sqrt{3}}\right)\approx35^\circ\ ,\ 
    \phi_g = 0\ .
\end{align}
The angles are semi-random but chosen to minimize accidental cancellations
and zeros.
The angles for $\tau^+$ (and hence $\nu_\tau$) 
are defined in the rest frame of the $W^{(*)}$.

Globally, we see in Fig.~\ref{ME1} and \ref{ME2}
the unpolarized and transverse cases are numerically similar,
while the longitudinal and interference contributions 
are both one or more orders of magnitude below.
For the full range of virtuality, 
the interference is negative and so its magnitude is plotted.
Qualitatively, all curves exhibit the Breit-Wigner line shape
since all matrix elements (and hence also the interference) 
carry propagator factors of $1/D_W(q^2)$.
Importantly, the interference does not vanish when $W$ goes on shell.

\begin{figure}[!t]
\subfigure[]{\includegraphics[width=0.48\textwidth]{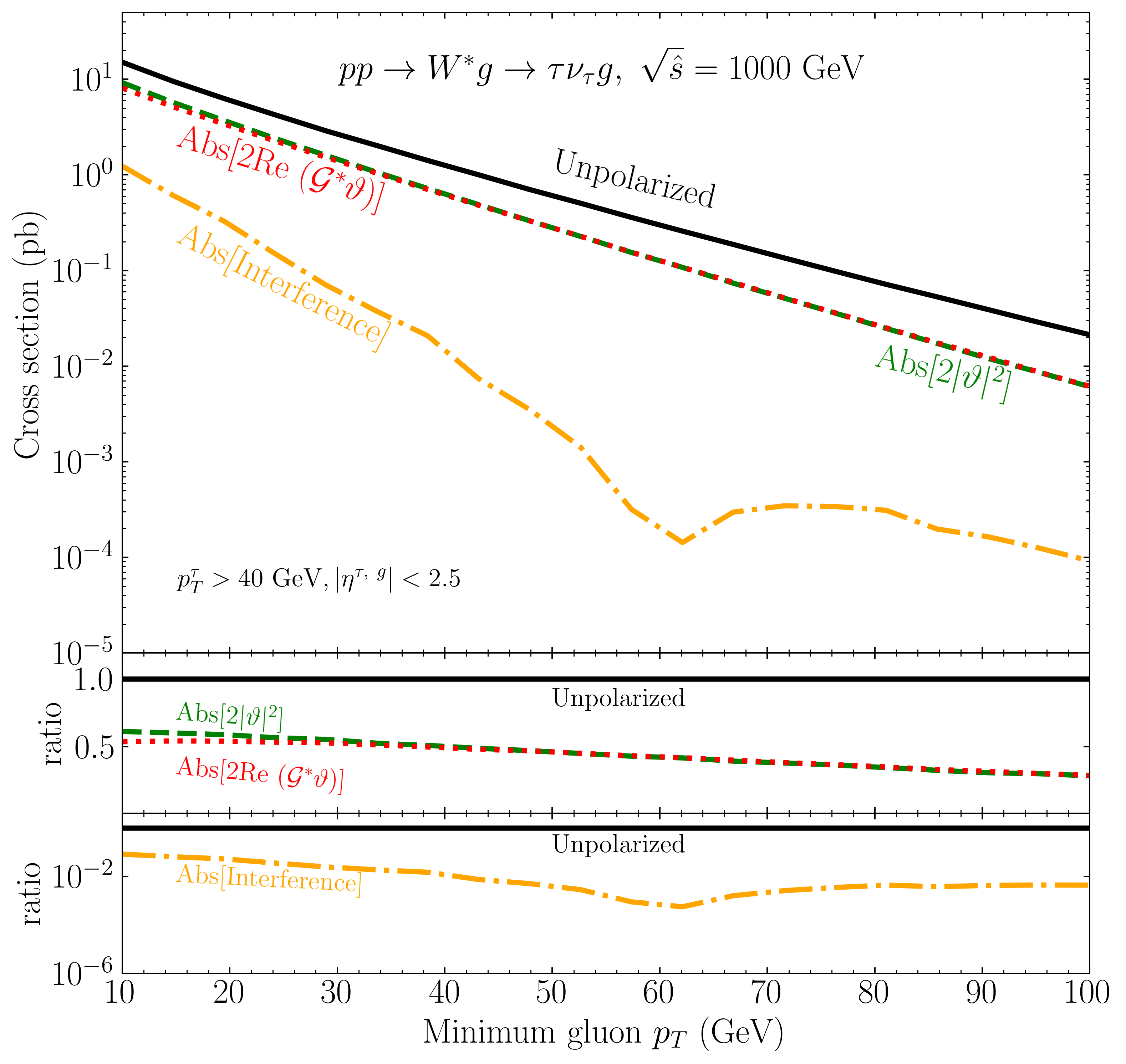}
\label{fig:Wjet_int1}}\hfill
\subfigure[]{\includegraphics[width=0.48\textwidth]{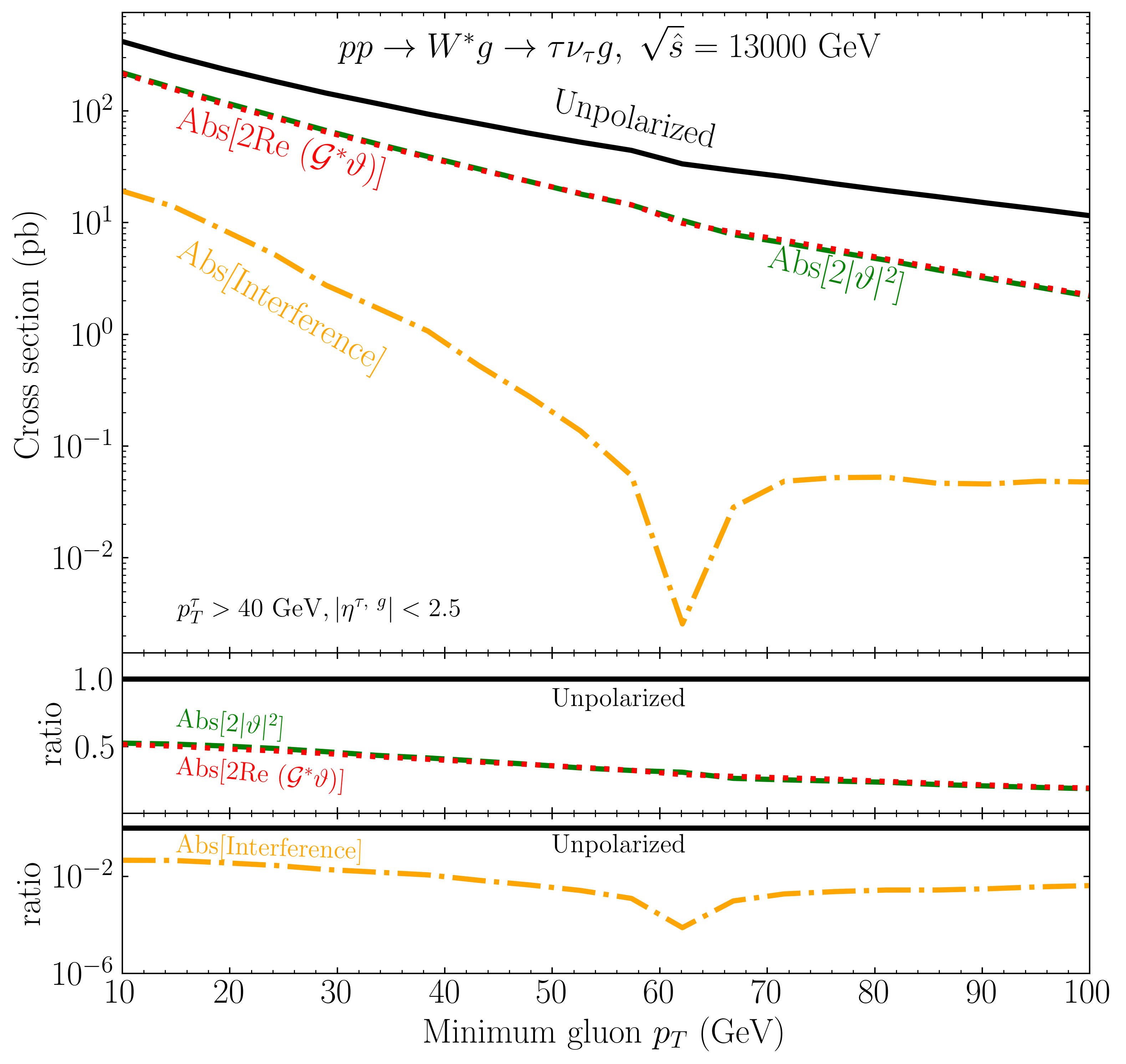}
\label{fig:Wjet_int2}}
\caption{Upper: 
For
(a) $\sqrt{s}=1\TeV$ and 
(b) $\sqrt{s}=13\TeV$,
the hadronic cross sections 
for the unpolarized process 
$pp\rightarrow W^{\pm (*)} g \to \tau^\pm \nu$
(solid)
as a function of the minimum gluon $p_T$,
as well as
the interference term $2|\vartheta|^2$  (dash),
the interference term $2\Re[\mathcal{G}^*\vartheta]$  (dot), and 
the total interference (dash-dot).
Middle and Lower: Ratio with respect to the unpolarized rate. 
}
\label{Wjet_int}
\end{figure}

Focusing on Fig.~\ref{ME1},
from low-to-high virtualities, the longitudinal (interference) 
contribution grows 
from about $\mathcal{O}(5\%)$ [$\mathcal{O}(10\%$] 
of the unpolarized case
to about $\mathcal{O}(40\%)$ [$\mathcal{O}(35\%$].
Below $\sqrt{q^2}\approx M_W$, 
the magnitude of the interference is larger than the 
longitudinal contribution while 
above $\sqrt{q^2}\approx M_W$
the longitudinal contribution is larger.
For our specific configuration, 
the size of the interference and 
longitudinal contribution are 
large, nearly equal, but have opposite signs and 
therefore and cancel strongly.

This similarity between the longitudinal polarization 
and polarization interference in Fig.~\ref{ME1} is not accidental.
The similarity is structural. 
As shown in Eq.~\eqref{eq:int_unitary_noq_w1g},
the interference scales with $\vartheta$, 
which is responsible for the $\mathcal{R}^{W+1g}_{\rm pol\ int}\sim1/\gamma_W^2$
scaling. 
A partonic center-of-mass energy 
of $\sqrt{\hat{s}}=250\GeV$ 
also does not induce a large boost 
to the $W^{(*)}$ 
system\footnote{We note that the $(\nu_\tau\tau^+)$ pair 
is the more physical system but our conclusions remain unchanged.}.
Meaning that leading 
$(q^2/E_W^2)$ and $(q^2/E_W^2)^2$ terms are comparable in size.
Numerically, for $\sqrt{\hat{s}}=250\GeV$ and 
virtuality $\sqrt{q^2}=60\GeV-100\GeV$, 
the energies and Lorentz boost factors 
carried by 
the $W^{+(*)}$ range $E_W\sim 130\GeV-145\GeV$ 
and $\gamma_W = E_W/\sqrt{q^2}\sim2.2-1.5$.
Complicating the matter is that spin correlation 
forces the $\tau^+$ lepton to be somewhat at rest 
in our configuration, with $E_\tau \sim m_\tau$.
This means that $1/\gamma_\tau=m_\tau/E_\tau$ factors 
are actually $\mathcal{O}(1)$ factors, 
and open several $(\nu_L\tau^+_L)$ terms in 
unpolarized and polarized squared matrix elements 
even though $(m_\tau/E_W)\ll1$.

In Fig.~\ref{ME2}, the partonic center-of-mass energy is increased
by fourfold. 
This causes $E_W/\sqrt{q^2}$ boost factors to increase 
by $(\gamma_W^{1000}/\gamma_W^{250})\sim 3.8\times$ to $3.5\times$ 
for $\sqrt{q^2}=60\GeV-100\GeV$.
Individual terms in longitudinal and interference 
contributions are then suppressed 
by at least a factor of $(\gamma_W^{1000}/\gamma_W^{250})^2\sim 10$,
with the longitudinal polarization decoupling more quickly
than the interference since $(\gamma_W^{1000}/\gamma_W^{250})^2\sim 150-200$.

In Figs.~\ref{ME3} and \ref{ME4}, we 
show the same curves as 
in Figs.~\ref{ME1} and \ref{ME2}
but for $\phi_\tau=0$.
This specific kinematical configuration 
forces a zero in the temporal component $(\beta=0)$
of the lepton current $J^\beta_{out}(\nu_{\tau L}\tau_R^+)$.
In other words, it forces $J^\beta_{out}(\nu_{\tau L}\tau_R^+)$ 
to be a purely transverse current.
Having no temporal component means 
that the current does not contribute to $\vartheta$
since $n_{\rm TL}\cdot J^\beta_{out}(\nu_{\tau L}\tau_R^+;\phi_g=0)=0$.
Consequentially, the entire amplitude for the 
longitudinal polarization and the polarization interference 
vanish for the $(\nu_{\tau L}\tau_R^+)$ configuration.
What survives is 
the $(\nu_{\tau L}\tau_R^+)$ configuration 
for the transverse polarization
and
the $(\nu_L\tau_L^+)$ configuration 
for the longitudinal polarization.
The latter is smaller by about $\mathcal{O}(10^{-3})$.
These are orthogonal helicity configurations 
and do not interfere.

To further explore the behavior 
of the polarized amplitudes 
and polarization interference,
we plot in Fig.~\ref{Wjet_int}
as a function of the minimum gluon $p_T$
the hadronic cross sections 
for the unpolarized process 
$pp\rightarrow W^{\pm (*)} g \to \tau^\pm \nu$
(solid)
for collider center-of-mass energy of 
(a) $\sqrt{s}=1\TeV$ and 
(b) $\sqrt{s}=13\TeV$.
Here, we sum over quark flavors and charge configurations,
keeping $m_\tau\neq0$.
To regulate infrared poles and reflect realistic detector thresholds,
we impose the following gluon and $\tau$ lepton rapidity cuts
and $\tau$ lepton $p_T$ cut
\begin{align}
    |\eta^{g,\tau}|<2.5\ \quad\text{and}\quad p_{T}^\tau> 40\GeV\ .
\end{align}

Our aim is to study the behavior of the interference
and its impact on LHC analyses.
Therefore, we show in  Fig.~\ref{Wjet}
(i) the $2|\vartheta|^2$ term (dash),
(ii) the $2\Re[\mathcal{G}^*\vartheta]$ term (dot), and 
(iii) the total interference (dash-dot).
These terms correspond to the contributions 
in Eq.~\eqref{eq:int_unitary_noq_w1g}
integrated over phase space.
Note that term (i) is a proxy for the longitudinal contribution.
In the middle panel, we show the ratio of 
terms (i) and (ii) relative to the total unpolarized rate.
In the lower panel, we show the ratio of 
the total interference relative to the total unpolarized rate.
Since interference is negative, 
we plot the absolute values of quantities.

For low (high) values of $p^g_{T\rm min}$ value
term (i) is slightly larger (smaller) than term (ii).
For the lowest $p^g_{T\rm min}$, individual 
interference terms reach $\mathcal{O}(50\%)$ 
of the unpolarized rate, but drop below $\mathcal{O}(10\%)$ 
for $p^g_{T\rm min}\gtrsim100\GeV$. 
The net longitudinal contribution is about half these values.
Importantly, for all $p^g_{T\rm min}$, 
the cancellation between (i) and (ii) is sizable,
leading to sub-percent interference,
$\mathcal{I}_{\rm pol}^{W+1g}\lesssim\mathcal{O}(1\%)$,
in accordance with Eq.~\eqref{eq:non_interference_integral_phi}. 

In practice, polarization interference in $W$+jets is small 
because TeV-scale collisions are very energetic compared to the masses of SM particles.
At TeV-scale colliders, 
weak bosons are produced with considerable transverse and longitudinal momenta.
They carry a lot of energy, even if slightly off shell.
$\mathcal{I}_{\rm pol}^{W+1g}(\lambda_\nu\lambda_\tau)$ in 
Eq.~\eqref{eq:int_unitary_noq_w1g_full} shows that 
polarization interference is tied to helicity inversion, 
which is evidenced by mass-over-energy factors.
However, matrix elements  for longitudinally polarized weak bosons
inherently carry mass-over-energy factors 
when coupling to SM fermions via SM interactions
[see, e.g., Eq.~\eqref{eq:w1g_current_projection_leptons}],
and therefore become strongly suppressed in the high-energy limit.

From Eq.~\eqref{eq:int_unitary_noq_w1g},
there is no uncanceled remainder from the Goldstone boson or scalar polarization.
Subsequently, the absence of $\lambda=0$ or $\lambda=T$ contributions 
for a fixed helicity configuration for $\nu_\tau$ and $\tau^+$ 
forces polarization interference for $W+1g$ to vanish.
This is the case here:
$\lambda=T$ amplitude is driven 
by the $(\nu_{\tau L}\tau_R^+)$ helicity configuration
while the $\lambda=0$ amplitude  
by the $(\nu_{\tau L}\tau_R^+)$ helicity configuration.

\subsection{\texorpdfstring{$W$}{W} 
Polarization in Top Quarks Decays}
\label{sec:top}

\begin{figure}[t!]
\begin{center}
\includegraphics[width=\textwidth]{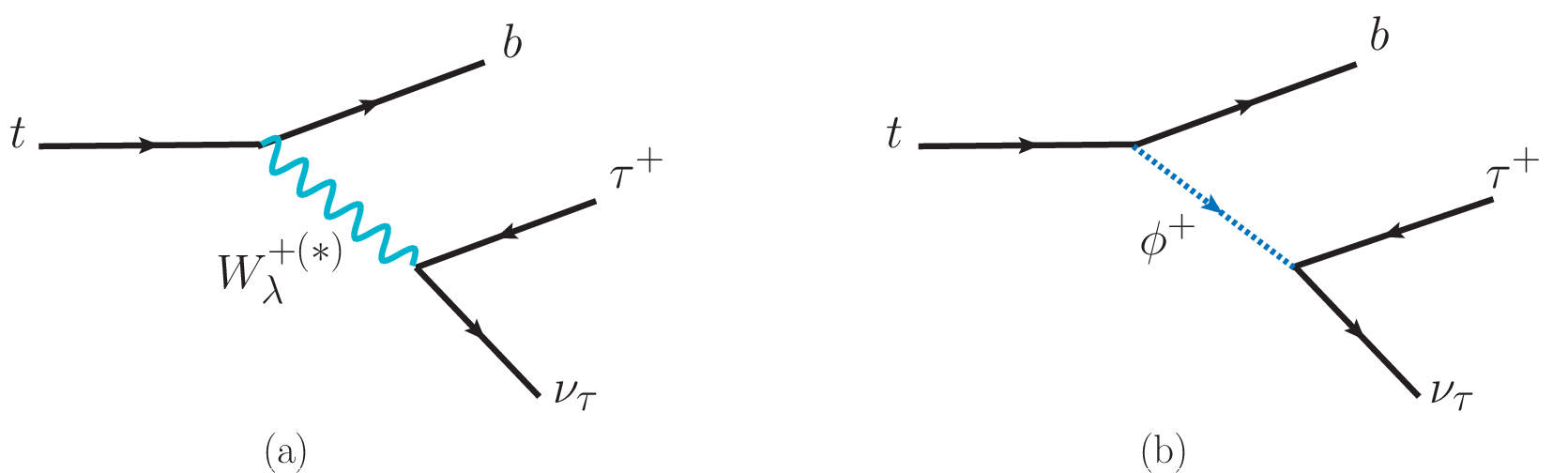}
\end{center}
\caption{Top quark decay to the $b\tau^+\nu_\tau$ system via a
(a) $W$ boson and (b) Goldstone boson.}
\label{fig:topdecay}
\end{figure}

Due to its heavy mass and short lifetime, 
the top quark can be produced resonantly and decay 
to on-shell $W$ bosons before undergoing hadronization.
For $W$s that decay to massive $\tau$ leptons,
as illustrated in Fig.~\ref{fig:topdecay}(a),
all elements of the longitudinal propagator become accessible.
This in contrast to the Drell-Yan (Sec.~\ref{sec:dy}) 
and $W+$jets (Sec.~\ref{sec:w1g}) processes, 
where $\mathcal{O}(q_\mu q_\nu)$ 
and $\mathcal{O}(q_\mu n_\nu, n_\mu q_\nu)$ terms 
in the longitudinal and scalar propagators vanish.
In the $R_\xi$ gauge, this also means that Goldstone exchanges, 
shown in Fig.~\ref{fig:topdecay}(b), also open.

In this section we apply our power counting  
to a situation 
where scalar and Goldstone contributions are 
both non-vanishing.
Following our strategy in Sec.~\ref{sec:interference_strategy},
the unpolarized, Goldstone, and polarized amplitudes  
in the $R_\xi$ gauge are
\begin{subequations}
\begin{align}
-i\mathcal{M}_{\rm unpol}\ =&\  -\mathcal{G} - \mathcal{Q}_\xi
\nonumber\\
= \frac{-ig^2}{2D_W(q^2)}&
\left[-J^\alpha_{tb}~g_{\alpha\beta}J^\beta_{\tau\nu}
-\frac{(\xi-1)}{D_W(q^2,\xi)}
(m_t J^R_{tb} - m_b J^L_{\tau\nu})(m_\nu J^L_{\tau\nu} -m_\tau J^R_{\tau\nu})
\right] ,
\label{UnPol}
\\
-i\mathcal{M}_{\rm Gold}\ =&\ \Xi
\nonumber\\
=& \frac{-ig^2}{2 D_W(q^2,\xi)} \frac{1}{\tilde{M}_W^2}  
\left(m_t J^R_{tb} - m_b J^L_{tb}\right)
\left(m_\nu J^L_{\tau\nu} - m_\tau J^R_{\tau\nu}\right)\ ,
\label{Goldstone}
\\
-i\mathcal{M}_{\lambda=T}\ =&\  -\mathcal{G} - \vartheta\ =\ \varphi
\nonumber\\
=& \frac{-ig^2}{2 D_W(q^2)}\ 
J^\alpha_{tb}~\Phi_{\alpha\beta}\ J^\beta_{\tau\nu}\ , 
\label{Trans}
\\
-i\mathcal{M}_{\lambda=0}\ =&\ \vartheta + \frac{\mathcal{Q}}{q^2}
\nonumber\\
=& \frac{-ig^2}{2 D_W(q^2)}
\left[J^\alpha_{tb} \Theta_{\alpha\beta}J^\beta_{\tau\nu}+\frac{1}{q^2}(m_t J^R_{tb} - m_b J^L_{\tau\nu})(m_\nu J^L_{\tau\nu} -m_\tau J^R_{\tau\nu})\right],
\label{LongME}
\\
-i\mathcal{M}_{\lambda=S}\ =&\ -\frac{\mathcal{Q}}{q^2} - \mathcal{Q}_\xi
\nonumber\\
=& \frac{-ig^2}{2 D_W(q^2)}
\left[\frac{-1}{q^2}-\frac{(\xi-1)}{D_W(q^2,\xi)}\right]
(m_t J^R_{tb} - m_b J^L_{\tau\nu})(m_\nu J^L_{\tau\nu} -m_\tau J^R_{\tau\nu})  .
\label{ScalarME}
\end{align}
\end{subequations}
Since we are working in the SM
we neglect neutrino masses $m_\nu$.
However, we write them here to show that 
the fully massive case does not 
significantly complicate our work.

For a particular helicity combination 
the incoming $t(p_t)\to b(p_b)$ vector current $J_{tb}^\alpha$ 
and outgoing $\tau^+(p_\tau)\nu_\tau(p_\nu)$ 
vector current $J_{\tau\nu}^\beta$ listed above are given by
\begin{align}
\label{eq:top_current_def}
J^\alpha_{tb} =  
\bar{u}(p_b,\lambda_b) \gamma^\alpha P_L u(p_t,\lambda_t) 
\quad\text{and}\quad
J^\beta_{\tau\nu} = 
\bar{u}(p_\nu,\lambda_\tau)\gamma^\beta P_L v(p_\tau,\lambda_\nu)\ .
\end{align}    
Via the Dirac equation, 
the contraction of these currents 
with the exchange momentum $q=(p_t-p_b)=(p_\nu+p_b)$
can be written in terms of the scalar currents $J^{L/R}_{tb}, J^{L/R}_{\tau\nu}$,
\begin{align}
    q_\alpha J^\alpha_{tb} = (m_t J^R_{tb} - m_b J^L_{\tau\nu})\ 
    &\quad\text{with}\quad\
    J^{L/R}_{tb} =\bar{u}_{\lambda_b}(p_b) P_{L/R} u_{\lambda_t}(p_t)\ , 
    \\
    q_\beta J^\beta_{\tau\nu} = (m_\nu J^L_{\tau\nu} -m_\tau J^R_{\tau\nu})\ 
    &\quad\text{with}\quad
    J^{L/R}_{\tau\nu} = \bar{u}_{\lambda_\nu}(p_\nu)P_{L/R} v_{\lambda_\tau}(p_\tau)\ .
\end{align}
Expressions for propagators $D_W(q^2)$ and $D_W(q^2,\xi)$ 
are given in Eq.~\eqref{eq:polprop_def}.

For the unpolarized case, 
the two amplitudes for Fig.~\ref{fig:topdecay},
which correspond to $\mathcal{M}_{\rm unpol}$ and $\mathcal{M}_{\rm Gold}$
respectively, 
must be summed to eliminate dependence on $\xi$ and obtain a gauge-invariant result.
For the polarized case, 
only the scalar and Goldstone contributions 
carry $\xi$ dependence.
Adding the amplitudes for the $\lambda=S$ and $\lambda=G$ amplitudes, one obtains 
\begin{align}
-i\mathcal{M}_{\lambda=S}&-i\mathcal{M}_{\rm Gold} =\ \frac{-ig^2}{2} 
\left(m_t J^R_{tb} - m_b J^L_{tb}\right)\left(m_\nu J^L_{\tau\nu} - m_\tau J^R_{\tau\nu}\right)
\nonumber\\
&\quad\times
\left[-\frac{1}{D_W(q^2)}\frac{1}{q^2}- \frac{1}{D_W(q^2)}\frac{(\xi-1)}{D_W(q^2,\xi)}
+\frac{1}{D_W(q^2,\xi)}\frac{1}{\tilde{M}_W^2}\right]
\\
&=\ \frac{-ig^2}{2} \left(m_t J^R_{tb} - m_b J^L_{tb}\right)
\left(m_\nu J^L_{\tau\nu} - m_\tau J^R_{\tau\nu}\right) 
\frac{1}{D_W(q^2)} \left[-\frac{1}{q^2}+\frac{1}{\tilde{M}_W^2}\right] ,
\label{ScalarGoldston}
\end{align}
which is independent of $\xi$.
In the Unitary gauge, 
the scalar amplitude is
\begin{align}
-i\mathcal{M}(\lambda=S)\Big\vert^{\rm Unitary} &=
\frac{-ig^2}{2 D_W(q^2)}
(m_t J^R_{tb} - m_b J^L_{\tau\nu})(m_\nu J^L_{\tau\nu} -m_\tau J^R_{\tau\nu}) 
\left[\frac{-1}{q^2}+\frac{1}{\tilde{M}_W^2}\right] 
\\
&=\ 
-i\mathcal{M}_{\lambda=S}\Big\vert^{R_\xi}-i\mathcal{M}_{\rm Gold}\Big\vert^{R_\xi}\ ,
\end{align}
in agreement with Eq.~\eqref{eq:ward_gauge_inv} 
and a manifestation of EW Ward identities.
The cancellation of $\xi$-terms highlights 
the relationship between Goldstone bosons and 
the scalar polarization state, and 
therefore the importance of treating gauge-fixing scalars on the same 
footing in practical calculations for polarization with massive external particles.

Turing to polarization interference, 
the leading behavior can be understood entirely 
from the $(tb)$ quark current itself without reference to the 
kinematics of the outgoing leptons.
We simplify the picture by taking 
the $b$ and $\tau$ massless and 
working in the top's rest frame.
Using Eq.~\eqref{eq:top_current_def},
the incoming currents 
for the top's two helicities are
\begin{subequations}
\begin{align}
J^\alpha(t_Lb_L)\ &=\  \sqrt{2 m_t E_b}
\left[\cos\frac{\theta_b}{2},e^{i\phi_b}\sin\frac{\theta_b}{2},
-ie^{i\phi_b}\sin\frac{\theta_b}{2},\cos\frac{\theta_b}{2}\right]\ ,
\label{Jin_topLL}
\\
J^\alpha(t_Rb_L)\ &=\  \sqrt{2 m_t E_b}
\left[-e^{i\phi_b}\sin\frac{\theta_b}{2},-\cos\frac{\theta_b}{2},
-i\cos\frac{\theta_b}{2},e^{i\phi_b}\sin\frac{\theta_b}{2}\right]\ ,
\label{Jin_topRL}
\end{align}    
\end{subequations}
and assume the following momentum assignments:
\begin{subequations}
\label{eq:top_kinematics_def}
\begin{align}
    p_t^\alpha\ &=\ (m_t,\vec{0})\ ,
    \quad 
     p_b^\alpha\ =\ \left(E_b, E_b \sin\theta_b \cos\phi_b, 
     E_b \sin\theta_b \sin\phi_b, E_b \cos\theta_b\right),\\
     q^\alpha\ &=\ p_t^\alpha-p_b^\alpha\ ,
     \quad
    n^\alpha_{\rm LL}\ =\ (1,-\hat{q}) = \left(1,  \sin\theta_b \cos\phi_b,  \sin\theta_b \sin\phi_b, \cos\theta_b\right)\ ,
    \\
    E_W\ &=\ \frac{m_t}{2}\left(1+\frac{q^2}{m_t^2}\right)\ ,
     \quad
    E_b\ =\ \vert\vec{q}\vert= \frac{m_t}{2}\left(1-\frac{q^2}{m_t^2}\right)\ .
\end{align}
\end{subequations}

For massless $\tau$ and $\nu$, 
outgoing current $J_{\tau\nu}^\beta$ is conserved
and the following hold:
\begin{align}
    \label{qJin_tb}
    q_\beta \cdot J^\beta_{\tau\nu} = 0
    \quad\text{and}\quad
    n_{\rm LL\beta} \cdot J^\beta_{\tau\nu} = 
    \left(\frac{E_W+\vert\vec{q}\vert}{\vert\vec{q}\vert}\right)
    n_{\rm TL\beta}\cdot J^\beta_{\tau\nu} = 
    \frac{m_t}{E_b}J^{\beta=0}_{\tau\nu}
\end{align} 
In the second expression we used the reference vector identities 
in Eq.~\eqref{eq:refvector_ident}.
Now, since $b$ is massless and $t$ is at rest, 
the current contraction $J_{tb}^\alpha\cdot q_\alpha$ simplifies.
In addition, for both top helicities, the incoming current $J_{tb}^\alpha$ 
is orthogonal to the light-like reference vector,
\begin{align}
J^\alpha_{tb}\cdot q_\alpha 
= J^\alpha_{tb} \cdot (p_{t\alpha} -p_{b\alpha}) = m_t J^0_{tb}
\quad\text{and}\quad
 J^\alpha_{tb} \cdot n_{\rm LL\alpha} = 0\ .   
\end{align}

Using these relationships
and defining $J_{\tau\nu}^\pm \equiv  (J^1_{\tau\nu} \pm iJ^2_{\tau\nu})/2$, 
the matrix elements 
for an unpolarized, longitudinal, and transverse $W$ 
and the $t_Lb_L$ helicity combination are
\begin{subequations}
\begin{align}
    -i\mathcal{M}_{\rm unpol}(t_L b_L) = &\ \frac{ig^2}{2}\frac{(-1)^2}{D_W(q^2)}\
    J^\alpha_{tb} (t_Lb_L) \  g_{\alpha\beta} \ J^\beta_{\tau\nu}
    \nonumber \\
        = &\ \frac{ig^2}{2}\frac{\sqrt{2 m_t E_b}}{D_W(q^2)} 
        \left[\cos\frac{\theta_b}{2}
        (J^0_{\tau\nu}-J^3_{\tau\nu}) 
        - \ 2 e^{i \phi_b} \sin\frac{\theta_b}{2}\ J^-_{\tau\nu}\ ,
        \right]\ ,
    \\
    -i\mathcal{M}_{\lambda=0}(t_L b_L) = & \ 
    \frac{ig^2}{2}\frac{(-1)^2}{D_W(q^2)}\ 
    \frac{1}{n_{\rm LL}\cdot q}\
    J^\alpha_{tb}(t_Lb_L)\  
    q_\alpha n_{\rm LL\beta}\
    J^\beta_{\tau\nu}
    \nonumber\\ 
    =&\ \frac{ig^2}{2}\frac{1}{D_W(q^2)} \ 
    \frac{2m_t^2}{\sqrt{2 m_t E_b}}\
    \cos\frac{\theta_b}{2} \ J^0_{\tau\nu}\ ,
    \\
    -i\mathcal{M}_{\lambda=T}(t_L b_L)  = & \ 
    -i\mathcal{M}_{\rm unpol}(t_L b_L) +i\mathcal{M}_{\lambda=0}(t_L b_L) 
    \nonumber\\ 
    = \  \frac{ig^2}{2}\frac{\sqrt{2 m_t E_b}}{D_W(q^2)} 
    &\left\{
    \cos\frac{\theta_b}{2}
    \left[\left(1-\frac{m_t}{E_b}\right)J^0_{\tau\nu} -J^3_{\tau\nu}\right] 
    - \ 2 e^{i \phi_b} \sin\frac{\theta_b}{2}\ J^-_{\tau\nu}
    \right\} .
\end{align}
\end{subequations}
For the transverse amplitude, we use completeness for clarity in intermediate steps.
Noting that $E_V J^0_{\tau\nu} 
= \vert \vec{q}\vert \hat{q}\cdot \vec{J}_{\tau\nu}
= -\vec{p}_b \cdot \vec{J}_{\tau\nu}$,
it is easy to confirm that this expression is equivalent 
to the one obtained from using $\varphi$.
For the $t_Rb_L$ helicity configuration, we similarly obtain
\begin{subequations}
\begin{align}
    -i\mathcal{M}_{\rm unpol}(t_R b_L) &=
    \frac{ig^2}{2}\frac{\sqrt{2m_tE_b}}{D_W(q^2)} 
    \left[-e^{i\phi_b}\sin\frac{\theta_b}{2}(J^0_{\tau\nu}+J^3_{\tau\nu}) 
    +2\cos\frac{\theta_b}{2}\ J^+_{\tau\nu} 
    \right], \label{TopUnpol_LL}
    \\
     -i\mathcal{M}_{\lambda=0} (t_Rb_L) &= 
     \frac{ig^2}{2}\frac{1}{D_W(q^2)} 
     \frac{2 m_t^2}{\sqrt{2m_t E_b}}
     \left(-e^{i\phi_b} \sin\frac{\theta_b}{2}\right) J^0_{\tau\nu}  , \label{TopLongLL}
     \\
     -i\mathcal{M}_{\lambda=T}(t_Rb_L) &= 
     \frac{ig^2}{2}\frac{\sqrt{2m_tE_b}}{D_W(q^2)}
     \nonumber\\
     &\times
     \left\{-e^{i\phi_b} \sin\frac{\theta_b}{2}
     \left[\left(1-\frac{m_t}{E_b}\right)J^0_{\tau\nu} +J^3_{\tau\nu}\right] 
     +2\cos\frac{\theta_b}{2}J^+_{\tau\nu} \right\}\ .\label{TopTransLL}
\end{align}
\end{subequations}

For each top quark helicity, the unintegrated polarization interference is then
\begin{subequations}
\label{eq:polint_top_nolep}
\begin{align}
\mathcal{I}_{\rm pol}^{t\to b\nu_\tau\tau^+}(t_Lb_L)\ 
&=\ 2\Re\left[\mathcal{M}_{\lambda=0}^*(t_Lb_L)\mathcal{M}_{\lambda=T}(t_Lb_L)\right], 
\\
\mathcal{I}_{\rm pol}^{t\to b\nu_\tau\tau^+}(t_Rb_L)\ 
&=\ 2\Re\left[\mathcal{M}_{\lambda=0}^*(t_Rb_L)\mathcal{M}_{\lambda=T}(t_Rb_L)\right], 
\end{align}    
\end{subequations}
where the products of polarized amplitudes are given by
\begin{subequations}
\label{eq:polint_top_nolep_me}
\begin{align}
\mathcal{M}_{\lambda=0}^*(t_Lb_L)&\mathcal{M}_{\lambda=T}(t_Lb_L) =\ 
\frac{g^4 m_t^2}{2\vert D_W(q^2)\vert^2}
\nonumber\\
\times\ &
    \left\{
    \cos^2\frac{\theta_b}{2}\left[\left(1-\frac{m_t}{E_b}\right)\vert J^0_{\tau\nu}\vert^2 -J^3_{\tau\nu}(J^0_{\tau\nu})^*\right] 
    - \  e^{i \phi_b} \sin\theta_b\ J^-_{\tau\nu}(J^0_{\tau\nu})^* 
    \right\}\ ,
\\
\mathcal{M}_{\lambda=0}^*(t_Rb_L)&\mathcal{M}_{\lambda=T}(t_Rb_L) =\ 
\frac{g^4 m_t^2}{2\vert D_W(q^2)\vert^2}
\nonumber\\
\times\ &
\left\{\sin^2\frac{\theta_b}{2}
     \left[\left(1-\frac{m_t}{E_b}\right)\vert J^0_{\tau\nu}\vert^2 
     +J^3_{\tau\nu}(J^0_{\tau\nu})^*\right] 
     - \  e^{-i \phi_b}\sin\theta_b J^+_{\tau\nu}(J^0_{\tau\nu})^* 
     \right\}\ .
\end{align}
\end{subequations}

Importantly, for decays of unpolarized top quarks, 
one sums over both top helicities.
The helicities of $b$, $\nu_\tau$, and $\tau^+$ are fixed since they are massless.
And since the real-part operator is linear, 
we can sum directly the product of polarized amplitudes, giving
\begin{align}
\label{eq:noninterference_top_nokin}
\mathcal{I}_{\rm pol}^{t\to b\nu_\tau\tau^+}(t_Lb_L&+t_Rb_L)\ 
=\ 2\Re\left[
\sum_{\lambda_t=L,R}
\mathcal{M}_{\lambda=0}^*(t_{\lambda_t}b_L)\mathcal{M}_{\lambda=T}(t_{\lambda_t}b_L)
\right]\ , \quad\text{where}
\\
\sum_{\lambda_t=L,R}
\mathcal{M}_{\lambda=0}^*(t_{\lambda_t}b_L)&\mathcal{M}_{\lambda=T}(t_{\lambda_t}b_L) =\ 
\frac{g^4 m_t^2}{2\vert D_W(q^2)\vert^2}\ 
\Big[
    \left(1-\frac{m_t}{E_b}\right)\vert J^0_{\tau\nu}\vert^2 
\nonumber\\
 &  - \ \cos\theta_b J^3_{\tau\nu}(J^0_{\tau\nu})^*
\nonumber\\ 
 &  - \  \sin\theta_b\cos\phi_b\ J^1_{\tau\nu}(J^0_{\tau\nu})^* 
    - \  \sin\theta_b\sin\phi_b\ J^2_{\tau\nu}(J^0_{\tau\nu})^* 
    \Big]\ ,
\\
&=\ \frac{g^4 m_t^2}{2\vert D_W(q^2)\vert^2 E_b}\ 
\left[
    \left(E_b-m_t\right)\ J^0_{\tau\nu}
    - \vec{p}_b\cdot \vec{J}_{\tau\nu}
    \right]\ (J^0_{\tau\nu})^*
=\ 0\ ,
\end{align}
with the last line following from current conservation,
$- \vec{p}_b\cdot \vec{J}_{\tau\nu} = \vec{q}\cdot \vec{J}_{\tau\nu} = E_V J^0_{\tau\nu}$.
In other words, the polarization interference for unpolarized top quarks that decay 
to massless fermions is zero at lowest order, 
independent of decay kinematics.
Moreover, this cancellation does not require $\sqrt{q^2}\approx M_W$ 
and is independent of the cancellations in Sec.~\ref{sec:noninterference}.

Despite the polarization interference being zero
in decays of unpolarized top quarks to massless fermions, 
we are still interested in finding the size of the interference 
for individual polarizations of the top.
To quantify this, we now evaluate the outgoing current $J^\beta_{\tau\nu}$.

To build $J^\beta_{\tau\nu}$ in the top's rest frame,
we note that the volume element for a three-body phase space
can always be decomposed into a convolution of two-body phase spaces,
\begin{align}
    dPS_3(t\rightarrow b \tau^+ \nu_\tau) = 
    dPS_2 (t\rightarrow b W^*)
    \times 
    dPS_2 (W^*\rightarrow \tau^+ \nu_\tau)
    \times 
    \frac{dq^2}{2\pi}\ .
\end{align}
This allows us to disentangle the angular variables entering 
the lepton current $J^\beta_{\tau\nu}$ 
from those entering the quark current $J^\alpha_{tb}$.

The leptonic current in the rest frame of $W(q)$,
denoted by $J^\beta_{\tau\nu}\vert_{W^*}$
and given in Eq.~\eqref{eq:dy_current_outgoing_LR},
is then related to $J^\beta_{\tau\nu}$ in the top's frame by 
a boost along the spin axis of $W(q)$ (the $\hat{z}$-direction)
and then a pair of rotations (defined by $\theta_b,\phi_b$).
For a Lorentz factor $\gamma = E_W/\sqrt{q^2} = 1/\sqrt{1-\beta^2}$
with $E_W$ given in Eq.~\eqref{eq:top_kinematics_def},
the boosted leptonic current is
\begin{align}
J_{\tau\nu}^\beta\big\vert_{\rm boost} =\  
[\Lambda(\gamma) \cdot J_{\tau\nu}\big\vert_{W^*}]^\beta\qquad\phantom{12} &
\\
= 2\sqrt{E_\nu E_\tau}\ \big[-\beta\gamma\sin\theta_\tau,\ 
&\cos\theta_\tau\cos\phi_\tau +i\sin\phi_\tau,\ 
\nonumber\\
&\cos\theta_\tau\sin\phi_\tau -i\cos\phi_\tau,\ -\gamma\sin\theta_\tau\big]\ 
\\
\equiv\ (J^0_B, J^1_B,J^2_B,J^3_B)\ .\qquad &
\end{align}
Again, $\theta_\tau$ and $\phi_\tau$ are defined in the rest frame of $W(q)$.
And after rotating, one has
\begin{align}
      J_{\tau\nu}^\beta\big\vert_{\rm top}\ =\ 
      & \left[R(z,\phi_b+\pi) 
      \cdot  R(y,\pi-\theta_b) 
      \cdot  J_{\tau\nu}\big\vert_{\rm boost}\right]^\beta\\
    = & 
\begin{bmatrix}
    J^0_B \\
    \cos\phi_b(J^1_B \cos\theta_b -J^3_B \sin\theta_b) + J^2_B \sin\phi_b\\
    -J^2_B \cos\phi_b + (J^1_B\cos\theta_b - J^3_
    B\sin\theta_b) \sin\phi_b\\
    -J^3_B\cos\theta_b -J^1_B\sin\theta_b
\end{bmatrix}\ .
 \end{align}

Including the lepton decay current in Eq.~\eqref{eq:polint_top_nolep}, 
the polarization interference is 
\begin{subequations}
\label{eq:polint_top_lep}
\begin{align}
\mathcal{I}_{\rm pol}^{t\to b\nu_\tau\tau^+}(t_Lb_L)\ 
&=\ 
\frac{g^4 m_t(m_t^2-q^2)\sqrt{q^2}}{4\vert D_W(q^2)\vert^2}
\sin\theta_b\ \cos\phi_\tau\ \sin\theta_\tau\ (1-\cos\theta_\tau)\ ,\
\\
\mathcal{I}_{\rm pol}^{t\to b\nu_\tau\tau^+}(t_Rb_L)\ 
&=\ -\mathcal{I}_{\rm pol}^{t\to b\nu_\tau\tau^+}(t_Lb_L)\ ,
\end{align}    
\end{subequations}
and we recover vanishing interference for unpolarized top quarks 
at the unintegrated level,
$\mathcal{I}_{\rm pol}^{t\to b\nu_\tau\tau^+}(t_Lb_L+t_Rb_L)=0$.
Notably, the interference vanishes as $q^2\to0$, 
which is consistent with massless vector bosons having no longitudinal polarization.

Following the same procedure as above, 
the squared matrix element for unpolarized $W$ bosons from 
unpolarized top quarks is
\begin{align}
 \vert\mathcal{M}_{\rm unpol}(t_L b_L)\vert^2 
 +   
 \vert\mathcal{M}_{\rm unpol}(t_R b_L)\vert^2 
  =\ & 
  \frac{g^4 m_t E_b}{2\vert D_W(q^2)\vert^2}\ 
  \left(\vert J^0_B+J^3_B\vert^2+\vert J^1_B-iJ^2_B\vert ^2\right) 
  \\
  =\ & 
  \frac{g^4 (m_t^2-q^2)}{4\vert D_W(q^2)\vert^2}\ 
  \left[m_t^2\sin^2\theta_\tau+q^2 (1-\cos\theta_\tau)^2\right]\ . 
  \label{eq:me2_top_unpolarized}
\end{align}
As a check, we recover the usual expression for the $1\to3$-body
partial decay width:
\begin{align}
\label{eq:top_decay_width_unintegrated}
    \frac{d\Gamma(t\to b\nu_\tau\tau^+)}{dq^2}\ =\ 
    &\frac{g^4 m_t^3}{2^{10}\ 3\ \pi^3 \vert D_W(q)\vert^2}
    \left(1-\frac{q^2}{m_t^2}\right)^2 
    \left(1+\frac{2q^2}{m_t^2}\right)\ 
    \\
     \overset{\rm NWA}{=} 
    &\frac{g^4 m_t^3}{2^{10}\ 3\ \pi^2\ M_W\ \Gamma_W}
    \left(1-\frac{q^2}{m_t^2}\right)^2 
    \left(1+\frac{2q^2}{m_t^2}\right)\ \delta(q^2-M_W^2)\ .
\end{align}
In the last line we used the narrow width approximation (NWA) to simplify $\vert D_W(q^2)\vert^2$.

\begin{figure}[!t]
\subfigure[]{\includegraphics[width=0.48\textwidth]{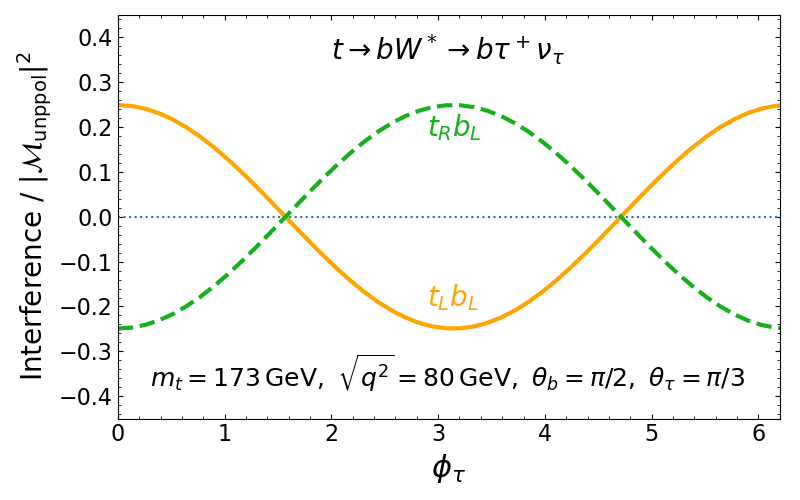}
\label{Int_phi80}}\hfill
\subfigure[]{\includegraphics[width=0.48\textwidth]{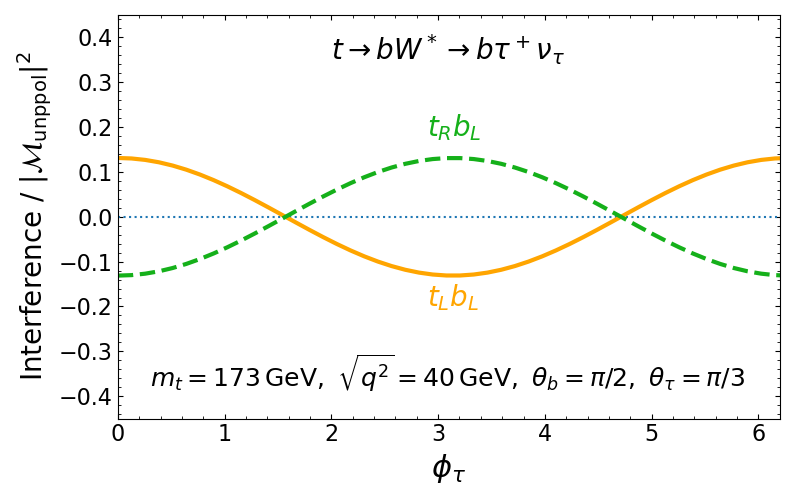}
\label{Int_phi40}}
\\
\subfigure[]{\includegraphics[width=0.48\textwidth]{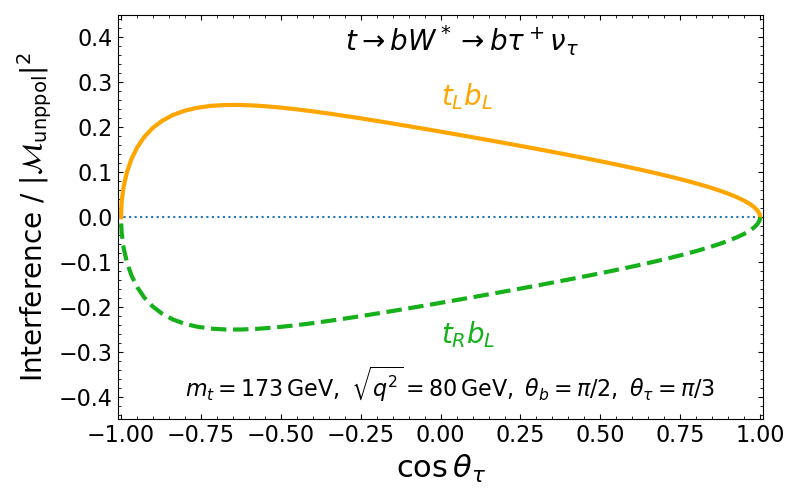}
\label{Int_theta80}}\hfill
\subfigure[]{\includegraphics[width=0.48\textwidth]{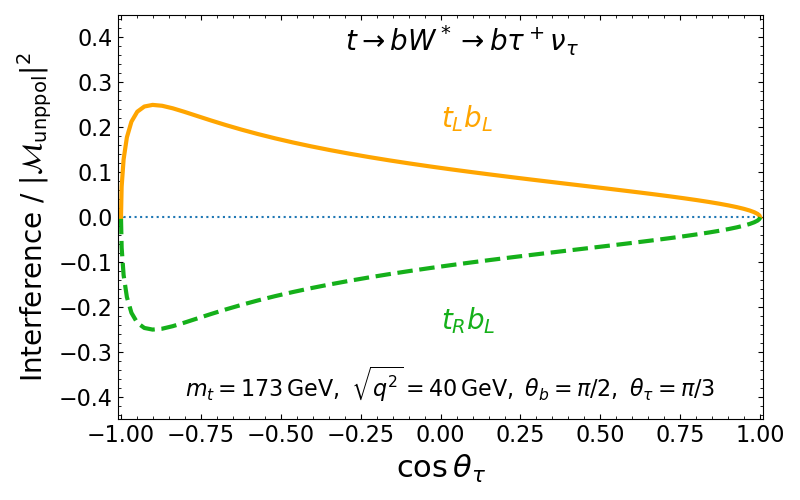}
\label{Int_theta40}}
\caption{Polarization in nterference ratios 
$\mathcal{R}^{t\to b\nu_\tau\tau^+}(t_Lb_L)$ (light curve)  and 
$\mathcal{R}^{t\to b\nu_\tau\tau^+}(t_Rb_L)$ (dark curve) 
as defined in Eq.~\eqref{eq:polint_ratio_top}
for top quark decays as a function of (a,b) $\phi_\tau$ and (c,d) $\cos\theta_\tau$
and $W$ virtuality (a,c) $\sqrt{q^2}=80\GeV$ and (b,d) $\sqrt{q^2}=40\GeV$.}
\label{Top_int_plots}
\end{figure}

Using the polarization interference in Eq.~\eqref{eq:polint_top_lep}
and the unpolarized squared matrix element in Eq.~\eqref{eq:me2_top_unpolarized},
the degree that polarization interference in top decays
cancels LH and RH top helicities can be quantified by the ratio 
\begin{align}
\mathcal{R}^{t\to b\nu_\tau\tau^+}(t_{\lambda_t}b_L)\ &\equiv\ 
\frac{\mathcal{I}_{\rm pol}^{t\to b\nu_\tau\tau^+}(t_{\lambda_t}b_L)}{
\vert\mathcal{M}_{\rm unpol}(t_L b_L)\vert^2 
 +   
 \vert\mathcal{M}_{\rm unpol}(t_R b_L)\vert^2
}
\\
&=\ 
(-2\lambda_t)\times
\frac{m_t\sqrt{q^2}\ \sin\theta_b\ \cos\phi_\tau\ \sin\theta_\tau\ (1-\cos\theta_\tau)}
{\left[m_t^2\sin^2\theta_\tau+q^2 (1-\cos\theta_\tau)^2\right]}\ ,
\label{eq:polint_ratio_top}
\end{align}
where $-2\lambda_t=\pm1$ for $t_L\ (t_R)$.
For representative inputs, we plot these ratios in Fig.~\ref{Top_int_plots}.

We find that the interference for individual helicity configurations 
depends considerably on angular variables,
except for $\phi_b$ due to rotational symmetry in 3-body decays.
Individual ratios vanish when the $b$ quark is emitted parallel or anti-parallel 
to the top's spin axis $(\theta_b = 0, \pi)$.
When $\tau^+$ is emitted in the direction of $W$ in the $W$'s rest frame $(\theta_\tau=0)$,
the unpolarized matrix element vanishes and the ratios become ill defined.
The ratio vanishes in the $q^2\to0$ limit, again due to a vanishing longitudinal polarization.
For an on-shell $W$, cancellations can be large, reaching $\mathcal{O}(25\%)$ 
for individual points of phase space.
This motivates further investigations into  polarization and  polarization interference 
in single top quark production, particularly in the context of new physics 
and quantum information.

In light of our findings, 
we revisit the impact of $W$ helicity polarization 
on the kinematics of final-state particles 
in the top quark decay chain 
\begin{align}
 t\ \to\ W^{+(*)}_\lambda b\ \to\ \tau^+\nu_\tau b \ .
\end{align}
Our discussion differs from past 
studies~\cite{Czarnecki:1990pe,Dalitz:1991wa,
Kane:1991bg,Stirling:2012zt}
in that the helicity of the $W$ is defined directly 
from polarization vectors / propagators,
and not via the injection of spin projectors,
which can hide intermediate cancellations.
We also allow the $W_{(\lambda)}^{+(*)}$ to go off-shell.
We focus on the (in)sensitivity 
of observables in different frames 
to different $W$ polarization states.
For concreteness, we work in the Unitary gauge,
use SM inputs listed in Sec.~\ref{sec:setup}.

\begin{table}[t!]
\centering
\resizebox{\columnwidth}{!}{
\begin{tabular}{|c|c|c|c|c|c|}
\hline\hline
Polarization &Unpolarized & Longitudinal & Transverse  & 
Scalar&
Auxiliary     \\ \hline
Decay width (GeV)& 0.163 & 0.114 & 0.0491
& 6.88$\times 10^{-6}$ 
& 6.83 $\times 10^{-6}$\\  
Polarization Fraction (\%) & 
100\% & 69.3\% & 29.9\% 
&4.18\%$\times 10^{-4}$ 
& 4.21\%$\times 10^{-4}$\\ 
\hline\hline
\end{tabular}
}
\caption{Top quark partial decay widths (top row) and branching rates (bottom row) 
for unpolarized (column 1) and polarized (columns 2-5) $W$ bosons.}
\label{tab:top_decays}
\end{table}

We start our numerical analysis with Table~\ref{tab:top_decays}, 
where we show the $t\to W^{+(*)}_\lambda b \to \tau^+\nu_\tau b$ 
partial decay width (row 1) 
for an unpolarized $W$ (column 1) and polarized $W_\lambda$ (columns 2-5).
We also show the polarization fraction $(f_\lambda)$ relative to the unpolarized case, 
\begin{align}
f_\lambda( t\ \to\ W^{+(*)}_\lambda b\ \to\ \tau^+\nu_\tau b)\ 
=\ \frac{\Gamma( t\ \to\ W^{+(*)}_\lambda b\ \to\ \tau^+\nu_\tau b)}{\Gamma( t\  \to\ \tau^+\nu_\tau b)}\ .
\end{align}

For the longitudinal and transverse cases, 
we observe the $70:30$ ratio that is the well-known
in the narrow width approximation,
and supports our expectation 
of vanishing polarization interference.
The split varies slightly with input masses.
We defer discussions of the scalar contribution 
to the end of this section, 
starting just above 
Eq.~\eqref{eq:top_decay_scalar_v_aux}.

\begin{figure}[!t]
\subfigure[]{\includegraphics[width=0.48\textwidth]{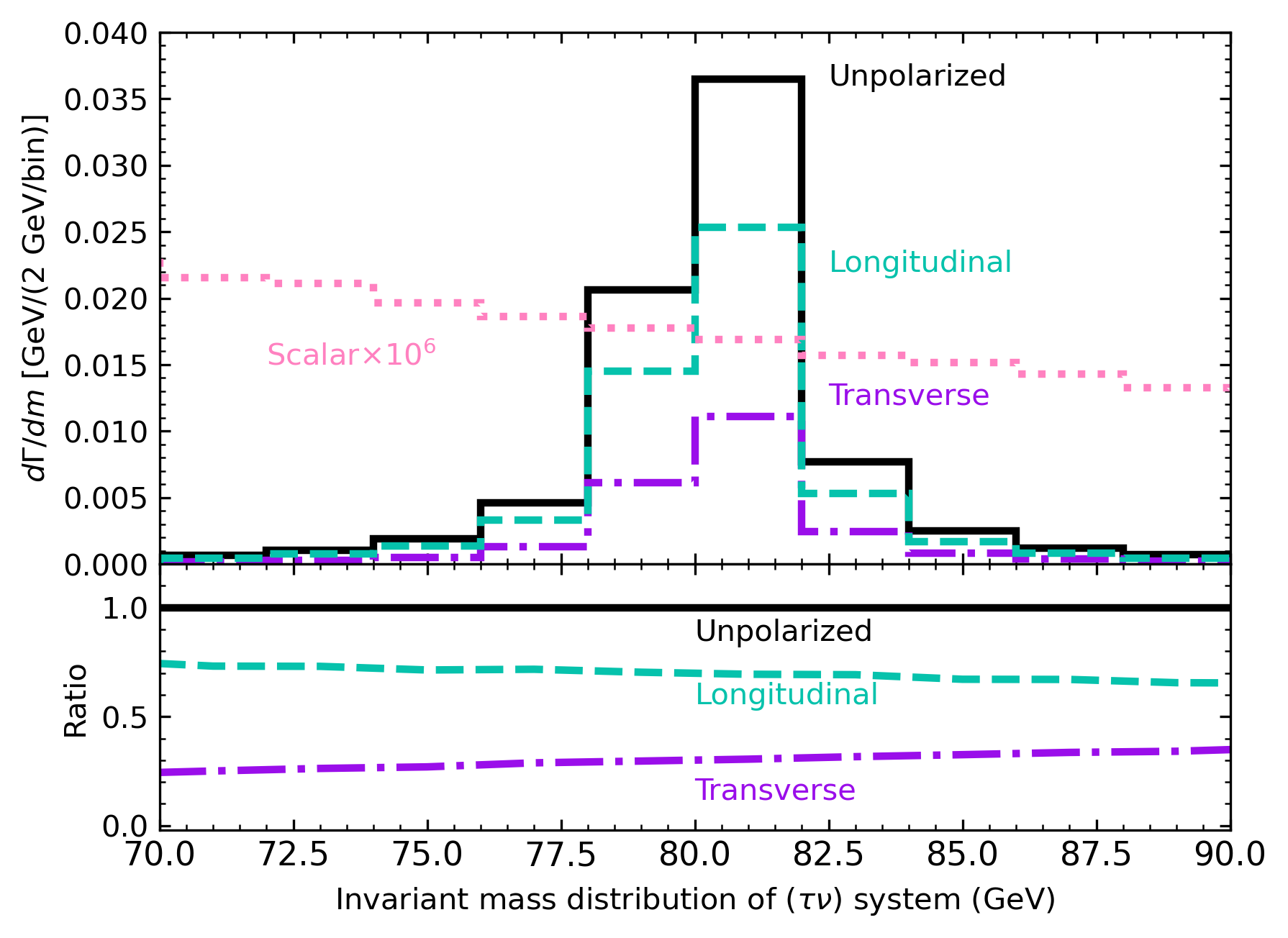}
\label{WMassRest}}\hfill
\subfigure[]{\includegraphics[width=0.48\textwidth]{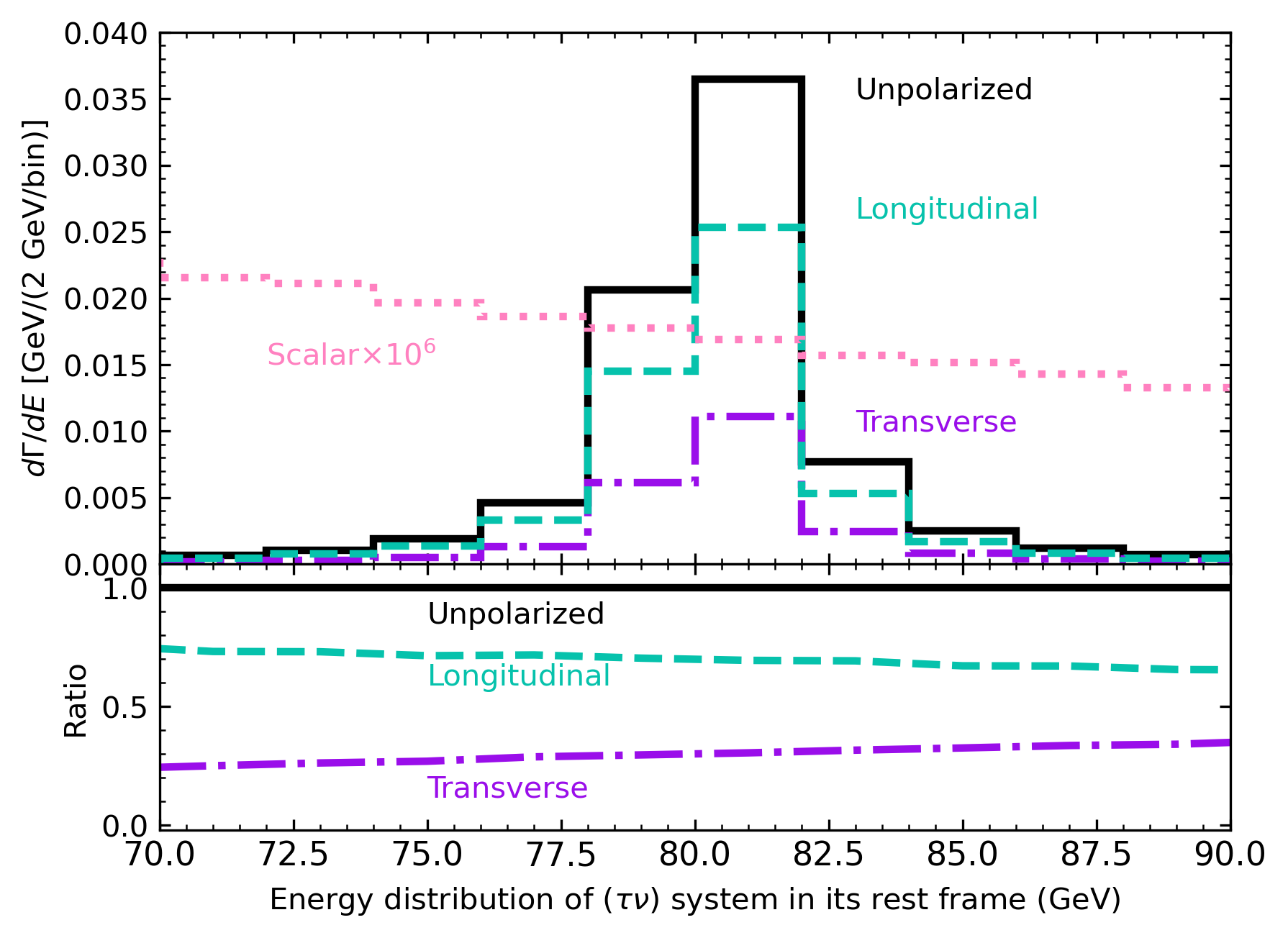}
\label{WenergyRest}}
\\
\subfigure[]{\includegraphics[width=0.48\textwidth]{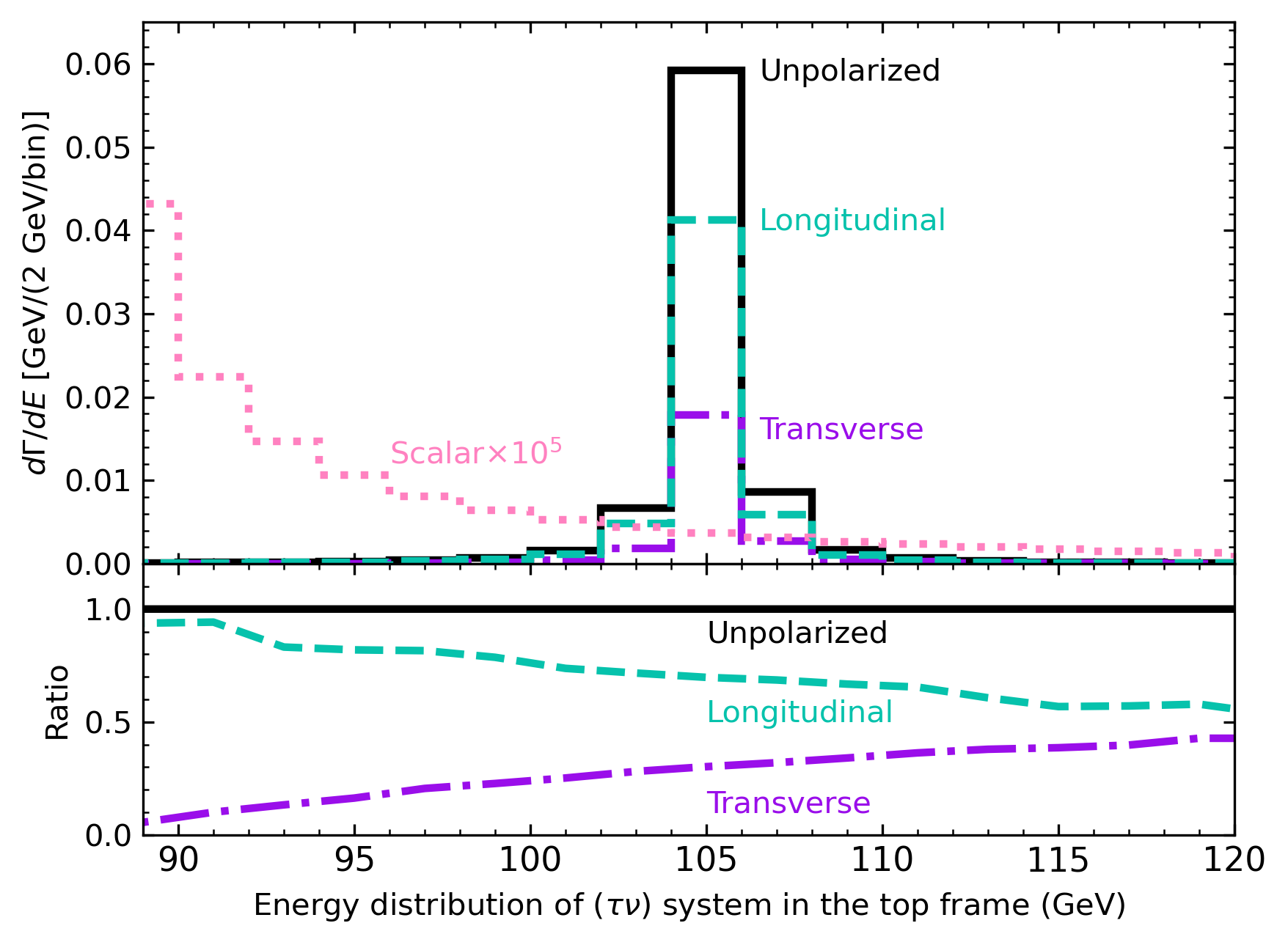}
\label{Wenergylab}}\hfill
\subfigure[]{\includegraphics[width=0.48\textwidth]{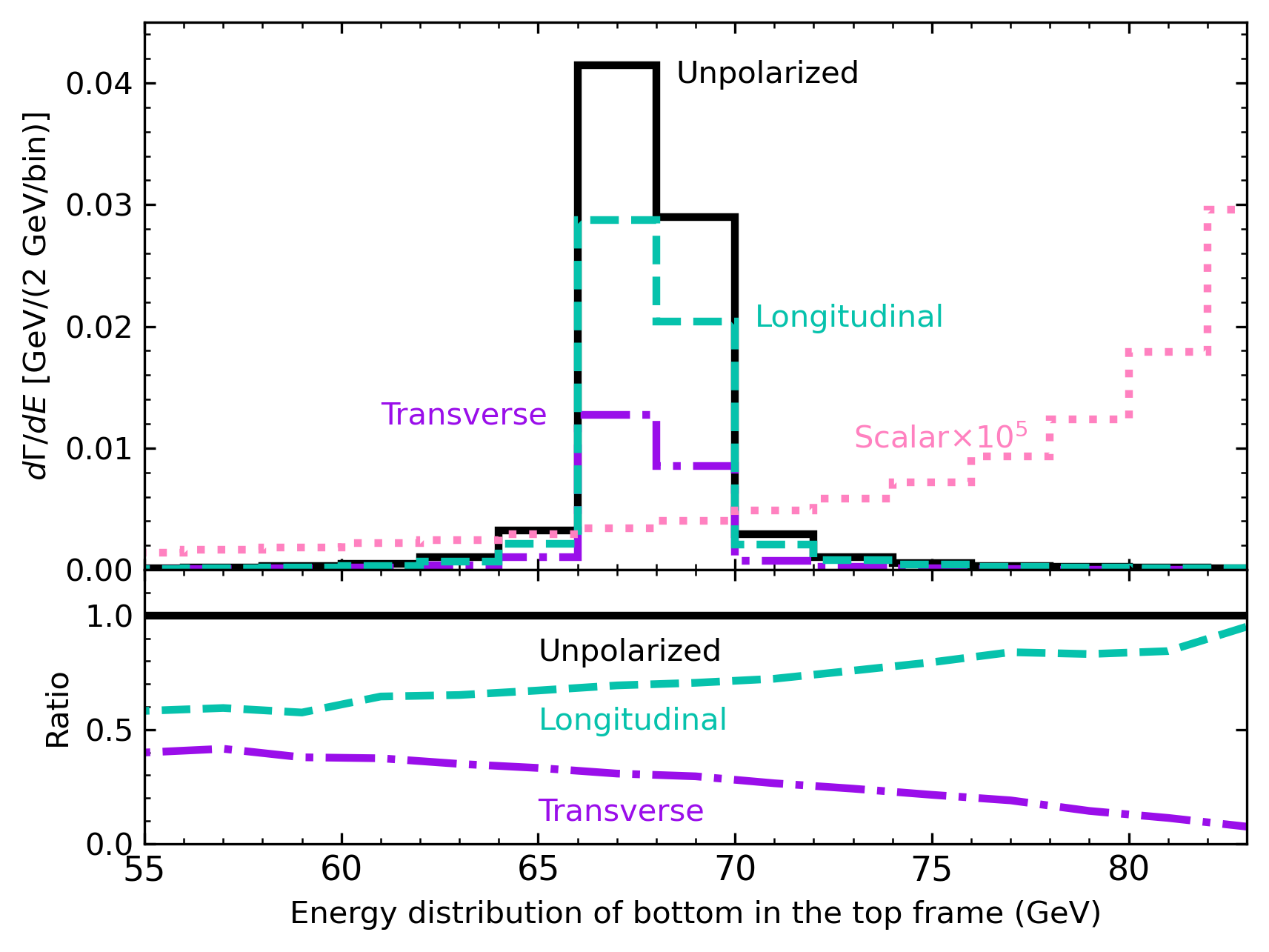}
\label{benergylab}}
\caption{Upper panel: For unpolarized (solid), longitudinal (dash),
transverse (dash-dot), and scalar (dot) $W_\lambda$ boson polarizations
in the $t\to W^{+(*)}_\lambda b \to \tau^+\nu_\tau b$ decay process,
(a) the invariant mass distribution of $(\tau\nu)$ system , 
(b) the energy of $(\tau\nu)$ system in its rest frame, 
(c) the energy of the $(\tau\nu)$ system in the lab frame, 
and
(d) the energy of the $b$ in the lab frame.
Lower panel: ratio with respect to the unpolarized case.}
\label{Wbplots}
\end{figure}

In Fig.~\ref{Wbplots} we show in the upper panels
various kinematical distributions in the top quark decay process
for unpolarized (solid), longitudinal (dash),
transverse (dash-dot), and scalar (dot) $W_\lambda$ boson polarizations.
In all the plots we scale the prediction for the scalar contribution 
by $10^5$ or $10^6$ to make it visible.
In the lower panels we show the ratio with respect to the unpolarized case
but omit the scalar polarization due to its smallness.

In Fig.~\ref{WMassRest} and Fig.~\ref{WenergyRest}, 
we plot as baselines 
the invariant mass of the composite system $(\tau\nu)$
and its energy in the $(\tau\nu)$ frame, respectively.
For different polarizations (except scalar) the matrix elements  all  depend 
on the invariant mass $\sqrt{q^2}$ through the Breit-Wigner 
propagator 
$\mathcal{M}_\lambda \sim D_W^{-1}(q^2) = [(q^2-M_W^2)^2+(\Gamma_W M_W)^2]^{-1}$,
which drives much of the kinematics.
When the $W$ goes on shell, the mass of the $(\tau\nu)$ system
is $\sqrt{q^2} = M_W \approx 80.4\GeV$,
which is clear in the plot. 
For the scalar polarization, 
the matrix element also depends on the 
Breit-Wigner propagator but 
factors in the polarization vector 
cause to the matrix element to scale as
$\mathcal{M}_{\lambda=S} \sim \Pi^W(q^2,\lambda=S) \sim 1/q^2$.
This pulls the distribution towards lower values 
of invariant mass (and energy).

In the lower panels we observe the $70:30$ split 
between the longitudinal and transverse polarizations.
We observe also 
that the longitudinal (transverse) contribution decreases (increases) 
with increasing mass and energy of the $(\tau\nu)$ system. 
We attribute this to a kinematical cancellation within the $\vartheta$ term.
In the absence of $b$ and $\tau$ masses, the longitudinal matrix element 
is given by $\vartheta$, with the nonzero terms in $\Theta_{\alpha\beta}$ scaling as
$\Theta_{\alpha\beta} \sim -q_\alpha n_\beta + q^2 n_\alpha n_\beta/(q\cdot n)$.
This means that $\vartheta$ and $\mathcal{M}_{\lambda=0}$ scale as
\begin{align}
    \mathcal{M}(\lambda=0)\ \sim\ \vartheta\ \sim\ -m_t + q^2/m_t\ .
\end{align}
In other words, as the invariant mass 
of the $(\tau\nu)$ system increases, 
the $t \to W_0^*$ decay mode turns off due 
to stronger phase-space suppression than present in the .
$t \to W_T^*$ decay mode.
Hence, the fraction of $t\to W_T^{*}$ increases with increasing $\sqrt{q^2}$.

Figure~\ref{Wenergylab} shows the energy distribution 
for the $(\tau\nu)$ system in the top's frame. 
The shapes for all polarizations follow na\"ive $1\to2$-body kinematics.
For the unpolarzied, transverse, and longitudinal cases, 
the intermediate $W$ is largely on shell, and
the energy of the $(\tau\nu)$ system is approximately
$E_{\tau\nu}^{\rm top}(\lambda={\rm unpol},0,T) \approx (m_t^2+ M_W^2-m_b^2)/2 m_t \approx 105\GeV$, as reflected in the plot.
For the scalar case, 
the propagator pole at $q^2 = 0$ 
favors $W^*_{\lambda=S}$ being nearly massless, and hence 
$E_{\tau\nu}^{\rm top}(\lambda=S) \approx m_t / 2 \approx 86\GeV$ 
(shown partially).

Similarly, Fig.~\ref{benergylab} shows the energy distribution 
of the bottom quark in the top frame.
For the unpolarzied, transverse, and longitudinal cases, 
the distributions has a peak around 
$E_b^{\rm top}(\lambda={\rm unpol},0,T) \approx m_t - E_W^{\rm top} \sim 68\GeV$,
as expected from energy conservation.
For the scalar case, 
$E_b^{\rm top}(\lambda=S) \approx m_t - E_W^{\rm top} \sim 86\GeV$ (shown partially).

\begin{figure}[!t]
\subfigure[]{\includegraphics[width=0.48\textwidth]{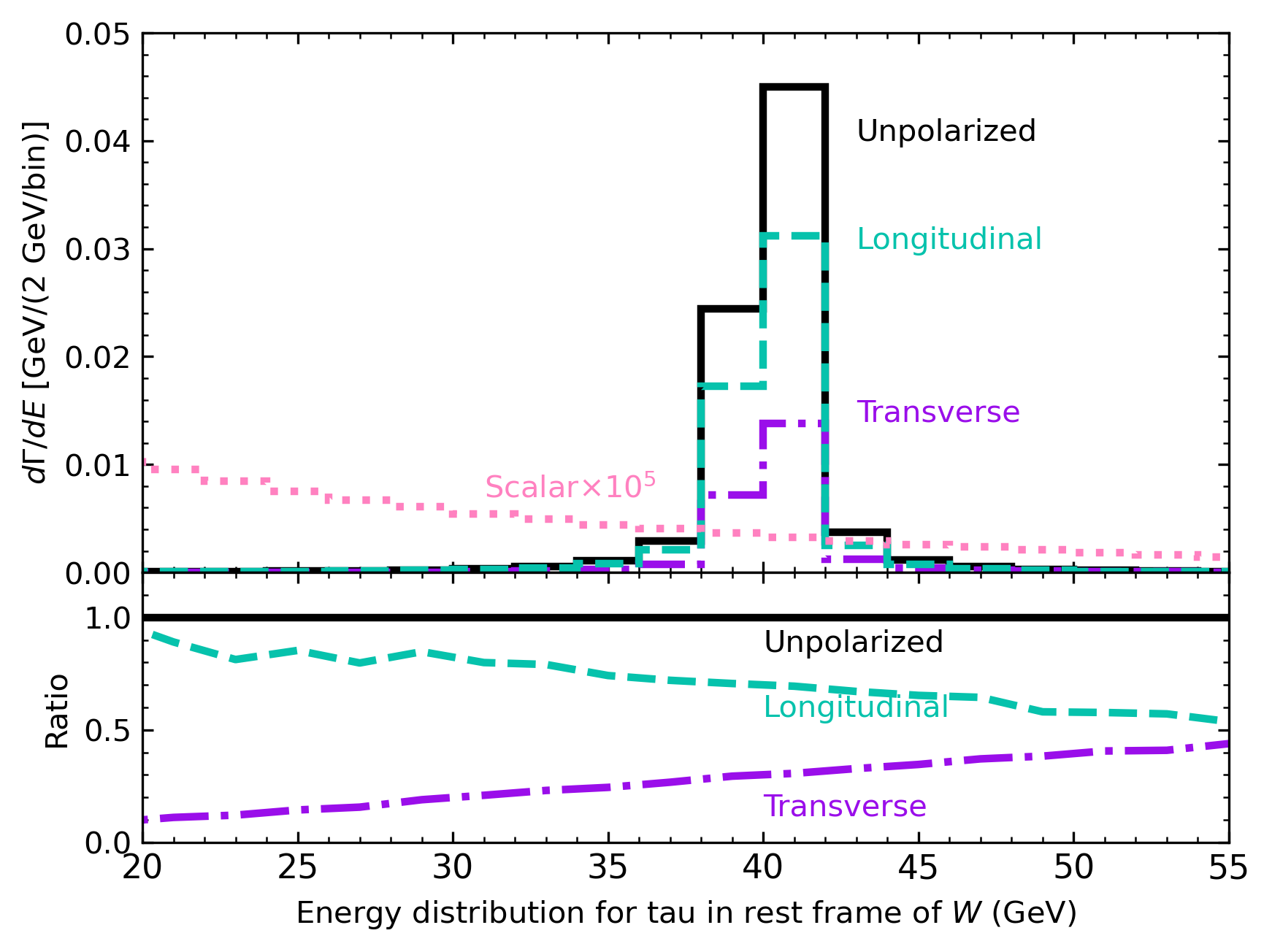}
\label{TauErest}}\hfill
\subfigure[]{\includegraphics[width=0.48\textwidth]{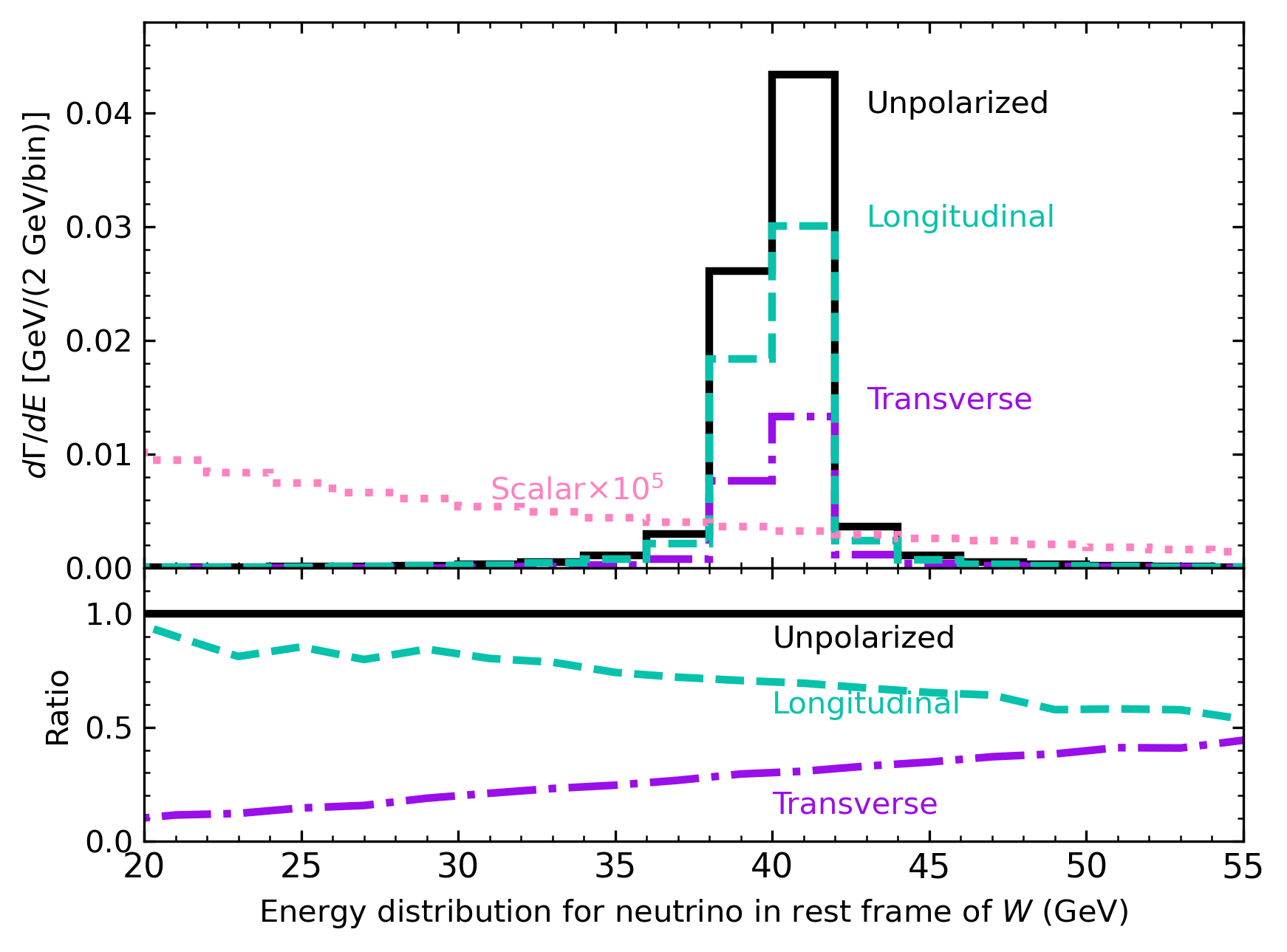}
\label{NeuEnergyrest}}
\\
\subfigure[]{\includegraphics[width=0.48\textwidth]{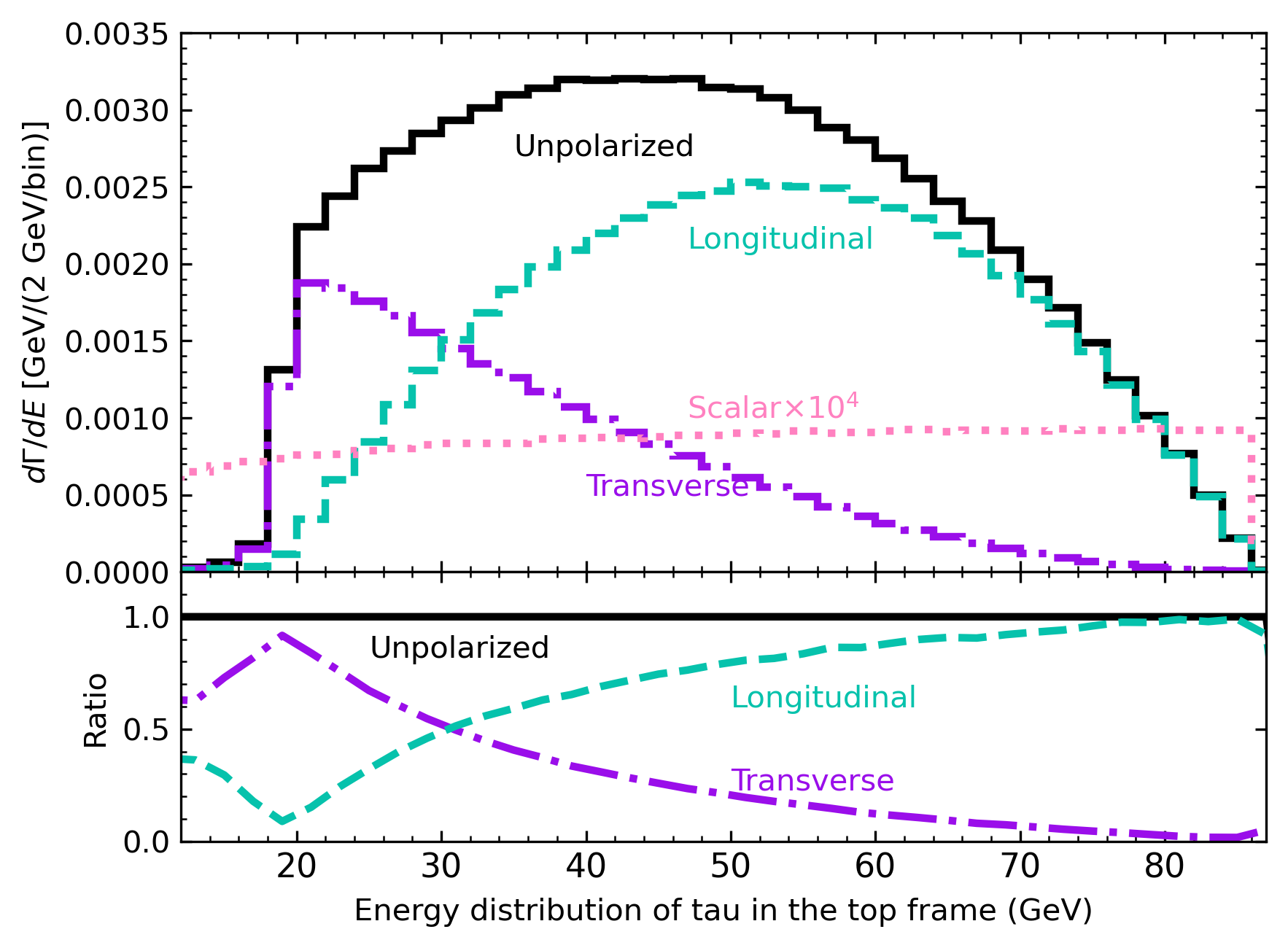}
\label{tauElab}}\hfill
\subfigure[]{\includegraphics[width=0.48\textwidth]{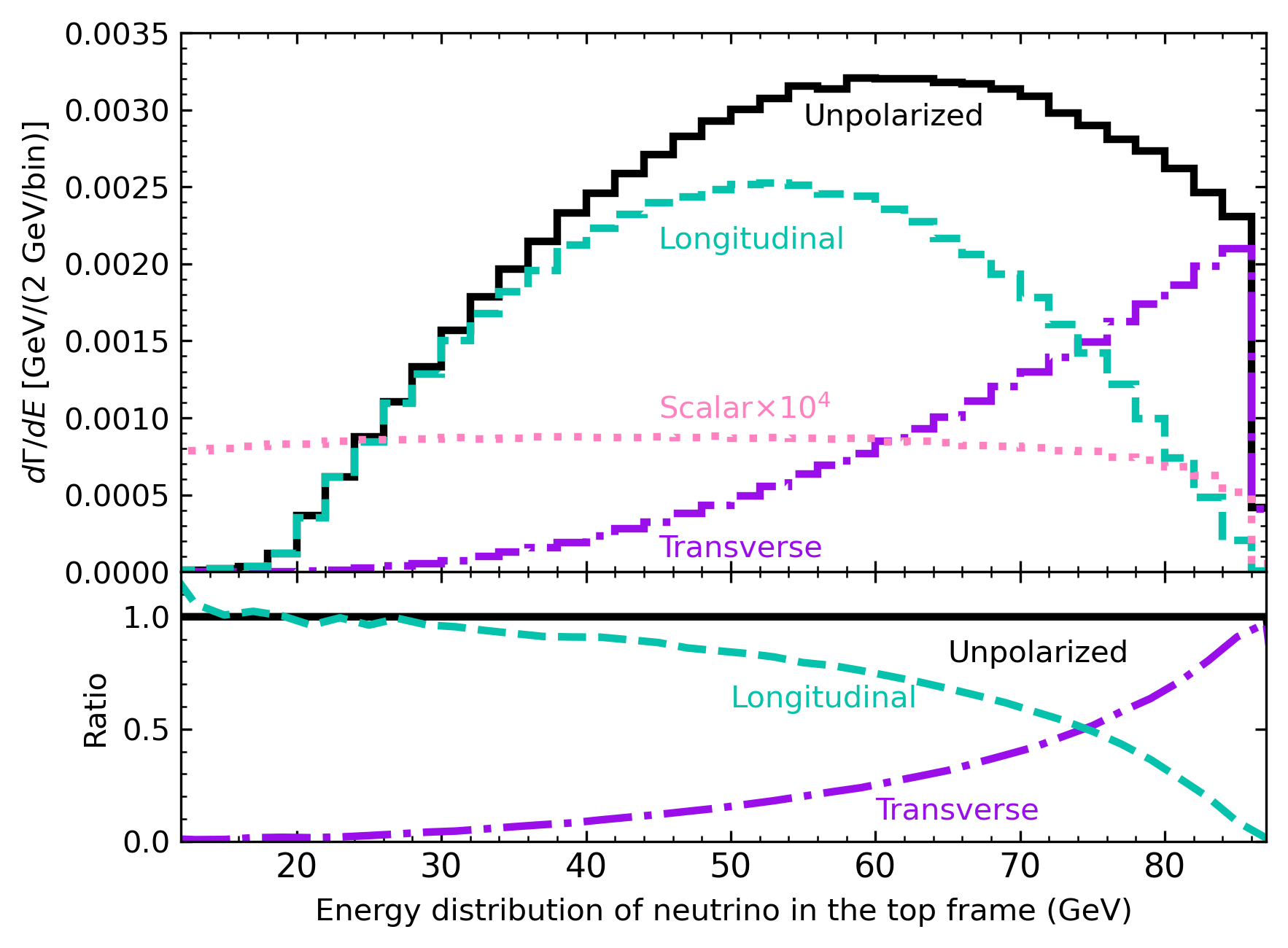}
\label{NeuElab}}
\caption{Same as Fig.~\ref{Wbplots} but for the energy distribution of 
(a) the $\tau^+$  and (b) the $\nu_\tau$ in the $(\tau\nu)$ frame.
(c,d) Same as (a,b) but in the top's frame.}
\label{fig:kinematics_nutau}
\end{figure}

We now turn to the kinematic distributions of the $\tau^+$ and $\nu_\tau$.
In Fig.~\ref{fig:kinematics_nutau} we show the 
energy distribution for (a,c) $\tau^+$ and (b,d) $\nu_\tau$ 
in (a,b) the  frame of the $(\tau\nu)$ system and 
in (c,d) the top's frame. 
In the $(\tau\nu)$ frame, we observe that both leptons carry an energy of about 
about $E_\tau^{\rm (\tau\nu)}\sim E_\nu^{\rm (\tau\nu)}\sim M_W/2\sim 40\GeV$,
which is consistent with the energy and invariant mass distributions
of the $(\tau \nu)$ system in Fig.~\ref{Wbplots}.
For the distributions in the top's frame, 
which are obviously more complicated, 
we turn to spin-correlation in decay chains.

\begin{figure}[t!]
\subfigure[]{\fbox{\includegraphics[width=0.31\textwidth]{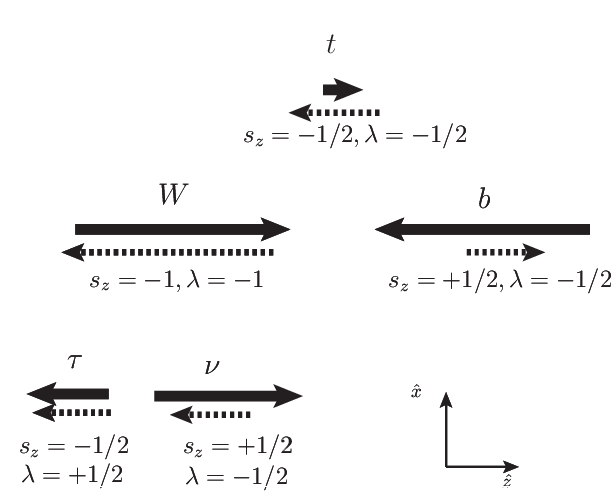}}
\label{tL1}}
\subfigure[]{\fbox{\includegraphics[width=0.275\textwidth]{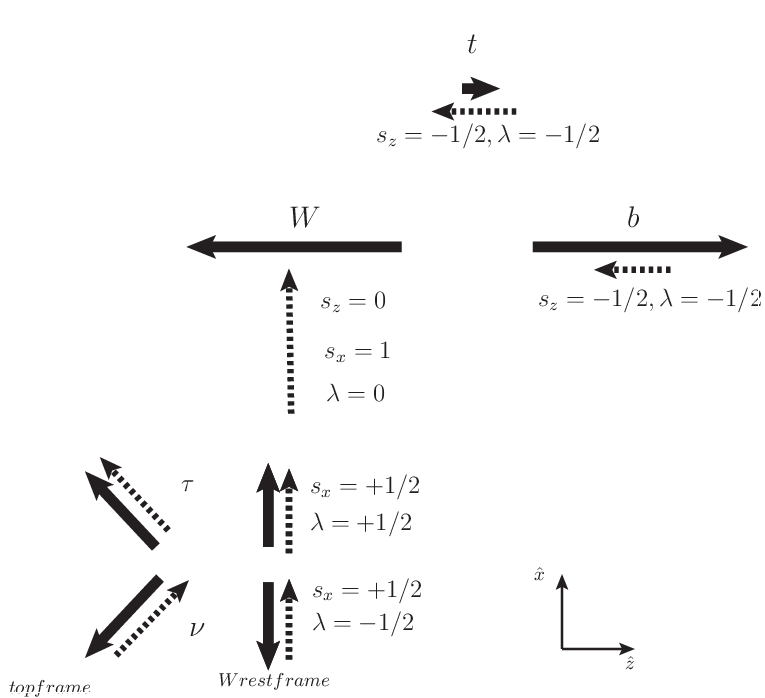}}
\label{tL2}}
\subfigure[]{\fbox{\includegraphics[width=0.31\textwidth]{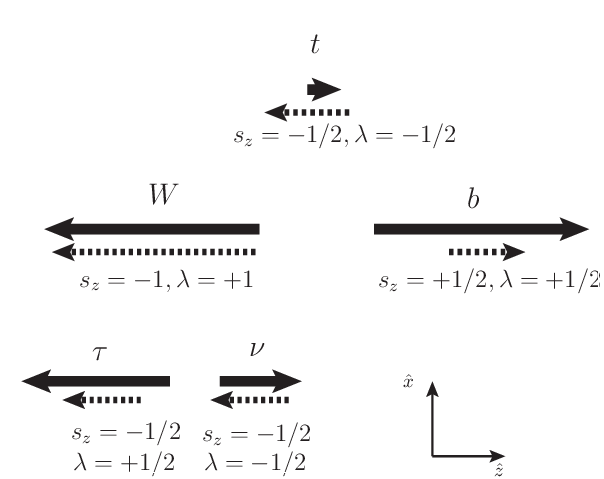}}
\label{tL3}}
\\
\subfigure[]{\fbox{\includegraphics[width=0.32\textwidth]{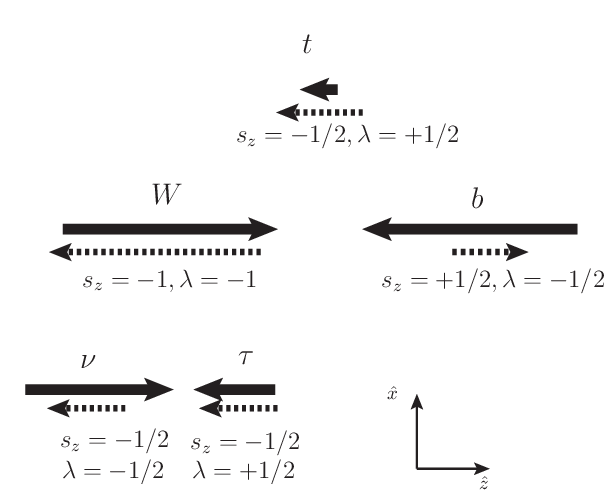}}
\label{tR3}}
\subfigure[]{\fbox{\includegraphics[width=0.26\textwidth]{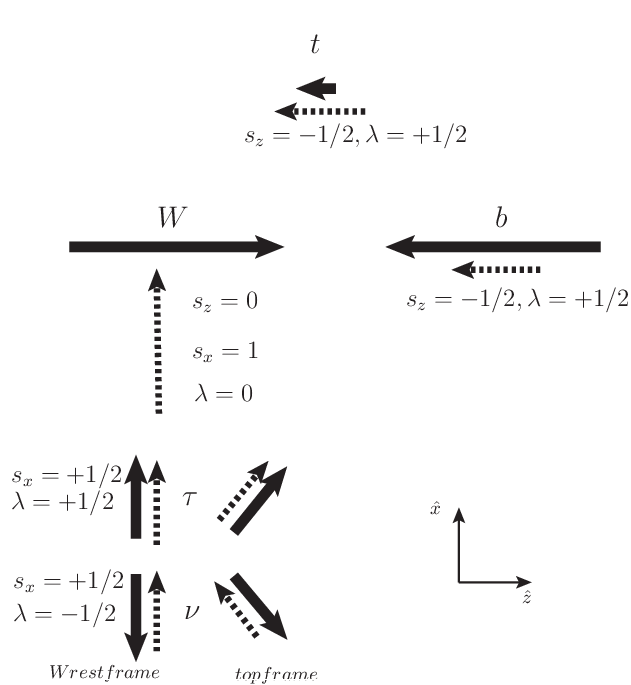}}
\label{tR2}}
\subfigure[]{\fbox{\includegraphics[width=0.34\textwidth]{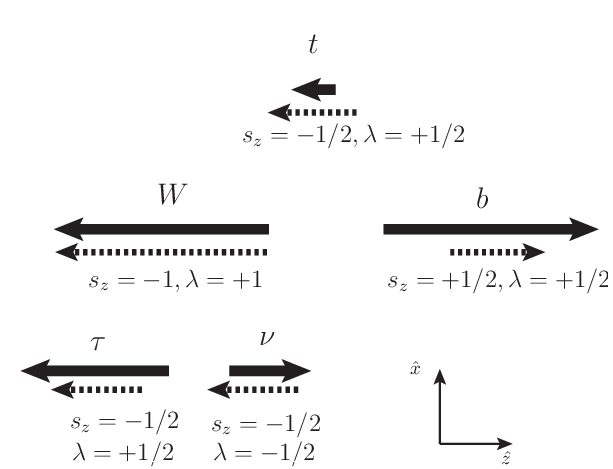}}
\label{tR1}}
\caption{Spin-correlation chains 
in $t({\lambda_t})\to W^+(\lambda_W) b(\lambda_b) \to 
\tau^+(\lambda_\tau) \nu_{\tau}(\lambda_\tau) b(\lambda_b)$
for 
(a,b,c) LH $(\lambda_t=-1/2)$ top quarks;
(d,e,f) RH $(\lambda_t=+1/2)$ top quarks;
(a,d) LH $(\lambda_W=-1)$ $W$ bosons;
(b,e) longitudinal $(\lambda_W=0)$ $W$ bosons; and
(c,f) RH $(\lambda_W=+1)$ $W$ bosons.
$s_z$ is the spin along $\hat{z}$, 
the momentum direction is given by a solid arrow,
and $\lambda$ is the helicity (dashed arrow).}    
\label{helicity}
\end{figure}

In Fig.~\ref{helicity} we show the possible helicity and spin configurations 
in the decay process $t({\lambda_t})\to W^+(\lambda_W) b(\lambda_b) \to 
\tau^+(\lambda_\tau) \nu_{\tau}(\lambda_\tau) b(\lambda_b)$,
assuming massless leptons and a massless bottom quark,
for (a,b,c) LH $(\lambda_t=-1/2)$ top quarks,
(d,e,f) RH $(\lambda_t=+1/2)$ top quarks,
(a,d) LH $(\lambda_W=-1)$ $W$ bosons,
(b,e) longitudinal $(\lambda_W=0)$ $W$ bosons, and
(c,f) RH $(\lambda_W=+1)$ $W$ bosons.
The solid arrows represent the direction of particle's 3-momentum and 
the dotted arrows represent the spin angular momentum direction $s_z$. 

For the decay of $t_L$, there are only two allowed helicity configurations for the $W$:
the LH transverse polarization $(\lambda_W = -1)$ as shown in Fig.~\ref{tL1}
and 
the longitudinal polarization $(\lambda_W = 0)$ as shown in Fig.~\ref{tL2}.
The RH transverse polarization of the $W$, 
shown in Fig.~\ref{tL3}, 
selects for a RH bottom quark.
However, RH fermions can only participate 
in LH chiral currents 
via helicity inversion. 
Since we assumed the $b$ to be massless, 
the $t_L\to b_R$ decay current vanishes, 
$J^\alpha_\text{in} \sim  \bar{u}_R(p_b) \gamma^\alpha P_L u_L(p_t) = 
\bar{u}_R(p_b) (P_L P_R) \gamma^\alpha u_L(p_t) = 0$.

For $\lambda_W=-1$ with LH tops [Fig.~\ref{tL1}] and 
RH tops [Fig.~\ref{tR3}], 
the LH transverse polarization of the $W$ is opposite to its motion.
This causes the neutrino (anti-tau) to move in the same (opposite) direction 
as the $W$'s boost to the top's frame.
This is why $\nu_\tau$s from  $W_{\lambda=T}$ decays 
acquire a higher energy compared to the $\tau^+$,
as reflected in Fig.~\ref{tauElab} and Fig.~\ref{NeuElab}.
Numerically, the energies of the $\tau^+$ and $\nu_\tau$ are related in the 
system and top frames by the $W$'s own boost 
$(\gamma_W = E_{W}^{\rm top}/M_W = 1/\sqrt{1-\beta_W^2}$)
from its rest frame:
\begin{subequations}
\begin{align}
    E_{\tau}^{\rm top}\ &=\ \gamma_W\ (1-\beta_W)\ E_{\tau}^{(\tau\nu)}\ \gtrsim\ 18\GeV\ ,
    \\
    E_{\nu}^{\rm top}\ &=\ \gamma_W\ (1+\beta_W)\ E_{\nu}^{(\tau\nu)}\ \lesssim\ 86\GeV\ . 
\end{align}
\end{subequations}
These are in agreement with the observed lower and upper 
values of the $\tau^+$ and $\nu_\tau$ energies.

For $\lambda_W = 0$ with LH tops [Fig.~\ref{tL2}]
and RH tops [Fig.~\ref{tR2}], 
the spin axis of longitudinal $W$s
is always perpendicular to its direction of motion.
However, as both leptons propagate 
preferentially \textit{along} 
the $W$'s spin axis in the $W$'s frame,
the boost for the leptons is along an axis 
that initially has no momentum. 
As a result, the momentum carried by the $W$ in the $t \to Wb$ decay
$(\vert \vec{p}_W^{\rm top}\vert\sim (m_t^2-M_W^2)/2m_t\sim 68\GeV)$
is largely split equally between the two leptons.
The resulting lepton energies in the top frame are 
\begin{align}
    E_\tau^{\rm top},\  E_\nu^{\rm top}  
    \sim \sqrt{\vert \vec{p}_\nu^{(\tau\nu)}\vert^2 + \left(\frac{\vert \vec{p}_W^{\rm top}\vert}{2}\right)^2} \approx \sqrt{(40\GeV)^2 + (34\GeV)^2}\ \approx\ 52\GeV\ ,
\end{align}
which is reflected in the peaks of the  $\tau^+$ and $\nu_\tau$ energies.

\begin{figure}[!t]
\subfigure[]{\includegraphics[width=0.32\textwidth]{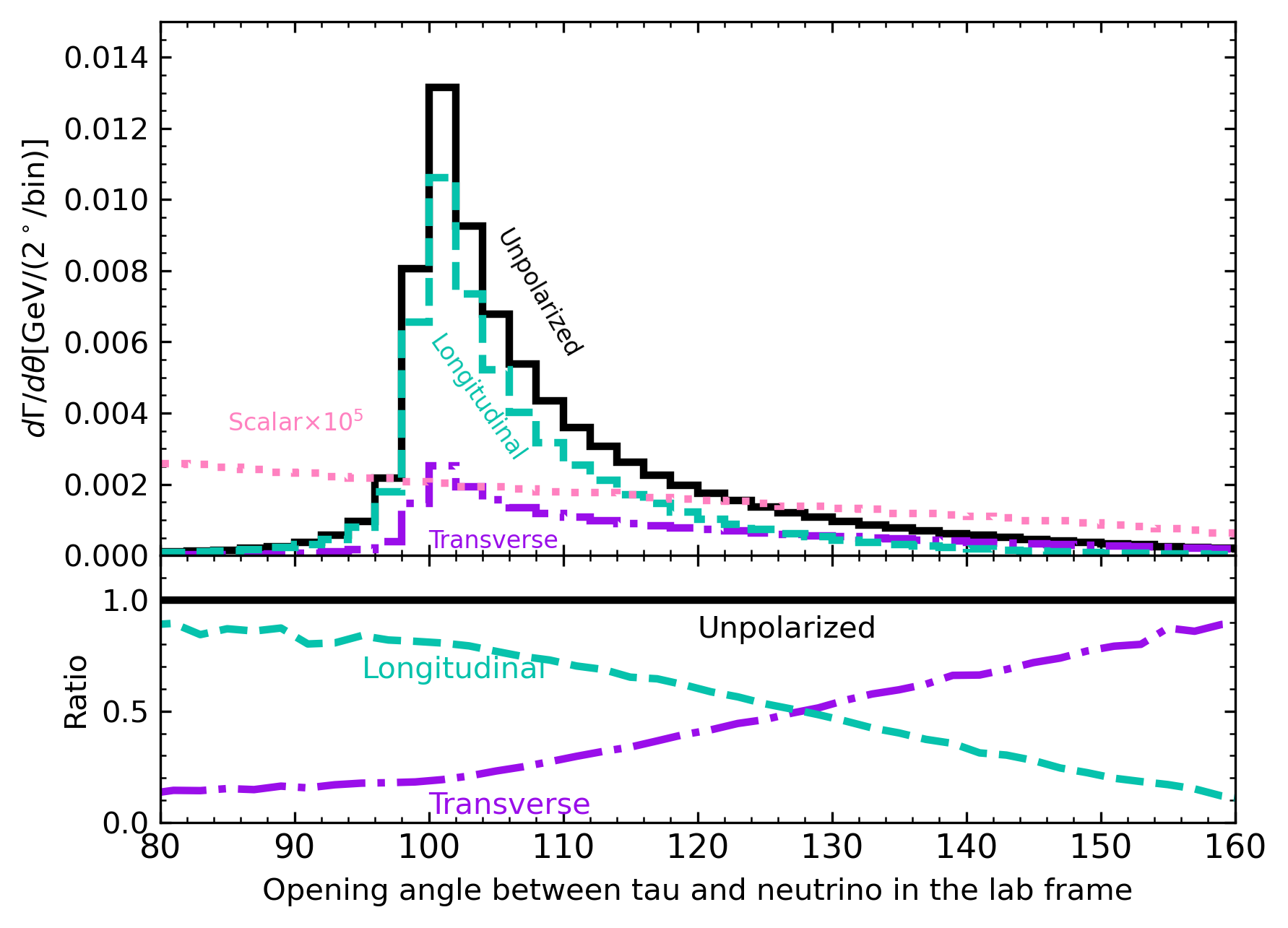}
\label{tauNeuAngle}}
\subfigure[]{\includegraphics[width=0.32\textwidth]{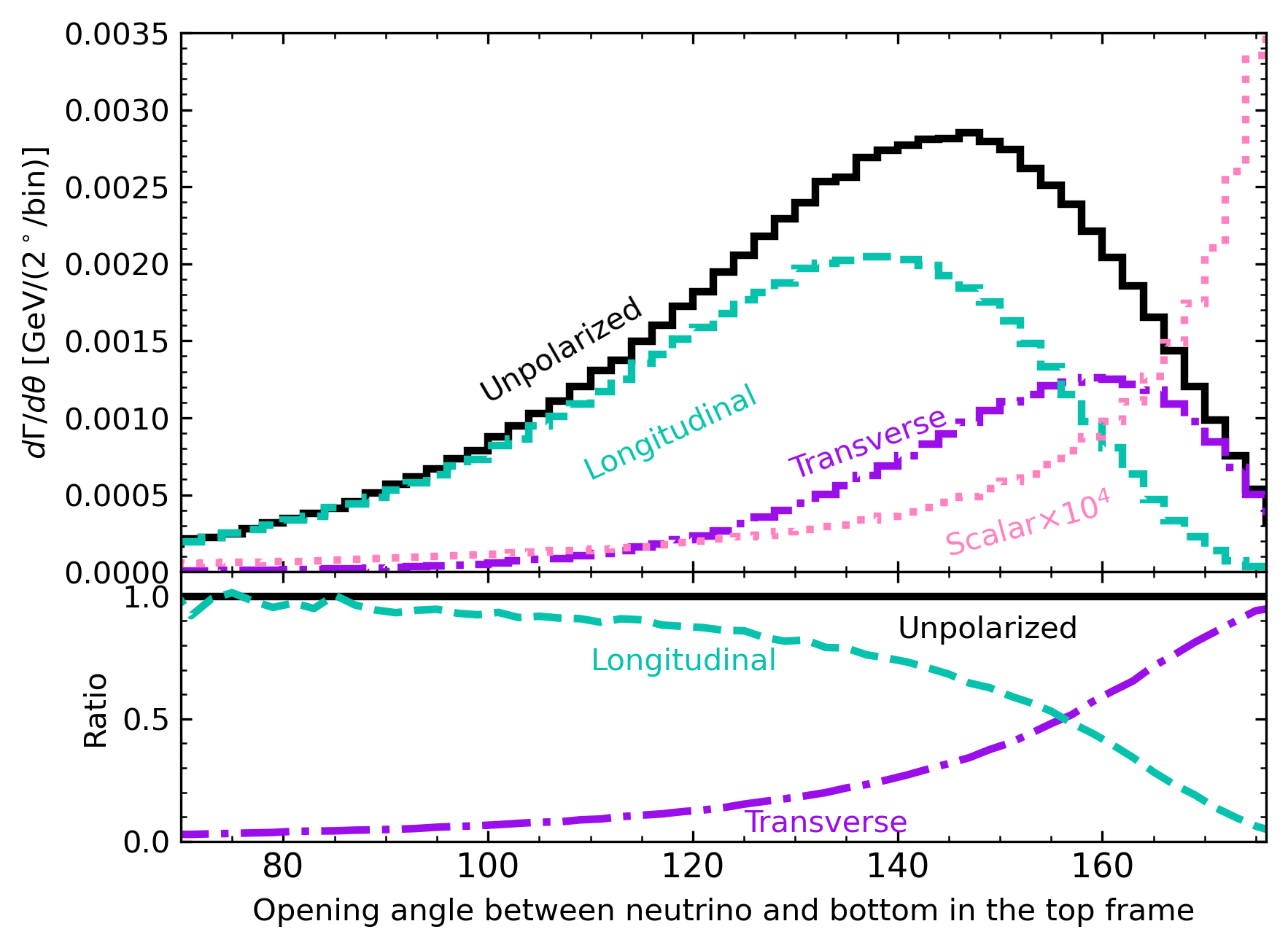}
\label{NeuBAngle}}
\subfigure[]{\includegraphics[width=0.32\textwidth]{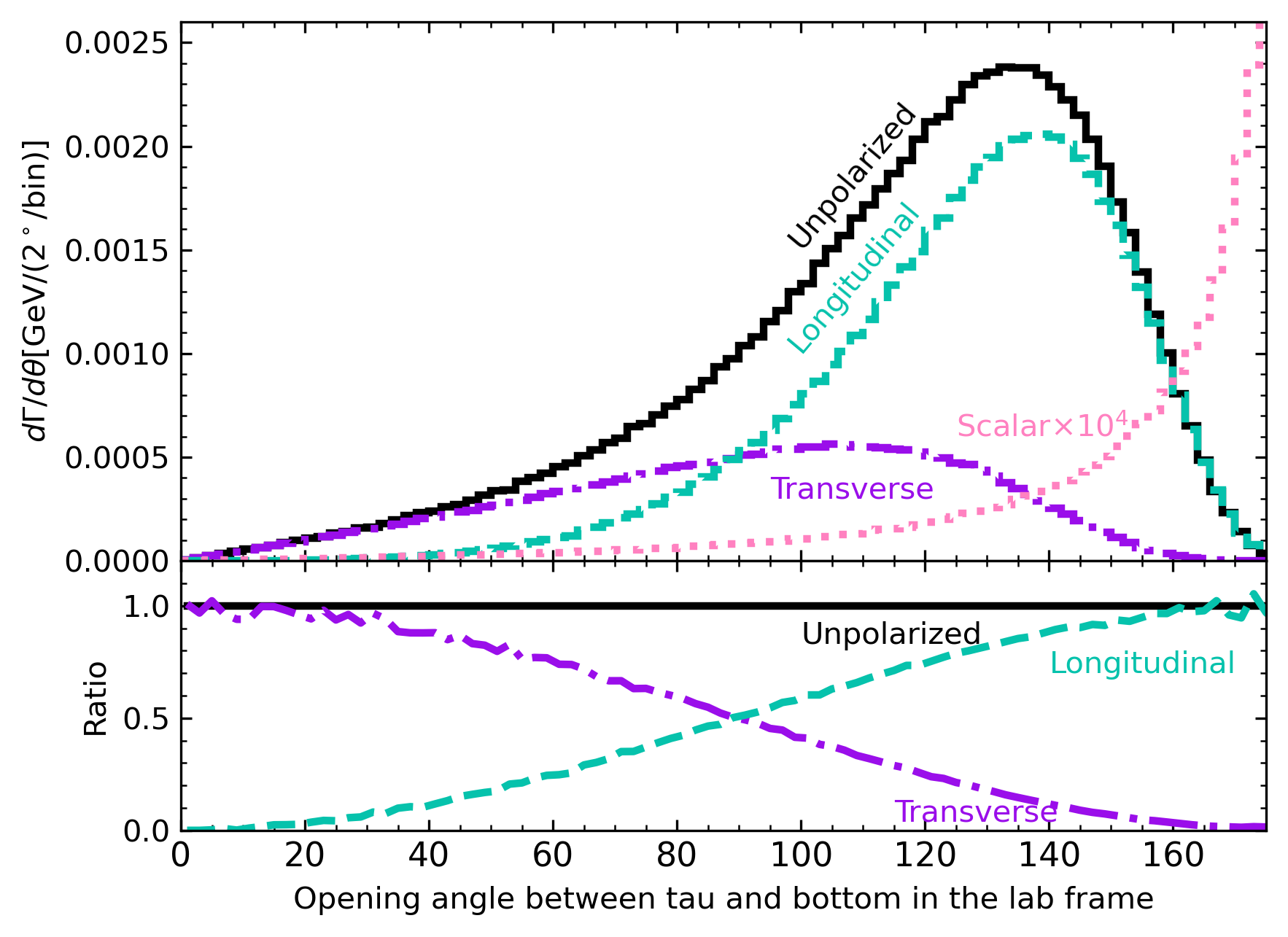}
\label{TauBtmAgle}}
\caption{Same as Fig.~\ref{Wbplots} but for the opening angles 
    in the top frame     between 
    (a) the $\tau$ and the $\nu_\tau$, 
    (b) the $\nu_\tau$ and the $b$, and 
    (c) the $\tau^+$ and the $b$.}
    \label{eq:kinematics_lepton_angles} 
\end{figure}

In Fig.~\ref{eq:kinematics_lepton_angles}
we show for unpolarized and polarized $W$
the opening angles 
in the top frame     between 
(a) the $\tau$ and the $\nu_\tau$, 
(b) the $\nu_\tau$ and the $b$, and 
(c) the $b$ and the $\tau^+$.
Using momentum conservation
we can roughly estimate the opening angles 
for each of these cases (assuming $m_\tau,m_b\approx 0$).
For the $\tau-\nu_\tau$ case, we have 
\begin{subequations}
\label{eq:angle_top}
\begin{align}
\label{eq:angle_tau_nu_top}
    q^2 = ~(p_\tau+p_\nu)^2 \approx ~ 2E_\tau E_\nu (1-\cos\theta_{\tau\nu})\implies
    \theta_{\tau\nu}^{\rm top} 
    \approx ~ \cos^{-1}\left[1- \frac{q^2}{2E_\tau^{\rm top} E_\nu^{\rm top}}\right] \ . 
\end{align}
In the on-shell limit and building on previous arguments (and distributions),
masses and energies are roughly 
$\sqrt{q^2} \approx M_W\approx 80\GeV$, $E_\tau^{\rm top}\approx 45\GeV$, and 
$E_\nu^{\rm top}\approx 60\GeV$. 
This gives a $\tau-\nu_\tau$ opening angle 
of about $\theta_{\tau\nu}^{\rm top}\approx 100^\circ$, 
in agreement with Fig.~\ref{tauNeuAngle}.

Similarly, for the $\tau-b$ and $b-\nu$ opening angles, we have the expressions
\begin{align}
    (p_\tau +p_b)^2 = &(p_t-p_\nu)^2 \implies 
    \theta_{\tau b}^{\rm top} 
    \approx \cos^{-1}
    \left[1-\frac{m_t^2-2m_t E_\nu^{\rm top}}{2 E_\tau^{\rm top} E_b^{\rm top}}\right],
    \\
   (p_b + p_\nu)^2 =   &(p_t-p_\tau)^2 
   \implies \theta_{\nu b}^{\rm top} \approx 
   \cos^{-1}\left[1-\frac{m_t^2-2m_t E_\tau^{\rm top}}{2 E_\nu^{\rm top} E_b^{\rm top}}\right],
\end{align}
\end{subequations}
where $E_\tau = m_t - E_b -E_\nu$.
For the range of $E_\nu^{\rm top} \sim 50-65$ GeV, 
we obtain 
$\theta_ {\nu b}^{\rm top}\sim 125^\circ - 143^\circ$ and 
$\theta_{\tau b}^{\rm top}\sim 111^\circ - 133^\circ$,
 in agreement with the distributions in 
Fig.~\ref{TauBtmAgle} and Fig.~\ref{NeuBAngle}.


Given the decay's sensitivity 
to the scalar polarization,
we now consider the impact of
including and neglecting 
the $\mathcal{O}(M_V\Gamma_V)$ term in 
the scalar helicity polarization vector 
of Eq.~\eqref{eq:polvec_scalar_uni}.
The term originates from demanding that 
the unpolarized propagator in Eq.~\eqref{eq:prop_unpol_rx}
respects Ward 
identities~\cite{LopezCastro:1991nt,LopezCastro:1995kg,Denner:1999gp,Denner:2005fg}
and enters the scalar polarization via the completeness relationship.
Unlike in Breit-Wigner propagators, 
the $\mathcal{O}(M_V\Gamma_V)$ term  
in scalar polarization vectors is often omitted 
in the literature~\cite{Ballestrero:2017bxn,Ballestrero:2019qoy,
BuarqueFranzosi:2019boy,Hoppe:2023uux,Javurkova:2024bwa}.
However, omitting this is justifiable only in 
$t$-channel exchanges 
or when $\mathcal{Q}$ terms can be neglected.

\begin{figure}[!t]
\subfigure[]{\includegraphics[width=0.48\textwidth]{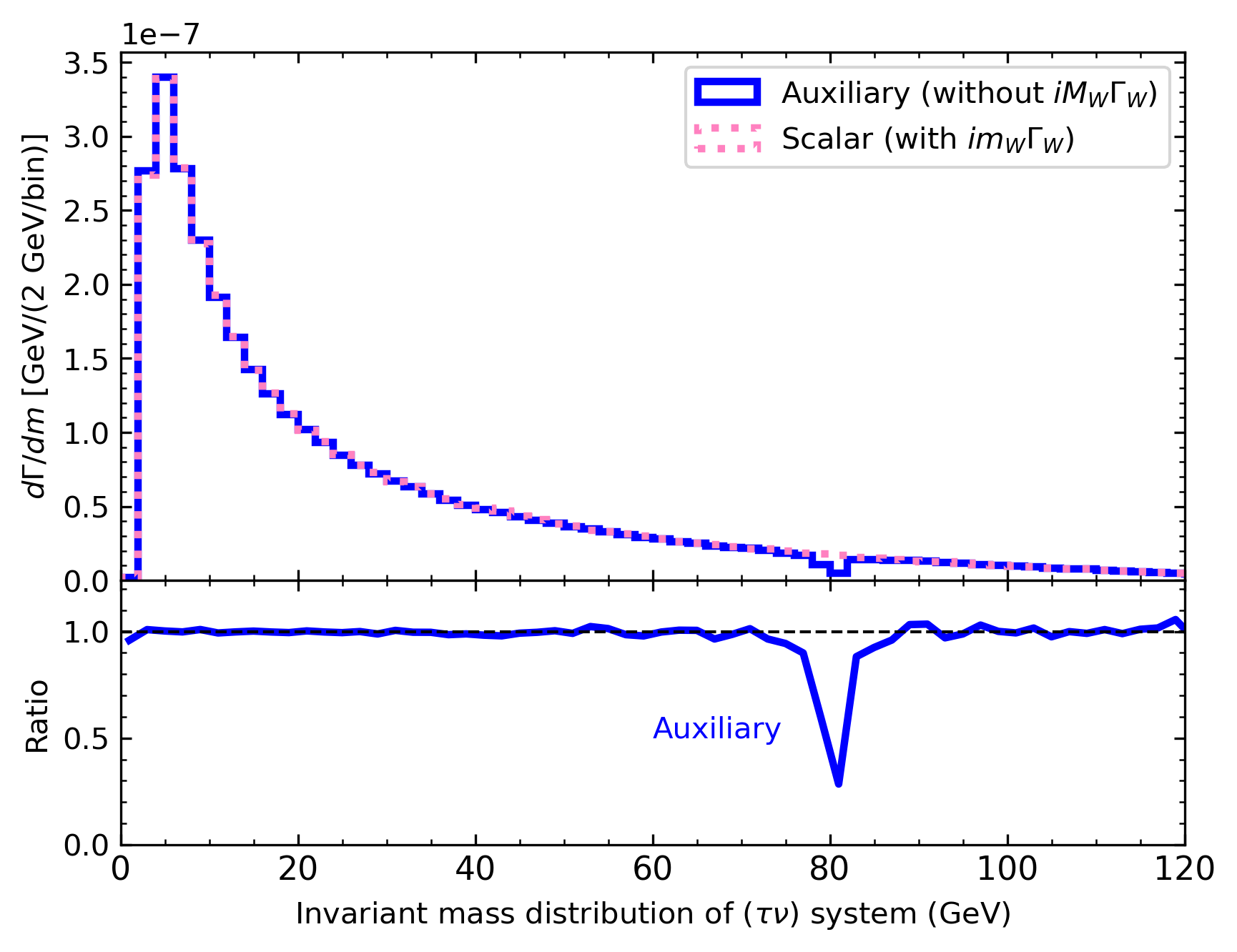}
\label{Wmass1}}\hfill
\subfigure[]{\includegraphics[width=0.48\textwidth]{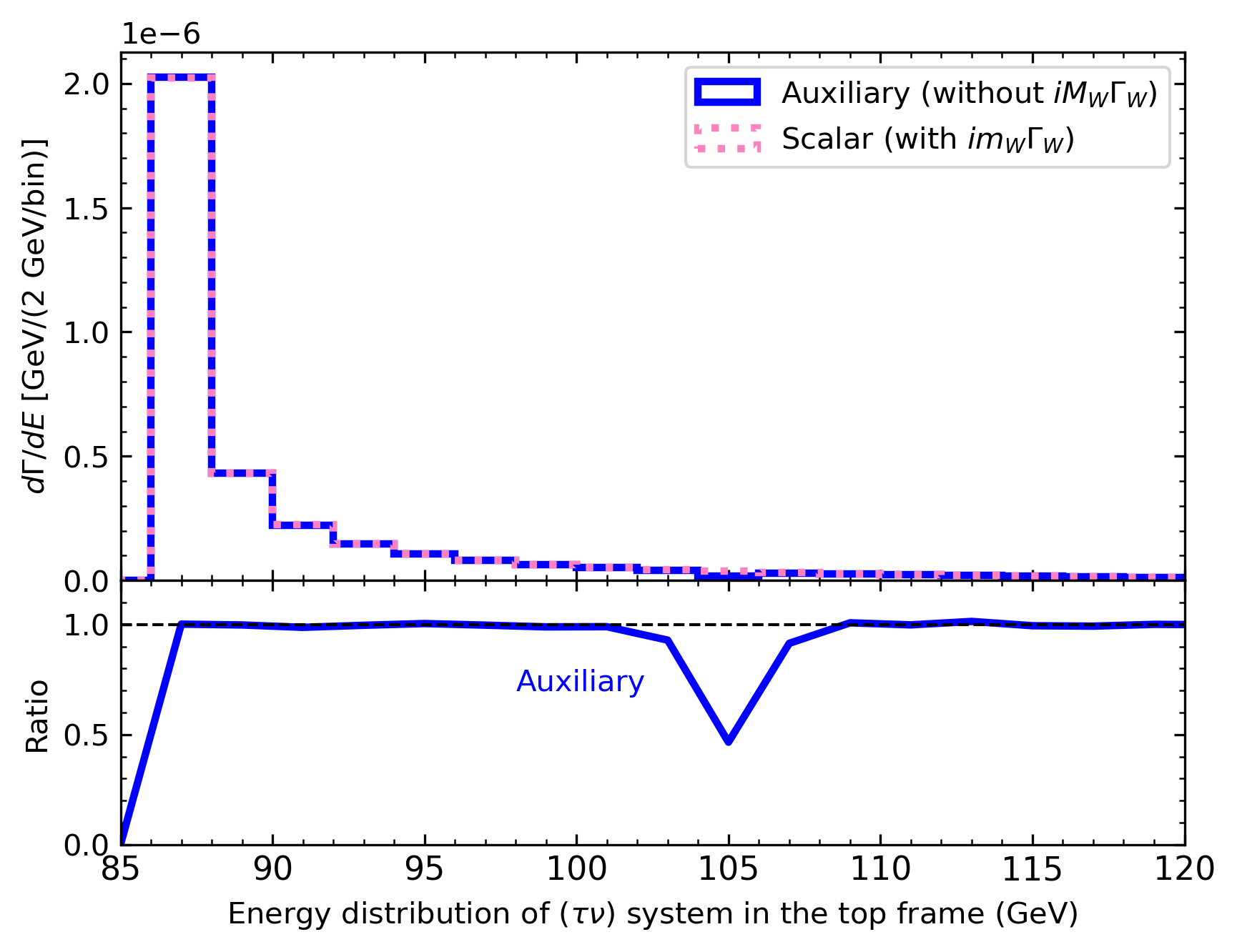}
\label{WElab1}}
\\
\subfigure[]{\includegraphics[width=0.48\textwidth]{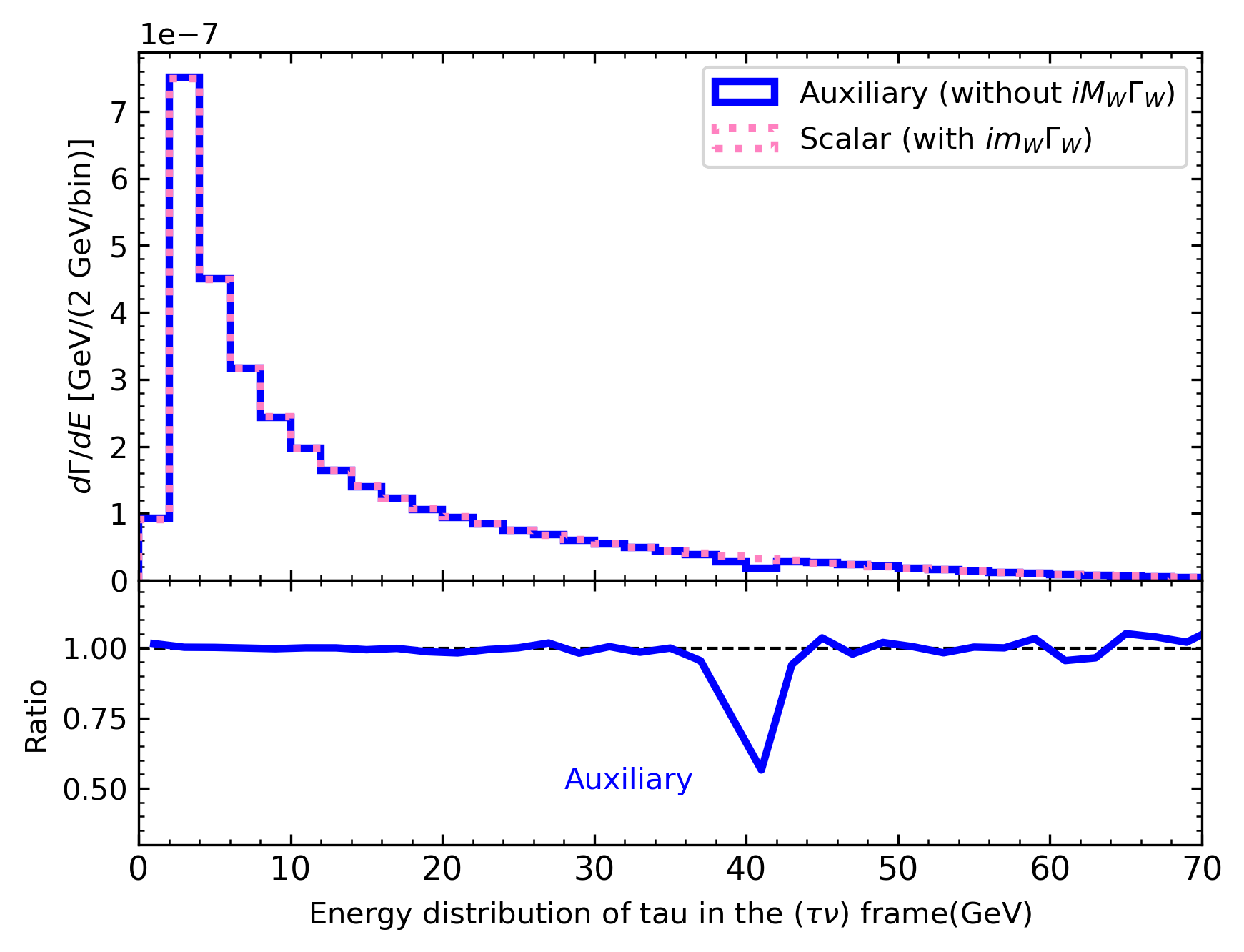}
\label{TauErest1}}\hfill
\subfigure[]{\includegraphics[width=0.48\textwidth]{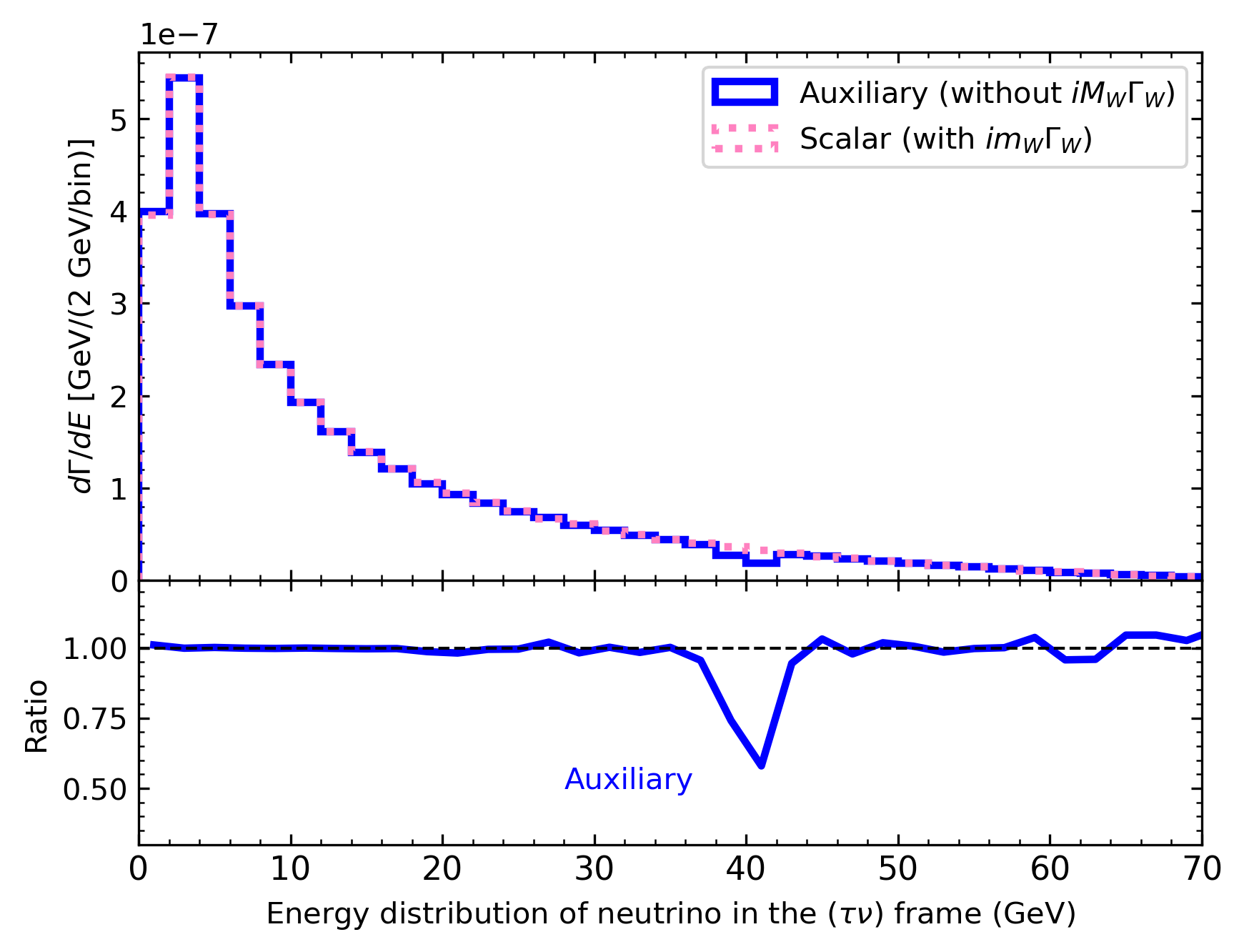}
\label{NeuErest1}}
\caption{Upper panel:
For the scalar (dash-dot) and ``auxiliary'' (solid) polarizations
in the $t\to W^{+(*)}_\lambda b \to \tau^+\nu_\tau b$ decay process,
(a) the invariant mass of the $(\tau^+\nu_\tau)$ system,
(b) the energy of the $(\tau^+\nu_\tau)$ system in the top's frame,
(c) the energy of the $\tau^+$ in the $(\tau^+\nu_\tau)$ frame, and
(d) same as (c) but for the $\nu_\tau$.
Lower panel: ratio to the ``scalar'' distribution.}
\label{fig:scalar_v_aux_scales}    
\end{figure}

For clarity, 
we refer to the scalar-helicity propagator  
with the $\mathcal{O}(M_V\Gamma_V)$ term 
as the ``scalar'' $(\lambda=S)$ polarization, 
while the scalar-helicity propagator without it 
is referred to as the  ``auxiliary'' $(\lambda=A)$ polarization.
The corresponding propagators are:
\begin{subequations}
\label{eq:top_decay_scalar_v_aux}
\begin{align}
\label{eq:top_decay_scalar_v_aux_s}
    \text{Scalar}\quad:\quad 
\Pi_{\mu\nu}^V(q,\lambda=S)\ &=\   
\cfrac{-i\ q_\mu q_\nu \left(\cfrac{1}{q^2} - \cfrac{1}{M_V^2 - i M_V\Gamma_V}\right)}{q^2 - M_V^2 + iM_V\Gamma_V} 
\nonumber\\
&=\ 
\frac{i\ q_\mu q_\nu}{(q^2)\ (M_V^2 - i M_V\Gamma_V)}\ ,
    \\
    \text{Auxiliary}\quad:\quad
\Pi_{\mu\nu}^V(q,\lambda=A)\ &=\   
\cfrac{-i\ q_\mu q_\nu \left(\cfrac{1}{q^2} - \cfrac{1}{M_V^2 }\right)}{q^2 - M_V^2 + iM_V\Gamma_V} 
\nonumber\\
&=\ \frac{i\ q_\mu q_\nu}{(q^2)\ (M_V^2)}
\frac{(q^2 - M_V^2)}{(q^2 - M_V^2 + iM_V\Gamma_V)}\ .
\label{eq:top_decay_scalar_v_aux_a}
\end{align}    
\end{subequations}

The subtle difference leads to 
significant qualitative differences.
When including the $\mathcal{O}(M_V\Gamma_V)$ term $(\lambda=S)$, 
the Breit-Wigner pole structure is cancelled,
leaving only a $1/q^2$ pole.
In other words, 
a scalar polarized $W_{\lambda=S}$ bosons 
behaves like a massless particle,
which is the expected behavior in the Unitary gauge~\cite{tHooft:1971qjg,
Becchi:1974md,Becchi:1974xu,Becchi:1975nq,Tyutin:1975qk}.
Neither the polarization vector nor the propagator 
for $\lambda=S$ vanish when $q^2\to M_V^2$, again consistent with expectations.
When omitting the $\mathcal{O}(M_V\Gamma_V)$ term $(\lambda=A)$,
one finds the $1/q^2$ pole and  
the original complex pole 
at $q=\sqrt{M_V^2 - i M_V\Gamma_V}$.
This second pole is typically obscured by 
a $(q^2- M_V^2)$ factor,
which causes the polarization vector and the 
propagator to vanish when $q^2\to M_V^2$.

To explore this behavior quantitatively, 
we implemented the $\lambda=S$ scalar polarization vector 
 in Eq.~\eqref{eq:top_decay_scalar_v_aux_s} into the 
simulation framework \texttt{MadGraph5\_aMC@NLO}.
Currently~\cite{BuarqueFranzosi:2019boy}, 
the framework supports the 
$\lambda=A$ ``auxiliary'' polarization vector
 in Eq.~\eqref{eq:top_decay_scalar_v_aux_a}.
In both cases, the $1/q^2$ pole is 
regulated\footnote{We also set \texttt{bwcutoff=100} to sample 
all momentum configurations allowed by momentum conservation.} 
by the $\tau^+$ mass since $q^2 > m_\tau^2$ must always hold.

In the last to columns of Table~\ref{tab:top_decays}, 
we show the ``scalar'' and ``auxiliary'' contributions
to the top quark's partial decay width.
We find that both reach the level of $\mathcal{O}(\text{several}\times10^{-4}\%)$.
This is consistent with $\mathcal{O}(m_\tau^2 m_b^2 M_W\Gamma_W /m_t^4 m_\tau^2)\sim 4\times10^{-6}$ that one estimates from power counting.
In absolute terms, 
the scalar partial width is $\mathcal{O}(5\%)$ larger than the auxiliary 
partial width, which we attribute to the scalar propagator 
not vanishing at $q^2 = M_W^2$.

\begin{figure}[!t]
\subfigure[]{\includegraphics[width=0.32\textwidth]{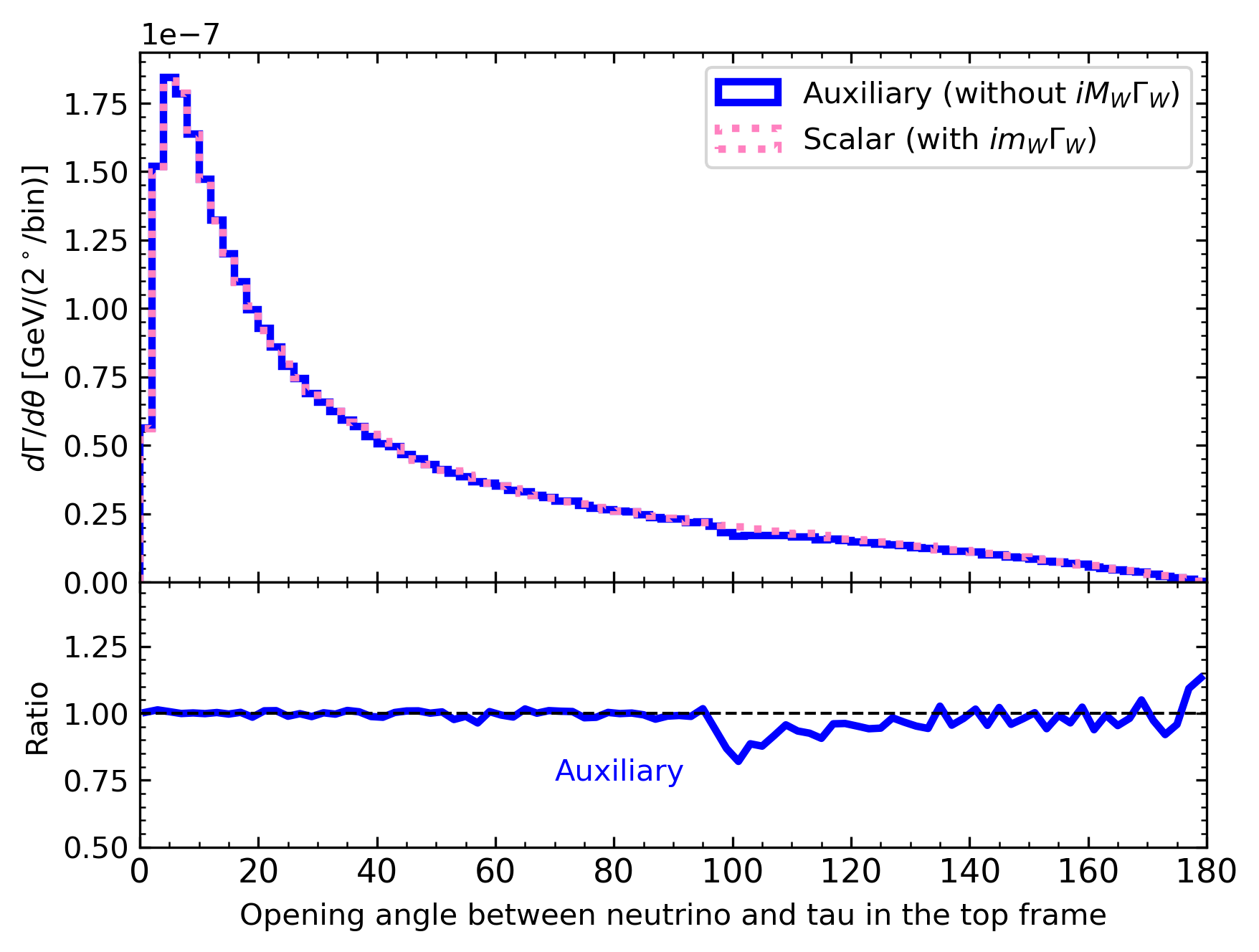}
\label{tauNeuAngle_scalar}}
\subfigure[]{\includegraphics[width=0.32\textwidth]{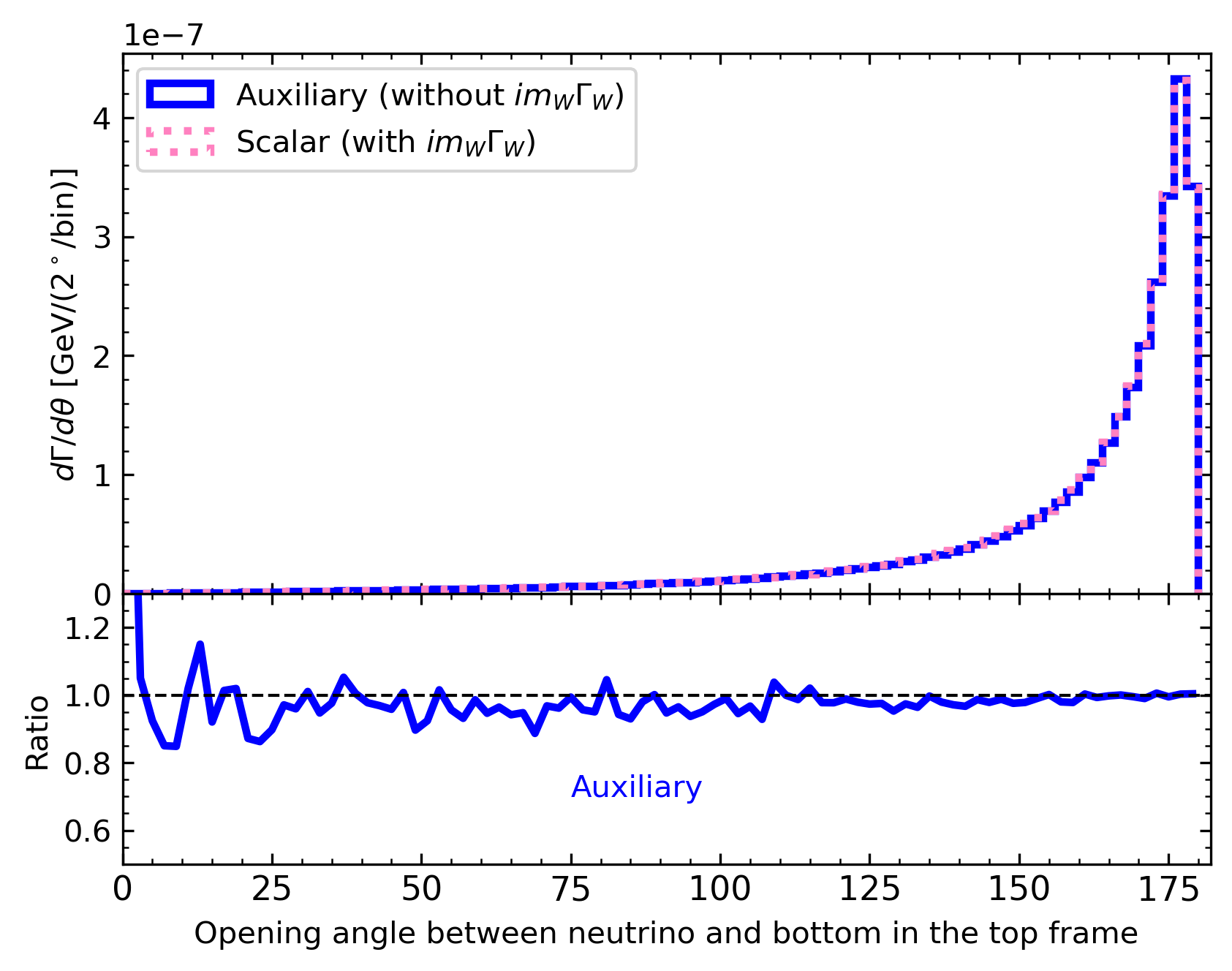}
\label{NeuBAngle_scalar}}
\subfigure[]{\includegraphics[width=0.32\textwidth]{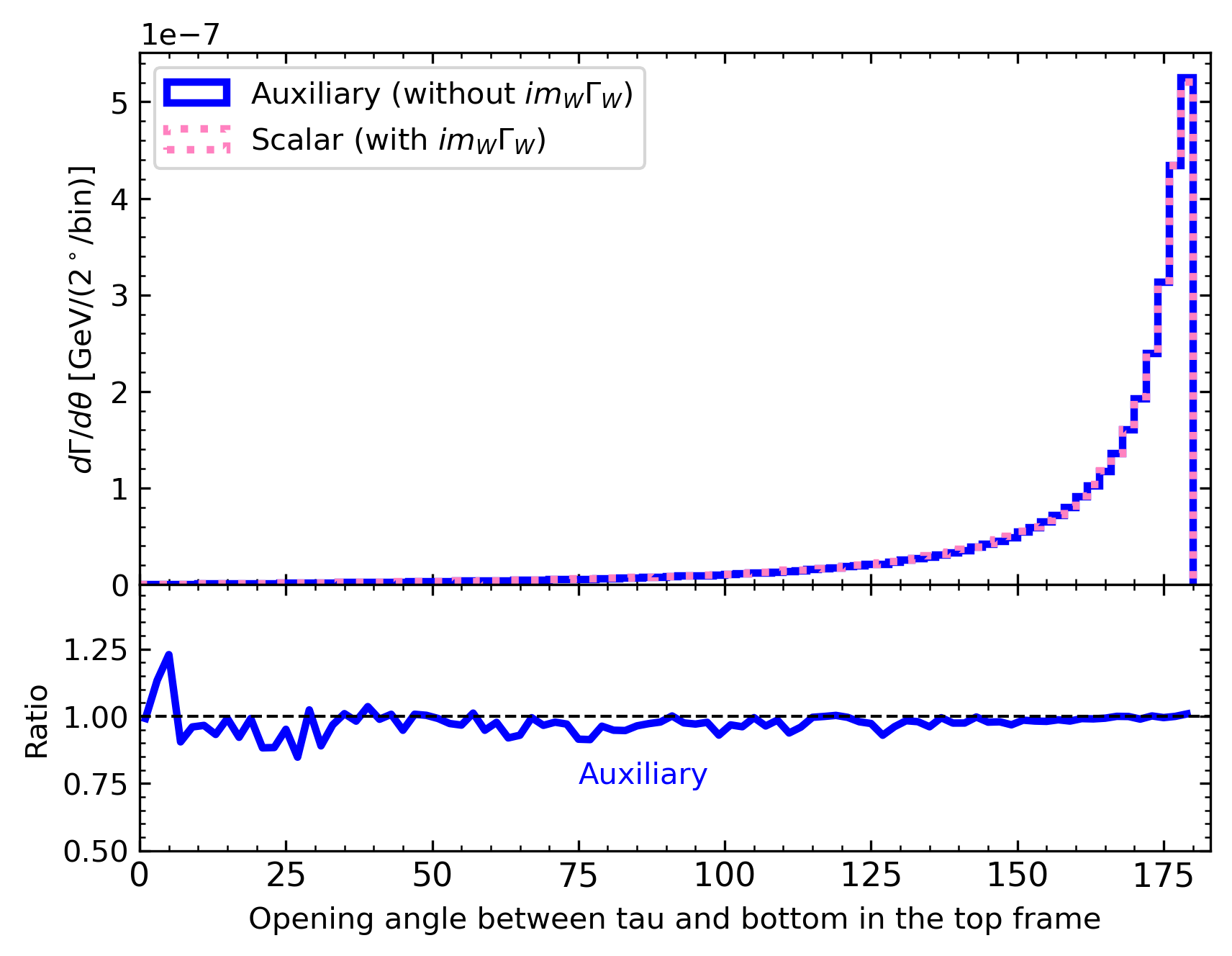}
\label{TauBtmAgle_scalar}}
\caption{
Distribution of opening angles between 
(a)  $\tau^+$ and $\nu_\tau$ in the lab frame, 
(b)  $\nu_\tau$ and $b$ in the lab frame, and
(c)  $b$ and $\tau^+$ in the lab frame.} 
\label{fig:scalar_v_aux_angles}
\end{figure}

In Fig.~\ref{fig:scalar_v_aux_scales} we plot 
for the scalar (dash-dot) and ``auxiliary'' (solid) polarizations
(a) the invariant mass of the $(\tau^+\nu_\tau)$ system,
(b) the energy of the $(\tau^+\nu_\tau)$ system in the top's frame,
(c) the energy of the $\tau^+$ in the $(\tau^+\nu_\tau)$ frame, and
(d) same as (c) but for the $\nu_\tau$.
In the lower panels we show the ratios relative to the ``scalar'' distributions.

In the invariant mass plot [Fig.~\ref{Wmass1}]
both curves have the expected $d\Gamma\sim1/q^4$ dependence
[see Eq.~\eqref{eq:top_decay_width_unintegrated}],
but with the auxiliary curve additionally showing a dip at $\sqrt{q^2}=M_W\approx80\GeV$.
For both cases, most of the phase space is restricted to $\sqrt{q^2}\ll M_W$.
Because of this, in the top's rest frame, 
the mass of the top quark ($m_t\sim 173$ GeV) is equally divided 
between the $(\tau\nu)$ system and the $b$, with 
$E_{(\tau\nu)}^{\rm top} \approx (m_t^2 + q^2)/2m_t \approx m_t/2$
and
$E_{b}^{\rm top} \approx (m_t^2 - q^2)/2m_t \approx m_t/2$.
This appears as a peak around $E_{\tau\nu}^{\rm top}\approx85\GeV$ 
in the Fig.~\ref{WElab1}.
Similarly, in the frame of $(\tau^+\nu_\tau)$ system,
the $\tau^+$ and the $\nu_\tau$ will each carry energies 
of around 
$E_{\tau/\nu}^{(\tau\nu)} = (q^2 \pm m_\tau^2)/2\sqrt{q^2} \approx m_\tau$ or $0$.
For the auxiliary polarization, 
we can also observe dips in the curves
at $E_{\tau/\nu}^{(\tau\nu)} \sim M_W/2\approx 40\GeV$,
mirroring the dip at $\sqrt{q^2}=M_W$ 
in the invariant mass of the $(\tau^+\nu_\tau)$ system.

Finally, in Fig.~\ref{fig:scalar_v_aux_angles},
we show for the scalar (dash-dot) and auxiliary (solid) modes
the opening angles between 
(a) the $\nu_\tau$ and $\tau^+$,
(b) the $\nu_\tau$ and $b$, and 
(c) the $\tau^+$ and $b$
in the top's frame.
The $\nu_\tau$ and $\tau^+$ pair are essentially collimated 
while the bottom is back-to-back with both the $\nu_\tau$ and $\tau^+$.
These distributions should be compared 
to the unpolarized, transverse, and longitudinal modes in 
Fig.~\ref{eq:kinematics_lepton_angles}.
Again, the behavior follows 
from the pole at $\sqrt{q^2}=0\GeV$
(which is regulated by $m_\tau$).
Taking $q^2 \gtrsim m_\tau^2$ and $E_{(\tau\nu)}^{\rm top}$, 
$E_b^{\rm top}\approx m_t/2$ 
as favored by Fig.~\ref{fig:scalar_v_aux_scales},
then by the relationship Eq.~\eqref{eq:angle_top}
one finds the $\tau^+-\nu_\tau$ opening angle to be 
$\theta_{\tau\nu}^{\rm top}\gtrsim 1^\circ$, 
in agreement with Fig.~\ref{tauNeuAngle_scalar}.
Likewise, taking $E_\tau^{\rm top}$, 
$E_\nu^{\rm top} \approx E_{(\tau\nu)}^{\rm top} /2 \sim m_t/4$,
we find with Eq.~\eqref{eq:angle_top} 
that the $\nu_\tau-b$ and $\tau^+-b$ opening angles 
are about $\theta_ {\nu b}^{\rm top}$,  
$\theta_{\tau b}^{\rm top}\approx \cos^{-1}[-1]=180^\circ$,
consistent with 
Fig.~\ref{NeuBAngle_scalar} and 
Fig.~\ref{TauBtmAgle_scalar}.

\subsection{\texorpdfstring{$W$}{W} 
Polarization in Neutrino Deep-Inelastic Scattering}
\label{sec:dis}

\begin{figure}[t!]
\begin{center}
    \includegraphics[width=.8\columnwidth]{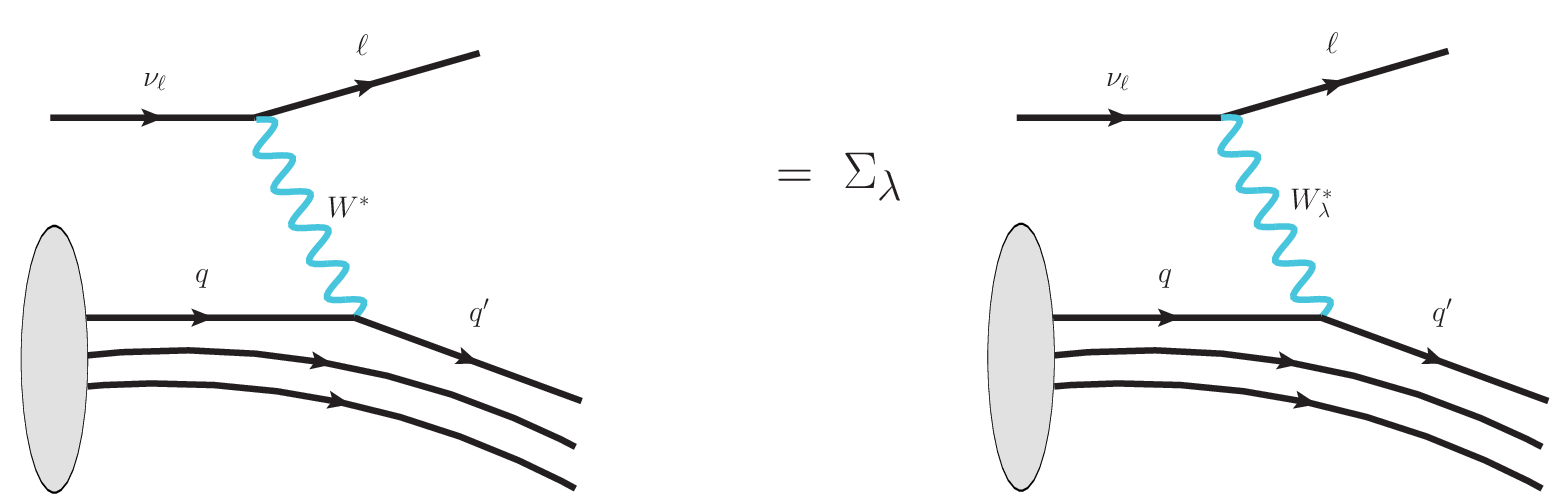}
\end{center}
    \caption{(L) Born-level diagram for the unpolarized, partonic process
    $\nu_\ell q\to \ell q'$ and its relationship to (R) the sum of helicity-polarized processes.}
    \label{fig:polexpand_dis}
\end{figure}

To briefly demonstrate that our power counting is also
applicable to $t$-channel exchanges, we consider as a 
 final case study 
inclusive, charged-current 
neutrino-hadron
deep-inelastic scattering ($\nu$DIS).
At lowest order, this is mediated by the partonic process
\begin{align}
\label{eq:proc_def_dis}
\nu_\ell(k_i)\ q(p_i)\ \xrightarrow{W^+(q)}\ \ell^-(k_f)\   q'(p_f)\ 
=\
\sum_{\lambda=T,0,S}\ 
\nu_\ell(k_i)\ q(p_i)\ \xrightarrow{W^+_\lambda(q)}\ \ell^-(k_f)\   q'(p_f)\ .
\end{align}
We can identify the unpolarized process 
as the sum over helicity-polarized processes,
as illustrated in Fig.~\ref{fig:polexpand_dis}.
For simplicity, 
we take both leptons 
and the incoming quark to be massless
and work exclusively at the partonic level.
An analysis with 
hadronic structure functions,
particularly those 
in the helicity
basis~\cite{Aivazis:1993kh,Ruiz:2023ozv},
is left to future work.

The construction of the polarization interference 
for $\nu$DIS
is similar to the Drell-Yan case in 
Sec.~\ref{sec:dy}.
The difference here is that
in the rest frame of the target hadron $A$ 
neither the $(\nu_\ell\ell)$ lepton current
nor the $(qq')$ quark current 
lies on a straight line;
the outgoing charged lepton (quark) is produced at some angle 
relative to the direction of the incoming neutrino (quark).
In the Drell-Yan case,
the $W$'s momentum is independent of the outgoing leptons
and is restricted to the $\hat{z}$ direction 
at lowest order in the partonic center-of-mass frame
because the incoming $(q\overline{q'})$
pair are traveling towards each other 
on a straight line.
In Eq.~\eqref{eq:proc_def_dis}, 
the $W$'s momentum 
is a function of $\ell$'s outgoing momentum.
Consequentially, both 
the transverse and longitudinal polarizations 
of $W^{(*)}$ contribute to the process.

In the $R_\xi$ gauge, 
the unpolarized, polarized, and Goldstone 
matrix elements are 
\begin{subequations}
\label{eq:dis_matrix_elements}
\begin{align}
    -i\mathcal{M}_{\rm unpol}\ &=\
    \frac{-ig^2}{2D_W(q^2)}
        J^\alpha_{\ell\nu} \ \left[-g_{\alpha\beta}-
    \frac{(\xi-1)q_\alpha q_\beta}{D_V(q^2,\xi)}\right] 
    J^\beta_{q'q}\ 
    \equiv\ -\mathcal{G}\ -\ \mathcal{Q}_\xi\ =\ -\mathcal{G}\ ,
    \\
    -i\mathcal{M}_{\lambda=T}\ &=\ 
    \frac{-ig^2}{2D_W(q^2)}
    J^\alpha_{\ell\nu}\ \left[\hat{q}_{\perp\mu}\hat{q}_{\perp\nu}\ 
    +\ 
    \hat{q}_{T\perp\mu}\hat{q}_{T\perp\nu}\right] 
    J^\beta_{q'q}\ \equiv\ \varphi\ ,
    \\
    -i\mathcal{M}_{\lambda=0}\ &=\ 
    \frac{-ig^2}{2D_W(q^2)}
    J^\alpha_{\ell\nu} \ \left[\Theta_{\alpha\beta}
    +\frac{q_\alpha q_\beta}{q^2}\right] 
    J^\beta_{q'q}\ \equiv\ 
    \vartheta\ +\ \frac{\mathcal{Q}}{q^2}\ =\ \vartheta\ ,
    \\
    -i\mathcal{M}_{\lambda=S}\ &=\ 
    \frac{-ig^2}{2D_W(q^2)}
    J^\alpha_{\ell\nu} \left[-\frac{q_\alpha q_\beta}{q^2}
    -
    \frac{(\xi-1)q_\alpha q_\beta}{D_V(q^2,\xi)}\right] 
    J^\beta_{q'q}\ 
    \equiv\ 
    -\frac{\mathcal{Q}}{q^2}\ -\ \mathcal{Q}_\xi\ =\ 0,
    \\
    -i\mathcal{M}_{G}\ &=\ 0\ .
\end{align}
\end{subequations}
Since the leptons and incoming quark are massless,
the Goldstone amplitude is zero.
Similarly, contracting the $W$'s momentum $q_\alpha$
with the lepton current $J^\alpha_{\ell\nu}$
vanishes by the Dirac equation:
$q\cdot J_{\ell\nu}=(k_\nu-k_\ell)\cdot J_{\ell\nu}=0$.
This means that $\mathcal{Q}$ and $\mathcal{Q}_\xi$
are zero, and hence the 
scalar polarization amplitude is also zero, 
$\mathcal{M}_{\lambda=S}=0$.
The unpolarized and longitudinal each reduce to a single term.
Consequentially, 
the net polarization interferences is
\begin{align}
\label{eq:int_unitary_noq_dis}
\mathcal{I}_{\rm pol}^{\nu\rm DIS}\ \overset{\mathcal{Q}\to0}{=}\ 
2\Re[\mathcal{M}_{\lambda=0}\mathcal{M}_{\lambda=0}^*]\
=\ 2\Re[\varphi\vartheta^*]\ .
\end{align}

After multiple applications of the Dirac equation,
the temporal/longitudinal term $\vartheta$ is 
\begin{align}
     \vartheta\ =&\ 
     \frac{-ig^2}{2D_W(q^2)}\ 
     J^\alpha_{\ell\nu}\ \Theta_{\alpha\beta}\ 
     J^\beta_{q'q}\
     \\
     =&\
    \frac{-ig^2}{2D_W(q^2)}\ 
    \frac{(n\cdot q)}{(n\cdot q)^2 - q^2 n^2}
    \left[
    J^\alpha_{\ell\nu} n_\alpha  q_\beta J^\beta_{q'q}
    + \frac{q^2}{(n\cdot q)}
    J^\alpha_{\ell\nu} n_\alpha    n_\beta J^\beta_{q'q}
    \right]\ 
    \\
    \overset{n\to n_{\rm TL}}{=}&\
    \frac{-ig^2}{2D_W(q^2)}\ 
    \frac{E_V}{(E_V^2 -q^2)}
    \left[
    J^{\alpha=0}_{\ell\nu}
    \sqrt{p_f^2}
    \tilde{J}_{q'q}
    + \frac{q^2}{E_V}
    J^{\alpha=0}_{\ell\nu} J^{\beta=0}_{q'q}
    \right]\ 
\end{align}
In the third line we fix the reference vector to be 
time-like, $n^\mu=(1,0)$. 
In this line we also 
reduce the $(qq')$ vector current $J^\beta_{q'q}$
into a scalar current $\tilde{J}_{q'q}$,
again with the Dirac equation
\begin{subequations}
\begin{align}
q_\beta\cdot J_{q'q}^\beta\ &=\ (p_f - p_i)_\beta \cdot J_{q'q}^\beta\ 
=\  \sqrt{p_f^2}\ \tilde{J}_{q'q}\ ,\ \text{where}
\\
J^\beta_{q'q}\  &=\ \overline{u}(p_f,\lambda_f) \gamma^\beta P_L u(p_i,\lambda_i)
\quad\text{and}\quad
\tilde{J}_{q'q}\ =\
\bar{u}(p_f,\lambda_f) P_L u(p_i,\lambda_i)\ .
\end{align}
\end{subequations}
Here and below we also use the DIS conventions in the rest frame of $A$:
\begin{subequations}
\begin{align}
    q\ &=\ k_i - k_f\ =\ p_f - p_i\ 
    \\
    Q^2\ &\equiv\ -q^2 > 0\ ,
    \quad 
    E_V = E_\nu - E_\ell\ ,
    \quad
    x_A\ =\ \frac{Q^2}{2 M_A E_V}\ .
\end{align}    
\end{subequations}
$x_A$ is the fraction of energy 
the incoming quark carries from $A$,
and the momentum of the outgoing quark $p_f$
is fixed by momentum conservation.
In terms of DIS variables, we have
\begin{align}
\vartheta\ =\     
\frac{-i}{(Q^2 + M_W^2)}\ 
    \frac{(J^{\alpha=0}_{\ell\nu})}{(E_V + 2x_A M_A)}\
    \left[
    \sqrt{p_f^2}
    \tilde{J}_{q'q}
    - 2 x_A M_A (J^{\beta=0}_{q'q})
    \right]\ .
\end{align}

Since the kinematics of outgoing leptons in $\nu$DIS  are generally known,
we can write 
\begin{align}
\label{eq:dis_lep_current}
J^\alpha_{\ell_L\nu_L}\ &= 
\bar{u}(k_f,\lambda_\ell) \gamma^\alpha P_L u(k_i,\lambda_\nu)
\\
&=
2\sqrt{E_\nu E_\ell}
\left[\cos\frac{\theta_\ell}{2},e^{i\phi_\ell}\sin\frac{\theta_\ell}{2},
-ie^{i\phi_\ell}\sin\frac{\theta_\ell}{2},\cos\frac{\theta_\ell}{2}\right]\ ,
\end{align}
which showcases the behavior 
of the polarized matrix element and interference.
While we do not explicitly construct the transverse amplitude, 
we note that $\hat{q}_{\perp}^\mu$ and $\hat{q}_{T\perp}^\mu$
are given by 
\begin{align}
\hat{q}_{\perp}^\mu\ =\
\frac{1}{q_T\vert\vec{q}\vert} (0, q_x q_z, q_y q_z, -q_T^2 ) 
&\quad\text{and}\quad
\hat{q}_{T\perp}^\mu\ =\ \frac{1}{q_T} (0, - q_y, q_x, 0 ) \ ,\quad
\text{where}
\\
\vert\vec{q}\vert^2\ =\ (E_\nu-E_\ell)^2 - 2k_i\cdot k_f
&\quad\text{and}\quad
q_T^2\ =\ q_x^2+q_y^2= E_\ell^2\sin^2\theta_\ell\ .
\end{align}

Focusing on $J^\alpha_{\ell_L\nu_L}$ in Eq.~\eqref{eq:dis_lep_current},
when $\theta_\ell\to0$, the lepton current 
becomes parallel to the spin axis of $W^*$ 
with vanishing transverse components,
while the longitudinal components of
$\hat{q}_{\perp}^\mu$ and $\hat{q}_{T\perp}^\mu$ vanish. 
In this limit, the transverse polarization amplitude 
and polarization interference vanish,
and the unpolarized amplitude is driven by the longitudinal 
polarization.
Conversely, when $\theta_\ell\to\pi$ 
the longitudinal polarization amplitude vanishes
since the temporal and longitudinal 
entries of the lepton current vanish,
leaving only the transverse components.
This means that the unpolarized matrix element 
for this kinematic configuration is determined  
by the transverse polarization amplitude,
and subsequently that 
the polarization interference also vanishes.

The scaling behavior of $\vartheta$ 
contains additional notable features.
For example: there is an interplay between the 
$\mathcal{O}(n_\alpha q_\beta)$ term,
which projects out 
the outgoing (virtual) quark mass $\sqrt{p_f^2}$,
and the $\mathcal{O}(q^2 n_\alpha n_\beta/E_V)$ term,
which scales 
as the momentum fraction and target mass, $x_A M_A$.
Another feature is the $\mathcal{O}[1/(n\cdot q)]\sim 1/E_V$
prefactor, which can control the relative importance of 
mass factors.
For example:
In the elastic scattering regime, 
$x_A$ tends towards unity 
while the mass of $q'$ tends towards zero,
causing the $\mathcal{O}(x_A M_A)$ term 
to dominate $\vartheta$.
Such terms are relevant at 
current accelerator neutrino facilities 
$(E_V^{\rm accelerator}\sim  M_A)$
but can be negligible 
when ultra high energy 
cosmic neutrinos are involved $(E_V^{\rm cosmic}\gg  M_A)$.
In the forward region of the deeply inelastic regime,
$x_A$ goes small and 
the outgoing quark mass $\sqrt{p_f^2}$
grows large.
Kinematics force $-\mathcal{G}\sim\vartheta$,
and hence suppress polarization interference.

\section{Outlook and Conclusion}\label{sec:conclusion}

Weak boson polarization in high-energy scattering 
remains an underexplored dimension of the SM paradigm. 
While studies exist, helicity polarization in particular is an
underutilized probe of new physics at the LHC.
The $W$ and $Z$ bosons differ from photons and gluons in that they have mass. 
Hence, the two have well-defined longitudinal polarizations 
when on shell.
However, like photons and gluons, the weak gauge bosons 
are still spin-one particles and therefore share many 
properties with off-shell photons and gluons. 

Inspired by power counting used in QCD
and building on new advances 
in understanding helicity polarization at a diagrammatic level, 
we introduced in Sec.~\ref{sec:polvector}
a covariant decomposition for helicity-polarized propagators 
of EW gauge bosons in terms of their momenta 
and light- and space-like reference vectors, 
both in covariant and axial gauges.
The decompositions are exact, 
applicable to other spin-1 particles,
including photons, gluons and new massive gauge bosons,
and make clearer mass-over-energy dependencies,
particularly the suppression of helicity inversion
in high-energy limits.

In Sec.~\ref{sec:interference}, we use our bookkeeping devices
to build a general expressions for polarization interference 
in different gauges.
Care is given to gauge cancellations and 
to the relationships between different 
polarized propagators in different gauges.

When working within polarized propagators,
realistic predictions are potentially sensitive to gauge choice 
since the number of helicity polarizations is gauge dependent.
Therefore, we introduced in Sec.~\ref{sec:interference_independence}
a scheme for combining contributions from 
longitudinally polarized gauge bosons 
with those from the gauge-fixing sector,
e.g., the scalar polarization and Goldstones bosons.
The scheme, motivated by BRST invariance,
is analogous to summing RH and LH polarizations
into a single ``transverse'' polarization.
The proposed scheme puts the covariant and EW axial gauges on closer footings, 
improves the robustness against gauge choice, 
is applicable to processes with massive external states,
and can be incorporated trivially into 
existing analysis frameworks within ATLAS and CMS.

In Sec.~\ref{sec:noninterference} and
Sec.~\ref{sec:case_studies} we considered several case studies 
that demonstrate the utility of our power counting.
In the general case with fully massive external states,
polarization interference in weak boson production
does not vanish, even when intermediate states are on-shell.
In practice, however, for LHC-like environments
polarization interference in inclusive, unpolarized processes   
is suppressed due to a combination of 
kinematical mass-over-energy effects,
rotation invariance,
and (partial) parity conservation,
in addition to usual arguments of spin correlation.

Importantly, cancellations are process dependent.
For example: in $V+{\rm jets}$, polarization interference 
cancels after integration over the angular kinematics 
of decay particles (see Sec.~\ref{sec:noninterference}),
while for decays of unpolarized top quarks, 
polarization interference cancels at the fully local level.
Likewise, the existence of new interactions,
such as those considered in Ref.~\cite{Panico:2017frx,Celada:2023oji},
can disrupt non-interference and we encourage explorations into this.

Finally, our work goes beyond contemporary analyses on polarization 
as we are able to compute polarization interference directly
while simultaneously keeping track of gauge artifacts
in on- and off-shell regimes.
The expressions throughout Sec.~\ref{sec:polvector}
make this procedure tractable.
While we focused on resonant single-boson exchanges,
the findings of Sec.~\ref{sec:noninterference} 
and Sec.~\ref{sec:case_studies}
tend to follow patterns and are likely to hold more generally.

We encourage the application
of our formalism to multiboson processes.
Many aspects of our power counting are expected to
hold at the loop level, including the 
``2P scheme'' in Sec.~\ref{sec:interference_independence},
but this should be investigated.
In kinematical regimes where EW radiation may be 
factorizable~\cite{Ruiz:2021tdt,Bigaran:2025rvb},
our work also suggests that polarization interference may 
be small.
Our work is also applicable to neutral-current exchanges
with $Z^{(*)}_{\lambda}/\gamma^*_{\lambda'}$
interference,
which is of broader interest~\cite{Chen:2016wkt,Marzocca:2024ica,Dittmaier:2025htf}.
Even in this situation, we find that power counting can
still be realized by applying partial fractions
to the product of $Z^{(*)}_{\lambda}/\gamma^*_{\lambda'}$ poles.

\section*{Acknowledgments}
The authors thank
A.\ Apyan,
I.\ Bigaran,
A.\ Denner,
S.\ Dittmaier,
J.\ Foreshaw,
S.\ Homiller,
A.\ Maas,
O.\ Mattelaer,
S.\ Pl\"atzer,
G.\ Pelliccioli,
R.\ Poncelet,
and
D.\ Wackeroth
for thoughtful and constructive discussions.
A.\ Denner,
M.\ Gallinaro,
G.\ Pelliccioli,
G.\ Marozzo, 
and
C.\ Tamarit are thanked 
for discussions 
that instigated parts 
of this work.
The authors acknowledge the support 
of Narodowe Centrum Nauki under Grant 
No.\ 2023/49/B/ST2/04330 (SNAIL). 
The authors acknowledge support 
from the COMETA COST Action CA22130.

\appendix

\section{Polarization Vectors for Intermediate Electroweak Bosons}
\label{app:polvectors}

In gauge quantum field theories, massive and massless spin-1 particles
are described by the 4-vector field $A^\mu(x)$,
which is characterized by the Fourier decomposition~\cite{Sterman:1993hfp,
Dreiner:2008tw,Coleman:2018mew,Weinberg:1995mt}
\begin{align}
    A^\mu (x) = \int \frac{d^3k}{(2\pi)^3 2 E_k} \sum_{\lambda=0}^{4} 
    \left[
    \varepsilon^\mu(k,\lambda) a(k,\lambda)e^{ik\cdot x}+
    \varepsilon^{*\mu}(k,\lambda) a^{\dagger}(k,\lambda)e^{-ik\cdot x}
    \right]\ .
\end{align}
For momentum $k$ and polarization $\lambda$, the $\varepsilon^\mu(k,\lambda)$ 
are the four helicity polarization vectors that enter scattering amplitudes. 
The fourth ``scalar'' helicity originates from embedding spin-1 objects 
(with two or three physical dof) into a 4-dimensional spacetime.
The purpose of gauge fixing is to remove this excess contribution from physical predictions.

The polarization vectors $\varepsilon^\mu(k,\lambda)$,
used throughout our study, 
can be built from a basis of orthonormal vectors $\epsilon^\mu(\tilde{\lambda})$. 
In the \textbf{Cartesian basis}, these are given by
\begin{subequations}
\begin{align}
    \epsilon^\mu(\tilde{\lambda}=t) =  \begin{pmatrix} 1\\ 0\\ 0\\ 0 \end{pmatrix},\
    \epsilon^\mu(\tilde{\lambda}=x) =  \begin{pmatrix} 0\\ 1\\ 0\\ 0 \end{pmatrix},\
    \epsilon^\mu(\tilde{\lambda}=y) =  \begin{pmatrix} 0\\ 0\\ 1\\ 0 \end{pmatrix},\
    \epsilon^\mu(\tilde{\lambda}=z) =  \begin{pmatrix} 0\\ 0\\ 0\\ 1 \end{pmatrix}\ .
\end{align}
\end{subequations}
The vectors manifestly recover the spacetime metric 
via the completeness relationship~\cite{Dreiner:2008tw}
\begin{align}
\label{eq:app_completion_cart}
    \sum_{\tilde{\lambda}\in\{t,x,y,z\}}\ 
    \eta_{\tilde{\lambda}}\ 
    \epsilon^\mu(\tilde{\lambda})\ 
    \epsilon^\nu(\tilde{\lambda})\ 
    =\ -g^{\mu\nu}\ ,\ 
    \text{where}\ (-\eta_t)=\eta_x=\eta_y=\eta_z=+1\ .
\end{align}
Starting from a spin-1 state with mass $\sqrt{k^2}$ and momentum 
$\tilde{k}^\mu = (\sqrt{k^2},0,0,0)^\mu$,
then for the Lorentz factor $\gamma = E_V/\sqrt{k^2}$,
$z$-boost $\Lambda^\mu_\nu(\gamma)$, 
and rotation matrices $R^\mu_\nu(i,\theta)$, 
\begin{subequations}
\begin{align}
\Lambda^\mu_\nu(\gamma) = \begin{pmatrix}
\gamma & 0 & 0 & \beta\gamma \\
0 & 1 & 0 & 0 \\
0 & 0 & 1 & 0 \\
\beta\gamma & 0 & 0 & \gamma \\
\end{pmatrix}\ ,\
&\quad
R^\mu_\nu(x;\theta) =  \begin{pmatrix}
1 & 0 & 0 & 0 \\
0 & 1 & 0 & 0 \\
0 & 0 & \cos\theta & -\sin\theta \\
0 & 0 & \sin\theta & \cos\theta \\
\end{pmatrix}   
\\
R^\mu_\nu(y;\theta) =  \begin{pmatrix}
1 & 0 & 0 & 0 \\
0 & \cos\theta & 0 & \sin\theta \\
0 & 0 & 1 & 0 \\
0 & -\sin\theta & 0 & \cos\theta \\
\end{pmatrix}\ ,\   
&\quad 
R^\mu_\nu(z;\theta) =  \begin{pmatrix}
1 & 0 & 0 & 0 \\
0 & \cos\theta & -\sin\theta & 0 \\
0 & \sin\theta & \cos\theta & 0\\
0 & 0 & 0 & 1 \\
\end{pmatrix} 
\end{align}
\label{Boost_rotation_matrices}
\end{subequations}
one can build the following momentum and polarization vectors in the 
this basis:
\begin{align}
k^\mu &=\ R^\mu_\nu(z,\phi)\ R^\nu_\rho(y,\theta)\ 
\Lambda^{\rho}_\sigma(\gamma)\ (\sqrt{k^2},0,0,0)^\sigma
\nonumber\\
&=\ (E_V, \vert\vec{k}\vert\sin\theta\cos\phi,\vert\vec{k}\vert\sin\theta\sin\phi,\vert\vec{k}\vert\cos\theta)\ 
\equiv\ (E_V, k_x, k_y, k_z),\ 
\label{eq:polvec_mom}
\\
\varepsilon^\mu(k,\tilde{\lambda}=t) 
&=\ 
R^\mu_\nu(z,\phi)\ 
R^\nu_\rho(y,\theta)\ 
\Lambda^{\rho}_\sigma(\gamma)\ 
\epsilon^\sigma(\tilde{\lambda}=t)
\nonumber\\
&=\ \frac{k^\mu}{\sqrt{k^2}}\ ,
\label{eq:polvec_cartT}
\\        
\varepsilon^\mu(k,\tilde{\lambda}=x)
&=\ R^\mu_\nu(z,\phi)\ 
R^\nu_\rho\left(y,\theta+\frac{\pi}{2}\right)\ 
\epsilon^\rho(\tilde{\lambda}=z)
\nonumber\\
&=\ (0, \cos\phi\cos\theta, \sin\phi\cos\theta, -\sin\theta)\ 
=\ \frac{1}{k_T\vert\vec{k}\vert } (0, k_x k_z, k_y k_z, -k_T^2 )
\label{eq:polvec_cartX}    
\\
\varepsilon^\mu(k,\tilde{\lambda}=y)
&=\ R^\mu_\nu\left(z,\phi+\frac{\pi}{2}\right)\ 
R^\nu_\rho\left(y,\frac{\pi}{2}\right)\ 
\epsilon^\rho(\tilde{\lambda}=z)
\nonumber\\
&=\ (0, -\sin\phi, ~\cos\phi, 0)\ 
=\ \frac{1}{k_T} (0, - k_y, k_x, 0 ) \ ,
\label{eq:polvec_cartY}
\\
\varepsilon^\mu(k,\tilde{\lambda}=z)
&=\ 
R^\mu_\nu(z,\phi)\ 
R^\nu_\rho(y,\theta)\ 
\Lambda^{\rho}_\sigma(\gamma)\ 
\epsilon^\sigma(\tilde{\lambda}=z)
\nonumber\\
&=\ \gamma ( \beta, \sin\theta \cos\phi, \sin\theta \sin\phi, \cos\theta )\ 
=\ 
\frac{E_V}{\sqrt{k^2} \vert\vec{k}\vert } 
\left( \frac{\vert\vec{k}\vert^2}{E_V}, k_x , k_y , k_z \right)\ .
\label{eq:polvec_cartZ}
\end{align}
Setting $q^2\to M_V^2$ in the above expressions 
recovers the polarization vectors in the Cartesian basis 
for on-shell massive spin-1 particles in the so-called 
\texttt{HELAS} convention~\cite{Hagiwara:1985yu,Murayama:1992gi}.

Alternative constructions of $\varepsilon^\mu(k,\tilde{\lambda})$ 
from different permutations of boosts and rotations are 
also possible~\cite{Dreiner:2008tw,Chen:2025dsi}.
Using $\epsilon^\mu(\tilde{\lambda}=z)$ to build 
$\varepsilon^\mu(k,\tilde{\lambda}=x)$ and 
$\varepsilon^\mu(k,\tilde{\lambda}=y)$
makes their orthogonality to $k^\mu$ explicit.
The boosts and rotations do not alter the 
original completeness relation
as explicit computation shows 
\begin{align}
\label{eq:polvectors_polsum}
    \sum_{\tilde{\lambda}=t,x,y,z}\  
    \eta_{\tilde{\lambda}}\ 
    \varepsilon_\mu(k,\tilde{\lambda})
    \varepsilon_\nu(k,\tilde{\lambda})\ 
    =\ - g_{\mu\nu}\ .    
\end{align}

\subsection*{Gauge Fixing}

When $V^{(a)}$ is the gauge field 
of an unbroken abelian or non-abelian gauge symmetry,
gauge fixing is necessary to help render the theory consistent.
In the $R_\xi$ gauge, this is done by introducing an unphysical 
gauge-fixing parameter $\xi$ and the gauge-fixing Lagrangian 
\begin{align}
\label{eq:polvectors_gf_lag}
    \mathcal{L}_{\rm GF}\ &=\ -\frac{1}{2\xi}\left(\partial_\mu A^{a\mu}\right)^2\ 
    \overset{\rm IBP}{=}\ -\frac{\delta^{ab}}{2\xi}A^{a\mu}\left(\partial_\mu\partial_\nu A^{b\nu}\right)\ 
    -\ \frac{\delta^{ab}}{2\xi}\partial_\mu\left(A^{a\mu} \partial_\nu A^{b\nu}\right)\ .
\end{align}
Here, $a,b=1,\dots$ run over the number of gauge fields in the 
non-Abelian theory. In Abelian theories, $a=b=1$.
The far-right term in Eq.~\eqref{eq:polvectors_gf_lag} is 
a total derivative and does not contribute to the theory
since fields are assumed to vanish at $x\to\infty$.

Taking the Fourier transform (FT) of $\mathcal{L}_{\rm GF}$ 
generates terms that scale as
\begin{align}
    {\rm FT}[\mathcal{L}_{\rm GF}]\ &\sim\ 
    \sum_{\lambda,\lambda'}\
    \frac{k^\mu k^\nu}{\xi}\varepsilon_\mu(k,\lambda)\varepsilon_\nu(k,\lambda')\ 
    =\ \frac{k^\mu k^\nu}{\xi}
    \varepsilon_\mu(k,\lambda=t)
    \varepsilon_\nu(k,\lambda'=t)\ .
\end{align}
Phases and permutations 
of creation and annihilation operators have been omitted in this expression.
Due to the orthogonality $k\cdot \varepsilon(k,\lambda=x,y,z)$, 
only the $\lambda=t$ polarization vector in the Cartesian basis 
contributes to gauge fixing.
This is the essence of covariant gauges: 
physical states obey the constraint equation $(\partial_\mu A^{\mu}_{\rm phys})=0$,
or equivalently $q_\mu \cdot \varepsilon^\mu_{\rm phys}=0$, while 
for gauge artifacts one has $(\partial_\mu A^{\mu}_{\rm unphys})\neq0$,
or equivalently $q_\mu \cdot \varepsilon^\mu_{\rm unphys.}\neq0$.

Importantly, $\xi$ is an artifact; it does not contribute 
to physical observables.
This is only possible if the introduction of 
$\mathcal{L}_{\rm GF}$ is accompanied by the redefinition
\begin{align}
    \varepsilon^\mu(k,\lambda=t)\ =\ \frac{k^\mu}{\sqrt{k^2}}\  \overset{\rm GF}{\longrightarrow}\ 
    \varepsilon^\mu(k,\lambda=t,\xi)\ =\ \sqrt{\frac{\xi}{k^2}}\ k^\mu\ .
\end{align}
Under this replacement, the FT of the gauge-fixing Lagrangian becomes
independent of $\xi$,
\begin{align}
    {\rm FT}[\mathcal{L}_{\rm GF}]\ &\sim\  \frac{k^\mu k^\nu}{\xi}
    \varepsilon_\mu(k,\lambda=t,\xi)
    \varepsilon_\nu(k,\lambda'=t,\xi)\ =\ \frac{\xi}{\xi}\frac{k^2k^2}{(k^2)^2}=1\ ,
\end{align}
and the associated completeness relationship 
of Eq.~\eqref{eq:polvectors_polsum} becomes
\begin{align}
\label{eq:polvectors_polsum_gf}
     \sum_{\lambda=t,x,y,z}\  
    \eta_\lambda\ \varepsilon_\mu(k,\lambda)\varepsilon_\nu(k,\lambda)\ 
    =\ - g_{\mu\nu}\ -\ (\xi-1)\frac{k_\mu k_\nu}{k^2}\ .
\end{align}
For related constructions, see Ref.~\cite{Dreiner:2008tw,Gallagher:2020ajd}.
For non-abelian theories, the gauge-fixing Lagrangian introduced above 
is augmented by Faddeev-Popov ghosts.

For the EW theory, gauge-fixing Lagrangian is more complicated 
due to spontaneous symmetry breaking and the presence of Goldstone bosons. 
In this case, we have 
\begin{align}
    \varepsilon^\mu(k,\lambda=t)\ =\ \frac{k^\mu}{\sqrt{k^2}}\  \overset{\rm GF}{\longrightarrow}\ 
    \varepsilon^\mu(k,\lambda=t,\xi)\ =\ \sqrt{\frac{1}{k^2}+\frac{(\xi-1)}{k^2-\xi M_V^2}}\ k^\mu\ .
\end{align}
Combining this with the other polarization vectors gives the 
completeness relationship 
\begin{align}
\label{eq:polvectors_polsum_gf_ssb}
     \sum_{\lambda=t,x,y,z}\  
    \eta_\lambda\ \varepsilon_\mu(k,\lambda)\varepsilon_\nu(k,\lambda)\ 
    =\ - g_{\mu\nu}\ -\ (\xi-1)\frac{k_\mu k_\nu}{k^2-\xi M_V^2}\ .
\end{align}

\subsection*{Helicity Basis with Gauge Fixing}
\label{app:polvectors_hel}
In the \textbf{helicity basis}, the transverse  $(\lambda=\pm1)$, 
longitudinal $(\lambda=0)$, and scalar $(\lambda=S)$ polarization vectors 
after gauge fixing are given by
\begin{subequations}
\label{eq:app_helicity_basis_def}
\begin{align}
\varepsilon^\mu (k,\lambda=\pm1) &=\ 
\frac{1}{\sqrt{2}}\left( -\lambda\varepsilon^\mu(k,x) - i\varepsilon^\mu(k,y)\right)\ ,
 \label{eq:polvec_hel}
 \\
 \varepsilon^\mu (k,\lambda=0) &=\ \varepsilon^\mu (k,\lambda=z)\ ,
 \\
 \varepsilon^\mu (k,\lambda=S) &=\ \varepsilon^\mu (k,\lambda=t,\xi)\ .
\end{align}
\end{subequations}
In this basis,
the polarization vectors for $t$-channel virtual photons 
used in Ref.~\cite{Aivazis:1993kh}
can be recovered by taking $\xi\to1$ (Feynman gauge),
$(\theta_V,\phi_V)\to (0,0)$,
and 
replacing $q^2 \to -q^2$ (since $q^2<0$ for $t$-channel exchanges).
The completeness relationships 
are those given in 
Eq.~\eqref{eq:polvectors_polsum_gf} for QED
and
Eq.~\eqref{eq:polvectors_polsum_gf_ssb} for the EW theory.

As the Cartesian and helicity bases are linear combinations of the other,
one has
\begin{align}
\label{eq:app_phi_matrix}
    \Phi_{\mu\nu}(\theta_V,\phi_V)\ =\ 
    \sum_{\lambda=\pm1}\  
    \eta_\lambda\ \varepsilon_\mu(k,\lambda)\varepsilon_\nu(k,\lambda)\ 
    =\ 
    \sum_{\lambda=x,y}\  
    \eta_\lambda\ \varepsilon_\mu(k,\lambda)\varepsilon_\nu(k,\lambda)\ ,
\end{align}
and recovers the expression for $\Phi_{\mu\nu}(\theta_V,\phi_V)$ in Eq.~\eqref{Phi_matrix}.

Finally, the generators of rotation for a spin-1 state are given 
by the tensor \cite{Weinberg:1995mt,Dreiner:2008tw}
\begin{align}
    \left(\mathcal{S}_{\rho\sigma}\right)^{\mu\nu} &=\ i\left(g_\rho^\mu g_{\sigma}^{\nu}-g_{\sigma}^\mu g_{\rho}^{\nu}\right)\ .
\end{align}
From this, one can define the spin operator $\mathcal{S}_i$
and the helicity operator $\hat{h}^{\mu\nu}$
that 
act on the polarization vectors. 
For a reference direction given by the 3-vector $\hat{k}=(\hat{k}_x,\hat{k}_y,\hat{k}_z)=(\sin\theta\cos\phi,\sin\theta\sin\phi,\cos\theta)$,
the spin and helicity operators are given by
\begin{align}
\label{eq:app_helicity_op_def}
     \left(\mathcal{S}_i\right)^{\mu\nu}\ &=\ \frac{1}{2}\ \epsilon^{ijk}\ \left(\mathcal{S}_{jk}\right)^{\mu\nu}\ ,\  
     \\
    \hat{h}^{\mu\nu}(\hat{k})\ &\equiv\ \left(\vec{\mathcal{S}}\cdot \hat{k}\right)^{\mu\nu}\ 
    =\ 
    \left(\mathcal{S}_x\hat{k}_x\right)^{\mu\nu}\ +\ 
    \left(\mathcal{S}_y\hat{k}_y\right)^{\mu\nu}\ +\ 
    \left(\mathcal{S}_z\hat{k}_z\right)^{\mu\nu}\
    \nonumber\\
    &=\ 
\begin{pmatrix}
0 & 0 & 0 & 0 \\
0 & 0 & i \hat{k}_z & -i \hat{k}_y \\
0 & -i \hat{k}_z  & 0 & i \hat{k}_x \\
0 & i \hat{k}_y & -i\hat{k}_x & 0 \\
\end{pmatrix}
= 
\begin{pmatrix}
0 & 0 & 0 & 0 \\
0 & 0 & i \cos\theta & -i \sin\theta\sin\phi \\
0 & -i \cos\theta  & 0 & i \sin\theta\cos\phi \\
0 & i \sin\theta\sin\phi & -i\sin\theta\cos\phi & 0 \\
\end{pmatrix} ,
\end{align}
where $\epsilon^{ijk}=+1$.
Using these operators, 
one finds the following eigenvalue relationships:
\begin{align}
\label{eq:polvectors_hel_helicity_op}
    \hat{h}^{\mu\nu}(\hat{k})\ \varepsilon_\nu (k,\lambda=\pm1) 
    =     \lambda\ \varepsilon^\mu (k,\lambda) 
    \quad\text{and}\quad
    \hat{h}^{\mu\nu}(\hat{k})\ \varepsilon_\nu (k,\lambda=0,S) 
    = 0^\mu\ .
\end{align}
In other words, the helicity polarization vectors of 
Eq.~\eqref{eq:app_helicity_basis_def} are the helicity 
eigenvectors of the operator $h^{\mu\nu}$, 
both for on-shell and off-shell momenta.

\section{Additional Properties of Polarized Propagators}
\label{app:properties}

In this appendix, 
we give additional properties, identities, and relationships
for the polarization polarized propagators 
introduced in Sec.~\ref{sec:polvector_covariant}
and Sec.~\ref{sec:polvector_axial}.

The inner products between $n^\mu$ in Eq.~\eqref{eq:ref_vector_def}  
and momentum  $q^\mu$ are
\begin{subequations}
\label{eq:refvector_dotprod}
\begin{align}
     n_{\rm LL}\cdot q &=\ E_V+\vert\vec{q}\vert\ =\ 
     E_V\left(1+\sqrt{1-q^2/E_V^2}\right)\ ,
    \\
     n_{\rm TL}\cdot q &=\ E_V\ ,
     \\
     n_{\rm SL}\cdot q &=\ \vert\vec{q}\vert\ =\ E_V\sqrt{1-q^2/E_V^2} \ .
\end{align}
\end{subequations}
Using these, 
different choices of $n^\mu$ 
are related to each other by the 
following identities:
\begin{subequations}
\label{eq:refvector_ident}
\begin{align}
 n_{\rm LL}^\mu   &=\ 
 n_{\rm TL}^\mu+ n_{\rm SL}^\mu\ ,\quad
 (n_{\rm LL}\cdot n_{\rm TL}) = 1\ ,\quad
 (n_{\rm LL}\cdot n_{\rm SL}) = -1\ ,\quad
 (n_{\rm TL}\cdot n_{\rm SL}) = 0\ ,
 \\
 n_{\rm LL}^\mu   &=\ 
 \frac{(n_{\rm LL}\cdot q)}{(n_{\rm SL}\cdot q)} n_{\rm TL}^\mu 
 -  \frac{q^\mu}{(n_{\rm SL}\cdot q)}\ 
 =\ 
 \left(\frac{E_V + \vert\vec{q}\vert}{\vert\vec{q}\vert}\right)n_{\rm TL}^\mu - \frac{q^\mu}{\vert\vec{q}\vert}\ ,
 \\
 n_{\rm SL}^\mu   &=\ 
 \frac{(n_{\rm TL}\cdot q)}{(n_{\rm SL}\cdot q)} n_{\rm TL}^\mu 
 - 
 \frac{q^\mu}{(n_{\rm SL}\cdot q)}\
 =\ 
 \left(\frac{E_V}{\vert\vec{q}\vert}\right)n_{\rm TL}^\mu - \frac{q^\mu}{\vert\vec{q}\vert}\ .
\end{align}
\end{subequations}

In other words, the SL and LL reference vectors,
which contain messy 3-momentum components,
can be decomposed into 
the momentum vector $q^\mu$ itself
and the simpler TL reference vector.
The TL reference vector 
 projects out temporal components $(\mu=0)$ from currents.
Momentum vectors  can then simplify currents 
via equations of motion, e.g., 
the Dirac equation.
When reference vectors are 
contracted with gamma matrices,
one obtains
\begin{align}
\label{eq:refvector_gamma}
    \not\! n_{\rm TL}\ =\ \gamma^0\ , \
    \not\! n_{\rm LL}\ =\ 
 \left(\frac{E_V + \vert\vec{q}\vert}{\vert\vec{q}\vert}\right)\ \gamma^0\ 
 -\ \frac{\not\! q}{\vert\vec{q}\vert}\ ,\
 \not\! n_{\rm SL}   =\ 
 \left(\frac{E_V}{\vert\vec{q}\vert}\right)\ \gamma^0\ -\ \frac{\not\! q}{\vert\vec{q}\vert}\ .
\end{align}
When contracting with the antisymmetric tensor 
($\epsilon^{\mu\nu\alpha\beta}=-\epsilon_{\mu\nu\alpha\beta}=+1$),
one has 
\begin{align}
   \epsilon_{\mu\nu\alpha\beta}\ q^\alpha n^\beta q^\nu\ =\ 0_\mu\  \quad\text{and}\quad\
    \epsilon_{\mu\nu\alpha\beta}\ q^\alpha n^\beta n^\nu\ =\ 0_\mu\ ,
    \label{eq:poltrans_inner_ident_antisymm}
\end{align}
and the identities of Eq.~\eqref{eq:refvector_ident}
further simplify possible contractions:
\begin{subequations}
\begin{align}
    \epsilon_{\mu\nu\alpha\beta}\ q^\alpha n^\beta_{\rm LL}\ &=\
    \epsilon_{\mu\nu\alpha\beta}\ q^\alpha n^\beta_{\rm TL}\ 
    \left(\frac{E_V + \vert\vec{q}\vert}{\vert\vec{q}\vert}\right)\ ,
    \\
    \epsilon_{\mu\nu\alpha\beta}\ q^\alpha n^\beta_{\rm SL}\ &=\ 
    \epsilon_{\mu\nu\alpha\beta}\ q^\alpha n^\beta_{\rm TL}\ 
    \left(\frac{E_V}{\vert\vec{q}\vert}\right)\ .
\end{align}
\end{subequations}

Like $\Theta_{\mu\nu}$, 
the $n^\mu$ are dimensionless.
Still, they still appear 
at the order of $\mathcal{O}(n^2/n^2)$
in Eq.~\eqref{Theta_matrix}.
This means that the decomposition of Eq.~\eqref{Theta_matrix}
also holds when any of the $n^\mu$ in Eq.~\eqref{eq:ref_vector_def}
is rescaled by a real number $a\neq0$, 
so that $n^\mu \to (n')^\mu = a n^\mu$.
If $a=\vert\vec{q}\vert$ or $a=E_V$, 
then we can also define 
the backwards-momentum, 
light-like vectors
\begin{subequations}
\label{eq:ref_vector_def_nq}
\begin{align}
    n^\mu_{\rm LLq}\ &=\ (\vert\vec{q}\vert,-\vec{q})\ 
    =\ (E_V+\vert\vec{q}\vert) n_{\rm TL}^\mu - q^\mu
    \quad\text{with}\quad 
     n_{\rm LLq}^2=\ 0\ \ , 
\\
n^\mu_{\rm LLE}\ &=\ (E_V,-E_V\hat{q})\ 
    =\ \frac{1}{|\vec{q}|}(E_V^2+ E_V\vert\vec{q}\vert) n_{\rm TL}^\mu - \frac{E_V}{\vert\vec{q}\vert}q^\mu
    \quad\text{with}\quad 
     n_{\rm LLE}^2=\ 0\ \  .
\end{align}
\end{subequations}

For the choices of $n^\mu$ in Eq.~\eqref{eq:ref_vector_def} 
the following identities and relationships hold:
\begin{subequations}
\begin{align}
   \varepsilon(q,\lambda=\pm1) \cdot n(\hat{q})\  =\ 0\ ,
   \quad&\quad
   \hat{h}_{\mu\nu}(\hat{q})\cdot n^\nu (\hat{q})\ =\ 0_\mu\ ,
   \\
   \hat{q}_{\perp\mu}\cdot n(\hat{q})\ =\ 0\ ,
   \quad&\quad
   \hat{q}_{T\perp\mu}\cdot n(\hat{q})\ =\ 0\ ,
   \\
   \Theta_{\mu\nu}\cdot n^\nu\ =\ -n_\mu\ , \quad&\quad
   \Pi_{\mu\nu}^V(q,\lambda=T)\cdot n^\nu\ =\ 0_\mu\ ,
   \label{eq:poltrans_inner_ident_theta_axial}
   \\
   \Theta_{\mu\nu}\cdot q^\nu\ =\ -q_\mu\ , \quad&\quad
   \Pi_{\mu\nu}^V(q,\lambda=T)\cdot q^\nu\ =\ 0_\mu\ ,
   \label{eq:poltrans_inner_ident_theta}
   \\
   \Pi_{\mu\nu}^V(q,\lambda=\pm1)\cdot n^\nu\ =\ 0_\mu\ , \quad&\quad 
   \Pi_{\mu\nu}^V(q,\lambda=\pm1)\cdot q^\nu\ =\ 0_\mu\ .   
\end{align}
\end{subequations}
Here, $\hat{h}^{\mu\nu}(\hat{q})$ is 
the helicity operator of Eq.~\eqref{eq:polvectors_hel_helicity_op}.
While the orthogonality conditions 
in Eq.~\eqref{eq:poltrans_inner_ident_theta}
are independent of our bookkeeping devices, 
using Eq.~\eqref{Theta_matrix} makes them clear.
Notably, if we interpret Eq.~\eqref{eq:poltrans_inner_ident_theta}
as the gauge-fixing condition $(\partial^\mu A_\mu)=0$
for physical gauge states in the $R_\xi$ gauge, 
then Eq.~\eqref{eq:poltrans_inner_ident_theta_axial}
can arguably be interpreted as the 
gauge-fixing condition $(n^\mu A_\mu)=0$ being satisfied 
for transverse polarizations. 
Again, the last expression is independent of our 
decomposition but is manifest through its adoption.

The longitudinal polarization vector 
and propagator obey the following 
relationships:
\begin{subequations}
\label{eq:pollong_inner_ident}
\begin{align}
 \varepsilon(q,\lambda=0) \cdot q = 0\ ,\
   &\quad\
 \Pi_{\mu\nu}^V(q,\lambda=0)\cdot q^\nu\ =\ 0_\mu\
\\
  \hat{h}^{\mu\nu}(\hat{q})\cdot 
   \varepsilon_\nu(q,\lambda=0)= 0^\mu\  ,\
  &\quad\
   \varepsilon(q,\lambda=0) \cdot n(\hat{q})\  =\ 
   \frac{\sqrt{(n\cdot q)^2 - q^2 n^2}}{\sqrt{q^2}}\ .   
\end{align}
\end{subequations}
While the orthogonality conditions in 
the first line of Eq.~\eqref{eq:pollong_inner_ident}
are independent of our decomposition,
they follow immediately from 
Eqs.~\eqref{eq:polvec_long_alt}
and \eqref{eq:poltrans_inner_ident_theta}.

We can expand the double-pole structure
in propagators with partial fractions
\begin{align}
\label{eq:partial_frac}
\frac{1}{(q^2-a^2)\ (q^2-b^2)}\ =\
\frac{1}{(a^2-b^2)\ (q^2-a^2)}\
-\
\frac{1}{(a^2-b^2)\ (q^2-b^2)}\ ,
\end{align}
for arbitrary complex $a^2,b^2$.
Using this,
the longitudinal propagator can be written as
\begin{align}
\label{eq:partial_frac_long}
    \Pi_{\mu\nu}^V(q,\lambda=0)\
=\
\cfrac{i\Theta_{\mu\nu}}{q^2 - M_V^2 + iM_V\Gamma_V}
&+
\cfrac{\cfrac{i q_\mu q_\nu}{M_V^2 - iM_V\Gamma_V}}{
q^2-M_V^2 + iM_V\Gamma_V}
\nonumber\\
&-
\cfrac{i q_\mu q_\nu}{q^2 (M_V^2 - iM_V\Gamma_V)}\ ,
\end{align}
showing that the longitudinal propagator
carries a massless, $1/q^2$ term
in covariant gauges.
While useful for insights,
we do not employ this expression
as applying the power counting
of Sec.~\ref{sec:interference}
to the $1/q^2$ term
is tantamount to undoing the
partial-fraction
decomposition.

With partial fractions
we can re-express the scalar propagator in the Unitary gauge,
\begin{align}
\label{eq:partial_frac_scalar}
 \Pi_{\mu\nu}^V(q,\lambda=S)\ &=\
\cfrac{i\ q_\mu q_\nu}{(q^2)\ (M_V^2 - iM_V\Gamma_V)}\
-\
\cfrac{\cfrac{i\ q_\mu q_\nu}{M_V^2 - i M_V\Gamma_V}}{q^2 - \xi M_V^2 + i\xi M_V\Gamma_V}\ .
\end{align}
When applying Eq.~\eqref{eq:partial_frac}
we find that two of the four terms cancel.
At the single-pole level, it is clear that
the first term in Eq.~\eqref{eq:partial_frac_scalar}
cancels the $1/q^2$ pole in the (expanded)
longitudinal propagator in Eq.~\eqref{eq:partial_frac_long}.
The second term
has the form but opposite sign of a Goldstone propagator.
As with Eq.~\eqref{eq:partial_frac_long},
we do not employ this expression
as consistent power counting
of the $1/q^2$ term
undoes the partial-fraction decomposition.

Interestingly, in the $R_\xi$ gauge
and at the double-pole level,
it is the \textit{longitudinal} propagator,
not the scalar propagator,
that is eliminated when summing over helicities. 
At the single-pole level,
it is the reverse:
the scalar propagator is eliminated
while the longitudinal propagator survives.
The exception to this is the Landau gauge, 
where one takes $\xi\to0$
and causes the scalar polarization vector
and the scalar propagator to vanish.

For all choices of $\xi$, 
the scalar polarization vector carries 
zero helicity; see Eq.~\eqref{eq:helicityOp_scalar_pol}.
\begin{align}
\label{eq:helicityOp_scalar_pol}
    \hat{h}^{\mu\nu}(\hat{q})\cdot \varepsilon_\nu (q,\lambda=S)\ 
    =\ 
    \hat{h}^{\mu\nu}(\hat{q})\cdot q_\nu\
    =\
    0^\mu\ .
\end{align}

The longitudinal polarization vector 
and propagator obey the following 
relationships:
\begin{subequations}
\begin{align}
\label{eq:pollong_inner_ident_axial}
 \varepsilon(q,\lambda=0) \cdot n(\hat{q}) = 0\ ,\
   &\quad\
 \Pi_{\mu\nu}^V(q,\lambda=0)\cdot n^\nu(\hat{q})\ =\ 0_\mu\ ,
\\
  \hat{h}^{\mu\nu}(\hat{q})\cdot 
   \varepsilon_\nu(q,\lambda=0)= 0^\mu\  ,\
  &\quad\
   \varepsilon(q,\lambda=0) \cdot q\  =\ 
   \frac{-\sqrt{q^2}}{(q\cdot n)}\ \sqrt{(n\cdot q)^2 - q^2 n^2}\ 
   .   
\end{align}
\end{subequations}
Note that the roles of $q^\mu$ and $n^\mu$ are inverted 
relative to Eq.~\eqref{eq:pollong_inner_ident}.

\section{Extended Support for Polarized Amplitudes in MadGraph5\_aMC@NLO}
\label{app:mgpolar}

Support for polarized propagators in \texttt{MadGraph5\_aMC@NLO}~\cite{Stelzer:1994ta,Alwall:2014hca} was introduced in
Ref.~\cite{BuarqueFranzosi:2019boy} 
for weak bosons and heavy quarks in the Unitary gauge,
and extended to loop-induced processes in Ref.~\cite{Javurkova:2024bwa}.
In this initial release, 
the polarized propagators for weak bosons include
\begin{subequations}
\label{eq:app_pol_prop_current}
\begin{align}
\label{eq:app_pol_prop_current_unpol}
\Pi_{\mu\nu}^V(q,\ {\rm unpol.})\ {\Big\vert}_{\rm Unitary}^{\rm MG5} &=\
\frac{-i}{D_V(q^2)}
\left[g_{\mu\nu} - \cfrac{q_\mu q_\nu}{M_V^2}\right]\ ,
\\
\Pi_{\mu\nu}^V(q,\lambda=\texttt{T})\ {\Big\vert}_{\rm Unitary}^{\rm MG5} &=\
\frac{-i}{D_V(q^2)}
\left[g_{\mu\nu}+ \Theta_{\mu\nu}\right]\ =\ {\rm Eq.}~\eqref{eq:prop_trans}\ ,
\\
\Pi_{\mu\nu}^V(q,\lambda=\texttt{0})\ {\Big\vert}_{\rm Unitary}^{\rm MG5} &=\
\frac{+i}{D_V(q^2)}
\left[\Theta_{\mu\nu} + \cfrac{q_\mu q_\nu}{q^2}\right]\ =\ {\rm Eq.}~\eqref{eq:prop_long}\ ,
\\
\Pi_{\mu\nu}^V(q,\lambda=\texttt{A})\ {\Big\vert}_{\rm Unitary}^{\rm MG5} &=\
\frac{+i}{D_V(q^2)}
\left[\cfrac{q_\mu q_\nu}{M_V^2} - \cfrac{q_\mu q_\nu}{q^2}\right]\ .
\end{align}    
\end{subequations}
The auxiliary propagator $\lambda=A$ can be obtained from the scalar 
propagator $\lambda=S$ in Eq.~\eqref{eq:prop_scalar_uni} 
by omitting the appropriate $\mathcal{O}(iM_V\Gamma_V)$ term.
As discussed in Sec.~\ref{sec:polvector_covariant}, 
the unpolarized propagator in Eq.~\eqref{eq:app_pol_prop_current_unpol}
does not respect QED Ward identities due 
to the same missing $\mathcal{O}(iM_V\Gamma_V)$ term.

As of \texttt{v3.7.1}, \texttt{MadGraph5\_aMC@NLO} supports the 
following polarized propagators 
\begin{subequations}
\begin{align}
\Pi_{\mu\nu}^V(q,\lambda=\texttt{W})\ {\Big\vert}_{\rm Unitary}^{\rm MG5} &=\
\frac{-i}{D_V(q^2)}
\left[g_{\mu\nu} - \cfrac{q_\mu q_\nu}{M_V^2-iM_V\Gamma_V}\right]\ =\ {\rm Eq.}~\eqref{eq:prop_unpol_unitary}\  ,
\\
\Pi_{\mu\nu}^V(q,\lambda=\texttt{S})\ {\Big\vert}_{\rm Unitary}^{\rm MG5} &=\
\frac{+i}{D_V(q^2)}
\left[\cfrac{q_\mu q_\nu}{M_V^2-iM_V\Gamma_V} - \cfrac{q_\mu q_\nu}{q^2}\right]\ =\ {\rm Eq.}~\eqref{eq:prop_scalar_uni}\ ,
\\
\Pi_{\mu\nu}^V(q,\lambda=\texttt{H})\ {\Big\vert}_{\rm Unitary}^{\rm MG5} &=\
\frac{+i}{D_V(q^2)}\ \Theta_{\mu\nu}\  ,
\\
\Pi_{\mu\nu}^V(q,\lambda=\texttt{Q})\ {\Big\vert}_{\rm Unitary}^{\rm MG5} &=\
\frac{+i}{D_V(q^2)}\ \cfrac{q_\mu q_\nu}{q^2}\ ,
\\
\Pi_{\mu\nu}^V(q,\lambda=\texttt{G})\ {\Big\vert}_{\rm Unitary}^{\rm MG5} &=\
\frac{-i}{D_V(q^2)}\ g_{\mu\nu}\ ,
\\
\Pi_{\mu\nu}^V(q,\lambda=\texttt{0,S})\ {\Big\vert}_{\rm Unitary}^{\rm MG5} &=\
\frac{+i}{D_V(q^2)}
\left[\Theta_{\mu\nu} - \cfrac{q_\mu q_\nu}{M_V^2-iM_V\Gamma_V}\right]\ =\ {\rm Eq.}~\eqref{eq:2pscheme_unitary}\ .
\end{align}
\end{subequations}
Each of the above propagators is defined in the Unitary gauge.
The new propagators are called following the same syntax as introduced
in Ref.~\cite{BuarqueFranzosi:2019boy}, i.e., 
with the bracket suffix $\texttt{\{X\}}$.
The propagator in Eq.~\eqref{eq:prop_unpol_unitary} 
respects QED Ward identities while the propagator in 
Eq.~\eqref{eq:2pscheme_unitary} 
is an implementation of the 2P scheme 
of Sec.~\ref{sec:interference_independence}.

In Table~\ref{tab:app_top_decay}, we list 
partial widths [GeV] (third and fifth columns)
for the unpolarized and polarized top quark decay process 
$t_{\lambda_t}\to W^+_{\lambda_W} b\to \nu_\tau \tau^+$,
with polarization defined in the top's frame,
assuming $W$ boson polarization $\lambda_W$ (first column),
along with their ratio 
relative to the unpolarized rate (fourth and sixth columns)
for massive (third and fourth columns)
and massless (fifth and sixth columns) $\tau$ leptons.

Widths were computed using the following steering:
\begin{verbatim}
set group_subprocesses False
import model loop_sm
generate t{X} > w+{Y} b, w+ > ta+ vt
output MyDir_tX_wY
launch MyDir_tX_wY
analysis=off
set no_parton_cut
set nevents 100k
set me_frame [1]
set bwcutoff 100
done    
\end{verbatim}

\begin{table}[!t]
\centering
\begin{tabular}{|m{0.5cm}|m{5.5cm}|m{2cm}|m{2cm}|m{1.5cm}|m{1.5cm}|}
\hline
\hline
& & \multicolumn{2}{c|}{$m_\tau \neq 0$} & \multicolumn{2}{c|}{$m_\tau = 0$}\\
\hline
$\lambda_W$ & Process Syntax 
& $\Gamma$ [GeV] & $\Gamma/\Gamma_{\rm unpol}$
& $\Gamma$ [GeV] & $\Gamma/\Gamma_{\rm unpol}$  \\
\hline
 \verb|| & \verb|t{L} > w+ b, w+ > ta+ vt| & \texttt{0.1631} & $-$ & \texttt{0.1633} & $-$ \\
\verb|0| & \verb|t{L} > w+{0} b, w+ > ta+ vt| & \texttt{0.1140} & \texttt{0.6992} & \texttt{0.1141} & \texttt{0.6988} \\
 \verb|0S| & \verb|t{L} > w+{0S} b, w+ > ta+ vt| & \texttt{0.1140} & \texttt{0.6992} & \texttt{0.1141} & \texttt{0.6988} \\
 \verb|A| & \verb|t{L} > w+{A} b, w+ > ta+ vt| & \texttt{6.830e-06} & \texttt{4.189e-05} & \texttt{<1e-21} & \texttt{<1e-21} \\
 \verb|G| & \verb|t{L} > w+{G} b, w+ > ta+ vt| & \texttt{0.1631} & \texttt{1.000} & \texttt{0.1633} & \texttt{1.000} \\
 \verb|H| & \verb|t{L} > w+{H} b, w+ > ta+ vt| & \texttt{0.2123} & \texttt{1.302} & \texttt{0.2124} & \texttt{1.301} \\
 \verb|Q| & \verb|t{L} > w+{Q} b, w+ > ta+ vt| & \texttt{9.051e-05} & \texttt{5.550e-04} & \texttt{<1e-21} & \texttt{1e-21} \\
 \verb|S| & \verb|t{L} > w+{S} b, w+ > ta+ vt| & \texttt{6.879e-06} & \texttt{4.219e-05} & \texttt{<1e-21} & \texttt{<1e-21} \\
 \verb|T| & \verb|t{L} > w+{T} b, w+ > ta+ vt| & \texttt{0.04913} & \texttt{0.3013} & \texttt{0.04919} & \texttt{0.3012} \\
 \verb|W| & \verb|t{L} > w+{W} b, w+ > ta+ vt| & \texttt{0.1631} & \texttt{1.000} & \texttt{0.1633} & \texttt{1.000} \\
\hline
\verb|| & \verb|t{R} > w+ b, w+ > ta+ vt| & \texttt{0.1632} & $-$ & \texttt{0.1633} & $-$ \\
\verb|0| & \verb|t{R} > w+{0} b, w+ > ta+ vt| & \texttt{0.1140} & \texttt{0.6987} & \texttt{0.1141} & \texttt{0.6986} \\
 \verb|0S| & \verb|t{R} > w+{0S} b, w+ > ta+ vt| & \texttt{0.1140} & \texttt{0.6989} & \texttt{0.1141} & \texttt{0.6986} \\
 \verb|A| & \verb|t{R} > w+{A} b, w+ > ta+ vt| & \texttt{6.823e-06} & \texttt{4.182e-05} & \texttt{<1e-21} & \texttt{<1e-21} \\
 \verb|G| & \verb|t{R} > w+{G} b, w+ > ta+ vt| & \texttt{0.1632} & \texttt{1.000} & \texttt{0.1633} & \texttt{1.000} \\
 \verb|H| & \verb|t{R} > w+{H} b, w+ > ta+ vt| & \texttt{0.2123} & \texttt{1.301} & \texttt{0.2124} & \texttt{1.300} \\
 \verb|Q| & \verb|t{R} > w+{Q} b, w+ > ta+ vt| & \texttt{9.056e-05} & \texttt{5.550e-04} & \texttt{<1e-21} & \texttt{1e-21} \\
 \verb|S| & \verb|t{R} > w+{S} b, w+ > ta+ vt| & \texttt{6.877e-06} & \texttt{4.215e-05} & \texttt{<1e-21} & \texttt{<1e-21} \\
 \verb|T| & \verb|t{R} > w+{T} b, w+ > ta+ vt| & \texttt{0.04911} & \texttt{0.3010} & \texttt{0.04919} & \texttt{0.3012} \\
 \verb|W| & \verb|t{R} > w+{W} b, w+ > ta+ vt| & \texttt{0.1632} & \texttt{1.000} & \texttt{0.1633} & \texttt{1.000} \\
\hline
 \verb|| & \verb|t > w+ b, w+ > ta+ vt| & \texttt{0.1633} & $-$ & \texttt{0.1633} & $-$ \\
\verb|0| & \verb|t > w+{0} b, w+ > ta+ vt| & \texttt{0.1140} & \texttt{0.6988} & \texttt{0.1141} & \texttt{0.6990} \\
 \verb|0S| & \verb|t > w+{0S} b, w+ > ta+ vt| & \texttt{0.1140} & \texttt{0.6988} & \texttt{0.1141} & \texttt{0.6990} \\
 \verb|A| & \verb|t > w+{A} b, w+ > ta+ vt| & \texttt{6.830e-06} & \texttt{4.186e-05} & \texttt{<1e-21} & \texttt{<1e-21} \\
 \verb|G| & \verb|t > w+{G} b, w+ > ta+ vt| & \texttt{0.1633} & \texttt{1.000} & \texttt{0.1633} & \texttt{1.000} \\
 \verb|H| & \verb|t > w+{H} b, w+ > ta+ vt| & \texttt{0.2123} & \texttt{1.301} & \texttt{0.2124} & \texttt{1.301} \\
 \verb|Q| & \verb|t > w+{Q} b, w+ > ta+ vt| & \texttt{9.054e-05} & {5.549e-04} & \texttt{<1e-21} & \texttt{1e-21} \\
 \verb|S| & \verb|t > w+{S} b, w+ > ta+ vt| & \texttt{6.875e-06} & \texttt{4.214e-05} & \texttt{<1e-21} & \texttt{<1e-21} \\
 \verb|T| & \verb|t > w+{T} b, w+ > ta+ vt| & \texttt{0.04910} & \texttt{0.3009} & \texttt{0.04917} & \texttt{0.3012} \\
 \verb|W| & \verb|t > w+{W} b, w+ > ta+ vt| & \texttt{0.1632} & \texttt{1.000} & \texttt{0.1632} & \texttt{1.000} \\
\hline
\hline
\end{tabular}
\caption{The partial widths [GeV] (third and fifth columns)
for the unpolarized and polarized top quark decay process 
$t_{\lambda_t}\to W^+_{\lambda_W} b\to \nu_\tau \tau^+$,
with polarization defined in the top's frame,
assuming $W$ boson polarization $\lambda_W$ (first column),
computed with in \texttt{MadGraph5\_aMC@NLO}
(syntax in the second column),
along with their ratio 
relative to the unpolarized rate (fourth and sixth columns)
for massive (third and fourth columns)
and massless (fifth and sixth columns) $\tau$ leptons.}
\label{tab:app_top_decay}
\end{table}

\newpage
\phantom{x}
\newpage

{\small
\bibliography{wPolar_refs}
}

\end{document}